\documentclass{jfm}
\usepackage{amsmath,bm}
\usepackage{graphicx}
\usepackage{newtxtext}
\usepackage{newtxmath}
\usepackage{natbib}
\usepackage{accents}
\usepackage{caption}
\usepackage{subcaption}
\usepackage{mathtools,tikz,caption}	

\usepackage{hyperref}
\usepackage{placeins}

\usepackage{xr}
\makeatletter

\newcommand*{\addFileDependency}[1]{
\typeout{(#1)}
\@addtofilelist{#1}

\IfFileExists{#1}{}{\typeout{No file #1.}}
}\makeatother

\hypersetup{
    colorlinks = true,
    urlcolor   = blue,
    citecolor  = black,
}

\newcommand{\be}{\begin{equation}}
\newcommand{\ee}{\end{equation}}
\newcommand{\bea}{\begin{eqnarray}}
\newcommand{\eea}{\end{eqnarray}}
\newcommand{\lb}{\label}
\newcommand{\black}[1]{\textcolor{black}{#1}}

\newcommand{\RomanNumeralCaps}[1]
\linenumbers

\title{Vorticity cascade and turbulent drag in wall-bounded flows: 
plane Poiseuille flow}

\author{Samvit Kumar\aff{1}
  \corresp{\email{skumar67@jh.edu}},
  Charles Meneveau\aff{1}
 \and Gregory Eyink\aff{1,2}}

\affiliation{\aff{1}Department of Mechanical Engineering, The Johns Hopkins University, Baltimore, Maryland, USA
\aff{2}Department of Applied Mathematics, The Johns Hopkins University, Baltimore, Maryland, USA}

\begin{document}
\maketitle

\begin{abstract}
Drag for wall-bounded flows is directly related to the spatial flux of spanwise vorticity outward from the wall. In turbulent flows a key contribution to this wall-normal flux arises from nonlinear advection and stretching of vorticity, interpretable as a cascade. We study this process using numerical simulation data of turbulent channel flow at $Re_\tau=1000$. The net transfer from the wall of spanwise vorticity created by downstream pressure drop is due to two large opposing fluxes, one which is ``down-gradient'' or outward from the wall, where most vorticity concentrates, and the other which is ``up-gradient'' or toward the wall and acting against strong viscous diffusion in the near-wall region. We present evidence that the upgradient/downgradient transport occurs by a mechanism of correlated inflow/outflow and spanwise vortex stretching/contraction that was proposed by Lighthill. This mechanism is essentially Lagrangian, but we explicate its relation to the Eulerian anti-symmetric vorticity flux tensor. As evidence for the mechanism, we study (i) statistical correlations of the wall-normal velocity and of wall-normal flux of spanwise vorticity, (ii) vorticity flux cospectra identifying eddies involved in nonlinear vorticity transport in the two opposing directions, and (iii) visualizations of coherent vortex structures which contribute to the transport. The ``D-type'' vortices contributing to down-gradient transport in the log-layer are found to be attached, hairpin-type vortices. However, the ``U-type'' vortices contributing to up-gradient transport are detached, wall-parallel, pancake-shaped vortices with strong spanwise vorticity, as expected by Lighthill's mechanism. We discuss modifications to the attached eddy model and implications for turbulent drag reduction.

\end{abstract}


\section{Introduction}\label{intro}

Most current approaches to wall-bounded turbulence are based on momentum conservation and the concept 
of ``momentum cascade'' to the wall \citep{tennekes1972first,jimenez2012cascades,yang2016moment}. However, vorticity conservation 
may arguably be of equal or even greater importance. One of the earliest 
advocates of this point of view was \cite{taylor1932}, although his ``vorticity transfer hypothesis''
was justly criticized for neglect of vortex-stretching. Nevertheless, Taylor arrived at an important exact  
result that pressure drop in turbulent flow down a pipe is directly related to flux of spanwise vorticity across 
the flow. \cite{lighthill1963} was an even more forceful champion for vorticity-based approaches, positing that 
``to explain convincingly the existence of boundary layers ...arguments concerning vorticity are needed'' 
and further that ``vorticity considerations ... illuminate the detailed development of the boundary layers 
just as clearly as do momentum considerations.'' In particular, \cite{lighthill1963} argued that vorticity 
is uniquely suited to a causal description of fluid flows, as it is the only variable whose effects propagate 
at finite speeds in the incompressible limit. 

Lighthill made in fact substantial concrete contributions to the program of explaining turbulent boundary layers
by means of vorticity dynamics. One key idea introduced by \cite{lighthill1963} which is now widely recognized is 
that vorticity generation at solid walls is due to tangential pressure gradients, with wall-normal 
vorticity flux given by 
\begin{align}\label{source} 
\bm{\sigma}&=\mathbf{n}\times (\nu\bm{\nabla}\times \bm{\omega})=-\mathbf{n}\times(\bm{\nabla}p+\partial_t \mathbf{u}),
\end{align}
where $\mathbf{n}$ is the unit normal vector at the boundary pointing inward to the fluid. The term $\partial_t\mathbf{u}$ 
which accounts for tangential acceleration of the wall was introduced by ~\cite{morton1984}, who emphasized further the 
inviscid character of such vorticity production. Although generally well accepted, the relations \eqref{source} have been 
the subject of some minor controversy, since they were first derived by \cite{lighthill1963} only for flat walls and 
were generalized subsequently to curved walls in the form \eqref{source} by ~\cite{lyman1990} and in an alternative form 
$\bm{\sigma}'=-\nu (\bm{n\cdot\nabla})\bm{\omega}$ by \cite{panton1984incompressible}. The subsequent debate
over which of these two forms is ``correct'' is reviewed by \cite{terrington2021generation}, who conclude that the 
two expressions measure slightly different things and have each their own (overlapping) domains of applicability. 
See also \cite{wang2022origin}. Lyman's version \eqref{source} uniquely describes 
the creation of circulation at the boundary \citep{eyink2008} and we use that form in our theoretical discussion here
(but note the two coincide in our concrete application to channel flow). In either guise, the Lighthill source reveals 
that the solid walls are the ultimate origin of all vorticity 
in the flow, whereas for momentum the walls act instead as the sink. In consequence, the profound sensitivity of fluid flows 
to the nature of the solid boundary is better revealed by vorticity considerations.


\cite{lighthill1963} made another essential contribution to wall-bounded turbulence which seems, however, 
to have been less appreciated. To introduce Lighthill's basic insight, we can do no better than to quote 
at length from his own paper: 
\begin{quotation}
``The main effect of a solid surface on turbulent vorticity close to it is to {\it correlate 
inflow towards the surface with lateral stretching.} Note that only the stretching of vortex lines 
can explain how during transition the mean wall vorticity increases as illustrated in Fig.II.21;
and only a tendency, for vortex lines to stretch as they approach the surface and relax as they 
move away from it, can explain how the gradient of mean vorticity illustrated in Fig.II.21 is maintained 
in spite of viscous diffusion down it --- to say nothing of any possible `turbulent diffusion'
down it, which the old `vorticity transfer' theory supposed should occur. It is relevant to both 
these points that Fig.II.21 relates to uniform external flow, which implies zero mean rate of 
production of vorticity at the surface; but, even in an accelerating flow, the rate of production 
$UU'$ is too small to explain either. 

A simplified illustration of how inflow towards a wall tends to go with lateral stretching, and how outflow 
with lateral compression, is given in Fig.II.22. Doubtless some longitudinal deformation is usually also
present, which reduces the need for lateral deformation (perhaps, on average, by half). However, there is 
evidence (from attempts to relate different types of theoretical model of a turbulent boundary layer 
to observations by hotwire techniques; see, for example, Townsend 1956) that the larger-scale motions
(which push out `tongues' of rotational fluid discussed above) are elongated in the stream direction,
as if their vortex lines had been stretched longitudinally by the mean shear; in such motions, the 
correlation between inflow and lateral stretching illustrated in Fig.II.22 would be particularly strong. 
We may think of them as constantly bringing the major part of the vorticity in the layer close to the wall,
while intensifying it by stretching and, doubtless, generating new vorticity at the surface; meanwhile,
they relax the vortex lines which they permit to wander into the outer layer. Smaller-scale movements 
take over from these to bring vorticity still closer to the wall, and so on. Thus, the `cascade process',
which in free turbulence (see, for example, Batchelor 1953) continually passes the energy of fluctuations
down to modes of shorter and shorter length-scale --- because at high Reynolds numbers motions in a whole
range of scales may be unstable, which implies that motions of smaller scale can extract energy from 
them---this cascade process has the additional effect in a turbulent boundary layer of bringing the 
fluctuations into closer and closer contact with the wall, while their vortex lines are more and 
more stretched.'' ---From \cite{lighthill1963}, pp.98-99. 
\end{quotation}

\begin{figure}
     \centering
     \begin{subfigure}[b]{0.9\textwidth}
         \centering
         \includegraphics[width=\textwidth]{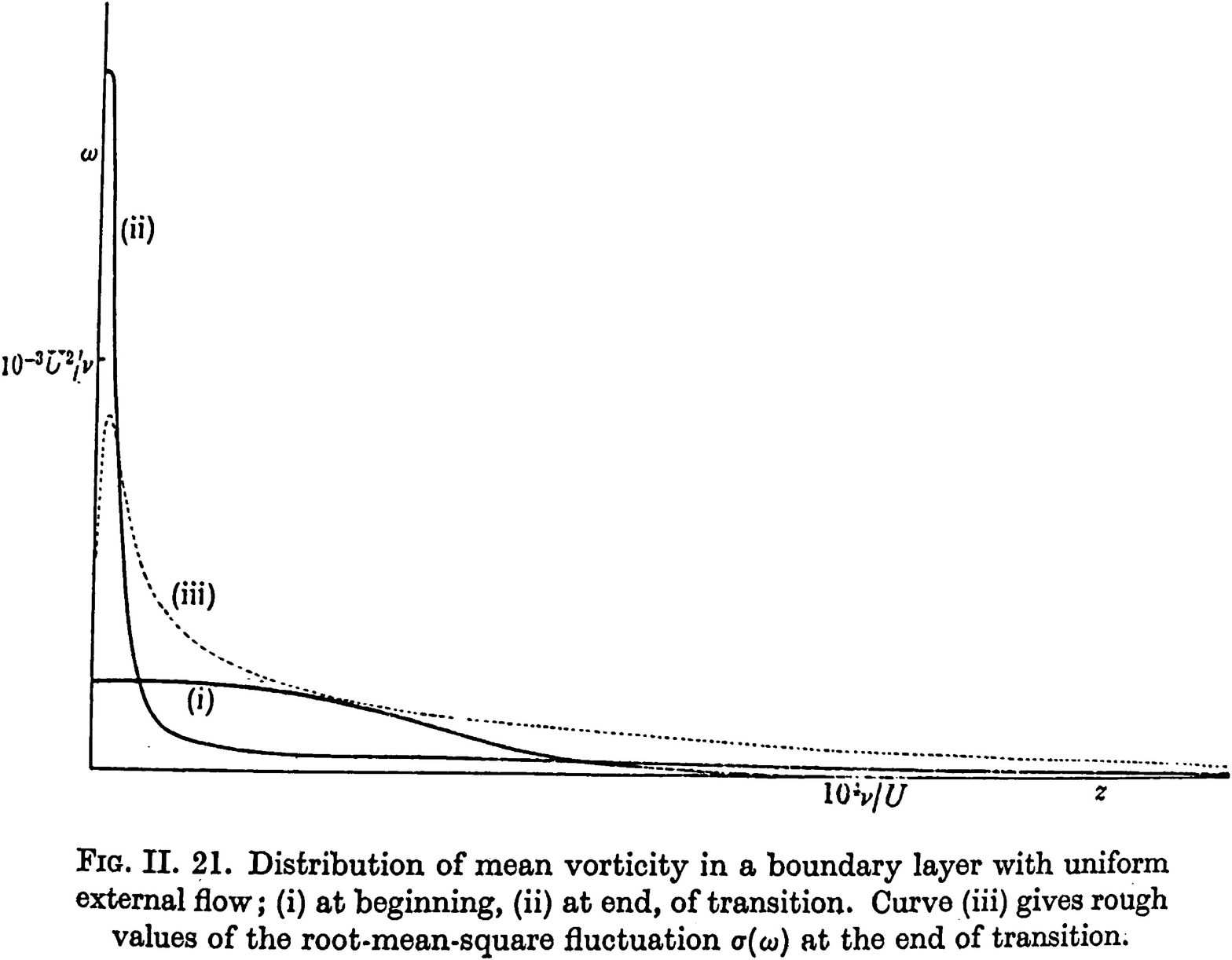}
         \label{concentrate}
         \caption{ }
     \end{subfigure}\\
     \hfill\\
     \begin{subfigure}[b]{.85\textwidth}
         \centering
         \includegraphics[width=\textwidth]{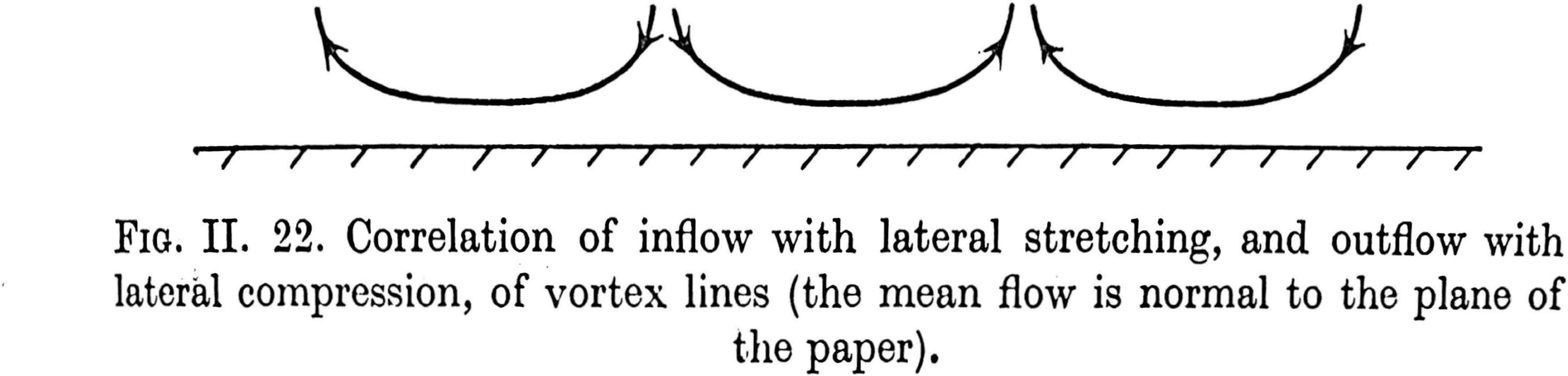}
         \label{correlate}
         \caption{}
     \end{subfigure}
             \caption{\black{From M. J. Lighthill, ``Introduction: Boundary layer theory,'' in: {\it Laminar Boundary Theory}, 
Ed. L. Rosenhead, pp. 46–113. Copyright \copyright 1963 by Oxford University Press. Reproduced with permission of the Licensor 
through PLSclear.}}
        \label{Lighthill_figs}
\end{figure}

We find in this remarkable passage four key ideas: (i) First, the correlation between turbulent inflow 
and lateral vortex stretching illustrated by \cite{lighthill1963} with his Figure II.22 acts to magnify 
principally spanwise vorticity and to drive it nearer the wall as shown in his Figure II.21 
(both reproduced here as our Figure \ref{Lighthill_figs}). \black{ Conversely, turbulent out-flow is 
correlated with lateral vortex compression, weakening the mainly spanwise vorticity as it is carried 
away from the wall. } 
Quoting again from \cite{lighthill1963}, p.96, ``it concentrates most of the vorticity much closer to the 
wall than before, although at the same time allowing some straggling vorticity to wander away from 
it farther.'' The validity of this mechanism for a transitional boundary-layer flow has been verified 
recently by \cite{wang2022origin}. One would expect on the basis 
of this argument to find spanwise-extended vortex structures as principal elements of wall-bounded 
turbulent flows.(ii) The mechanism of correlated inflow\black{/outflow} and 
vortex-stretching\black{/contraction} is powerfully {\it up-gradient}, acting against both viscosity 
and ``eddy-viscosity'' effects which attempt to smooth the very sharp gradients of vorticity created 
near the wall. As remarked by \cite{lighthill1963},p.96, ``turbulence redistributes the vorticity in such 
a way that viscous diffusion becomes more effective in countering the amplitude of the disturbances.'' 
(iii) This intense competition between up-gradient vortex-stretching on the one hand and diffusion 
away from the wall by molecular and turbulence effects on the other is narrowly won by the latter, 
since the net vorticity produced at the wall by the Lighthill source \eqref{source} must be transferred 
away under statistically steady conditions. Note that Lighthill's argument presumes that the total pressure 
$p+(1/2)|\mathbf{u}|^2$ is continuous across accelerating turbulent boundary layers, 
so that the mean vortex production at the wall with steady, external mean velocity $U=\langle u\rangle$ 
is given by $U\partial_x U.$ (iv) Finally, Lighthill saw this up-gradient transport of vorticity toward 
the wall as a scale-by-scale cascade process, proceeding by the successive stretching, straightening, and 
strengthening of spanwise vorticity through a hierarchy of eddy scales. Among the chief results 
of the present work will be direct empirical confirmation of these insights of Lighthill. 

Closely related ideas were developed somewhat after Lighthill's work in the adjacent field of quantum superfluids,
where \cite{Josephson65} for superconductors and \cite{Anderson66} for neutral superfluids recognized 
the relation between drops of voltage/pressure in flow through wires/channels and the cross-stream flux 
of quantized magnetic-flux/vortex lines. Their ideas closely mirror those of \cite{taylor1932}
and \cite{lighthill1963} for classical fluids, but \cite{Josephson65} and \cite{Anderson66} were 
seemingly unaware of those earlier works and the two literatures have subsequently developed in parallel. 
In quantum superfluids the Josephson-Anderson relation has become the paradigm to explain 
drag and dissipation in otherwise ideal superflow \citep{packard1998role,varoquaux2015anderson}.
This understanding is based in particular on the work of \cite{Huggins1970a}, who derived a 
``detailed Josephson-Anderson relation'' that exactly relates energy dissipation to vortex motions.
Interestingly, although the target application of \cite{Huggins1970a} was quantum superfluids, 
his mathematical model was the incompressible Navier-Stokes equation for a classical viscous fluid. 
In fact, somewhat later \cite{huggins1994vortex} applied his ideas to classical turbulent channel flow. 

More precisely, \cite{Huggins1970a,huggins1994vortex} considered a classical incompressible fluid at 
constant density $\rho$ and with kinematic viscosity $\nu$ flowing in a channel with accelerations due 
both to a conservative force $-\mathbf{\nabla} Q$ and to a non-conservative force $-\mathbf{g}$ 
(with $\bm{\nabla \times}\mathbf{g} \neq \bm{0}),$ described by
\begin{align}
\partial_t\mathbf{u}&=\mathbf{u}\times \bm{\omega} -\nu\bm{\nabla\times\omega}-
\bm{\nabla}(p/\rho+|\mathbf{u}|^2/2+Q )-\mathbf{g},\label{eq_mom}
\end{align}
For example, $Q$ might be the gravitational potential energy density $\rho g y $ for vertical height $y$ (with acceleration due to gravity $g$), and $\mathbf{g}$ might 
be $-\bm{\nabla\cdot\tau}_p$ with $\bm{\tau}_p$ the stress of a polymer additive. 
\cite{Huggins1970a,huggins1994vortex} noted that this 
equation for the momentum balance may be rewritten as 
\begin{align}
\partial_tu_i&=(1/2)\epsilon_{ijk}\Sigma_{jk}-\partial_i h,\label{JAsimple} \end{align} 
with anti-symmetric {\it vorticity flux tensor} 
\begin{align} \Sigma_{ij} &= u_i\omega_j-u_j\omega_i 
+\nu\left(\frac{\partial\omega_i}{\partial x_j}-\frac{\partial\omega_j}{\partial x_i}\right)-\epsilon_{ijk}g_k, 
\label{Sigma} \end{align} 
and {\it total pressure} or enthalpy 
\begin{align} h&=p/\rho+|\mathbf{u}|^2/2+Q. \end{align}
Here the total pressure \textit{h} includes both the hydrostatic and the dynamic pressures, 
and the tensor $\Sigma_{ij}$ represents the flux of the \textit{j}th vorticity component in the \textit{i}th 
coordinate direction. The latter interpretation is made clear by taking the curl of the momentum equation \eqref{eq_mom}, 
which yields a local conservation law for vector vorticity: 
\begin{align}
\partial_t\omega_j+\partial_i\Sigma_{ij} =0. \lb{vort-cons} 
\end{align}
The first term in \eqref{Sigma} for $\Sigma_{ij}$ represents the advective transport of vorticity, the second represents 
transport by 
nonlinear stretching and tilting, the third represents viscous transport, and the fourth represents 
transport of vorticity perpendicular to an applied, non-conservative body-force $\mathbf{g}$
akin to the Magnus effect. The stretching\black{/tilting} term \black{$(\bm{\omega\cdot\nabla}){\bf u}$} in the Helmholtz equation \black{violates material conservation, so that $D_t\bm{\omega}\neq 0,$ but it can nevertheless be written 
as a total divergence $\bm{\nabla\cdot}(\bm{\omega}{\bf u})$ and thus interpreted as} a space transport of vorticity (see section \ref{sec:lighthuggins}). 
The equation \eqref{JAsimple} thus shows the deep connection between momentum balance and vorticity transport, 
and this equation is the most elementary version of the classical Josephson-Anderson relation. 
See also \black{the insightful study of \cite{brown2012turbulent} in the context of flow past a cylinder and}  
the more recent work of \cite{terrington2021generation}, who discuss at length the meaning and 
applications of the anti-symmetric vorticity flux tensor \eqref{Sigma}, which they call the ``Lyman-Huggins tensor''.

A first attempt was made by \cite{eyink2008} to unify these parallel theories in the context 
of two canonical turbulent flows, plane-parallel channels and straight pipes. He discussed the physical 
significance of the observation by \cite{taylor1932} and by \cite{huggins1994vortex} that there is a 
mean cross-stream vorticity flux driven by the downstream pressure gradient. In the context of 
channel flow, with $x$ the streamwise direction, $y$ the wall-normal direction, and $z$ the spanwise 
direction, this average relation takes the form 
\be \langle \Sigma_{yz}\rangle =\langle v\omega_z- w\omega_y-\nu(\partial_y\omega_z-\partial_z\omega_y)\rangle= \partial_x \langle p\rangle= -u_\tau^2/h, \lb{const-flux} \ee 
where $\Sigma_{yz}$ is the wall-normal flux of 
spanwise vorticity, $h$ is the channel half-height, and $u_{\tau}$ is the friction velocity.  
The standard result that $\partial_y\partial_x \langle p\rangle=0$ (\cite{tennekes1972first}, section 5.2) 
is seen to be a consequence of vorticity conservation $\partial_y\langle \Sigma_{yz}\rangle =0.$ 
\cite{eyink2008} noted that Huggins' vorticity flux tensor \eqref{Sigma} and Lighthill's vorticity source \eqref{source} 
in the form of \cite{lyman1990} are simply related by $\bm{\sigma}=
\mathbf{{n}}
\cdot \bm{\Sigma},$ so that 
the origin of the constant mean flux is the vorticity created at the wall, which flows toward the 
channel center to be annihilated by opposite-sign vorticity from the facing wall. \cite{eyink2008}
referred to this phenomenon as an ``inverse vorticity cascade'', but note that the vorticity transport
involved here is down-gradient, opposed by the up-gradient cascade mechanism proposed by \cite{lighthill1963}.
The term ``inverse cascade'' was used by \cite{eyink2008} because the \black{mean} vorticity flux is in the opposite 
direction as the \black{mean} flux of momentum, that is, out from the wall and via eddies of increasing size 
at further distances from the wall. \black{ Just as spatial momentum flux in wall-bounded turbulence can be interpreted as a stepwise cascade (\cite{tennekes1972first,jimenez2012cascades}), so also spatial 
vorticity transport can be understood as a cascade through a hierarchy of eddies whose sizes scale with distance to the wall. }

A possible mechanism for this cascade is lifting and growing hairpin-like vortex structures 
in the inertial sublayer of the channel flow. It was shown by \cite{eyink2008} that constant mean 
down-gradient flux of vorticity via the nonlinear dynamics can in fact be explained by the attached-eddy model (AEM) 
of \cite{townsend1976structure} (see \cite{marusic2019attached} for a recent comprehensive review).  
Since the AEM is not designed to describe the statistics and dynamics of the 
fine-grained vorticity 
(\cite{marusic2019attached}, section 4.1), it is not entirely trivial that the model should 
account for the mean vorticity flux. However, this flux can be deduced from the Reynolds stress 
by the standard relation \citep{taylor1915eddy,tennekes1972first,klewicki1989velocity} 
\be  \langle v \omega_z- w\omega_y\rangle = -\partial_y\langle u'v'\rangle, \lb{vortRey} \ee 
from which it can be shown that the AEM implies $\langle v\omega_z-w\omega_y\rangle \sim -u_\tau^2/h$ for 
$y\gtrsim y_p,$ where $y_p$  is the wall distance of the peak Reynolds stress \citep{eyink2008}. On the contrary, for
$y<y_p$ it follows directly from \eqref{vortRey} that $\langle v\omega_z- w\omega_y\rangle>0,$ whose positive 
sign indicates up-gradient nonlinear transport of (negative) spanwise vorticity. It was noted by \cite{eyink2008} that this 
up-gradient transport is not obviously explained by attached eddies and we shall present here strong evidence 
that the underlying mechanism is in fact that of \cite{lighthill1963}. 
A further impetus to our investigation comes from recent work of \cite{eyink2021}, who showed that 
the ``detailed relation'' of \cite{Huggins1970a} for energy dissipation in channel flows holds also 
for flow around a uniformly moving solid body. In fact, this result holds much more generally 
for bodies that are moving non-uniformly and even changing shape and volume (Eyink, unpublished)
and also for channel flows with periodic boundary conditions \citep{kumar2023turbulent}.
In all of these situations, there is flux of vorticity away from the solid surface and net drag is given 
instantaneously by the spatial integral of spanwise vorticity flux across the streamlines of a background 
potential Euler flow.

To gain further insight into the underlying fluid-dynamical mechanisms of turbulent vorticity cascade,  we here carry out a detailed investigation of the turbulent vorticity dynamics in the simplest case of turbulent channel flow. Although viscous diffusion plays a dominant role in the mean vorticity transport out to the wall distance $y_p$ \citep{klewicki2007physical,eyink2008,brown2015vorticity}, its properties follow directly 
from the mean velocity profile and are thus relatively easy to understand. We shall therefore be more 
concerned with the  nonlinear vorticity dynamics and the resulting statistics of the velocity-vorticity correlations 
$\langle v \omega_z\rangle,$ $\langle w\omega_y\rangle$ at various wall distances. We employ data for our study from the Johns Hopkins Turbulence database (JHTDB) which stores the output of a direct numerical simulation (DNS) of turbulent channel flow at $Re_{\tau}=1000$ \citep{jhtdb1,jhtdb_channel}.  This simulation was performed using the petascale DNS channel flow code {\it PoongBack} \citep{lee2013petascale} with driving force provided by a constant applied pressure gradient. The resulting online database archives full space-time fields of velocity and pressure throughout the channel domain and for about one flow-through time. The archived data permit us to calculate not only the velocity-vorticity correlations 
but also their Fourier cospectra in streamwise wavenumber $k_x,$ spanwise wavenumber $k_z$ and 2D wavenumber 
$(k_x,k_z),$  which 
prove particularly illuminating of the physics. 


\black{It is important to emphasize once again the dual role of the quantity $(\mathbf{u}\times \bm{\omega})_x
=v \omega_z- w\omega_y$ that is studied in this work. On the one hand, it is the streamwise 
component of the ``vortex force'' which appears in the momentum balance \eqref{eq_mom}, while 
on the other hand it is the inertial contribution to the component $\Sigma_{yz}$ of the conserved vorticity 
current. Much prior work has focused on the role of the mean vortex force 
$\langle v\omega_z-w\omega_y \rangle=\langle v'\omega_z'-w'\omega_y' \rangle$, interpreted as a ``turbulent inertia'' (TI) 
term through equation \eqref{vortRey}.
Separate contributions $\langle v'\omega_z'\rangle$ and $\langle w'\omega_y'\rangle$ were measured experimentally by~\cite{klewicki1989velocity} and weighted joint PDFs of $v',\omega_z'$ and of $u',\omega_z'$ were obtained by~\cite{klewicki1994vortical}. 
A four-layer structure for wall-bounded flows was proposed by~\cite{wei_fife_klewicki_mcmurtry_2005}, based on the relative magnitude of the viscous and TI term in the mean momentum equation (see also~\cite{klewicki2007physical}). The wall-normal derivatives of streamwise spectra of the Reynolds' shear stress, equal to the nonlinear flux co-spectra for periodic flows, were studied as ``Net Force Spectra'' for pipe flows~(\cite{guala_hommema_adrian_2006}) and for channel flows and zero pressure-gradient boundary layers~(\cite{balakumar_adrian_2007}). An in-depth study of the statistics and streamwise spectral behavior of the velocity-vorticity products for turbulent boundary layers, at several $Re$ values, was carried out by~\cite{priyadarshana2007statistical}. They compared streamwise spectra for velocity and vorticity with the corresponding co-spectra and plotted profiles of the velocity-vorticity products and correlation coefficients. The correlations between velocity and vorticity were seen to arise from a ``scale selection'' associated with peaks in the velocity and vorticity streamwise spectra.}
\black{~\cite{monty2011characteristics} interpret the TI term as a momentum source/sink (depending upon the sign) and carried out detailed calculations of the streamwise and spanwise two-point correlations of $v$ with $\omega_z$ and $w$ with $\omega_y$ in a DNS of channel flow. They drew an important conclusion, which anticipates our own, that ``the mean Reynolds stress gradient at any wall-normal location is a direct result of a slight asymmetry in the characteristic vortical motions of the flow.'' 
The work of~\cite{wu_baltzer_adrian_2012}, primarily studied streamwise very large-scale motions (VLSMs) and their relations to shear-stress in DNS 
of pipe flows, but they computed as well 2D ``net force spectra'' jointly in streamwise and spanwise wavenumbers at four wall-normal locations, scaled with outer units. ~\cite{morrill2013influences} carried out experimental investigations for flat plate boundary layers, studying streamwise co-spectra, scale selection, two-point correlations, and Reynolds number effects. 
~\cite{chin_et_al_2014} made a detailed analysis of the TI term for DNS of pipe flows, decomposing it into advective transport and vorticity stretching/tilting and studying wall-normal variation of the respective streamwise co-spectra and the combined ``net force spectrum'', as well as joint PDF's of $v',\omega_z'.$ 
}

\black{
A smaller body of work has studied velocity-vorticity correlations instead as the nonlinear contribution to mean vorticity flux, 
following the early work of ~\cite{taylor1915eddy,taylor1932}, including the DNS studies of \cite{bernard1990turbulent,crawford1997reynolds,vidal_etal_2018}. In DNS of channel flows over a range of Reynolds numbers, 
\cite{brown2015vorticity} studied vorticity flux, highlighting the fact that its mean is constant across every wall-parallel plane for a turbulent pressure-driven channel flow and evaluating the two nonlinear contributions. Of particular interest is their calculation of the p.d.fs of $w\omega_y$ close to the wall and their visualization of the vortex lines passing through such a region.  
 Experimental measurements for an open channel flow by~\cite{chen2014experimental} focused on the contributions of spanwise vortex filaments to Reynolds shear stress and to $v\omega_z$. Apart from measuring separate contributions to the advection term from prograde and retrograde vortices, they observed that the movement away from the wall yields the significant contribution of spanwise vortex filaments (identified by a swirling strength-based criterion) to the ``net force''. These ideas were further explored in \cite{chen2018contributions} where flow structures were classified into four groups based on vorticity and swirling strength, and contributions made by these structures to the nonlinear vorticity fluxes were measured.}
\\



\black{
A key contribution of our work, which distinguishes it from all of the previous studies cited above, is to make a clear 
connection of our numerical results with the ideas of \cite{lighthill1963} which focus on vorticity dynamics.  
Based on new theoretical insights and novel analysis of data using conditional averaging,  
we shall argue that Lighthill's theory provides
a compelling explanation of many prior observations. Furthermore, by a targeted filtering 
based upon joint velocity-vorticity co-spectra, we show 
that Lighthill's ``up-gradient'' vorticity cascade involves a previously unidentified hierarchy of non-attached, 
near-wall eddies, with important implications for theory and modelling.}
The detailed contents of this paper are outlined as follows. In section \ref{sec:lighthuggins} we discuss how 
Lighthill's Lagrangian mechanism is represented by the Eulerian vorticity flux tensor, a necessary theoretical 
prelude so that our subsequent numerical results can be appropriately interpreted. The main section \ref{sec:numerics}
of the paper presents our numerical study. In section \ref{sec:means} we study the mean vorticity flux
and its component velocity-vorticity correlations, validating our own numerical results against previously 
published results and illustrating the mean flow of vorticity along isolines of total pressure. The next
section  \ref{sec:correlations} presents results on conditional averages of fluxes given the direction of 
the wall-normal velocity as inward or outward, in order to investigate the proposed strong correlation. 
Section \ref{sec:cospectra} presents results on cospectra of the nonlinear vorticity flux 
and velocity-vorticity correlations, both 1D spectra in the streamwise and spanwise wavenumbers and 
joint 2D spectra. Then in section \ref{sec:coherent} we use the 2D cospectra to divide the velocity
and vorticity fields into ``down-gradient'' and ``up-gradient'' eddy contributions and we visualize 
the coherent vortices which dominate transport in both of these components. Finally, in the conclusion section
\ref{sec:conclusion} we review our main results, draw relevant lessons and discuss important future directions. 
Incidental numerical results of various sorts are presented in Supplementary Materials.

\section{Lighthill Mechanism and Huggins' Vorticity Flux Tensor}\label{sec:lighthuggins}

    

In order to properly interpret the results of our numerical study, we must first discuss carefully the physical and 
mathematical meaning of Lighthill's arguments. This is necessary especially because Lighthill's dynamical picture 
is essentially Lagrangian whereas Huggin's vorticity flux tensor \eqref{Sigma} is Eulerian. Thus, the relation 
between Lighthill's mechanism and the predicted behaviour of Huggins' flux can be somewhat subtle and even 
counter-intuitive.

The basic picture behind Lighthill's argument is sketched as a cartoon in Fig.~\ref{fig_cartoon}. Shown there 
is a representative vortex line carrying spanwise vorticity and also wall-normal vorticity 
associated to a lifted arch. If the flow is inward toward the wall ($v<0$) at this location, then, 
by incompressibility, there must be diverging flow in the spanwise and/or streamwise directions. 
\black{See Fig.~\ref{fig_cartoon1}.}  
This divergent flow should generate corresponding velocity gradients in those directions which 
Lighthill argued should be strongest spanwise because the well-known longitudinal organization 
of the near-wall structures would tend to reduce streamwise gradients.  According to the Helmholtz laws 
of ideal vortex dynamics, one would therefore expect the vortex line to be, first, carried down by the flow closer 
to the wall, and, second, flattened and stretched out, principally in the spanwise direction. This action 
of the Lagrangian flow is indicated by the blue dashed arrows in Fig.~\ref{fig_cartoon1} which depict typical particle 
trajectories. Since the stretched vortex line should intensify, Lighthill suggested that the plausible 
result would be increased spanwise vorticity concentrated closer to the wall. The opposite effect should occur 
in regions of flow outward from the wall ($v>0),$ which corresponds to the same cartoon but reversing all velocity 
vectors given by red arrows and all Lagrangian trajectories given by blue dashed lines. \black{See Fig.~\ref{fig_cartoon2}.}
In this case the vortex line according to ideal laws would be lifted away from the wall, compressed in the 
spanwise direction and correspondingly weakened. Such motions, according to Lighthill, would reduce 
the spanwise vorticity at further distances from the wall, so that the net effect of both types of motion 
would be an increase in the intensity of vorticity at the wall and a steepening of its wall-normal gradient.  
\black{However, note that according to the Kolmogorov theory of local isotropy (\cite{tennekes1972first}), streamwise and spanwise velocity gradients may be of a similar magnitude at small enough scales. Therefore, the association of inflow/outflow with stretching/relaxing of spanwise aligned vortex lines is expected to be primarily valid at scales that are large enough to possess the streamwise organization associated with stronger spanwise gradients.}

{ \begin{figure}
 \centering
 \begin{subfigure}[b]{0.49\textwidth}
\centering
\includegraphics[width=\textwidth]{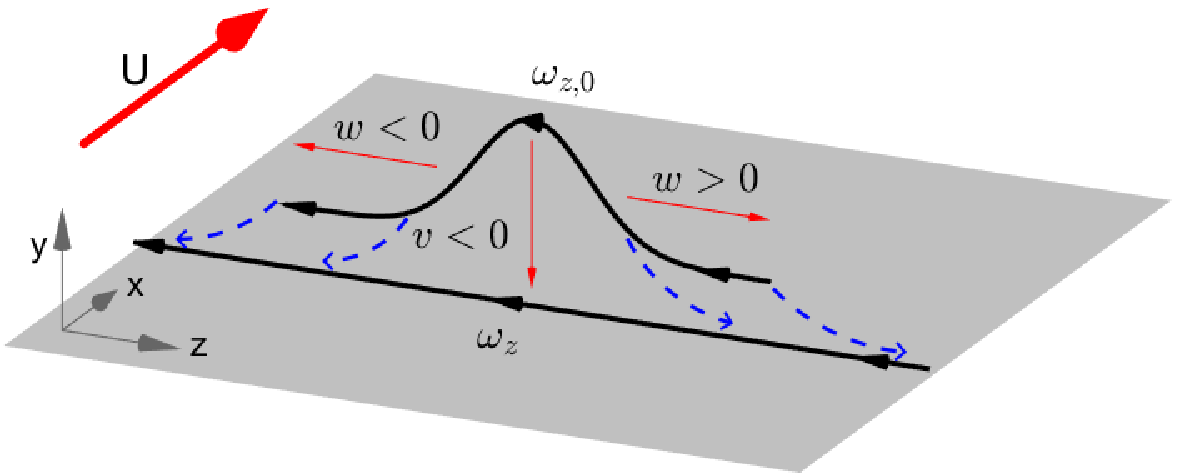}
\caption{$|\omega_{z}|>|\omega_{z,0}|$}
\label{fig_cartoon1}
\end{subfigure}
\hfill
\begin{subfigure}[b]{0.49\textwidth}
\centering
\includegraphics[width=\textwidth]{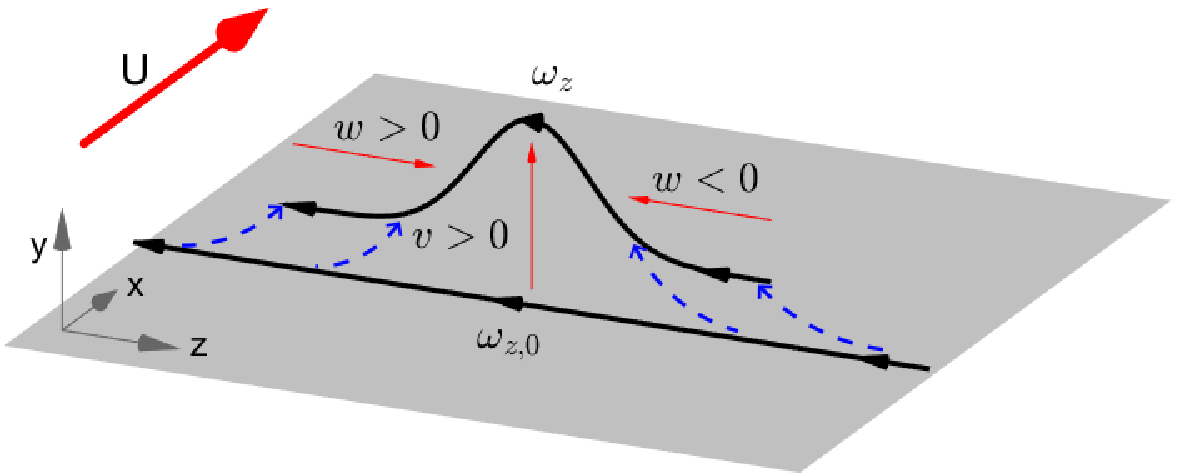}
\caption{$|\omega_{z}|<|\omega_{z,0}|$}
\label{fig_cartoon2}
\end{subfigure}
\caption{Cartoon of \black{of Lighthill's Lagrangian mechanism (``vortex lines ... stretch as they approach the surface and relax 
as they move away from it") for typical vortex lines in the buffer and log layers.
Panel (a): inflow and vortex stretching, $|\omega_{z}|>|\omega_{z,0}|$; panel (b): outflow and vortex compression, $|\omega_{z}|<|\omega_{z,0}|$. 
In both cases, initial vorticity is given by $\omega_{z,0}$ and final vorticity by $\omega_z$.}}
\label{fig_cartoon}
\end{figure}}

An obvious concern with this picture is its neglect of viscous diffusion effects, which certainly 
must be substantial in the near-wall buffer layer and viscous sublayer. In fact, as noted above,
viscous diffusion dominates the average wall-normal flux of spanwise vorticity out to the location
$y_p$ of peak Reynolds stess \citep{klewicki2007physical,eyink2008,brown2015vorticity}. The viscous 
modifications of ideal vortex 
dynamics can be incorporated by means of a stochastic Lagrangian formulation of incompressible Navier-Stokes 
in vorticity-velocity representation \citep{constantin2011stochastic,eyink2020stochastic}, which represents 
viscous diffusion of vorticity by an average over stochastic Brownian perturbations of Lagrangian 
particle motions.  This approach was exploited by \cite{wang2022origin} to investigate the 
origin of the enhanced wall vorticity and skin friction in a transitional zero pressure-gradient boundary layer, 
as discussed in the passage from \cite{lighthill1963} quoted in the Introduction. This study used the Lighthill 
vorticity source $\bm{\sigma}$ in \eqref{source} as Neumann boundary conditions, so that the wall vorticity 
at points of local maximum amplitude could be expressed in terms of two contributions: (i) the Lighthill source 
integrated over earlier times and (ii) the initial conditions for the vorticity as modified by subsequent
advection, stretching and viscous diffusion. It was found that the dominant source of the high wall-vorticity 
is the spanwise stretching of pre-existing spanwise vorticity, exactly as argued by \cite{lighthill1963}. 
In particular, as also suggested by Lighthill, the rate of production by the vorticity source $\bm{\sigma}$
was ``too small to explain'' the maxima. This contribution was found in general to be about an order of magnitude smaller 
than that from spanwise stretching and also found to give vorticity contributions of both 
signs with about equal probability, thus often reducing the magnitude. The conclusion of  \cite{wang2022origin} 
from their analysis of the numerical data was that, despite strong viscous effects in the near-wall region, 
the theory of \cite{lighthill1963} explained well the origin of high wall-stress events observed in transitional flow.  

{ \begin{figure}
 \centering
\begin{subfigure}[b]{0.49\textwidth}
\centering
\includegraphics[width=\textwidth]{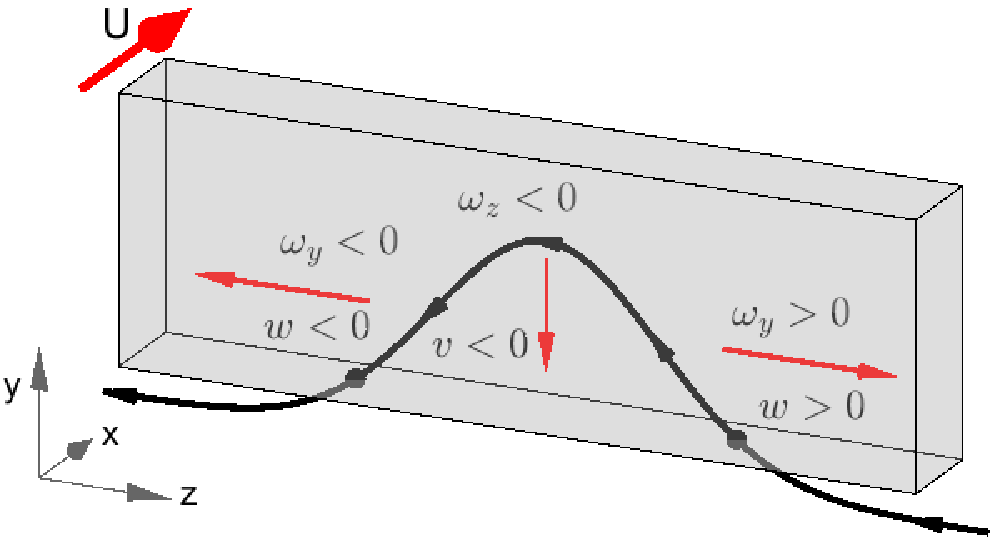}
\caption{\black{$v<0$}}
\label{fig_control}
\end{subfigure}
\hfill
\begin{subfigure}[b]{0.49\textwidth}
\centering
\includegraphics[width=\textwidth]{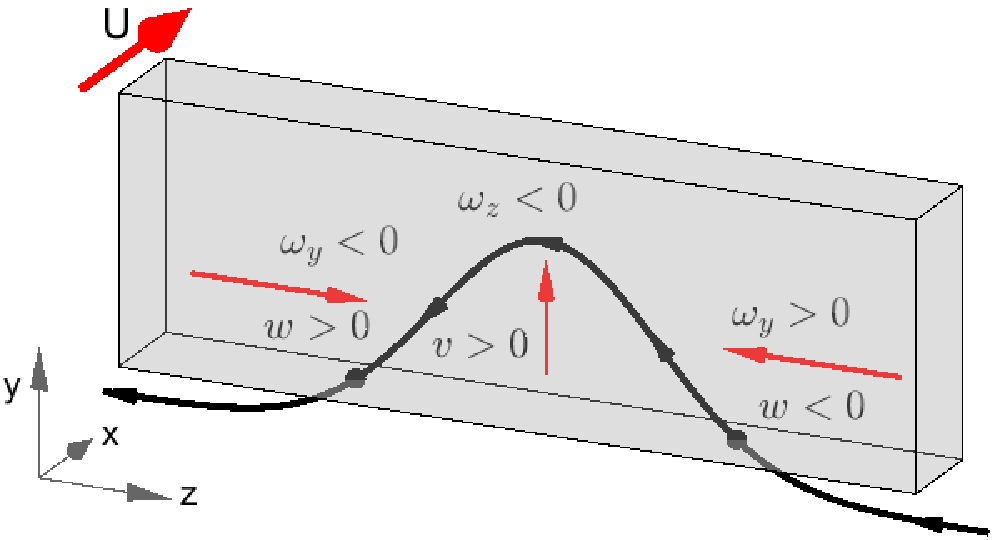}
\caption{ \black{$v>0$}}
\label{fig_control2}
\end{subfigure}
\caption{\black{Eulerian control-volume analysis of the same vorticity dynamics illustrated in Fig.~\ref{fig_cartoon}.
Panel (a): inward flow ($v<0$) and panel (b): outward flow ($v>0$). The control volumes are chosen to highlight 
the contribution of the stretching/tilting term to the change of integrated $\omega_z$ in the volume by flux across the bottom face.}}
\label{fig_cv_updown}
\end{figure}}

The same stochastic Lagrangian methods can be applied also to fully-developed turbulent channel flow, 
but here, motivated by recent work of \cite{eyink2021,kumar2023turbulent} on the Josephson-Anderson 
relation, we instead aim to understand vorticity dynamics in terms of Huggins' vorticity current 
tensor \eqref{Sigma}. Because the vorticity conservation law \eqref{vort-cons} is expressed 
in Eulerian form, its relationship to Lighthill's Lagrangian picture is not entirely self-evident. 
The physical meaning of Huggins' tensor $\Sigma_{ij}$ has previously been discussed by 
\cite{terrington2021generation,terrington2022vorticity} using control volumes and 
control surfaces. To adapt their arguments, we integrate \eqref{vort-cons} over a volume $V$ to obtain 
\be \frac{d}{dt} \int_V \omega_j d^3x = -\oint_{\partial V} n_i\Sigma_{ij} dA\, \lb{vort-int} \ee
where $\bm{n}$ is the outward-pointing normal at the boundary $\partial V$ and each individual term
in $n_i\Sigma_{ij}$ should represent a rate of change of $\omega_j$ integrated over $V.$ The meaning 
of the nonlinear term $u_i\omega_j$ is transparent, as $(\bm{n\cdot u})\omega_j$ just represents the advection of 
$\omega_j$ across the boundary $\partial V.$ The other nonlinear contribution $-u_j\omega_i$ to the flux 
upon taking taking its divergence yields the term $-(\bm{\omega\cdot\nabla})\bm{u}$ in the Helmholtz equation 
associated to vortex stretching and tilting, so that it must somehow express that physics. It is worth 
remarking that \cite{huggins1994vortex}, p.326 concluded that this term ``does not appear to have a particularly 
simple interpretation'' but suspected that it is ``a vortex stretching term''. The relation with stretching\black{/tilting} 
is clarified by Fig.~\ref{fig_control} which plots schematically the \black{first} ``up-gradient'' configuration 
considered by \cite{lighthill1963} with flow inward to the wall advecting and stretching\black{/tilting} a hairpin-like 
vortex. We have drawn as control volume a rectangular box selected so that only the term $-w\omega_y$ 
in the flux $\Sigma_{yz}$ contributes to growth of spanwise vorticity $\omega_z$ in the volume,
whereas the advection term $v\omega_z$ does not contribute. Because of the diverging flow, 
two contributions with signs $\omega_y>0,$ $w>0$ and $\omega_y<0,$ $w<0$ occur at the bottom face of the 
box and these correspond indeed to an increase of (negative) spanwise vorticity in the pictured control volume, due to the lengthening of the spanwise vortex line segment \black{and the tilting of the wall-normal vortex line segments}. 
It is notable that the contribution $-w\omega_y<0$ at the bottom face in Fig.~\ref{fig_control} 
corresponds to an {\it outward} flux of spanwise vorticity into the control volume, with a sign which is 
formally ``down-gradient'' or away from the wall. By contrast, the advection term has sign $v\omega_z>0$ 
which is ``up-gradient'' \black{and from Lighthill's Lagrangian
argument we may expect that the net nonlinear transfer 
is ``up-gradient''} for this flow configuration with $v<0.$ 
If we consider instead
the outward flow configuration with $v>0$ \black{ illustrated in Fig.~\ref{fig_control2}}, which compresses and weakens the spanwise vorticity, then the signs 
would both reverse to $v\omega_z<0$ and $-w\omega_y>0$ (because of converging flow), but would remain 
opposite.\black{ The advective term is now making a ``down gradient" contribution and the stretching/tilting term is making an ``up gradient" contribution, with the net nonlinear flux expected to be ``up-gradient''.} 

To determine whether the net vorticity flux from nonlinearity is ``up-gradient'' or ``down-gradient'', 
it is important consider the combined term  $u_i\omega_j-u_j\omega_i,$ which is anti-symmetric. 
As stressed by \cite{terrington2021generation}, the anti-symmetry $\Sigma_{ji}=-\Sigma_{ij}$ 
expresses the fundamental property that vortex lines cannot end in the fluid so that flux 
of $\omega_j$ in the $i$th direction is necessarily associated with an equal and opposite
flux of $\omega_i$ in the $j$th direction. This relation of flux anti-symmetry and 
non-termination of vortex lines is clearly illustrated in Fig.~\ref{fig_control}, \black{for example.} In the 
flow sketched there, the depicted spanwise flux $\Sigma_{zy}$ of $\omega_y$-vorticity implies that the 
$\omega_z$-line must lengthen, because the vortex line which enters at one $z$-location in the 
bottom face must exit at the other. The resulting growth of $\omega_z$-vorticity in the interior 
\black{by stretching and tilting} corresponds 
to a wall-normal flux $\Sigma_{yz}=-\Sigma_{zy}$ outward into the control volume. To determine from our 
numerical data whether nonlinear flux of spanwise vorticity is ``down-gradient'' or ``up-gradient'' 
it will therefore be crucial to consider the combined quantity $v\omega_z-w\omega_y$ that contains
both advection and stretching\black{/tilting}, since these two effects cannot be separated physically without violating 
the kinematic condition of non-terminating vortex lines. 

The anti-correlated sign of the two \black{separate} flux contributions\black{ from advection and stretching/tilting will be 
crucial, on the other hand,} in interpreting our numerical results below, \black{since this anti-correlation is
a key Eulerian signature of Lighthill's mechanism. Fig.~\ref{fig_control} shows that the strengthening 
of spanwise vorticity during an inflow is represented in the Eulerian flux by a ``down-gradient'' stretching/tilting term,
even though the net flux is ``up-gradient''.\, Similarly, Fig.~\ref{fig_control2} shows that the weakening 
of spanwise vorticity during an outflow is represented in the Eulerian flux by an ``up-gradient'' 
stretching/tilting term, even though the net flux is ``down-gradient''. 
This anti-correlation of the stretching/tilting term with the net flux is thus a direct manifestation of Lighthill's mechanism.}
It should be clear that this anti-correlation of signs depends upon the geometry of the vortex line.
For example, if the vortex line in Figure \ref{fig_control} were instead bent inward into a U-shape
and entered the control volume from the top face, then the sign of the stretching\black{/tilting} term would have 
become $-w\omega_y>0.$ \black{This inward  flux into the control volume would again correspond to vortex-strengthening,
but it would now represent formally ``up-gradient'' transport} positively correlated with the advection term $v\omega_z>0.$ 
{Because of the assumption of a specific vortex line geometry in Fig.\ref{fig_cv_updown}, the suggested anti-correlation between the signs of the advection and stretching/tilting terms can be claimed only to be consistent 
with Lighthill's ideas, which should be further investigated. More positively,} the relative sign of the advection term and of the stretching/tilting term potentially contains some information about the typical geometry of vortex lines.

%

\section{Numerical Study of Vorticity Flux in Pressure-Driven Channel Flow}\label{sec:numerics} 


We now report on our empirical study of the flux of spanwise vorticity,  hereafter referred to simply 
as ``vorticity flux''. This component of the vorticity is crucial to drag and energy dissipation, since its flux is directly related 
to streamwise pressure drop. As already mentioned in the Introduction (Sec ~\ref{intro}), we employ direct numerical simulation data of channel flow at 
$Re_{\tau}=1000$ from the Johns Hopkins Turbulence Database (JHTDB) (see ~\cite{jhtdb1,jhtdb_channel}). The right-handed Cartesian 
coordinate system for this data is the same as shown in Figures \ref{fig_cartoon} \& \ref{fig_control}, with $x$ streamwise, $y$ wall-normal
and $z$ spanwise.  Although the database 
provides built-in tools to calculate velocity and vorticity gradients from Lagrange interpolants, our study of vorticity 
dynamics required greater accuracy for these crucial quantities. We have thus used the database cut-out service to download
time snapshots of data for the entire channel. Gradients in the spanwise and streamwise directions are then calculated spectrally
by FFT, and wall-normal gradients are calculated using seventh-order basis-splines based on the collocation points of the 
original simulation (~\cite{jhtdb_channel}). All statistics are thereafter calculated by averaging over wall-parallel planes 
in the $x-$ and $z-$directions of homogeneity, as well as over multiple snaphots. The steady-state statistics presented here 
were calculated with 38 time snaphots. We shall generally plot our results only for the bottom half of the channel, with 
reflected results from the top half included to double the sample-size of our averages. 

\subsection{Mean vorticity flux and flow-lines}\label{sec:means}  

{ \begin{figure}
 \centering
\includegraphics[width=0.6\textwidth]{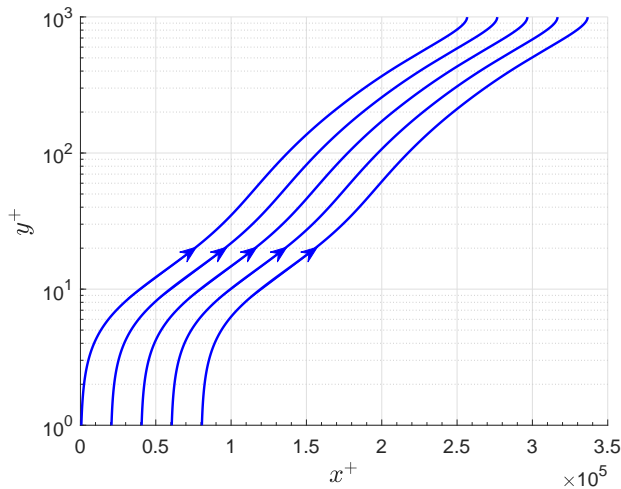}
\caption{ Flowlines of mean spanwise vorticity flux $(\langle\Sigma_{xz}\rangle,\langle\Sigma_{yz}\rangle)$ 
obtained as the isolines of mean total pressure $\langle h\rangle$.}
\label{fig_flowlines}
\end{figure}}

To provide an intuitive understanding of Huggins' vorticity flux tensor \eqref{Sigma} and of the mean vorticity
dynamics in turbulent channel flow, we first present numerical results on the average flux $\langle\Sigma_{ij}\rangle.$
An important theoretical result which follows directly by averaging the momentum balance equation 
\eqref{JAsimple} is the steady-state relation between vorticity flux and the gradients of the total pressure: 
\be \langle \Sigma_{ij}\rangle = \epsilon_{ijk}\partial_k\langle h\rangle. \ee 
This general result implies immediately for channel flow that 
\be \langle\Sigma_{yz}\rangle=\partial_x\langle h\rangle, \quad 
\langle\Sigma_{xz}\rangle=-\partial_y\langle h\rangle, \quad \langle \Sigma_{xy}\rangle=0, \lb{meanJA} \ee 
with all other components given by anti-symmetry. Since it is the flux of spanwise vorticity only which enters into the 
Josephson-Anderson relation for plane-parallel channel flow \citep{kumar2023turbulent} we shall 
focus on its dynamics in the $(x,y)$-plane (since $\Sigma_{zz}=0$). It can be very 
instructive about the physics to trace the integral flow lines of mean fluxes for 
transported quantities such as energy and momentum \citep{meyers_meneveau_2013} and we carry out this construction for the conserved spanwise vorticity. Here there is a substantial simplification because, as a simple consequence of \eqref{meanJA}, the integral lines of the mean flux vector $(\langle\Sigma_{xz}\rangle,\langle\Sigma_{yz}\rangle)$ coincide 
with the isolines of mean total pressure $\langle h\rangle= P + \frac{1}{2}U^2 + \frac{1}{2}\langle |u'|^2+|v'|^2+|w'|^2\rangle.$
We follow the usual notations, $U=\langle u\rangle,$ $u'=u-U,$ $P=\langle p\rangle,$ etc. In Figure \ref{fig_flowlines} we plot
these isolines resulting from numerical computation of $\langle h\rangle.$ Consistent with the remark of \cite{lighthill1963}
that ``tangential vorticity created is in the direction of the surface isobars,'' the mean vorticity generated at the 
wall is spanwise and flows outward from points of constant $\langle h\rangle=P$ at $y=0.$ The vorticity flux 
is about three orders of magnitude larger streamwise than wall normal, mainly because of the large term $U\Omega_z$
contributing to $\langle\Sigma_{xz}\rangle,$ with $\Omega_z=-\partial_yU.$ Thus, the mean vorticity flow lines extend about 250 channel 
half-widths downstream as they cross from the wall to the channel center, reflecting the strong 
streamwise advection of vorticity. It is, however, the much smaller wall normal vorticity flux which is directly 
related to drag and energy dissipation, since $\langle\Sigma_{yz}\rangle=\partial_xP$
by \eqref{meanJA}. As earlier remarked by \cite{eyink2008}, the latter takes 
on the $y$-independent value $\langle\Sigma_{yz}\rangle=-u_\tau^2/H$ because 
of the conservation relation $\partial_y\langle\Sigma_{yz}\rangle=
\partial_y\langle\Sigma_{yz}\rangle+\partial_x\langle\Sigma_{xz}\rangle=0$
and the \cite{lighthill1963} relation for wall-generation of vorticity 
by tangential pressure gradients. This argument assumes as well the $x$-independence of steady-state averages such as $\langle\Sigma_{xz}\rangle$, which is evident in the parallel vorticity flux lines of Figure ~\ref{fig_flowlines}. 


{ \begin{figure}
 \centering
\includegraphics[width=0.6\textwidth]{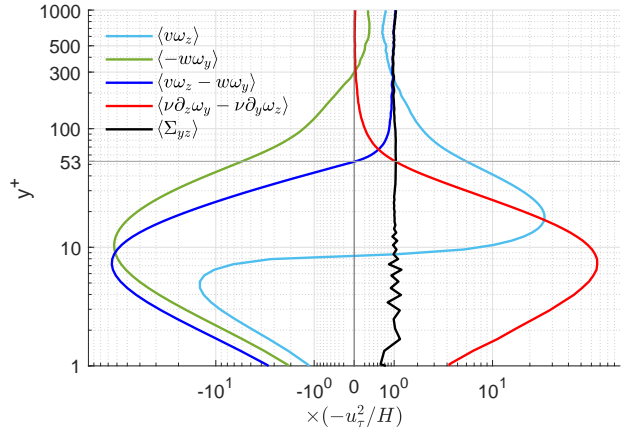}
\caption{Vorticity flux in a pressure driven periodic channel, averaged in time as well as in the streamwise and spanwise directions. A good agreement is found with expected constant behaviour across the height of the channel.  For $y^+<y_p^+(=53)$, the nonlinear term gives a strong up-gradient contribution to the flux which is balanced by a larger down-gradient viscous contribution. As $y$ increases above $y_p$, the nonlinear term contributes an increasing share of the down-gradient flux while the viscous contribution decreases as $y$ grows so that for $y^+ \gtrsim 250$, the nonlinear contribution carries nearly the entire vorticity flux.} 
\label{fig_conv_visc}
\end{figure}}

Exact results of \cite{klewicki2007physical,eyink2008,brown2015vorticity} for the 
nonlinear $\langle v\omega_z -w\omega_y  \rangle$ and viscous $-\nu\langle \partial_y\omega_z-\partial_z\omega_y\rangle$ contributions to the mean vorticity flux 
 $\langle\Sigma_{yz}\rangle$ provide a further check of the reliability of our numerics. 
Our numerical data are plotted in Fig~\ref{fig_conv_visc} and show good agreement with 
the theoretically required behavior. First, we observe that the magnitude of the mean vorticity flux is constant in $y$ 
to a very good approximation, except for small numerical oscillations very close to the wall ($y^+<10$), and it matches 
quite well the average streamwise pressure gradient. This means that, on average, negative spanwise vorticity (the same sign 
as the mean vorticity) is being transported away from the wall and that overall, the flux is down-gradient. As in the Introduction, 
we shall refer to flux of vorticity away from the wall as ``down-gradient,'' since the vorticity is already highly concentrated at the wall
\citep{lighthill1963}, and flux in the opposite direction will be referred to as ``up-gradient". It was also shown 
by \cite{klewicki2007physical,eyink2008,brown2015vorticity} that, while viscous flux should be expected to be always down-gradient, the net 
nonlinear flux will be down-gradient above the height of the peak Reynolds' stress ($y^+=y_p^+=53$ for the
data at $Re_\tau = 1000$)
but up-gradient below that height and opposing the large viscous flux there. These theoretical results are well confirmed 
by our empirical data in Fig~\ref{fig_conv_visc}. In addition, we have calculated the separate contributions of the 
advective ($v\omega_z$) and stretching\black{/tilting} ($-w\omega_y$) terms to the nonlinear flux, for which no exact predictions exist.
However, our results for these two quantities plotted also in Fig.~\ref{profile_comp} agree well with those 
earlier reported by \cite{monty2011characteristics,brown2015vorticity,chen2018contributions}
from the channel-flow simulation of \cite{delalamo2004scaling} at $Re_\tau=934$ and also with the experimental results of \cite{chen2014experimental}  for an open channel flume  at somewhat lower $Re_\tau=740.$  
Similar observations have been made both at lower and at higher Reynolds numbers, but we postpone until 
our conclusions section the discussion of the important issue of $Re$-dependences. Confirming those earlier studies, 
we find that both contributions are down-gradient in the outer range ($y^+\gtrsim 300$) and both up-gradient 
in the near-wall ($y^+\lesssim 10$), but have opposite signs in the intermediate range corresponding roughly to the 
logarithmic mean velocity profile ($30\lesssim y^+\lesssim 300$). Note that we use the term ``log layer'' 
for this range rather than the frequent term ``inertial sublayer'' because, as pointed out 
by \cite{klewicki2007physical,eyink2008,brown2015vorticity}, viscous diffusion dominates mean vorticity 
transport at least up to $y_p.$ In this logarithmic layer the nonlinear advection term is also down-gradient but  
the stretching\black{/tilting} term up-gradient, with advection dominating for $y\gtrsim y_p$ and stretching\black{/tilting} dominating 
for $y\lesssim y_p.$ The importance of the stretching\black{/tilting} term in the range $y\lesssim y_p$ is suggestive of Lighthill's 
mechanism and the anti-correlation there is reminiscent of the opposite signs found in the control volume 
analysis of \black{Figure \ref{fig_cv_updown}}. However, to identify precisely whether the vorticity up-gradient transport occurs 
by Lighthill's mechanism, we must study the crucial question of correlation with motion inward \black{($v<0$)} or outward \black{($v>0$)}
from the wall. 

{ \begin{figure}
 \centering
\includegraphics[width=0.7\textwidth]{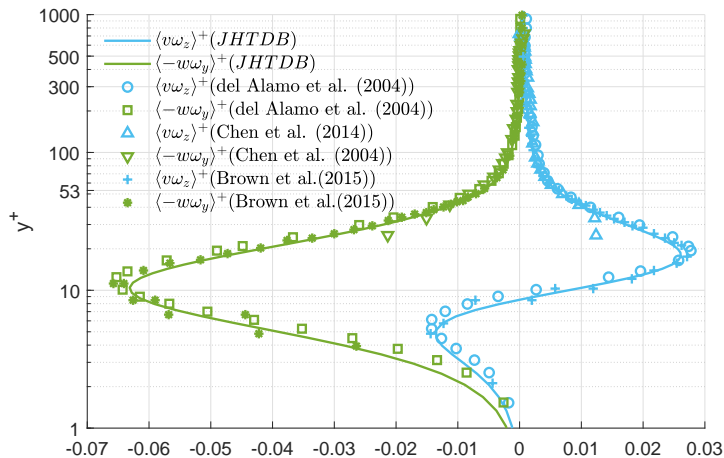}
\caption{Contributions to nonlinear vorticity flux averaged over time and wall parallel planes, computed from channel flow data at $Re_{\tau}=1000$ from JHTDB(~\cite{jhtdb_channel}) and from the earlier numerical simulation of channel flow at $Re_{\tau}=1000$ by ~\cite{brown2015vorticity},  $Re_{\tau}=934$ 
by ~\cite{delalamo2004scaling}, as reported in \cite{monty2011characteristics}, and from experimental measurements for an open channel flume at $Re_{\tau}=740$ by ~\cite{chen2014experimental}.}
\label{profile_comp}
\end{figure}}

\subsection{Evidence for Lighthill's mechanism from flux-velocity correlations}\label{sec:correlations}

\begin{figure}
      \centering
     \begin{subfigure}[b]{0.32\textwidth}
         \centering
         \includegraphics[width=\textwidth]{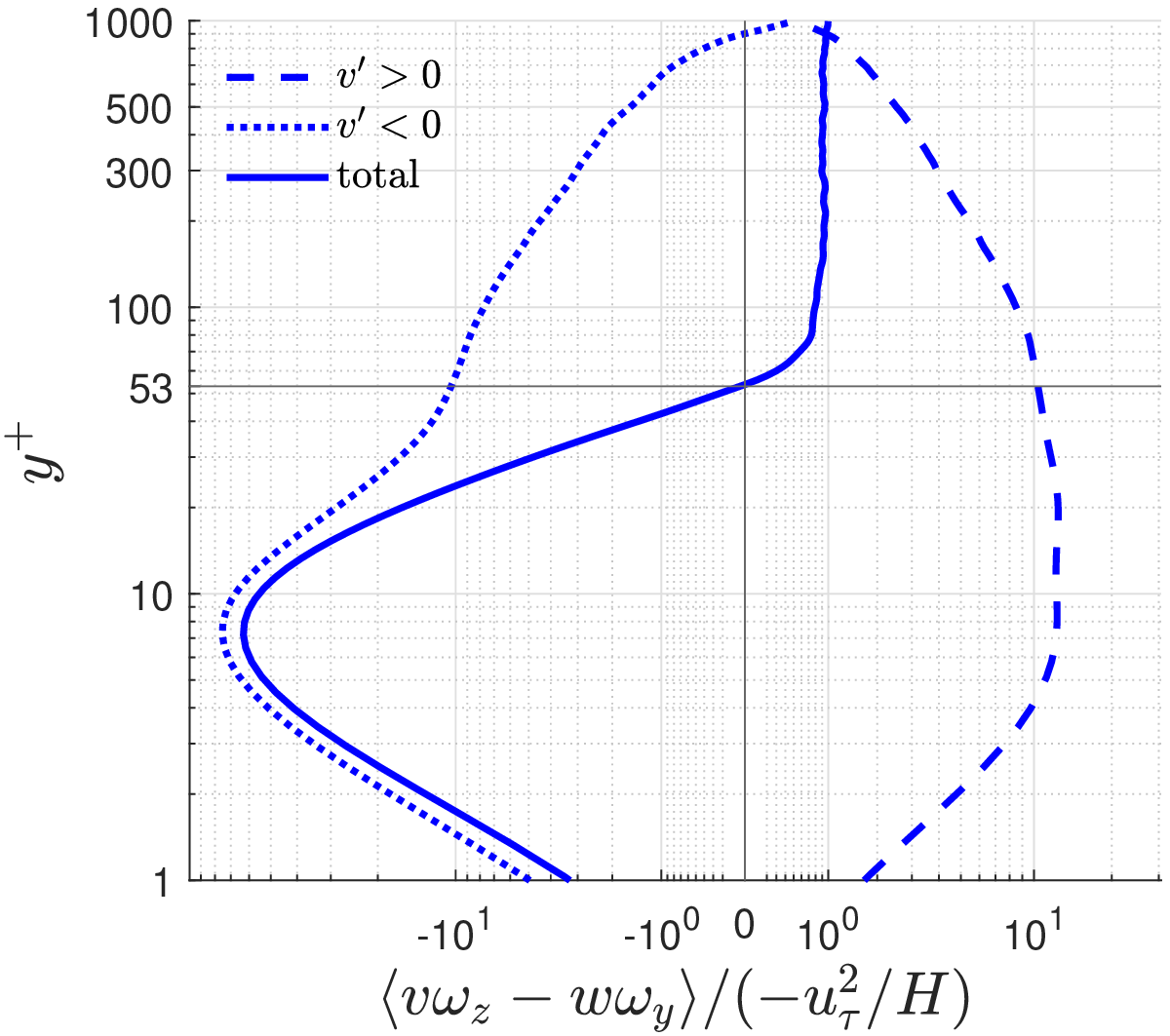}
        \caption{}
         \label{quad_nonlinear_v}
     \end{subfigure}
     \hfill
     \begin{subfigure}[b]{0.32\textwidth}
         \centering
         \includegraphics[width=\textwidth]{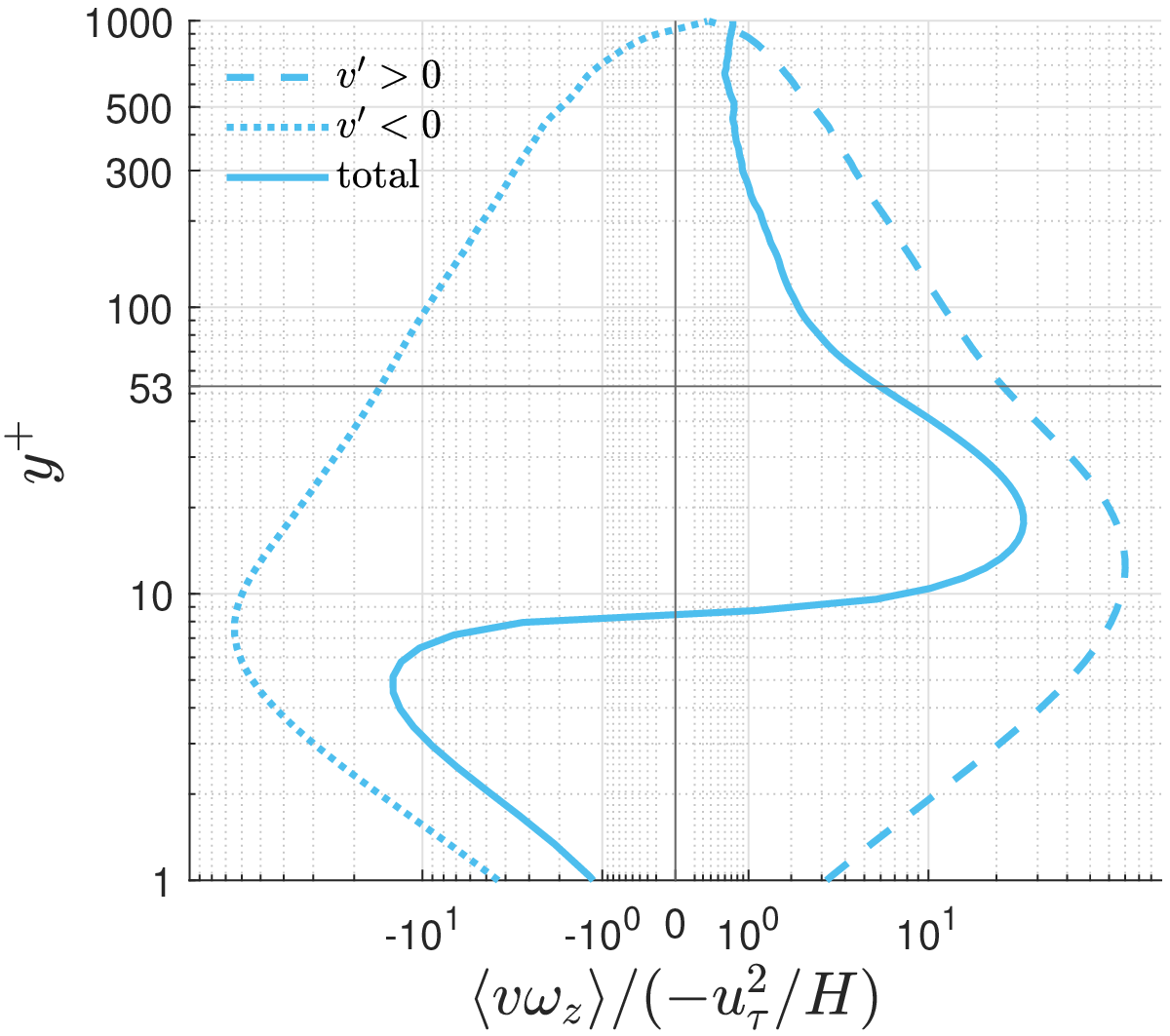}
         \caption{}
         \label{quad_voz_v}
     \end{subfigure}
     \hfill
     \begin{subfigure}[b]{0.32\textwidth}
         \centering
         \includegraphics[width=\textwidth]{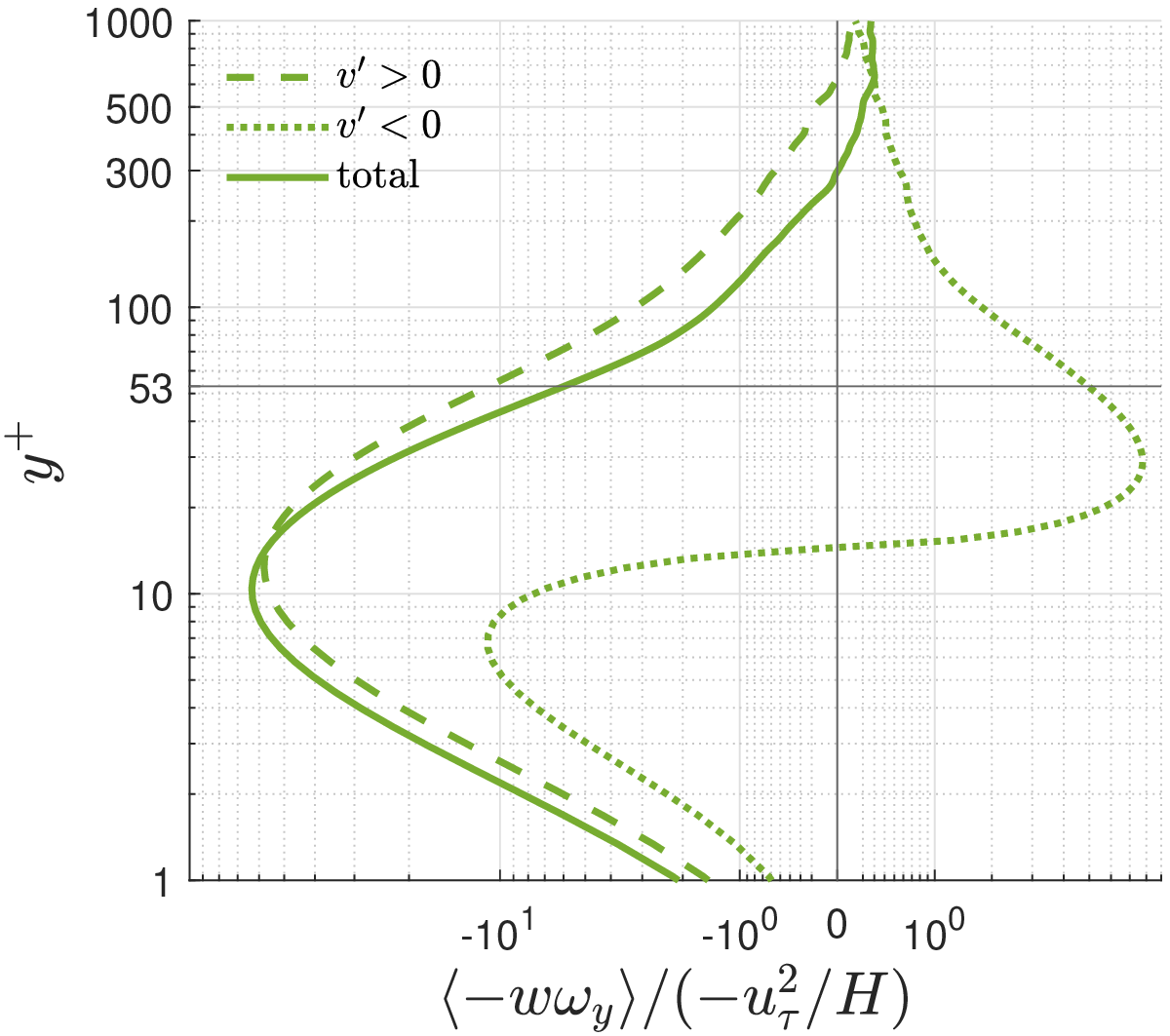}
         \caption{}
         \label{quad_woy_v}
     \end{subfigure}
        \caption{Contributions from regions where turbulent flow is outward $(v'>0)$ and inward $(v'<0)$, to the (a) nonlinear flux , (b) convection/advection and (c) stretching/tilting , averaged over time and wall parallel planes, plotted as a function of wall distance.   }
        \label{quad_contri_v}
\end{figure}

A crucial feature of the theory of \cite{lighthill1963} is the proposed correlation between 
vorticity-strengthening and inward motion toward the wall, and likewise vorticity-weakening and 
outward motion away from the wall. To test for this key correlation we consider partial averages 
depending upon the two conditions $v'>0$ and $v'<0,$ where the wall-normal velocity fluctuation 
is $v'=v$ since $V=0.$ Note that by ``partial average subject to $X$'' we mean the average 
conditioned upon the event $X$ but further multiplied by the probability of $X.$ Defined in this manner,
the sum of the partial averages for the two exclusive events $v'>0$ and $v'<0$ yields the total average.
\black{In Fig.~\ref{quad_nonlinear_v} we plot }these partial averages for the total nonlinear flux 
$v\omega_z-w\omega_y.$ Although Lighthill's proposed mechanism is essentially Lagrangian,
we see a clear correlation in the Eulerian vorticity flux, with outflow \black{($v>0$) }associated at all wall distances 
to down-gradient mean vorticity flux and inflow \black{($v<0$)} associated to up-gradient 
mean flux, except possibly very near the center of the channel. The inflow contribution \black{appears to prevail for $y<y_p$, where the net nonlinear flux is up-gradient.} 
To gain further insight into the vorticity dynamics, we 
consider next the partial averages of the separate flux contributions from advection and stretching\black{/tilting}. 

The partial averages of advective vorticity flux $v\omega_z$ are plotted in Fig.~\ref{quad_voz_v},
which exhibit the same correlation as the total: down-gradient flux is associated to outflow
and up-gradient flux to inflow (except at the channel center). 
This correlation is essentially obvious in the near-wall region since instantaneous vorticity has the sign 
$\omega_z<0$ nearly always there, the same sign as the mean vorticity $\Omega_z<0.$ Note that the 
partial average from outflow prevails above the buffer layer, where the combined average is down-gradient. 
The dominance of outflow over inflow can be easily understood if the wall-normal velocity magnitudes 
$|v|$ are roughly comparable for the two events, but if spanwise vorticity magnitudes $|\omega_z|$
are generally greater for flows originating near the wall. We can thus have net advection of 
vorticity away from the wall, even though there is no net advection of mass. In contrast to this down-gradient 
behaviour in the log layer, we note that net mean advection is up-gradient for $y^+<10$. This result 
is plausibly explained if inflow events are correlated
with locally higher spanwise vorticity \black{due to stretching of spanwise aligned vortex lines} in the near-wall region and outflow events with locally weaker vorticity \black{due to relaxation and attenuation of vortex lines.} 
These conclusions and interpretations 
agree with earlier work of \cite{klewicki1994vortical}, who measured experimentally the joint PDF 
$P(v',\omega_z')$ of the fluctuations of wall-normal velocity $v'$ and spanwise vorticity $\omega_z'$ in a turbulent zero
pressure-gradient boundary-layer. Their results showed for $y^+=5.3$ that the mean $\langle v'\omega_z'\rangle$
gets most of its contribution from quadrants Q1 and \black{Q3} where $v'$ and $\omega_z'$ are positively correlated,
consistent with Lighthill's mechanism. \cite{klewicki1994vortical} interpreted this result apparently 
somewhat differently in terms of low-speed streaks with $u'<0,$ $\omega_z'>0$ moving upward with $v'>0,$ 
and high-speed streaks with $u'>0,$ $\omega_z'<0$ moving downward with $v'<0.$ However, these 
observations are not inconsistent with Lighthill's argument (see especially the remarks at the end of this 
subsection and Figs.~\ref{streak_contri}-\ref{quad_contri} in the Supplementary Materials). 
\cite{klewicki1994vortical} observed also that $\langle v'\omega_z'\rangle<0$ for $y^+>10$ with 
the main contributions from quadrants $Q4$ and especially $Q2$ of the PDF $P(v',\omega_z'),$ 
and they explained this down-gradient transport in essentially the same manner as we have. 

The partial averages of the stretching contribution $-w\omega_y$ to vorticity flux plotted in Fig.~\ref{quad_woy_v}
show diametrically opposite correlations with wall-normal velocity as those for the advection term \black{in the buffer layer, the log layer, and some of the outer layer}. The plot shows
that \black{in and above} the buffer layer inflow is associated with down-gradient flux, while outflow is associated with 
up-gradient flux (except very near the channel center). This opposite correlation may be explained 
by the control-volume analysis in section \ref{sec:lighthuggins}, which associates down-gradient 
flux with \black{the} line-stretching mechanism for $v'<0$ \black{(Fig.~\ref{fig_control})} and up-gradient flux 
with the corresponding line-compressing mechanism for $v'>0$ \black{(Fig.~\ref{fig_control2}).
The observed anti-correlation between the advection term in Fig.~\ref{quad_voz_v} and the stretching term 
in Fig.~\ref{quad_woy_v} is thus one of our key pieces of evidence in favor of Lighthill's mechanism, 
since ``down-gradient'' stretching flux during downflow is consistent with strengthening spanwise vorticity and 
``up-gradient'' stretching flux during upflow is consistent with weakening spanwise vorticity.}
\black{We see that the outflow contribution ($v'>0$) prevails in the combined average, implying that the 
stretching/tilting term contributes a net up-gradient flux across the buffer and log layers. }
We may explain 
this again as a consequence of the near-equality of spanwise velocity magnitudes $|w|$ for 
the two conditions but with wall-normal vorticity magnitude $|\omega_y|$ larger nearer the wall and 
smaller further away. 

For $y^+<15,$ \black{by contrast}, inflow and outflow \black{in Fig.~\ref{quad_woy_v}} are both associated with up-gradient vorticity flux 
\black{from the stretching/tilting term}.
\black{The outflow term makes the larger contribution, which is now augmented by a smaller contribution from the inflow. 
A different mechanism than Lighthill's seems to be at play here.}
It may be relevant that all nonlinear vorticity flux terms 
vanish identically at the channel wall. Thus, growth on average of spanwise vorticity in this 
very near-wall region must be due to nonlinear flux of vorticity inward, because the nonlinear 
flux outward must be small. The vortex lines in this region tend also to be strongly aligned in the 
spanwise direction, so that the geometry assumed in the control-volume argument is not typical here. 
\cite{brown2015vorticity} investigated the buffer-layer statistics of $w\omega_y$ by a DNS 
nearly identical to ours and their Figure 4 plots vortex lines originating at $y^+=10$ 
in the vicinity of a quasi-streamwise vortex. They argued that such coherent streamwise vortices 
are responsible for creating the ``up-gradient'' correlation $\langle w\omega_y\rangle<0$
in the buffer layer, because vortex lines on the side where $v>0$ are lifted into $\Lambda$-shape
with converging legs, while lines on the $v<0$ side are depressed into $U$-shape with diverging legs.
Note that this explanation agrees with our observation in Fig.~\ref{quad_woy_v} that partial averages of $w\omega_y$ 
are negative both for $v>0$ and for $v<0$ in the buffer layer. \cite{brown2015vorticity} anticipated also our result that the 
partial average for $v>0$ should exceed that for $v<0,$ arguing that $\omega_y$ magnitudes for 
$v<0$ will be weakened by creation of image vorticity at the wall. It is worth emphasizing a 
further numerical finding of \cite{brown2015vorticity} that the pointwise values of $w\omega_y$
at $y^+=5$ have a strongly non-Gaussian PDF giving high probability to events with magnitudes $\sim 25$ 
larger than the mean, so that the overall negative value $\langle w\omega_y\rangle<0$ in the 
buffer layer results from near-cancellation between much larger contributions of opposite signs. 


An \black{important} conclusion of all three of our Figs.~\ref{quad_nonlinear_v}-\ref{quad_woy_v} is that the
net effects seen in the partial averages of the total nonlinear flux in Fig.~\ref{quad_nonlinear_v} 
are due to the dominance of the advection term. Thus,  of the two 
effects considered in the control-volume analysis in section \ref{sec:lighthuggins}, 
the advective contribution generally outweighs the stretching\black{/tilting} contribution. 
It might 
appear paradoxical at first glance that inflow dominates in the total nonlinear flux 
for the region $y<y_p,$ while outflow dominates in the two separate flux contributions 
from advection and stretching\black{/tilting} 
\black{in and above the buffer layer, including a region where $y<y_p$}. However, the advection and stretching\black{/tilting} 
contributions oppose each other in this region and the outflow contributions suffer more 
cancellation in the combined flux than do the inflow contributions. In particular, 
the ``down-gradient'' flux from the stretching term during inflow is comparatively weak. 
In the 
viscous sublayer, \black{where the control volume analysis of section \ref{sec:lighthuggins} does not apply,} the inflow contribution to the stretching\black{/tilting} term
is again weaker than the outflow contribution but both are now ``up-gradient'', 
together with the net advective flux. 
\black{The opposing nature of the partial averages for the advection and stretching/tilting terms in the buffer and log layers, in addition to the correlations with inflow and outflow, strongly support Lighthill's idea of inflow being correlated with stretching/strengthening and outflow with compression/weakening of vortex lines, thereby providing \textit{a posteriori} validation of his theory.}

In order to make an objective assessment of the evidence we have 
investigated other possible correlations as well. However, none of these alternative 
correlations presented such a clear picture as the correlations with outflow/inflow.
Thus, we present these alternative correlations in the Supplementary Materials for completeness.
For example, we considered partial averages of the three flux terms $v\omega_z-w\omega_y,$
$v\omega_z,$ and $-w\omega_y$ conditioned on ``low-speed'' ($u'<0$) and ``high-speed'' 
($u'>0$) events \citep{meinhart1995existence,kim1999very,hwang2016inner,hwang2018wall}, 
as shown in Fig.~\ref{streak_contri} of the Supplementary Materials. 
While there is a clear correlation of the advective term (down-gradient for low speed
and up-gradient for high speed, consistent with \cite{klewicki1994vortical}), the stretching term 
is found to be insensitive to the conditions $u'>0$ and $u'<0.$ Perhaps the most interesting of these additional
correlation studies involved the standard quadrant analysis of the Reynolds stress \citep{willmarth_lu_1972,lu_willmarth_1973,bogard_tiederman_1986,pope2000turbulent,lozanoetal2012}. 
As shown in Fig.~\ref{quad_contri} of the Supplementary Materials, the partial averages 
of the flux terms conditioned on $Q2$-events $(u'<0,v'>0)$ or ``ejections'' are very similar to 
those conditioned on $v'>0$ alone, and those conditioned on $Q4$ events $(u'>,v'<0)$ or ``sweeps'' are
very similar to those for $v'<0$ alone. By contrast, the partial averages for $Q_1$ and $Q_3$ events 
are distinctly smaller in magnitude. The relevant conclusion is that much of the vorticity flux 
correlations observed in this section with outflow and inflow events arise from the corresponding 
``active'' regions of the flow, $Q2$ and $Q4,$ which contribute also to the Reynolds shear stress. 
See \cite{vidal_etal_2018} for related results discussed more in the Supplemental Materials, 
section \ref{sec:quad}.

\subsection{Velocity-vorticity cospectra}\label{sec:cospectra}

The main prediction of the \cite{lighthill1963} theory is up-gradient vorticity transport toward a solid wall, 
with eddy contributions naturally depending upon the size of eddies relative to the wall distance $y.$ Since the theory 
posits strong spanwise stretching, the vorticity transport effects should be particularly sensitive to the spanwise 
extent of eddies. This motivates us to consider the 1-D  spanwise vorticity-flux co-spectrum $\phi_{v \omega_z-w\omega_y}(k_z)$ 
which gives the net contribution of eddies with spanwise wavenumber magnitude $k_z$ or corresponding wavelength
$\lambda_z=2\pi/k_z.$ Such a co-spectrum may defined for any direction of homogeneity ($x,$ $z,$ or a linear combination 
thereof) by taking FFTs of velocity and vorticity, followed by an inner product, and averaging in time over snapshots
and along the orthogonal homogeneous direction. The cospectrum so defined yields a spectral decomposition of the nonlinear 
vorticity flux:  
\be \int_0^{\infty} \phi_{v\omega_z-w\omega_y}(k_i,y)  dk_i=\langle v\omega_z - w\omega_y\rangle(y),\  k_i=k_x \ or \ k_z. \lb{int_prop} \ee
\black{These velocity-vorticity co-spectra are identical to the ``net force spectra", defined as the wall-normal derivative of the Reynolds shear stress co-spectra, discussed in prior works of  ~\cite{guala_hommema_adrian_2006}, \cite{balakumar_adrian_2007} and \cite{wu_baltzer_adrian_2012}.}
Similar cospectra $\phi_{v\omega_z}(k_i,y),$ $\phi_{w\omega_y}(k_i,y)$ can be defined for the individual velocity-vorticity
correlations $\langle v\omega_z\rangle(y)$ and $\langle w\omega_y\rangle(y),$ with prior empirical studies 
of \cite{priyadarshana2007statistical} and \cite{morrill2013influences} having calculated \black{individual} streamwise cospectra.
\black{We are aware of no prior studies which computed spanwise cospectra for channel flows, although analogous co-spectra 
were calculated by \cite{wu_baltzer_adrian_2012} for pipe flows as a function of azimuthal angle. 
Therefore, we }have validated our calculations \black{of spanwise cospectra} by comparing with the corresponding spatial two-point velocity-vorticity correlations in the spanwise direction obtained from channel 
flow DNS at $Re_{\tau}=934$ by ~\cite{delalamo2004scaling}, as reported in ~\cite{monty2011characteristics}. This comparison, 
shown in Fig~\ref{cor_comp} in the Supplementary Materials, confirms our own data presented here.  

\begin{figure}
     \centering
     \begin{subfigure}[b]{0.32\textwidth}
         \centering
         \includegraphics[width=\textwidth]{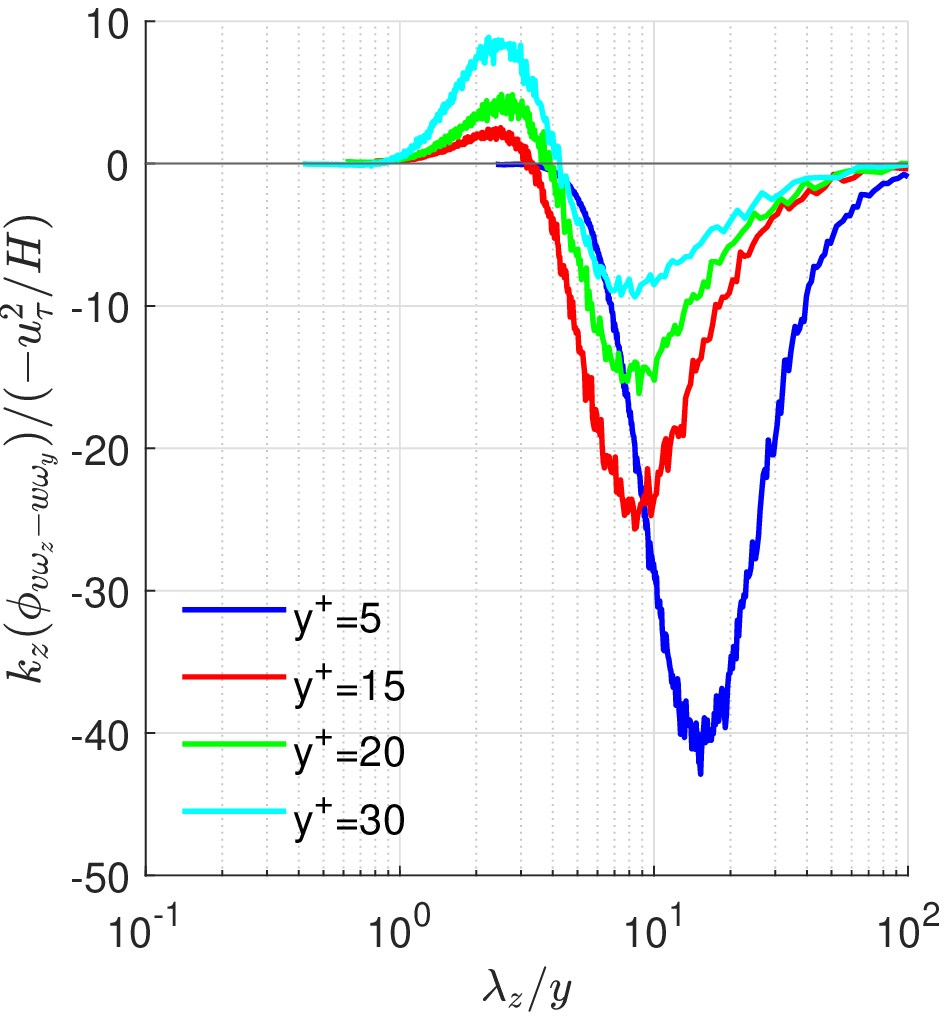}
        \caption{Spanwise co-spectra: $\mbox{Viscous \& Buffer Layers}$}
         \label{fig_bufferz}
     \end{subfigure}
     \hfill
     \begin{subfigure}[b]{0.32\textwidth}
         \centering
         \includegraphics[width=\textwidth]{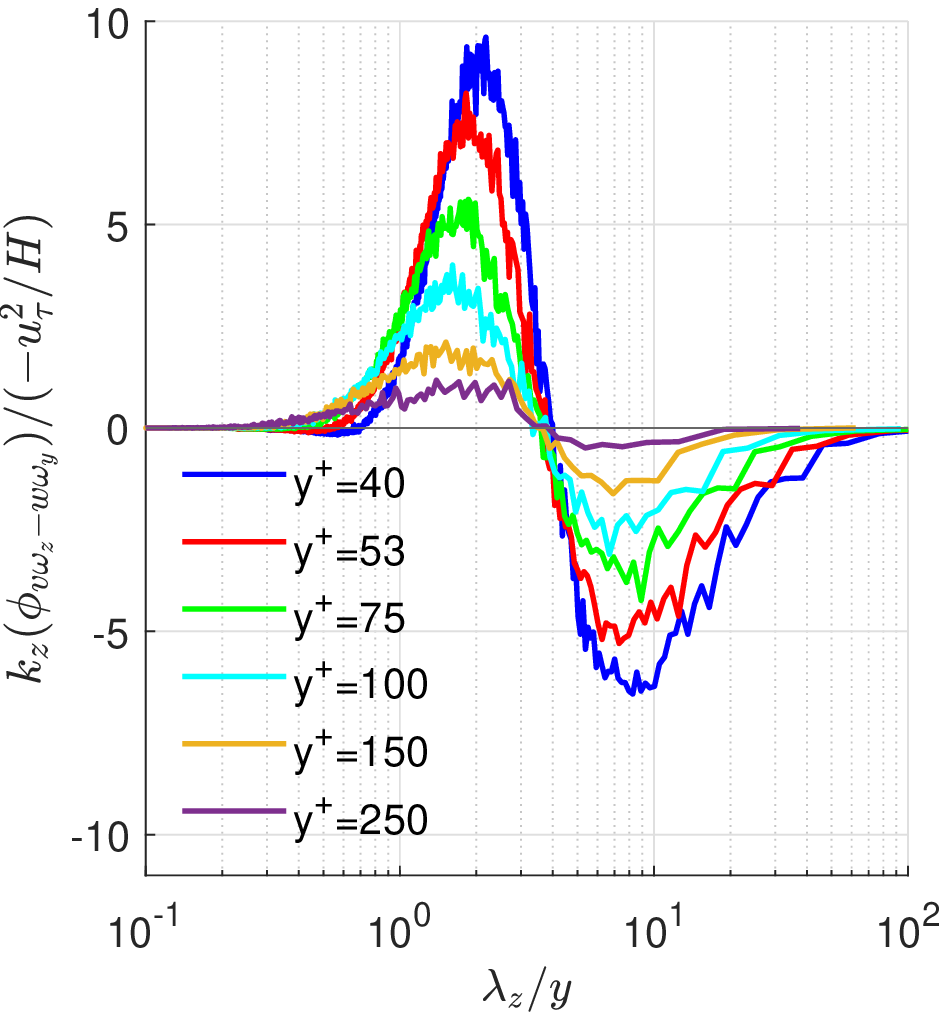}
         \caption{ Spanwise co-spectra: Log Layer}
         \label{fig_inertialz}
     \end{subfigure}
     \hfill
     \begin{subfigure}[b]{0.32\textwidth}
         \centering
         \includegraphics[width=\textwidth]{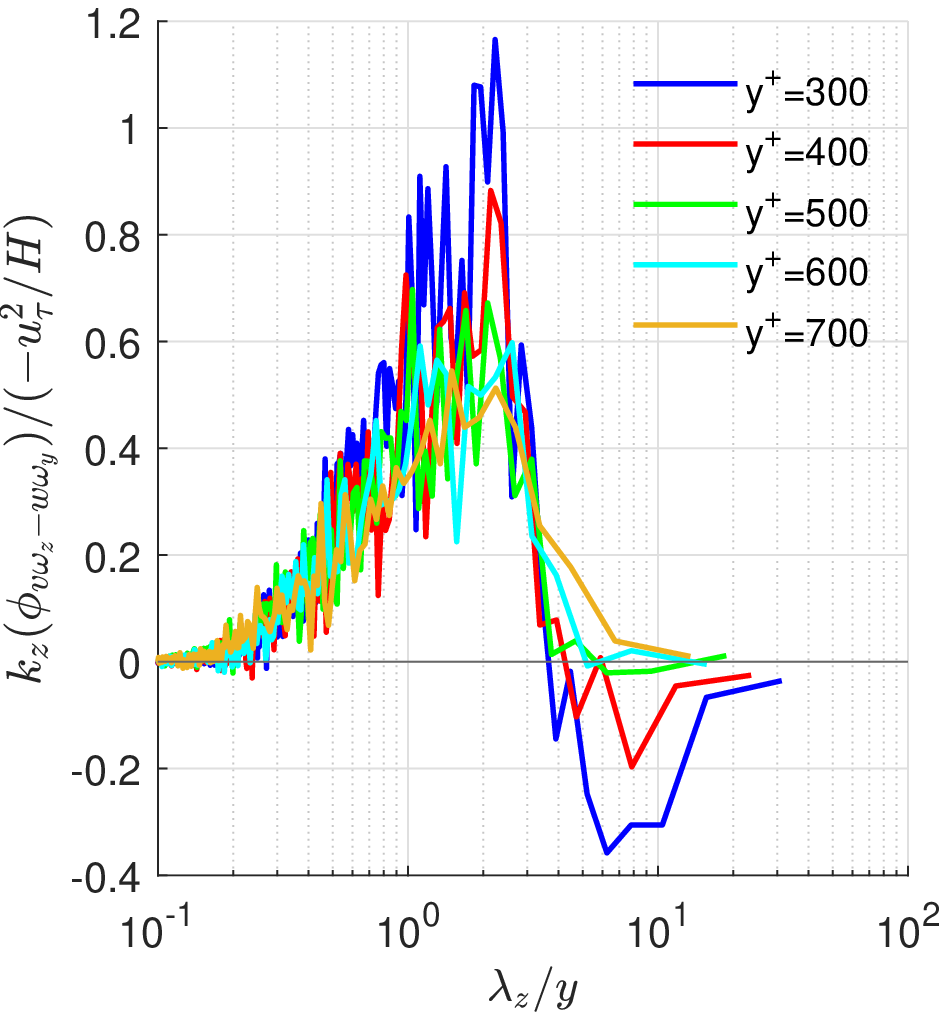}
         \caption{Spanwise co-spectra:Outer Layer}
         \label{fig_outerz}
     \end{subfigure}
             \caption{Normalized spanwise cospectra of the nonlinear term, in the  (a) viscous \& buffer layers, 
             (b) log layer and  (c) outer layer. Here, the abscissa shows wavelength divided by $y$.}
        \label{fig_1dcospec_z}
\end{figure}

\begin{figure}
    \begin{subfigure}[b]{0.49\textwidth}
         \centering
         \includegraphics[width=\textwidth]{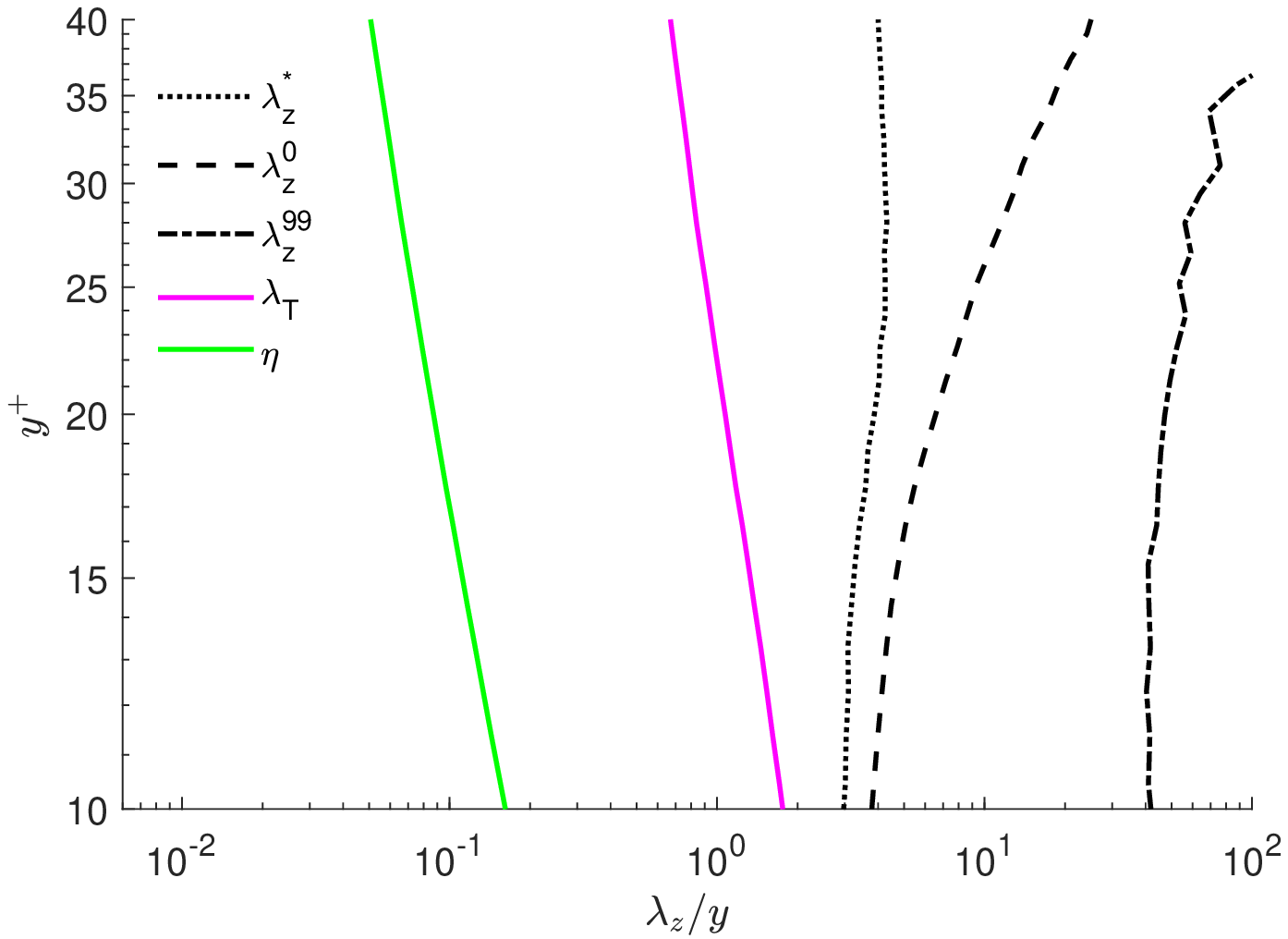}
         \caption{$y<y_p$}
         \label{fig_kzvals_40}
     \end{subfigure}
     \hfill
     \begin{subfigure}[b]{0.49\textwidth}
         \centering
         \includegraphics[width=\textwidth]{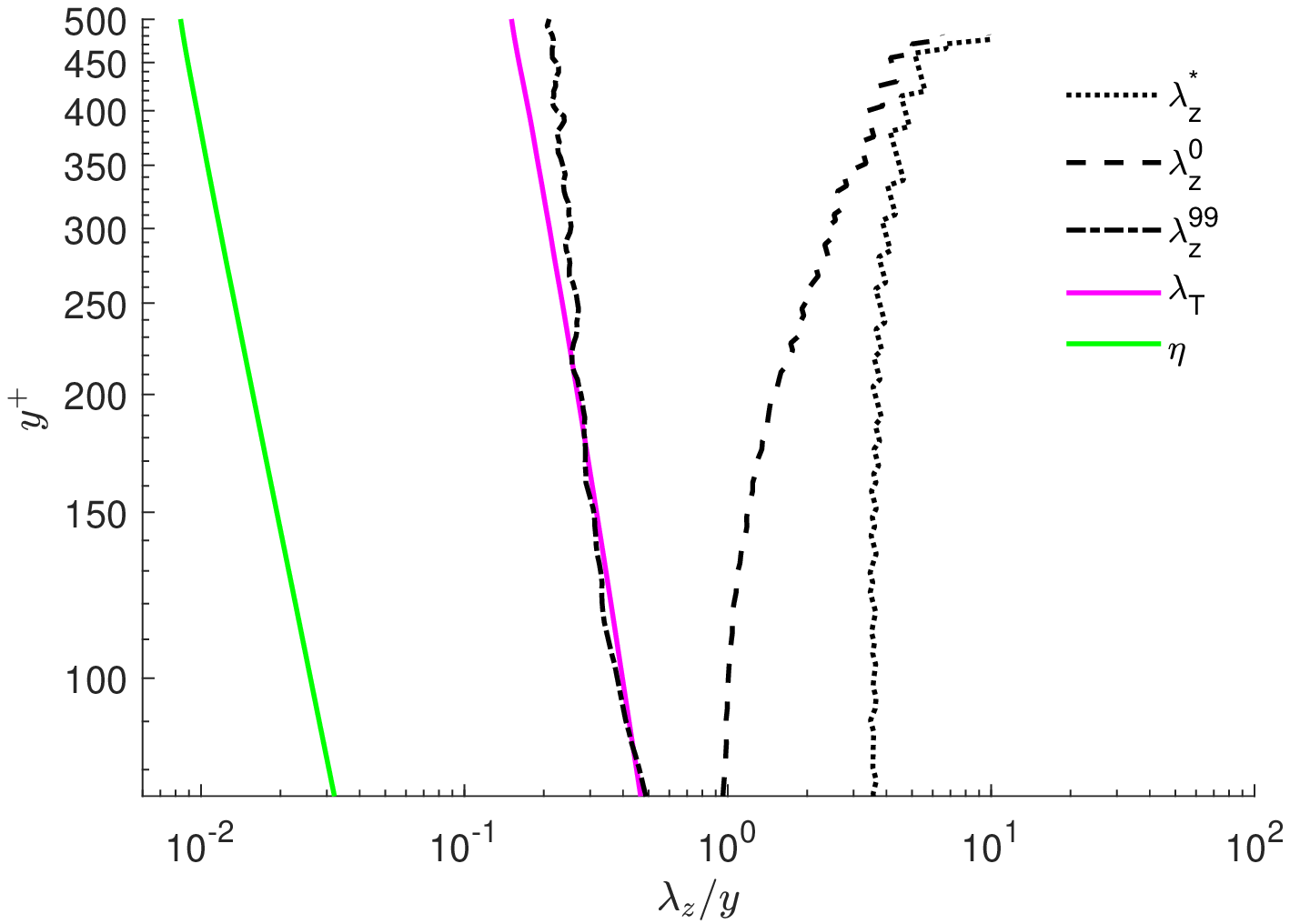}
         \caption{$y>y_p$}
         \label{fig_kzvals_75}
     \end{subfigure}
             \caption{\black{Characteristic spanwise wavelengths relevant to nonlinear vorticity transport, defined based on Eq.~\ref{scales-z}, for (a) $y<y_p$ and (b) $y>y_p$.} }
        \label{fig_kzvals}
\end{figure}

We plot in Fig.~\ref{fig_1dcospec_z} the spanwise flux cospectra for several $y$ values each in the buffer layer, 
log-layer and outer layer. To make the results more physically intuitive, we have plotted the cospectra versus 
$\lambda_z/y$ in log-scale and then compensated by the factor $k_z=2\pi/\lambda_z$ necessary to yield the correct total 
integrals. We have further normalized by the asymptotic value $-u_\tau^2/H$ achieved by the mean nonlinear flux 
$\langle v\omega_z-w\omega_y \rangle(y)$ for $y\gg y_p.$ With the latter normalization, positive values of the 
cospectra count as ``down-gradient'' and negative as ``up-gradient.'' The most striking feature of the 
results plotted in Fig.~\ref{fig_1dcospec_z} is the existence of a precise spanwise length scale 
$\lambda_z^*$ such that the cospectrum is ``down-gradient'' for $\lambda_z<\lambda_z^*$ and ``up-gradient'' 
for $\lambda_z>\lambda_z^*.$ Furthermore, $\lambda_z^*$ is proportional to $y,$ $\lambda_z^*\doteq \gamma y$ 
with a prefactor $\gamma=3\sim 4$ over the entire range of $y$-values; see also the plot of $\lambda_z^*$ vs. $y$
in Figure \ref{fig_kzvals}. Physically, it is the eddies with spanwise wavelengths $\lambda_z$ greater 
than $\gamma y$ that are subjected to the correlated inflow and spanwise stretching proposed by 
\cite{lighthill1963}, whereas the eddies with $\lambda_z$ less than $\gamma y$ instead transport vorticity 
down-gradient away from the wall. 
To compare $\lambda_z^*$ with other relevant length scales in 
wall-bounded flow, note that the integral length given nominally by $\ell=\kappa y$ for von K\'arm\'an constant
$\kappa\doteq 0.4$ is about 10 times smaller. \cite{nickels2005evidence} have estimated that the 
``production range'' of attached eddies with $k_x^{-1}$ energy spectrum occurs for $15.7y<\lambda_x<0.3H.$
If we adopt the relation $\lambda_z\sim \lambda_x/7$ suggested by results for 2D energy spectra 
in $(k_x,k_z)$ \citep{chandran2017two}, then this production range corresponds to $\lambda_z>2.24 y$ and includes
the ``up-gradient'' scales $\lambda_z>(3\sim 4)y$ identified by our results. However, we shall present 
concrete evidence later that the up-gradient vorticity flux is not associated to wall-attached eddies. 

Although the normalized flux cospectra plotted in Figure~\ref{fig_1dcospec_z} all pass through zero at 
$\lambda_z/y\doteq 3\sim 4,$ their integrals over $\log(\lambda_z/y)$ must shift from negative values 
for $y$ in the buffer layer, pass through zero at $y=y_p,$ and then approach 1 for $y\gg y_p,$
consistent with the results for $\langle v\omega_z-w\omega_y\rangle(y)$ plotted in Figure~\ref{fig_conv_visc}.
This change in the integrated values occurs via a shift between the two halves of the cospectrum with increasing $y,$ 
whereby the negative ``up-gradient'' branch at $\lambda_z>\lambda_z^*$ dominates in the buffer layer but 
diminishes with increasing $y$ as the positive ``down-gradient'' branch at $\lambda_z<\lambda_z^*$ increases. 
This increase of the down-gradient branch relative to the up-gradient branch continues in the log-layer, 
with the two coming into exact balance at $y=y_p\doteq 53.$ However, unlike the buffer layer, increasing 
distance from the wall in the log layer sees a decrease in the magnitude of both down-gradient and up-gradient 
branches. For  $y$ further increasing into the outer layer, the up-gradient branch at $\lambda_z>\lambda_z^*$ 
continues to diminish and the down-gradient branch at $\lambda_z<\lambda_z^*$ stabilizes to a positive 
cospectrum independent of $y$. At the extremes, for $y^+\lesssim 10$ there is essentially no down-gradient branch 
and for $y^+\gtrsim 500$ no up-gradient branch. Physically, the eddies for $y^+\gtrsim 500$ (or $y\gtrsim 0.5H)$ 
no longer feel the effect of the wall. These results for the flux cospectrum highlight the delicate balance between competing 
fluxes proposed by ~\cite{lighthill1963} and restated in Sec~\ref{intro}. The remarkable persistence of 
$\lambda_z^*/y=3\sim 4$ lends credence to Lighthill's idea that the up-gradient transport of vorticity 
towards the wall is a scale-by-scale cascade process (see Sec ~\ref{intro}), since $\lambda_z^*$ can be viewed 
as the ``smallest" up-gradient spanwise scale which gets smaller $\propto y$ as vorticity is transported nearer to the wall. 




The intense competition between nonlinear vorticity transport in opposite directions arising from different 
scales of motion, vividly illustrated in Figure~\ref{fig_1dcospec_z}, implies that the net ``down-gradient'' 
transport for $y>y_p$ must arise from scales much smaller than $y$ and likewise the net ``up-gradient transport'' 
for $y<y_p$ must arise from scales much larger than $y$. 
\black{In order to quantify the extent of cancellation between co-spectral regions with opposing fluxes, we define the }fractional 
cumulative flux
\begin{equation}\label{scales-z}
f(y,\Lambda)= 
    \begin{cases}
        \frac{1}{\langle v \omega_z -w\omega_y \rangle(y)} \int_{2\pi/\Lambda}^\infty 
        \phi_{v\omega_z-w\omega_y}(k_z,y) dk_z, & \text{if } y<y_p \\
        \frac{1}{\langle v \omega_z -w\omega_y \rangle(y)} \int_0^{2\pi/\Lambda} 
        \phi_{v\omega_z-w\omega_y}(k_z,y) d k_z, & \text{if } y > y_p
    \end{cases}
\end{equation}
which for $y<y_p$ measures the fraction of nonlinear vorticity flux arising from wavelengths 
$\lambda_z<\Lambda$ and for $y>y_p$ measures the fraction arising from $\lambda_z>\Lambda.$
One important spanwise scale which may be defined in the range $10\lesssim y^+\lesssim 500$
where two opposite-signed branches of the cospectrum co-exist is the ``break-even'' 
wavelength $\lambda_z^0$ satisfying $f(y,\lambda_z^0)=0.$ For $y<y_p$ spanwise scales as large as 
$\lambda_z^0>\lambda_z^*$ must be included to get exactly cancelling flux, and for $y>y_p$ spanwise 
scales as small as $\lambda_z^0<\lambda_z^*$ must be included to get cancelling flux. Perhaps even 
more relevant is the ``99\%'' wavelength $\lambda_z^{99}$ satisfying $f(y,\lambda_z^{99})=0.99.$
According to this definition, spanwise scales as large as $\lambda_z^{99}>\lambda_z^0$ must be included 
to get 99\% of the net up-gradient flux for $y<y_p$, and spanwise scales as small as 
$\lambda_z^{99}<\lambda_z^0$ must be included to get 99\% of the net down-gradient flux for $y>y_p$. \black{Note that at $y=y_p$ the co-spectrum integrates to zero, and hence $\lambda_z^0\rightarrow \infty$ and $\lambda_z^{99}\rightarrow\infty$ as $y\uparrow y_p$, while
$\lambda_z^0\rightarrow 0$ and $\lambda_z^{99}\rightarrow 0$ as $y\downarrow y_p$. Therefore, these length-scales provide useful 
information about the sizes of eddies yielding the net nonlinear flux, but they are exaggerated to absurdity for $y$ 
too close to $y_p.$}
All of the length-scales $\lambda_z^*,$ $\lambda_z^0$ and $\lambda_z^{99}$ are plotted 
together versus $y^+$ \black{for $10<y^+<40$ in Fig.~\ref{fig_kzvals_40} and for $75<y^+<500$ in Fig.~\ref{fig_kzvals_75}, corresponding 
to $y\lesssim y_p$ and $y\gtrsim y_p$ respectively}. For \black{$y\lesssim y_p$ in Fig.~\ref{fig_kzvals_40}} the plotted results show that 
$\lambda_z^{99}$ begins as $40y$ in the buffer layer and increases to more that \black{$100 y$ 
approaching $y^+=40,$} making manifest the very large spanwise scales involved in Lighthill's 
``up-gradient'' mechanism. 
\black{We} have \black{also} added for reference to Fig.~\ref{fig_kzvals} 
two characteristic turbulent small scales, the Kolmogorov scale $\eta$ and the Taylor scale $\lambda_T.$
Here we have followed standard definitions $\eta=\nu^{3/4} \epsilon^{-1/4}$ and 
$\lambda_{T}=[5\nu(u_{rms}^2+v_{rms}^2+w_{rms}^2)/\epsilon]^{1/2},$ estimating energy dissipation as 
$\epsilon = P_K/\alpha$ in terms of turbulence production $P_K=- \langle u' v'\rangle \partial_y U$ 
and the factor $\alpha$ from Figure 7 of ~\cite{lee_moser_2015}.
\black{Note, that the scales associated with up-gradient flux $(\gtrsim \lambda_z^*)$ are at least one order of magnitude larger than the Kolmogorov scale, where streamwise organization associated with stronger spanwise scales can be expected.}
Remarkably, for \black{wall distances $y\gtrsim y_p$ in Fig.~\ref{fig_kzvals_75}}
in the log layer we see that $\lambda_z^{99}\doteq \lambda_T$ to a very good approximation 
and in the outer layer \black{$\lambda_z^{99}$}
is only a factor of a few times larger than $\lambda_T.$ 
Thus, we conclude that fine-scale eddies with spanwise wavelength $\lambda_z$ down to nearly the Taylor 
microscale contribute significantly to the down-gradient transport of vorticity at $y>y_p.$ Similar observations 
were made previously by \cite{priyadarshana2007statistical} based on streamwise 
cospectra $\phi_{v\omega_z}(k_x,y),$ $\phi_{w\omega_y}(k_x,y)$ (see further below). The sensitivity
of the nonlinear vorticity transport to such small scales for $y>y_p$ has important implications 
for physical phenomena such as polymer drag-reduction since modifications of the flux cospectrum 
at very small scales can alter the delicate balance between down-gradient and up-gradient transport
and lead to a drastic reduction of drag (cf. \cite{crawford1997reynolds,monty2011characteristics}).

We have calculated as well the separate spanwise cospectra for the advective flux $\phi_{v\omega_z}(k_z,y),$
and the stretching flux $-\phi_{w\omega_y}(k_z,y).$ These contain little new information beyond the cospectrum 
of the total nonlinear flux, so that we just briefly summarize here the key observations and relegate the plots 
of those cospectra to Fig~\ref{fig_1dcospec_voz} and Fig~\ref{fig_1dcospec_woy} in the Supplementary Materials. Most 
significantly, the stretching cospectrum is predominantly ``up-gradient'' for all $y^+$ values 
and the advective cospectrum predominantly ``down-gradient'' for $y^+>5.$ Intriguingly, $\phi_{v\omega_z}(k_z,y),$
switches sign for $y^+\lesssim 10,$ where it becomes almost entirely ``up-gradient". These signs are 
all consistent with those of the mean values $\langle v\omega_z\rangle(y)$ and $-\langle w\omega_y\rangle(y)$ 
plotted in Fig.~\ref{fig_conv_visc} and the underlying physical mechanisms are presumably the same as discussed
in that connection. A relevant conclusion is that in Fig.~\ref{fig_1dcospec_z} the ``down-gradient" branch 
in the nonlinear flux cospectrum $\phi_{v\omega_z-w\omega_y}(k_z,y)$ arises mainly from advection, whereas the 
``up-gradient'' branch arises mainly from stretching.

\begin{figure}
     \centering
      \begin{subfigure}[b]{0.32\textwidth}
         \centering
         \includegraphics[width=\textwidth]{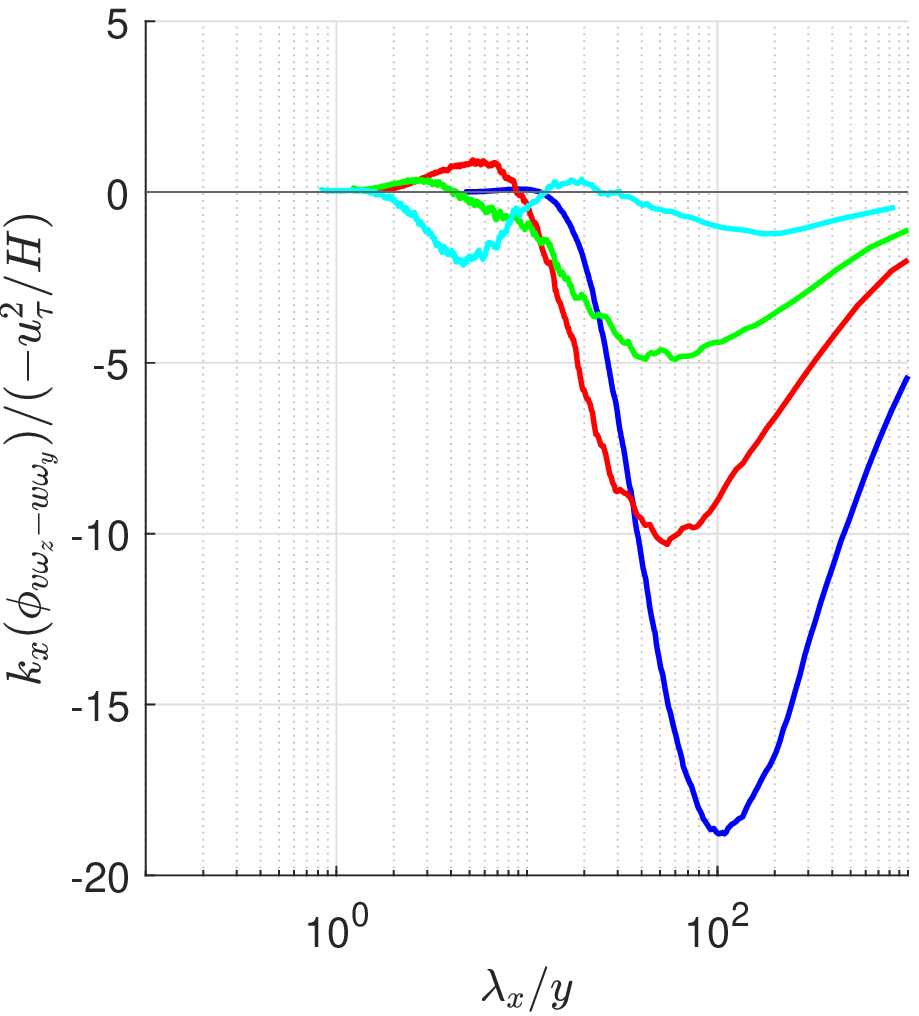}
        \caption{Streamwise co-spectra: $\mbox{Viscous \& Buffer Layers}$}
         \label{fig_bufferx}
     \end{subfigure}
     \hfill
     \begin{subfigure}[b]{0.32\textwidth}
         \centering
         \includegraphics[width=\textwidth]{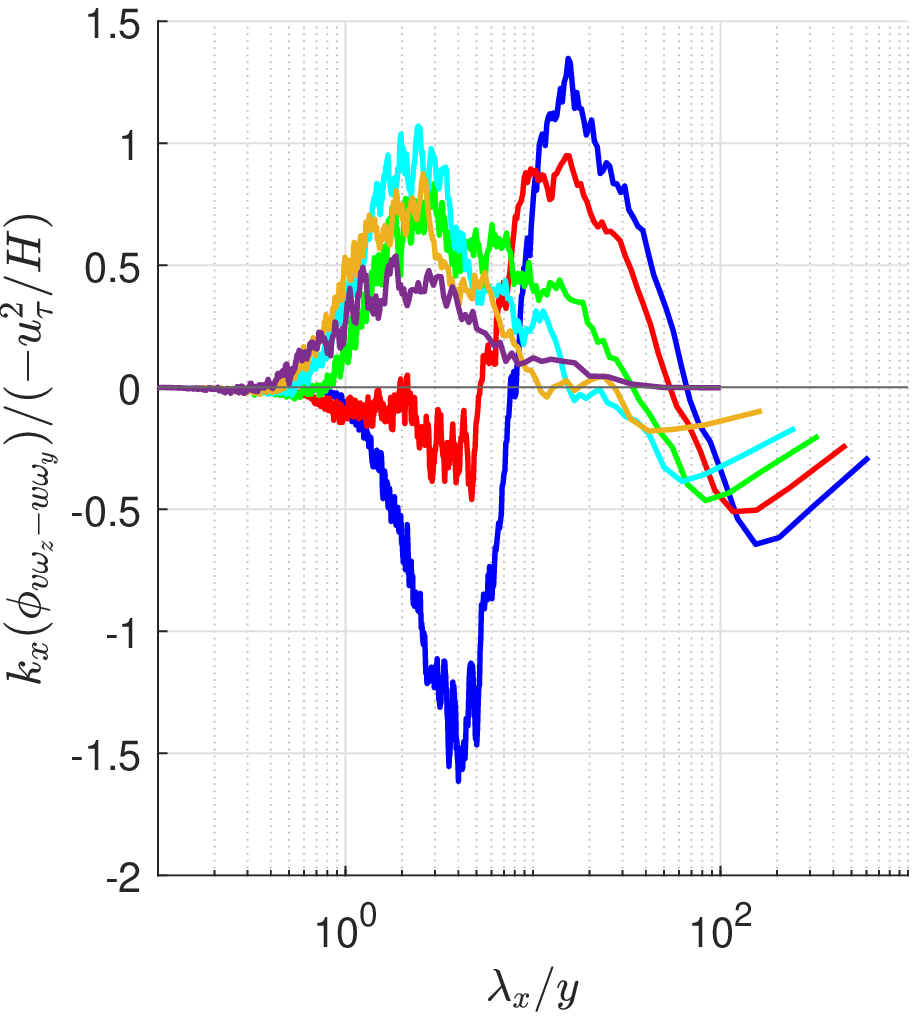}
         \caption{Streamwise co-spectra: Log Layer}
         \label{fig_inertialx}
     \end{subfigure}
     \hfill
     \begin{subfigure}[b]{0.32\textwidth}
         \centering
         \includegraphics[width=\textwidth]{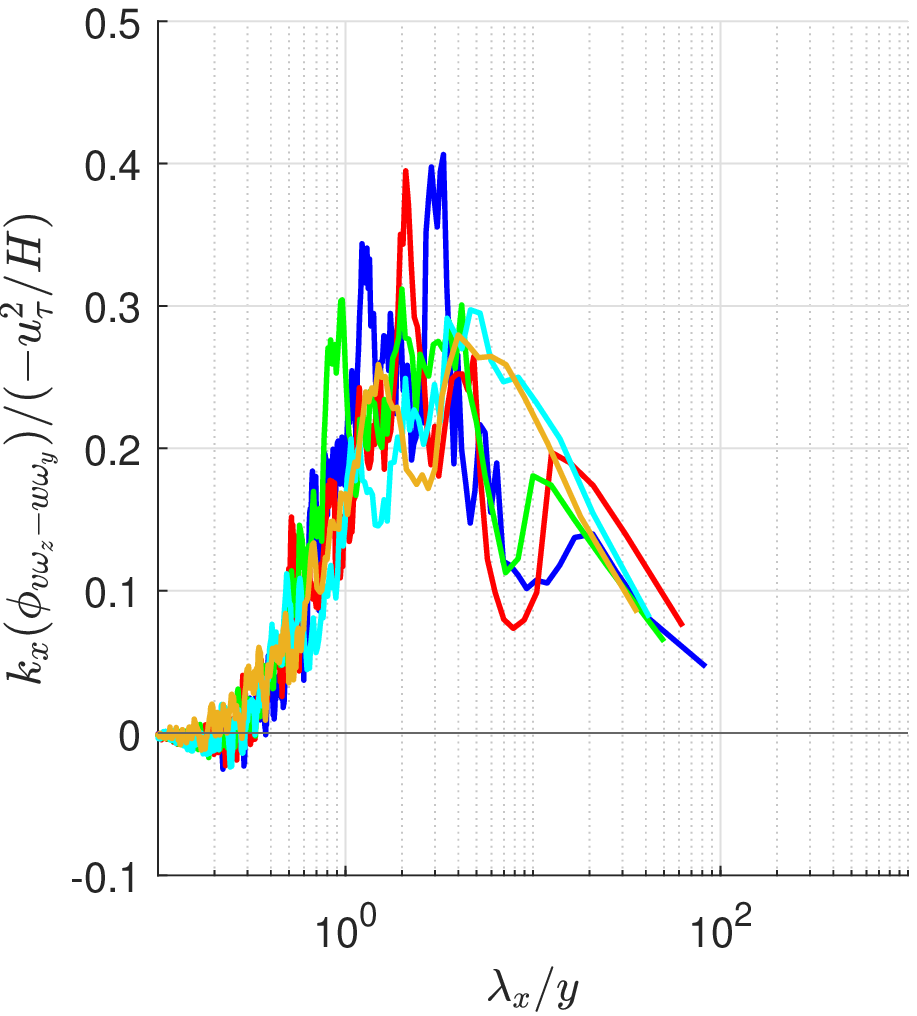}
         \caption{Streamwise co-spectra: Outer Layer}
         \label{fig_outerx}
     \end{subfigure}
     
        \caption{Normalized streamwise co-spectra of the nonlinear term ($\phi_{v\omega_z}-\phi_{w\omega_y}$), in (a) the viscous \& buffer layers , (b) log layer and (c) outer layer. Different curve colors represent different distances from the wall within each of the three layers, as in corresponding plots in Fig~\ref{fig_1dcospec_z}.}
        \label{fig_1dcospec}
        
\end{figure}

We have studied in addition the streamwise cospectra $\phi_{v\omega_z}(k_x,y),$ $-\phi_{w\omega_y}(k_x,y),$ 
and $\phi_{v\omega_z-w\omega_y}(k_x,y),$ but we have found that these present a much less clear physical 
picture and are not as easily interpretable as the spanwise cospectra. We thus present here in 
Fig.~\ref{fig_1dcospec} only the streamwise cospectra for the total nonlinear flux, \black{which we 
compare briefly} with the spanwise cospectra in Fig.~\ref{fig_1dcospec_z}. \black{More detailed 
discussion is given in section \ref{spec2d} of the Supplementary Materials, along with comparison 
to prior results of \cite{guala_hommema_adrian_2006,balakumar_adrian_2007,wu_baltzer_adrian_2012}
for ``total force spectra'' in channel flow, pipe flow and boundary layers. We also} relegate 
to the Supplementary Materials in Figs.~\ref{fig_1dcospecx_voz},~\ref{fig_1dcospecx_woy} \black{our results for} 
the cospectra for advective and stretching fluxes. The latter are shown 
to agree qualitatively with prior experimental results of \cite{priyadarshana2007statistical,morrill2013influences}
at somewhat different Reynolds numbers and for boundary layers. Referring briefly to the results 
plotted in Fig.~\ref{fig_1dcospec}, we remark that the streamwise cospectra at $y^+=5$ and $y^+=15$ shown in Fig ~\ref{fig_bufferx} are qualitatively similar to the corresponding spanwise cospectra.  However, as $y^+$ increases to 30, the streamwise cospectrum develops a region of down-gradient behaviour (at $\lambda_x\sim 10y$) sandwiched between regions of up-gradient behaviour (at $\lambda_x\sim4y$ and $\lambda_x\sim 200y$). Such behaviour persists into the log layer until reaching $y^+=53$ (shown in Fig~\ref{fig_inertialx}) whereupon the behaviour changes again to qualitatively resemble spanwise spectra with down-gradient contributions from $\lambda_x<\lambda_x^*$ and up-gradient contributions from $\lambda_x>\lambda_x^*$ (as seen at  $y^+=75-250$). Here, $\lambda_x^*$ is the streamwise wavelength at which the co-spectrum crosses the x-axis. However, streamwise cospectra do not possess the persistent and sharp boundary between competing fluxes across the log layer seen in the spanwise case and $\lambda_x^*/y$ varies significantly with $y$. At $y^+=250$ and into the outer layer (Fig ~\ref{fig_outerx}),the streamwise cospectra are purely down-gradient. The streamwise and spanwise cospectra are naturally similar near the center line due to the larger degree of component isotropy there.

\begin{figure}
     \centering
      \begin{subfigure}[b]{0.32\textwidth}
         \centering
         \includegraphics[width=\textwidth]{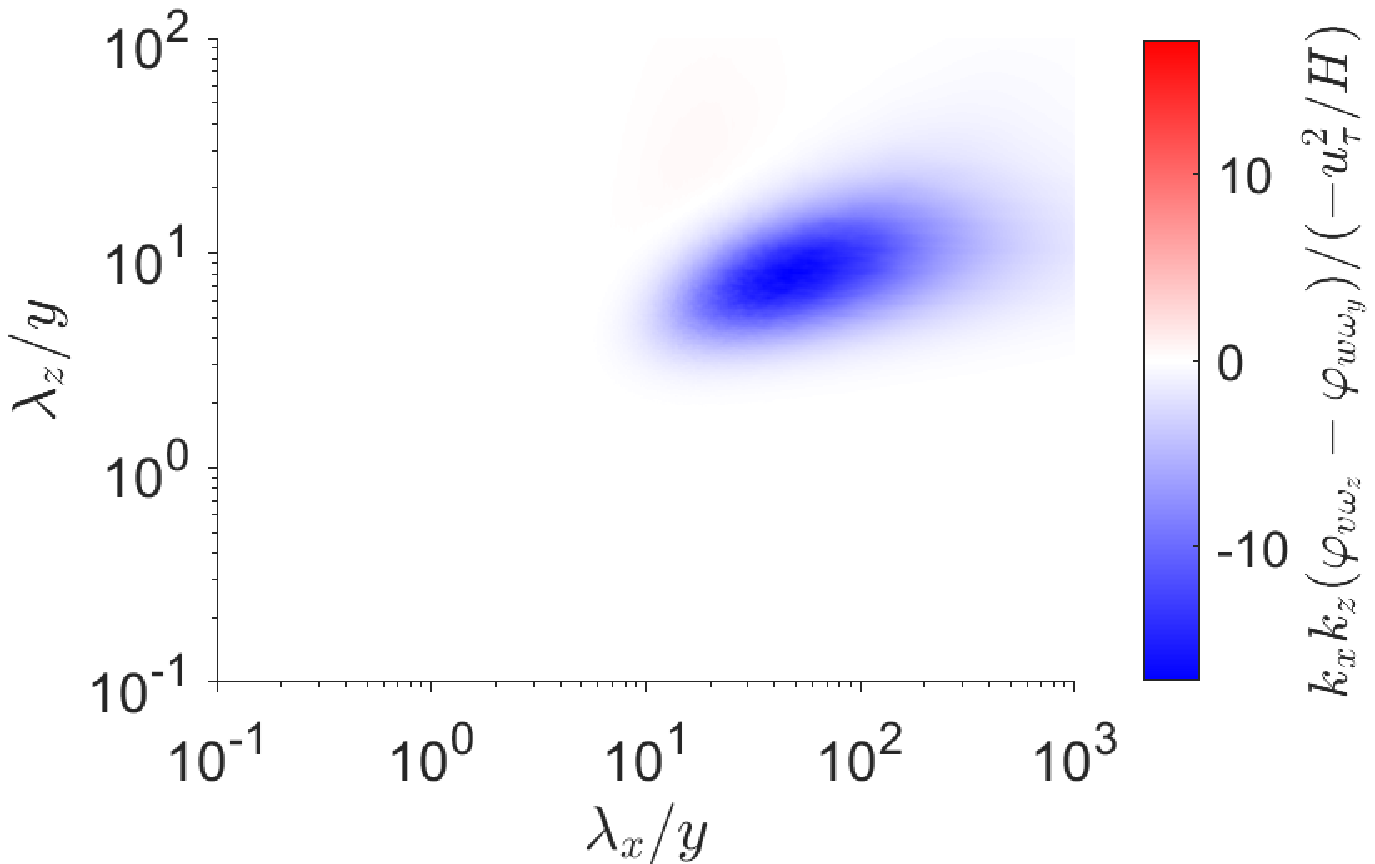}
        \caption{$y^+=5$}
         \label{y5}
     \end{subfigure}
     \hfill
     \begin{subfigure}[b]{0.32\textwidth}
         \centering
         \includegraphics[width=\textwidth]{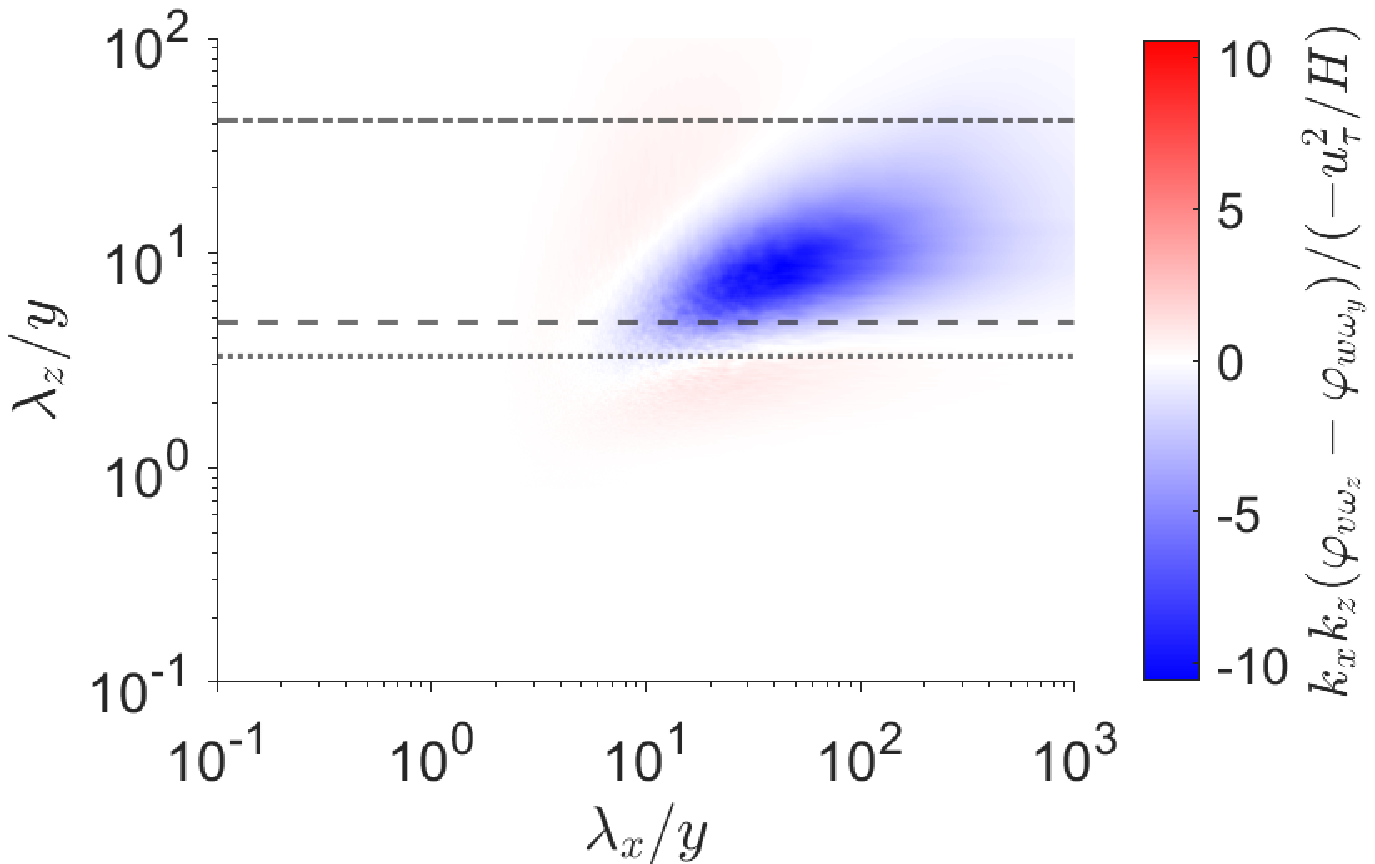}
         \caption{$y^+=15$}
         \label{y15}
     \end{subfigure}
     \hfill	
     \begin{subfigure}[b]{0.32\textwidth}
         \centering
         \includegraphics[width=\textwidth]{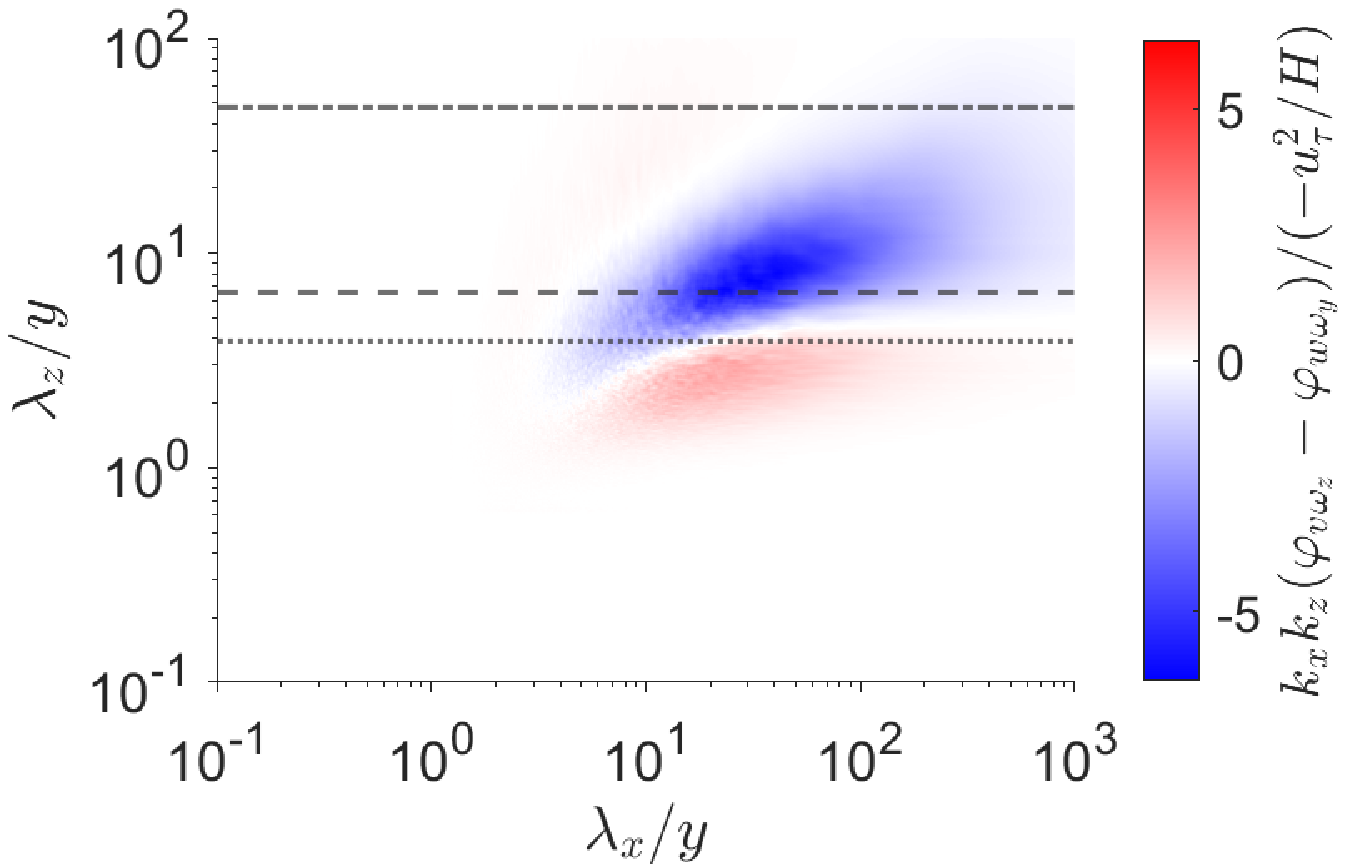}
        \caption{$y^+=20$}
         \label{y20}
     \end{subfigure}
     \vfill
          \begin{subfigure}[b]{0.32\textwidth}
         \centering
         \includegraphics[width=\textwidth]{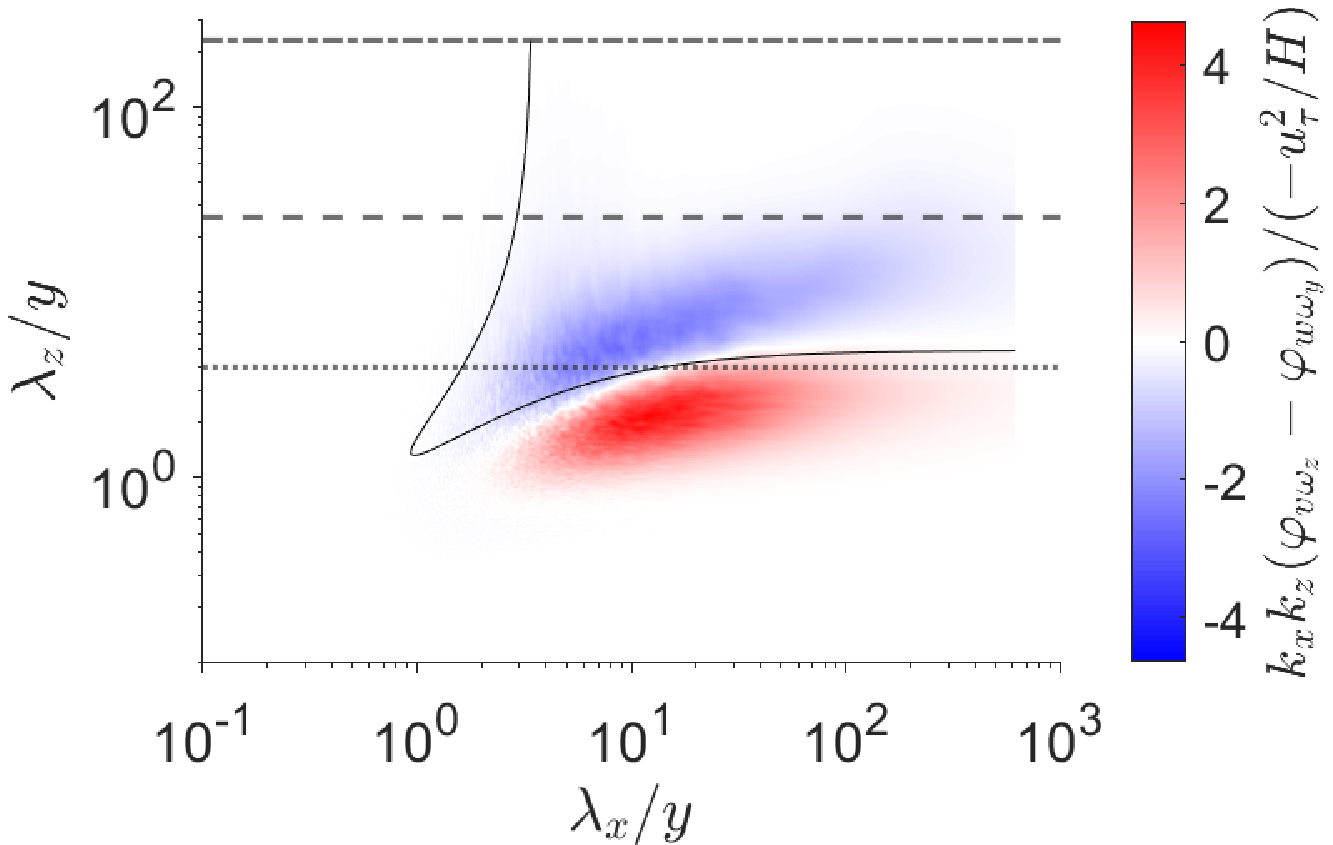}
        \caption{$y^+=40$}
         \label{y40}
     \end{subfigure}
     \hfill
     \begin{subfigure}[b]{0.32\textwidth}
         \centering
         \includegraphics[width=\textwidth]{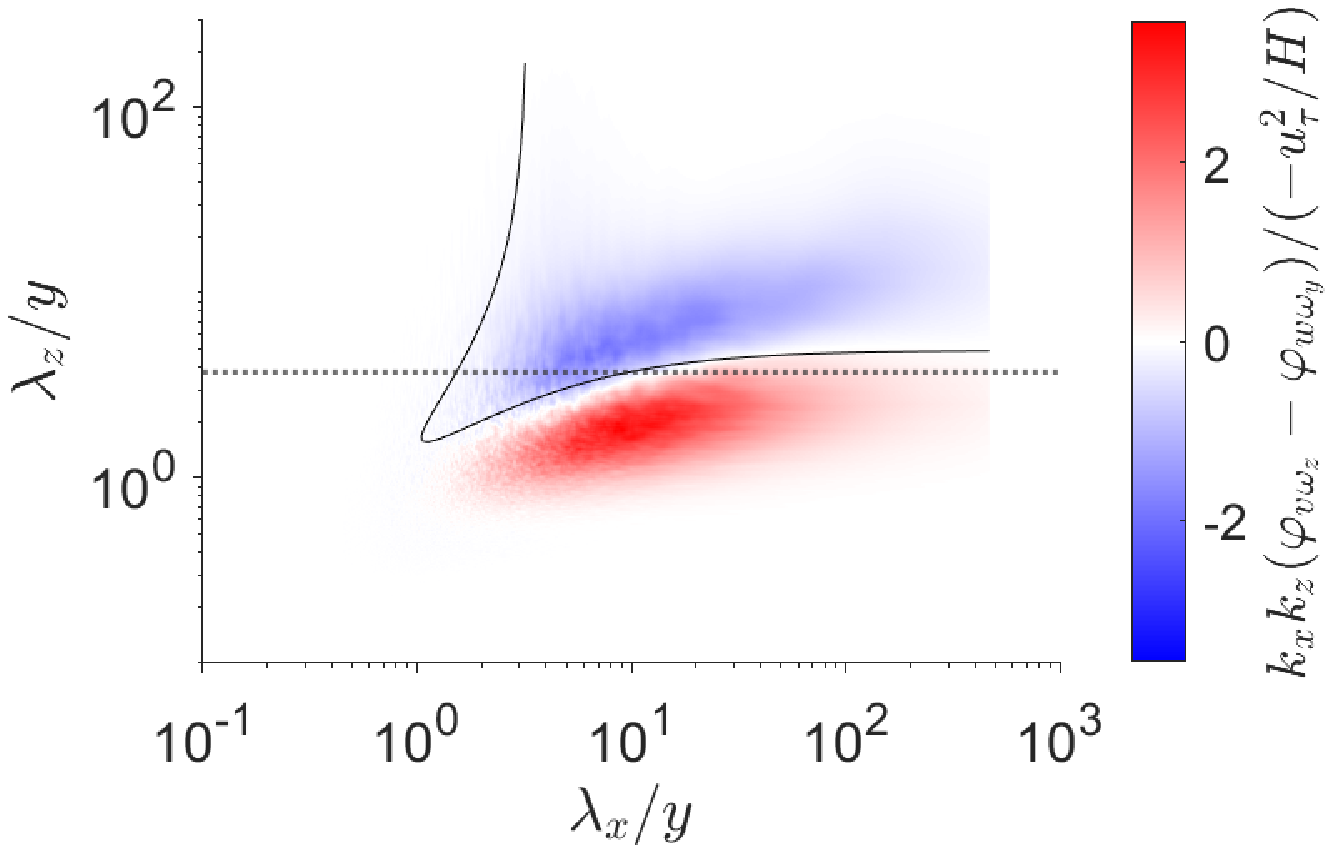}
         \caption{$y^+=53$}
         \label{y53}
     \end{subfigure}
     \hfill
     \begin{subfigure}[b]{0.32\textwidth}
         \centering
         \includegraphics[width=\textwidth]{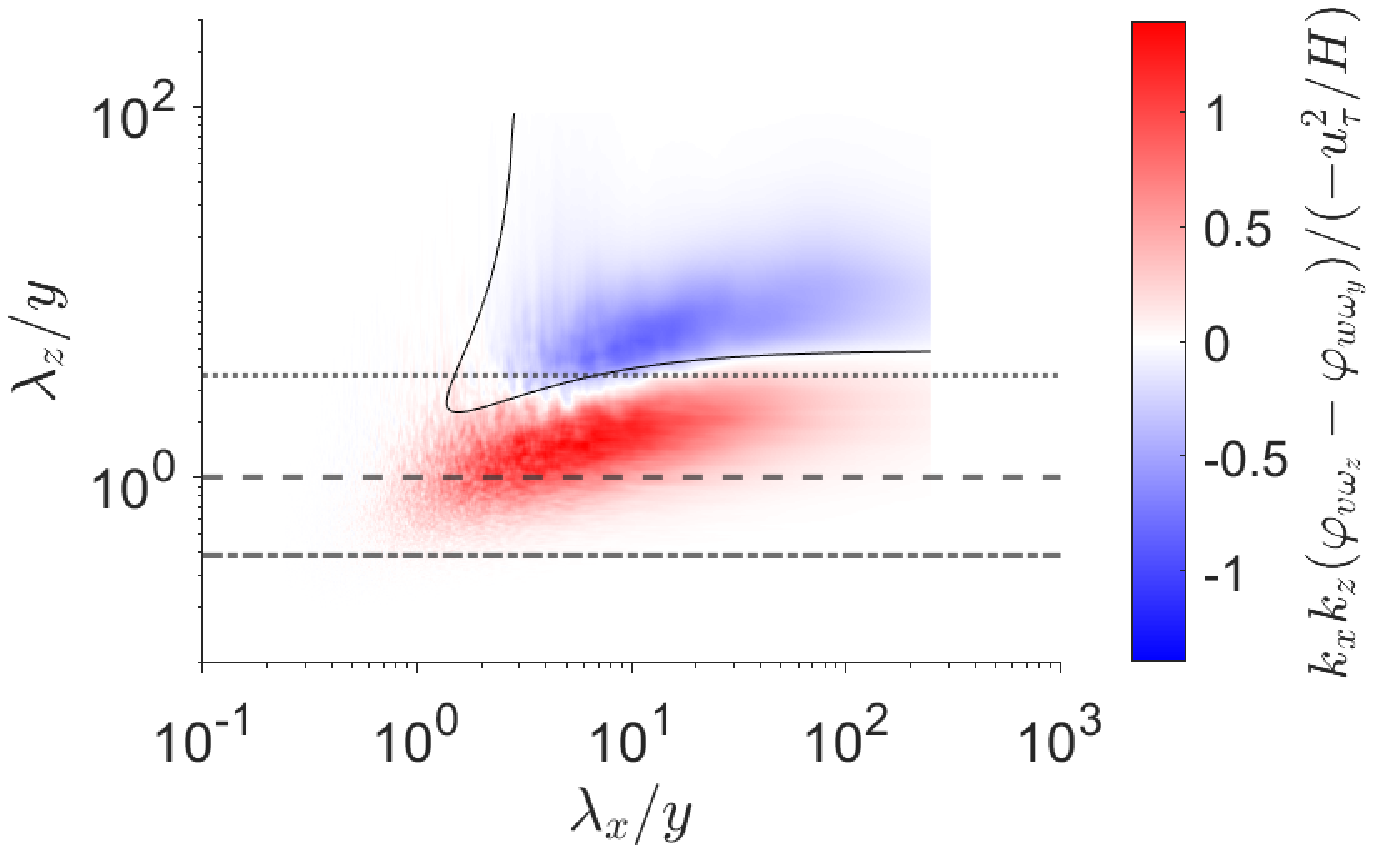}
         \caption{$y^+=100$}
         \label{y100}
     \end{subfigure}
     \vfill
     \begin{subfigure}[b]{0.32\textwidth}
         \centering
         \includegraphics[width=\textwidth]{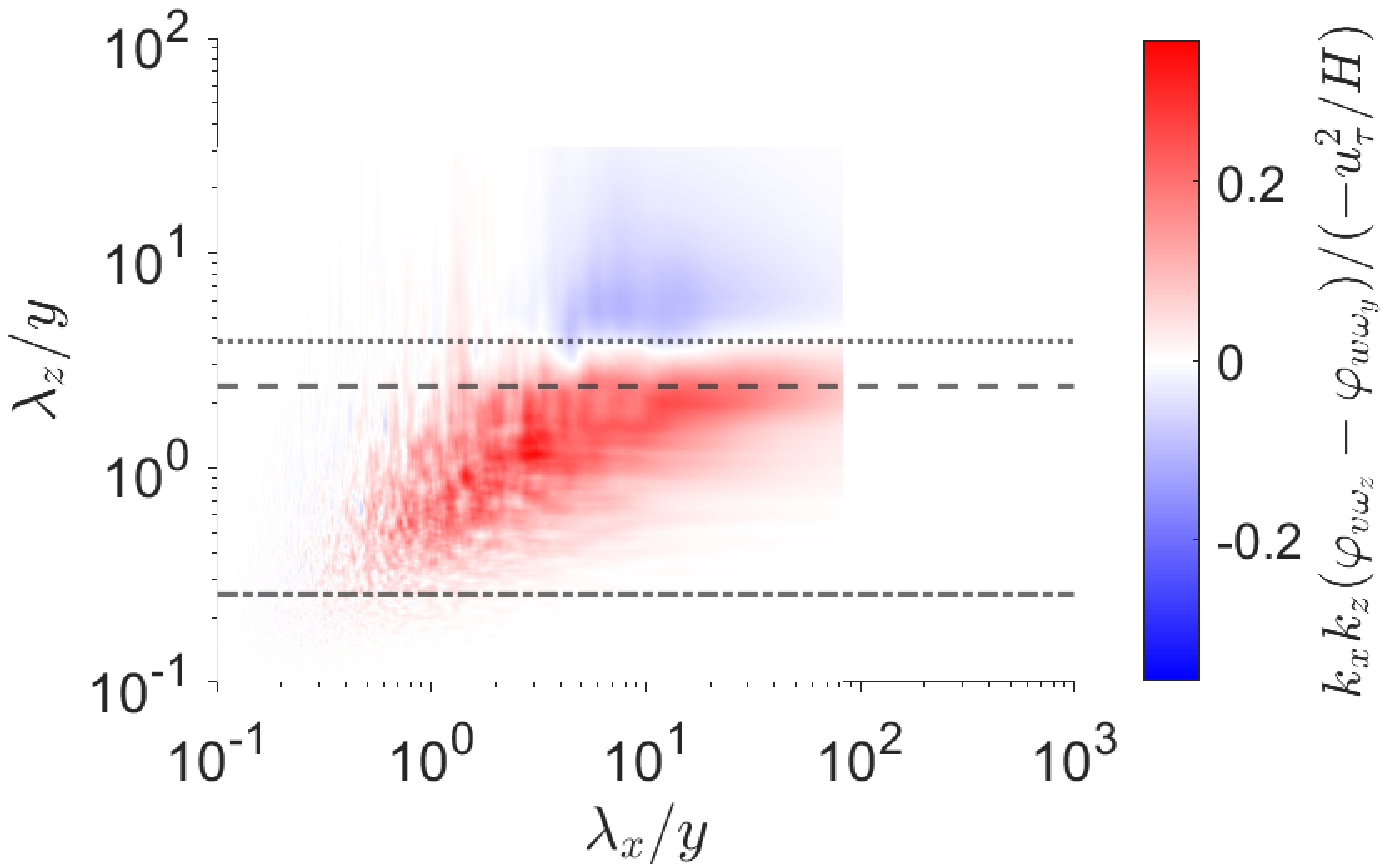}
         \caption{$y^+=300$}
         \label{y300}
     \end{subfigure}
     \hfill
     \begin{subfigure}[b]{0.32\textwidth}
         \centering
         \includegraphics[width=\textwidth]{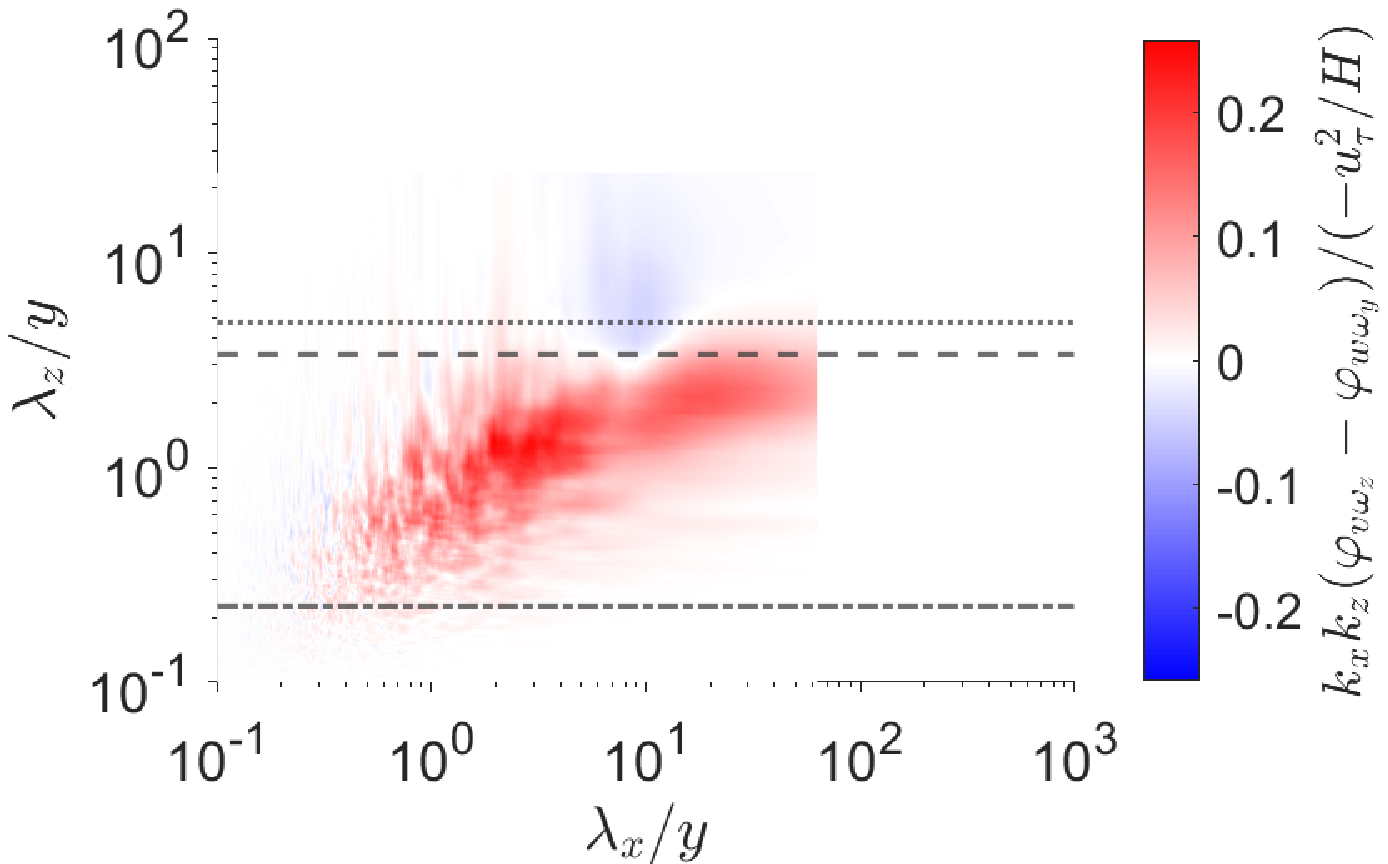}
         \caption{$y^+=400$}
         \label{y400}
     \end{subfigure}
     \hfill
          \begin{subfigure}[b]{0.32\textwidth}
         \centering
         \includegraphics[width=\textwidth]{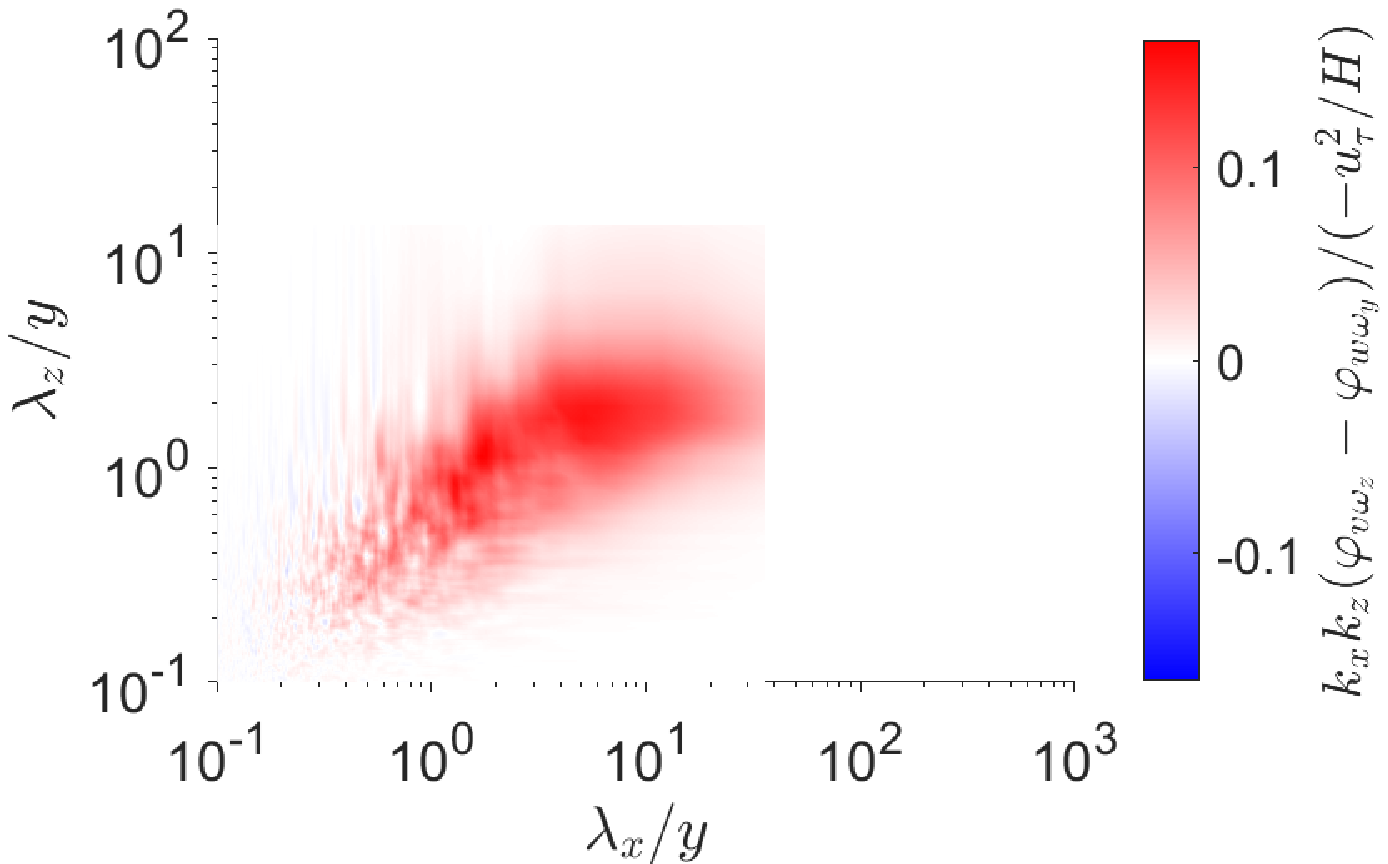}
         \caption{$y^+=700$}
         \label{y700}
     \end{subfigure}
        \caption{Normalized compensated 2D co-spectra of the nonlinear flux $\varphi_{v\omega_z-w\omega_y}(k_x,k_z,y)$, in the viscous \& buffer layers (a,b,c), log layer (d,e,f), and outer layer (g,h).The \black{solid} curves mark the iso-contour of the filter $\mathcal{D}(k_x,k_z,y)=0.5$, described in Supplementary Materials~\ref{dragonfly}. \black{The dotted, dashed and dashed-dot lines represent $\lambda_z^*$, $\lambda_z^0$ and $\lambda_z^{99}$ respectively, as in Fig~\ref{fig_kzvals}.Note that the range of a color-bar does not reflect the actual range of values of the spectrum since only one color achieves saturation while the other does not. For example, spectrum (c) ranges from -6.4 to 2.5 while (g) ranges from -0.11 to 0.36. } }
        \label{fig_2dcospec}
        
\end{figure}

\black{The somewhat complicated}
picture arising from the streamwise cospectra \black{is clarified by studying the} 
2D cospectra in $(k_x,k_z)$, which provide more detailed information about the different 
scales of motion in streamwise and spanwise directions simultaneously.
\black{The paper of \cite{wu_baltzer_adrian_2012} previously studied analogous 2D cospectra for pipe flows. Our} 
cospectra are calculated by computing 2D FFTs of velocity and vorticity, followed by taking appropriate 
inner products and averaging in time. We also added contributions reflected in the $x$- and $z$-axes so that the spectra depend only on wavelength 
magnitudes $k_x\geq 0,$ $k_z\geq 0,$ yielding the following spectral decomposition of the nonlinear vorticity flux: 
\be
\int_0^{\infty}\int_0^{\infty} \varphi_{v\omega_z-w\omega_y}(k_x,k_z,y) \ dk_x\,dk_z = \langle {v\omega_z}-{w\omega_y} \rangle(y)
\label{2D-decomp} 
\ee
Because the 2D cospectra were obtained by averaging over only 38 snapshots, rather sizable fluctuations remained in the results. 
Thus, to obtained more converged results, we smoothed these cospectra using 2D running averages in Fourier space. The smoothing based on the Principle of Minimal Sensitivity \citep{Stevenson1981} is discussed fully in the Supplementary Materials, section \ref{smoothing}, but we note here that the smoothing employed preserves the total integral 
in Eq.\eqref{2D-decomp} and that the cospectra plotted in Fig.~\ref{fig_2dcospec} are relatively insensitive to the exact choice 
of filter width $\Delta k$ in the range considered. \black{Whereas \cite{wu_baltzer_adrian_2012} normalized the wavelengths
in their 2D cospectra by the pipe radius $R$, we have found it more illuminating to plot our}
2D cospectra versus normalized 
wavelengths $\lambda_x/y$ and $\lambda_z/y,$ both in log-scale and compensated by the factor $k_xk_z$ to yield the 
correct double integral over $(\log(\lambda_x/y),\log(\lambda_z/y)).$ We have again normalized by the factor 
$-u_\tau^2/H,$ so that positive values represent down-gradient transport and negative values up-gradient. We plot 2D cospectra 
for three $y$-values each in the viscous sublayer and buffer layer (Fig.~\ref{y5}-\ref{y20}), the log layer (Fig.~\ref{y40}-\ref{y100}), 
and outer layer (Fig.~\ref{y300}-\ref{y700})). The most important feature of the 2D spectra is that, like the 1D spanwise spectrum but unlike 
the streamwise spectrum, there is a clear bipartite structure, with two distinct branches  or ``lobes", with clear spectral 
boundaries, corresponding to the competing down-gradient and up-gradient transport. In fact, these boundaries are mainly 
along the line $\lambda_z/y\doteq 3\sim 4$ but with also another boundary depending upon $\lambda_x$ for $\lambda_z/y\gtrsim3\sim 4.$
These results illuminate why the 1D streamwise cospectra do not yield a clear bipartite structure after integration over $k_z,$
while integration over $k_x$ preserves such structure. A main new implication of the 2D cospectra is that up-gradient
transport requires not only $\lambda_z/y\gtrsim 3\sim 4$ but also $\lambda_x/y\gtrsim 3\sim 4,$ or even larger.
\black{Most importantly, the boundaries clearly seem to scale with wall-distance $y.$}  

\black{In their prior work on pipe flow, ~\cite{wu_baltzer_adrian_2012} observed a similar bipartite structure in their plots of 2D net force spectra, shown in their Figure 18. They obtained spectra similar to ours at each considered $y$ value, with an ellipse of negative net force (decelerating or with a down-gradient flux contribution) at shorter azimuthal wavelengths and an ellipse of positive net force (accelerating or with an up-gradient flux contribution) at longer azimuthal wavelengths. 
We can thus determine from their data the azimuthal wavenumber $\lambda_{\theta}^*$ at which their net force spectrum changes sign and 
the corresponding wavelength defined based on the azimuthal arclength, $\lambda_{s}^*=r\lambda_{\theta}^*=(R-y)\lambda_{\theta}^*$. The resulting values of $\lambda_{s}^*/y$ are compared in Table~\ref{tab:wu} with our previously determined $\lambda_z^*/y$ . It is interesting to note that both quantities have only small variations across the different $y$ values considered, especially within the log layer ($y^+=50,101,200$). The values of $\lambda_z^*/y$ are at most 22\% smaller than $\lambda_z^*/y$. This is a high degree of agreement, considering that they are for different flow configurations and two different Reynolds numbers. This reanalysis of the results of ~\cite{wu_baltzer_adrian_2012} confirms our finding that the spectral 
boundary between up-gradient and down-gradient vorticity transport scales with wall-distance $y,$
as expected from the \cite{lighthill1963} theory.} 
\begin{table}
  \begin{center}
\def~{\hphantom{0}}
  \begin{tabular}{lccc}
      $y^+$& $\lambda_{\theta}^*$  & $\lambda_s^*/y$   &   $\lambda_z^*/y$ \\[3pt]
       20  &0.109 & 3.624 & 3.866\\
       50  & 0.234 & 2.972 & 3.852\\
       101 &0.526 & 3.040 & 3.596\\
       200 & 1.156 & 2.803 & 3.621\\
  \end{tabular}
  \caption{\black{Wavenumber $\lambda_{\theta}^*$ (based on azimuthal angle) and wavelength $\lambda_s^*$ (based on azimuthal arclength) 
  at which the net force spectra in Figure 18 of ~\cite{wu_baltzer_adrian_2012} change sign, compared with the 
  $\lambda_z^*$ from our Figure~\ref{fig_kzvals}.}  }
  \label{tab:wu}
  \end{center}
\end{table}

We now comment 
briefly on significant features and consequences of our 2D cospectra in Fig.~\ref{fig_2dcospec} for the various wall distances.

In the viscous sublayer and buffer layer, the dominant up-gradient contribution to the nonlinear flux corresponds 
to the blue region in Fig.~\ref{y5}-\ref{y20}, which is characterized roughly by $\lambda_z/y\gtrsim 3$ 
and $\lambda_x\gtrsim \lambda_z.$ The competing down-gradient contribution indicated by red color is weaker and in the viscous sublayer at $y^+=5$ 
it is almost entirely negligible. The peak negative value of the cospectra associated to up-gradient transport 
occurs around wavelengths $(\lambda_x,\lambda_z)\sim (30y,8y),$ whose ratio is indicative of sublayer streaks. These are 
the type of flow structures ``elongated in the stream direction'' mentioned by \cite{lighthill1963} and whose relevance 
to near-wall vorticity transport has been discussed in several previous studies 
(\cite{klewicki1994vortical,brown2015vorticity,Arosemena_et_al_2021}).

In the log-layer, the contributions to down-gradient and up-gradient transport from the 2D cospectra plotted in Fig.~\ref{y40}-\ref{y100} 
are more nearly in balance, with an exchange of dominance at $y=y_p.$ The blue portion associated with up-gradient transport 
occupies very crudely the region specified by the two constraints $\lambda_z/y\gtrsim 4$ and $\lambda_x/y\gtrsim 4,$
requiring large wavelengths in both spanwise and streamwise directions. 
(The black curves in Fig.~\ref{y40}-\ref{y100}
plot more precise boundaries of this region.
The aim to separate the flow into contributions on either side of these lines motivates us to define a filter, termed the ``dragonfly filter'' due to its shape once the black line is mirrored to all four quadrants. The dragonfly filter is discussed in more detail in \S \ref{sec:coherent}  and  
in the Supplementary Materials \ref{dragonfly}.)
The red portion of the cospectrum associated to down-gradient 
transport obtains most of its contribution, on the other hand, 
from the region with $\lambda_z/y\lesssim 4$ but with $\lambda_x/y$ ranging over values both much smaller and larger than 
unity. As $y$ increases, this down-gradient spectral region extends to smaller $\lambda_x/y$ and $\lambda_z/y,$ and 
simultaneously the position of peak positive cospectrum shifts to smaller wavelengths. The negative, up-gradient portion 
of the cospectrum instead peaks around $(\lambda_x/y,\lambda_z/y)\sim (10,10)$ for all $y$. These observations 
imply not only that the down-gradient transport becomes more dominant with increasing $y$ but also that it originates 
from an increasing range of spanwise and streamwise scales, down to the Taylor microscale
\citep{priyadarshana2007statistical}. Meanwhile, the up-gradient contributions arise only from spanwise and 
streamwise scales larger than the integral scale and with dimishing magnitudes at greater wall distances. 

These trends with increasing $y$ continue for the 2D cospectra in the outer layer plotted in Fig.~\ref{y300}-\ref{y700}.
The up-gradient spectral region continues to be specified roughly by $\lambda_z/y\gtrsim 4$ and $\lambda_x/y\gtrsim 4,$
but the cospectrum magnitudes in this sector drop rapidly with $y.$ For $y^+\gtrsim 500$ the up-gradient contribution 
is essentially negligible and only the down-gradient contribution from the small scales remains. The cospectra for the 
latter have furthermore shifted to even smaller values of $\lambda_z/y$ and $\lambda_x/y.$ 

\black{Additionally, we have marked on the co-spectra in Fig.~\ref{fig_2dcospec} the lengths $\lambda_z^*,\lambda_z^0$ and $\lambda_z^{99}$ previously plotted in Fig.~\ref{fig_kzvals}. For $y<y_p$ (Fig.\ref{y15}-\ref{y40}), the part of the co-spectrum below the dashed line ($\lambda_z^0$) integrates to zero, and 99\% of the up-gradient flux comes from the region between the dashed and the dot-dashed line ($\lambda_z^{99}$). Similarly, for $y>y_p$ (Fig.\ref{y100}-\ref{y400}), the part of the co-spectrum above the dashed line ($\lambda_z^0$) integrates to zero, and 99\% of the down-gradient flux comes from the region between the dashed and the dot-dashed line ($\lambda_z^{99}$). 
Strikingly, these regions between dashed and dot-dashed lines which contribute most of the net flux may not correspond to the 
regions with the largest magnitude of the cospectrum, at least for $y$ near $y_p.$ In fact, the contributions of the large positive 
and large negative values of the cospectra nearly cancel each other in much of the log-layer and the net contribution arises from much lower-magnitude regions of the cospectrum. This observation highlights the delicate balance between up-gradient and down-gradient transport 
and the sensitivity of the net flux to contributions from very large scales for $y<y_p$ and from very small scales for $y>y_p.$}

We have calculated also the separate 2D cospectra for the advective flux $\varphi_{v\omega_z}(k_x,k_z,y)$ and 
stretching flux $-\varphi_{w\omega_y}(k_x,k_z,y).$ These are plotted in Fig.~\ref{fig_2dvoz} and Fig.~\ref{fig_2dwoy},
respectively, in the Supplementary Materials and here we just summarize their main features. These cospectra have 
the same bipartite structure as the cospectra for the total nonlinear flux plotted in Fig.~\ref{fig_2dcospec}.
The most important difference is that the down-gradient contribution is greatly reduced for the 
stretching cospectrum in Fig.~\ref{fig_2dwoy} and likewise the up-gradient contribution is greatly reduced 
for the advective cospectrum in Fig.~\ref{fig_2dvoz}. Thus, the stretching cospectrum contributes primarily 
up-gradient transport and the advective cospectrum primarily down-gradient transport. The only exceptions to 
the latter statements are for $y^+\lesssim 10$ where the advective cospectrum plotted in Fig.~\ref{y5voz} 
is almost entirely up-gradient and for $y^+\gtrsim 500$ where the stretching cospectrum plotted in 
Fig.~\ref{y700woy} is almost entirely down-gradient. The other general change in the separate
2D cospectra is that the boundaries between up-gradient and down-gradient transport are slightly shifted,
upward to $\lambda_z/y\simeq 6\sim 8$ for the advective cospectra in Fig.~\ref{fig_2dvoz} and downward 
to $\lambda_z/y\simeq 1\sim 2$ for the stretching cospectra in Fig.~\ref{fig_2dwoy}.  The relevant conclusion 
for the competing contributions to the nonlinear flux cospectra plotted in Fig.~\ref{fig_2dcospec} is that 
the down-gradient contribution arises mainly from advection and the up-gradient contribution mainly 
from stretching/tilting. 

\begin{figure}
     \centering
     \begin{subfigure}[b]{0.329\textwidth}
         \centering
         \includegraphics[width=\textwidth]{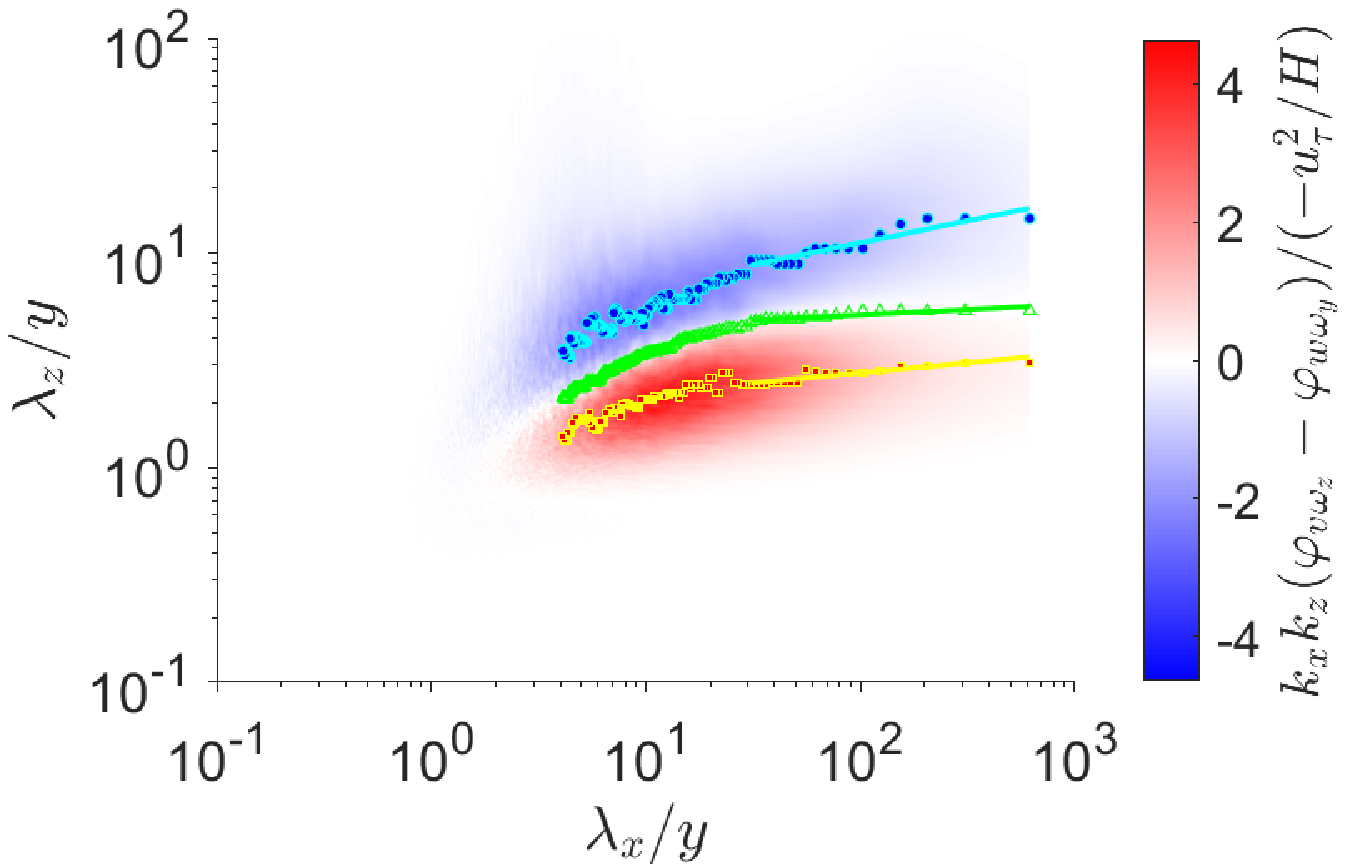}
        \caption{$y^+=40$}
         \label{log_lessyp}
     \end{subfigure}
     \hfill
     \begin{subfigure}[b]{0.329\textwidth}
         \centering
         \includegraphics[width=\textwidth]{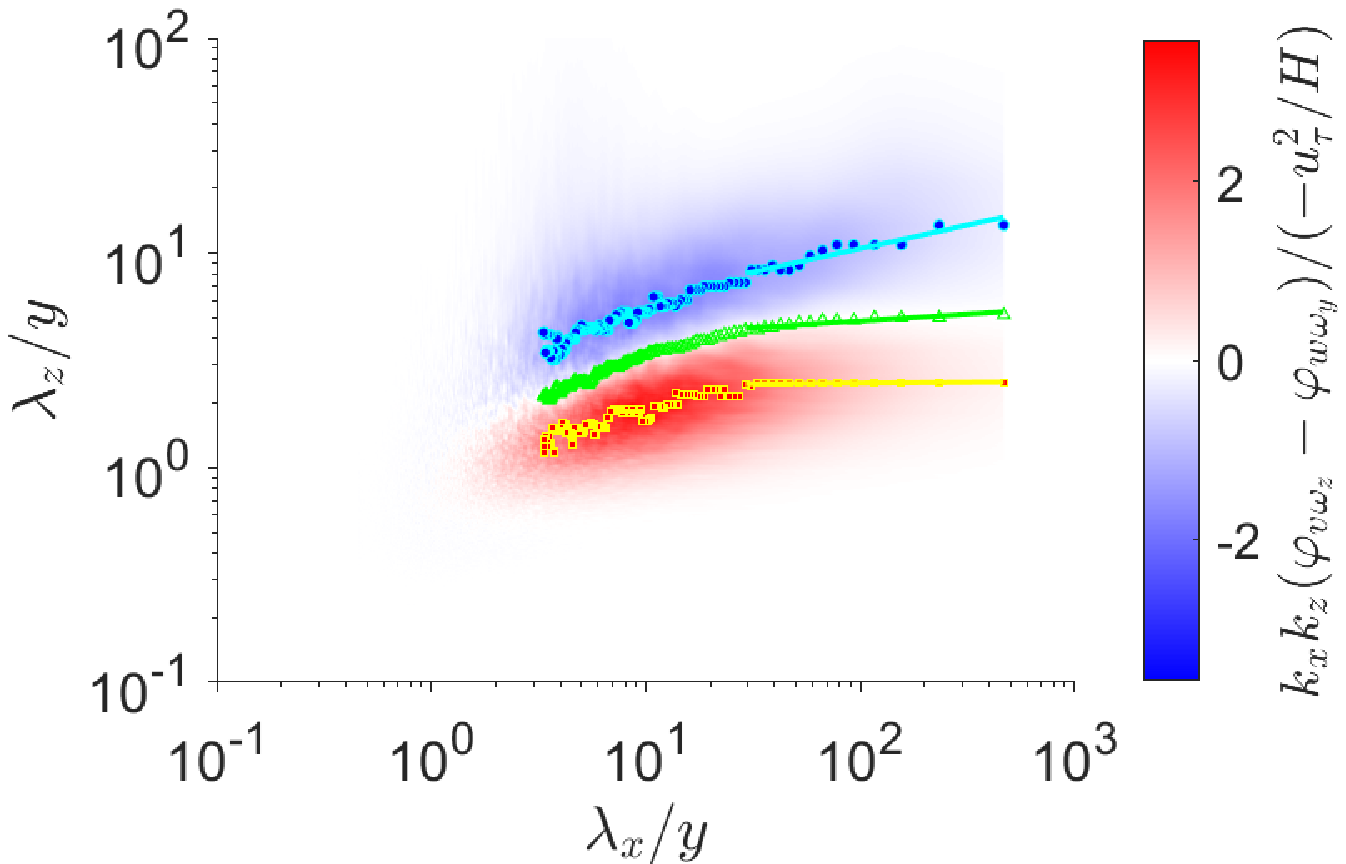}
         \caption{ $y^+=53$}
         \label{log_equalyp}
     \end{subfigure}
     \hfill
     \begin{subfigure}[b]{0.329\textwidth}
         \centering
         \includegraphics[width=\textwidth]{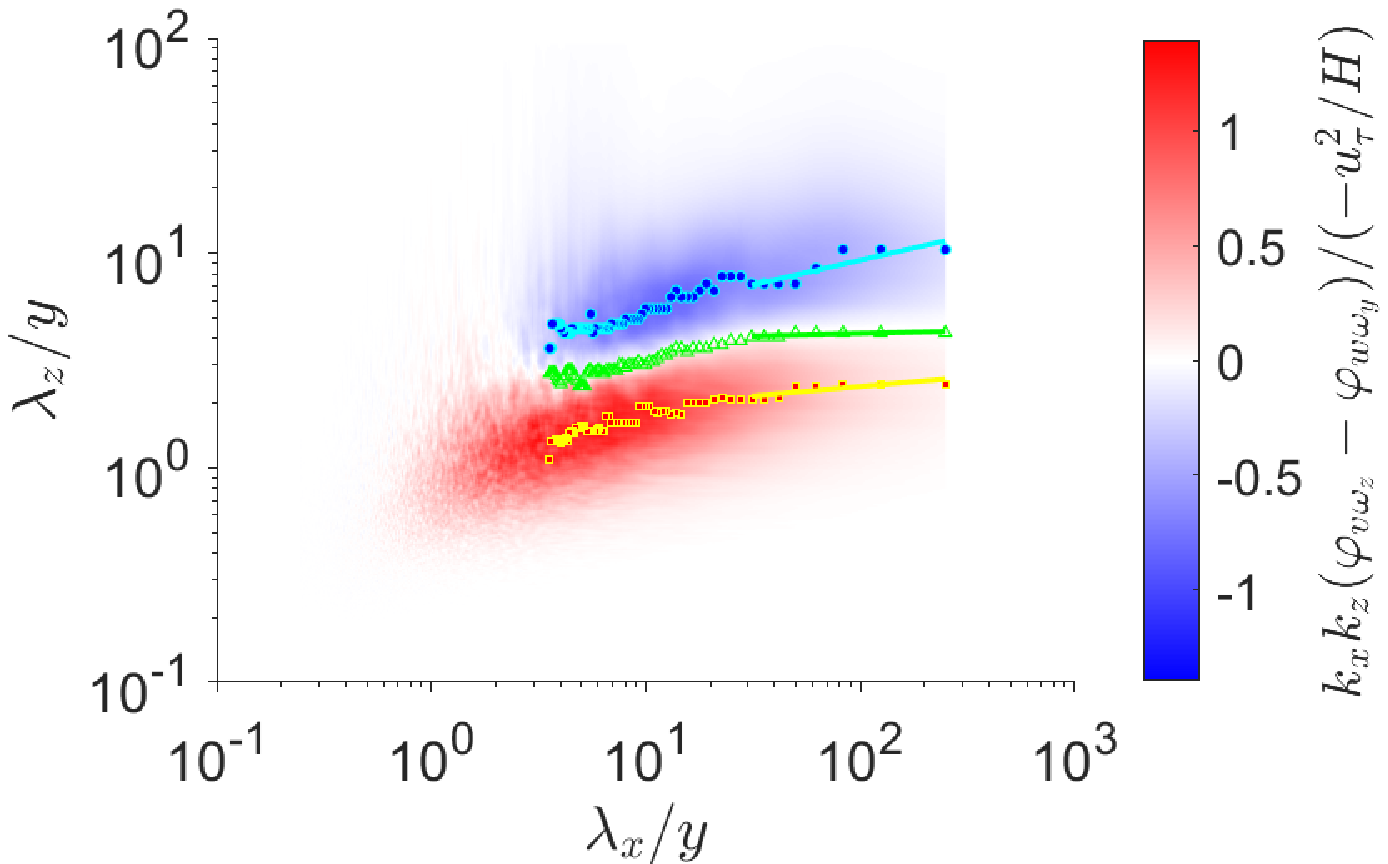}
         \caption{$y^+=100$}
         \label{log_moreyp}
     \end{subfigure}
             \caption{Ridges of spanwise wavelengths $\lambda_z^u/y$ at minimum cospectrum (blue-filled cyan circles), 
             $\lambda_z^d/y$ at maximum cospectrum (red-filled yellow squares), and $\lambda_z^*/y$ at zero cospectrum
             (green triangles), plotted log-log versus streamwise wavelength $\lambda_x/y$, together with the normalized 2D cospectra $\phi_{v\omega_z-w\omega_y}(k_x,k_z;y)$ for $y$ in the log layer. Power-law fits for the minimum, maximum, and zero cospectral ridges at $\lambda_x  \gtrsim 30y$, are given by cyan, yellow and green lines respectively.}
        \label{fig_2D_ridge}
\end{figure}

It is informative to compare our results for the 2D flux cospectra in the log layer with those for 2D energy 
spectra obtained from channel-flow DNS \citep{delalamo2004scaling} in the range of Reynolds numbers $Re_\tau=547\sim 1901,$ 
comparable to ours, and also from boundary-layer experiments \citep{chandran2017two,chandran2020spectral}
at much higher Reynolds numbers $Re_\tau=2430\sim 26,090.$ It was found by \cite{delalamo2004scaling}
that the $\lambda_z$ at the maximum of the 2D energy spectrum for each $\lambda_x$ corresponded to a ridge 
given by a power-law scaling 
\be \lambda_z/y\sim (\lambda_x/y)^p \label{plaw} \ee 
with an exponent $p\doteq 1/2$ that differed from the value $p=1$ corresponding to the self-similar 
structures assumed in the AEM. \cite{chandran2017two,chandran2020spectral} verified this result at their 
lowest Reynolds numbers but found that for higher $Re_\tau$ the spectral ridge is better fit by a 
power-law with $p>1/2,$ especially in the large-scale range $\lambda_x\gtrsim 20y.$ In the limit of 
very large Reynolds numbers, they found that $p\to 1$ and that $\lambda_z\sim \lambda_x/7,$ consistent 
with the spectra arising from streamwise elongated but self-similar structures, such as hairpin packets, 
as assumed in the AEM. Our results for the 2D flux cospectra are strikingly different, becoming almost 
independent of $\lambda_x$ for $\lambda_x/y\gtrsim 10.$ We have quantified this independence by 
calculating three spanwise wavenumbers $k_z^u,$ $k_z^d,$ and $k_z^*$ for each streamwise wavelength $k_x,$
corresponding respectively to the wavenumber where the pre-multiplied cospectrum normalized by $-u_\tau^2/H$
has its minimum (most negative) value, its maximum value, and its zero-crossing, respectively, for that 
$k_x$-slice. The corresponding ``ridges'' are plotted in Fig.~\ref{log_lessyp}-\ref{log_moreyp} for $y=40,\, 53,\, 100$ 
in the log-layer and we find that these are fit reasonably well for $\lambda_x/y>30$ by power-laws of the form \eqref{plaw},
with linear best-fit values given in Table~\ref{tab:p}. The small values of $p$ quantify how the cospectra 
become nearly independent of $\lambda_x$ for wavelengths $\lambda_x/y\gtrsim 10.$ This finding seems to indicate 
that the eddies contributing to nonlinear vorticity transport in the log layer are strongly non-self-similar,
with $\lambda_z$ nearly independent of $\lambda_x.$ This is one piece of evidence that the up-gradient 
transport proposed by \cite{lighthill1963} does not arise from attached eddies that are usually assumed to be self-similar. In addition, despite 
the arguments of \cite{eyink2008} to the contrary, the results in Fig.\ref{log_moreyp} suggest that for $y>y_p$ 
the down-gradient vorticity transport is as well not provided by the self-similar attached eddies assumed in the AEM,
at least for $Re_\tau=1000.$ In the next section we shall try to clarify this issue by identifying the 
vortex structures in the flow which are most relevant to the nonlinear vorticity transport.

\begin{table}
  \begin{center}
\def~{\hphantom{0}}
  \begin{tabular}{lccc}
      $y^+$  & $p_u$   &   $p_d$ & $p_*$ \\[3pt]
       40   & 0.2025 & 0.0936 & 0.0491\\
       53   & 0.2176 & 0.0078 & 0.0629\\
       100  & 0.2291 & 0.0894 & 0.0236\\
  \end{tabular}
  \caption{Linear fits for $(\lambda_x/y)\gtrsim30$, $\lambda_z^u/y=a_u(\lambda_x/y)^{p_u}$, $\lambda_z^d/y=a_d(\lambda_x/y)^{p_d}$, $\lambda_z^*/y=a_*(\lambda_x/y)^{p_*}$. Here, $a_u, \ a_d$ and $a_*$ are constants.}
  \label{tab:p}
  \end{center}
\end{table}

\subsection{Coherent vortices and their vorticity flux contributions}\label{sec:coherent} 

There has long been interest in the role of coherent vortex structures in turbulent flows 
\citep{cantwell1981organized,hussain1986coherent,robinson1991coherent,adrian2007hairpin}, with new insights related 
to exact Navier-Stokes solutions that represent such organized states \citep{mckeon2017engine,graham2021exact}.
One of the drivers of this sustained attention is the empirical fact that coherent vortices contribute 
a disproportionate amount to turbulent transport, outsize relative to their small volume fraction in the flow. 
These considerations have motivated extensive efforts over many years to develop vortex identification methods,
with classical methods discussed in prominent reviews \citep{chakraborty2005relationships,kolavr2007vortex}
and with improved methods continuing to be developed \citep{haller2016defining}
The bipartite 
character of the 2D flux cospectrum discussed in the previous subsection provides the basis to investigate the 
coherent vortex structures which contribute separately to up-gradient and down-gradient vorticity transport.  
The straightforward idea is to decompose the flow at each $y$-level into the contributions of two sets of eddies 
characterized by their support in the 2D Fourier space $(k_x,k_z),$ with one field coming from eddies
with negative (up-gradient) sign of the normalized flux cospectrum and the other from eddies with positive 
(down-gradient) sign. Methods of vortex visualization that have been applied to the full fields can then 
be applied to the two (nearly) orthogonal fields in order to identify the coherent vortices that contribute most 
significantly to up-gradient and down-gradient transport. 

In Supplementary Materials, section~\ref{dragonfly} we devise a convenient low-pass filter function $\mathcal{D}(k_x,k_z,y)$ 
that selects the spectral region of up-gradient flux, while its complement high-pass function $\mathcal{D}^c(k_x,k_z,y)=1-\mathcal{D}(k_x,k_z,y)$ 
selects the region of down-gradient flux. The filter functions were chosen to be graded to avoid Gibbs-type 
oscillations in physical space due to sharp spectral cutoffs. The particular filter function $\mathcal{D}(k_x,k_z,y)$ that 
we employ is a Gaussian function with elliptical level sets and with rotation angle relative to the Cartesian 
axes that depend upon the wall distance $y.$ When extended to the space of signed wavenumbers by reflections 
in the Cartesian coordinate axes, the levels of this function (see Fig.~\ref{flower}) resemble the crossed wings of a 
dragonfly and hence we have dubbed this function the ``dragonfly filter''. In the other panels of Fig.~\ref{fig:three graphs}
in the Supplementary Materials we illustrate how this filter selects regions of negative normalized flux cospectrum. 
All spatial fields $q(x,y,z)$ such as velocity and vorticity are then filtered by taking 2D FFT's, multiplying by 
$\mathcal{D}(k_x,k_z,y)$ or $\mathcal{D}^c(k_x,k_z,y),$ and then taking a 2D inverse FFT to obtain two fields, 
the contributions $q^U(x,y,z)$ of ``U-type eddies'' and $q^D(x,y,z)$ of ``D-type eddies''. We can then calculate 
separate nonlinear fluxes $\Sigma_{yz}^F:=v^{F}\omega_z^{F}-w^{F}\omega_y^{F}$ 
for both $F=U,D.$  Since off-diagonal terms such as 
$v^{U}\omega_z^{D}$ are small after averaging over both $x$ and $z,$ this yields a nearly additive decomposition
for averages $\langle\Sigma_{yz}^{nlin}\rangle\doteq \langle\Sigma_{yz}^U\rangle+\langle\Sigma_{yz}^D\rangle.$

When can then visualize vortices for the two velocity fields ${\bf u}^U$ and ${\bf u}^D.$ We present results 
here for the $\lambda_2$-criterion of \cite{jeong_hussain_1995} which is based on the intermediate 
eigenvalue $\lambda_2({\mathbf \nabla}{\bf u})$ of the symmetric matrix ${\bf S}^2+{\mathbf \Omega}^2$
where ${\bf S}$ and ${\mathbf \Omega}$ are, respectively, the symmetric and anti-symmetric parts of 
${\mathbf \nabla}{\bf u}.$ We can then define $\lambda_2^U:=\lambda_2({\mathbf \nabla}{\bf u}^U)$ and 
$\lambda_2^D:=\lambda_2({\mathbf \nabla}{\bf u}^D)$ and visualize vortices by negative levels of these 
scalar fields. We have also considered other vortex visualization schemes such as the $Q$-criterion 
of \cite{hunt1988eddies} but, consistent with other works \citep{chakraborty2005relationships}, we obtain 
very similar results from the  different visualization criteria when applied to turbulent fields and 
we thus present here our results only for the $\lambda_2$-criterion. We follow the suggestion of  
\cite{wu_christensen_2006} to visualize structures in inhomogeneous wall-bounded turbulence based
on levels of the vortex discriminant function normalized by its variance, here $\lambda_2/\lambda_2^{rms}=-\beta.$
Since the vorticity magnitudes decrease rapidly with $y$, this normalization permits uniform visualization 
of coherent vortices at all wall distances. The imposed level is somewhat arbitrary but we choose 
here $\beta=1$ which is in the range of earlier related studies 
\citep{wu_christensen_2006,chen2014experimental,chen2018comparison,chen2018contributions}, 
which we discuss at length below.

\begin{figure}
        \centering
        \includegraphics[width=\textwidth]{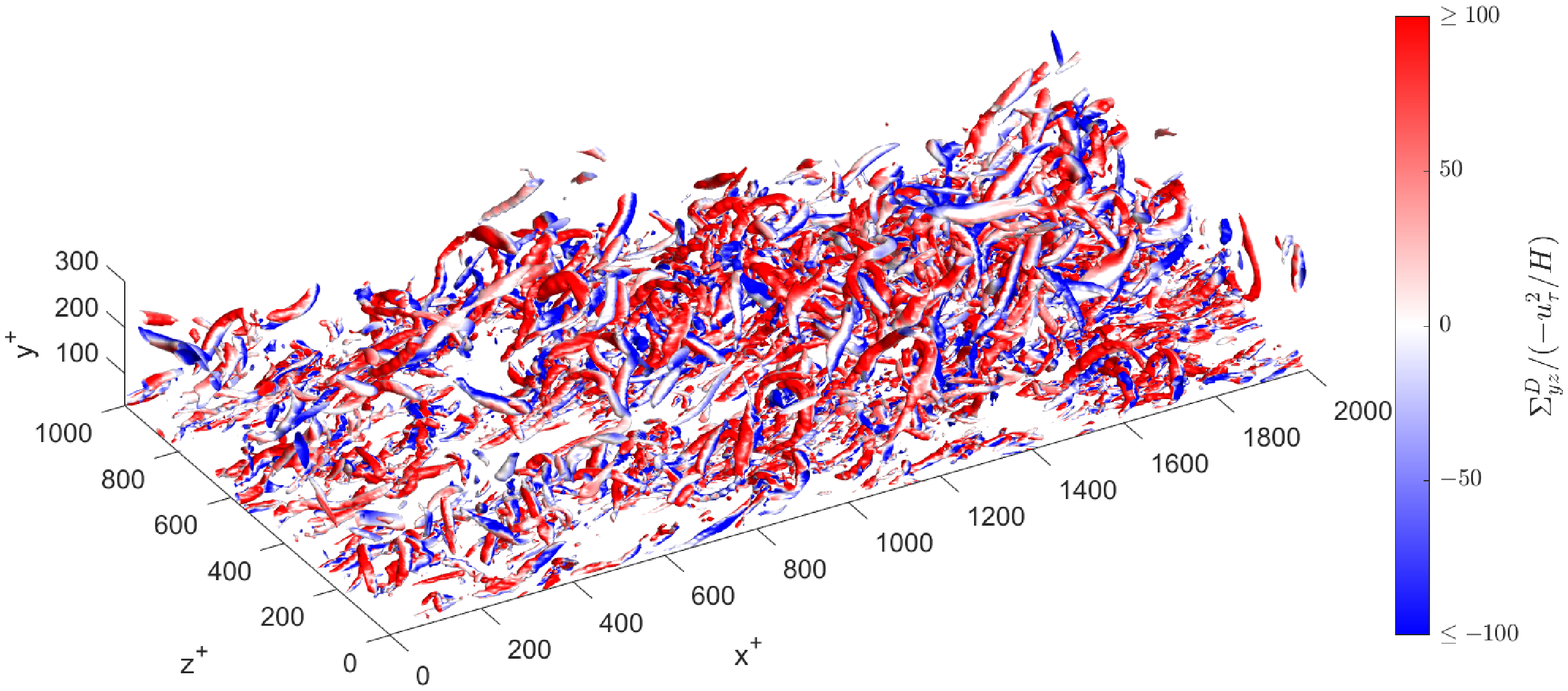}
        \caption{Vortices identified using the $\lambda_2$-criterion for the velocity field ${\bf u}^D$ filtered using $\mathcal{D}^c.$ Isosurfaces are plotted for $\lambda_2^D=-\lambda_2^{D,rms}$ and coloured by the nonlinear flux $\Sigma_{yz}^D$. The three dimensional figure is available at \href{https://protect-eu.mimecast.com/s/oHEhCwm99i43jGVc98kd8?domain=cocalc.com}{http://cocalc.com/.../D-vortices.html}. The data and code to generate such 3D vortices are available at \href{https://protect-eu.mimecast.com/s/mLweCvoqqFPz97LsXfK_g?domain=cocalc.com}{http://cocalc.com/.../3D-D-vortices/}. }
        \label{vortices_hp}
\end{figure}

         We begin with a discussion of the coherent vortices for the high-pass field ${\bf u}^D$, identified based on the discriminant 
function $\lambda_D.$ These structures are plotted in Fig~\ref{vortices_hp} and shall be referred to here as ``$D$-type vortices'' 
since they arise from the field ${\bf u}^D$ which accounts for the down-gradient nonlinear vorticity flux away from the wall. 
We visualize here only the vortices in the log-layer of the simulation, corresponding to $30<y^+<300.$
The vortex surfaces in the figure are colored based on the pointwise values of the down-gradient flux $\Sigma_{yz}^D$ 
The most immediate observation about the $D$-type vortices is that they have a very similar morphology to the 
well-known ``hairpin vortices'' that have been frequently visualized in the full velocity field ${\bf u}$ of turbulent wall-bounded flows, not only by the $\lambda_2$-criterion \citep{jeong1997coherent} but also by alternative methods such as 
swirling strength \citep{adrian2002observation,alfonsi2011hairpin} and the $Q$-criterion \citep{wu2009direct}.
We note that most of these vortices are ``prograde'' with $\omega_z<0$ and 
``retrograde'' vortices with $\omega_z>0$ are greatly outnumbered. 
Many of the hairpins also appear to be strongly asymmetric, with one much weaker leg, in agreement with some previous observations
\citep{adrian2007hairpin}. Hairpin vortices or packets of hairpins are often considered to be  plausible candidates for the ``attached eddies"  in the AEM \citep{adrian2007hairpin,marusic2019attached}. Although we attempt here no detailed statistical analysis, 
the $D$-type vortices pictured in Fig~\ref{vortices_hp} appear indeed to be wall-attached structures, with 
feet of one or both legs planted in the near-wall region. These observations agree with the suggestion of \cite{eyink2008}
and others \citep{chen2014experimental,chen2018contributions} that the down-gradient transport of vorticity should 
be supplied by attached hairpin-type structures, although we recall the evidence from the previous subsection 
that D-type vortices are not self-similar at $Re_\tau=1000.$ An interesting fact that may be inferred from the color 
plot in Fig~\ref{vortices_hp} is that $\Sigma_{yz}^D$ is not down-gradient (red color) at every point on the 
$D$-vortices. In fact, there are points also of very large up-gradient transport (blue color), which is permitted 
because the filter function $\mathcal{D}^c$ selects positive flux only in Fourier space not in  physical space. 
We can see furthermore that the regions of the two different signs of transport are organized, with red
(down-gradient) generally on the upstream side of the vortex and blue (up-gradient) generally on the downstream side.
This tendency is even more obvious in an interactive 3D version of figure \ref{vortices_hp}
that can be found at \href{https://protect-eu.mimecast.com/s/oHEhCwm99i43jGVc98kd8?domain=cocalc.com}{http://cocalc.com/.../D-vortices.html}
made available in the JFM Notebook. 
This observation can be easily understood in terms of the direction of the Lamb vector calculated 
from the vorticity vector and the local vortex-induced velocity in the vicinity of a hairpin-type vortex.

    \begin{figure}
    
     \begin{subfigure}[b]{0.329\textwidth}
         \centering
         \includegraphics[width=\textwidth]{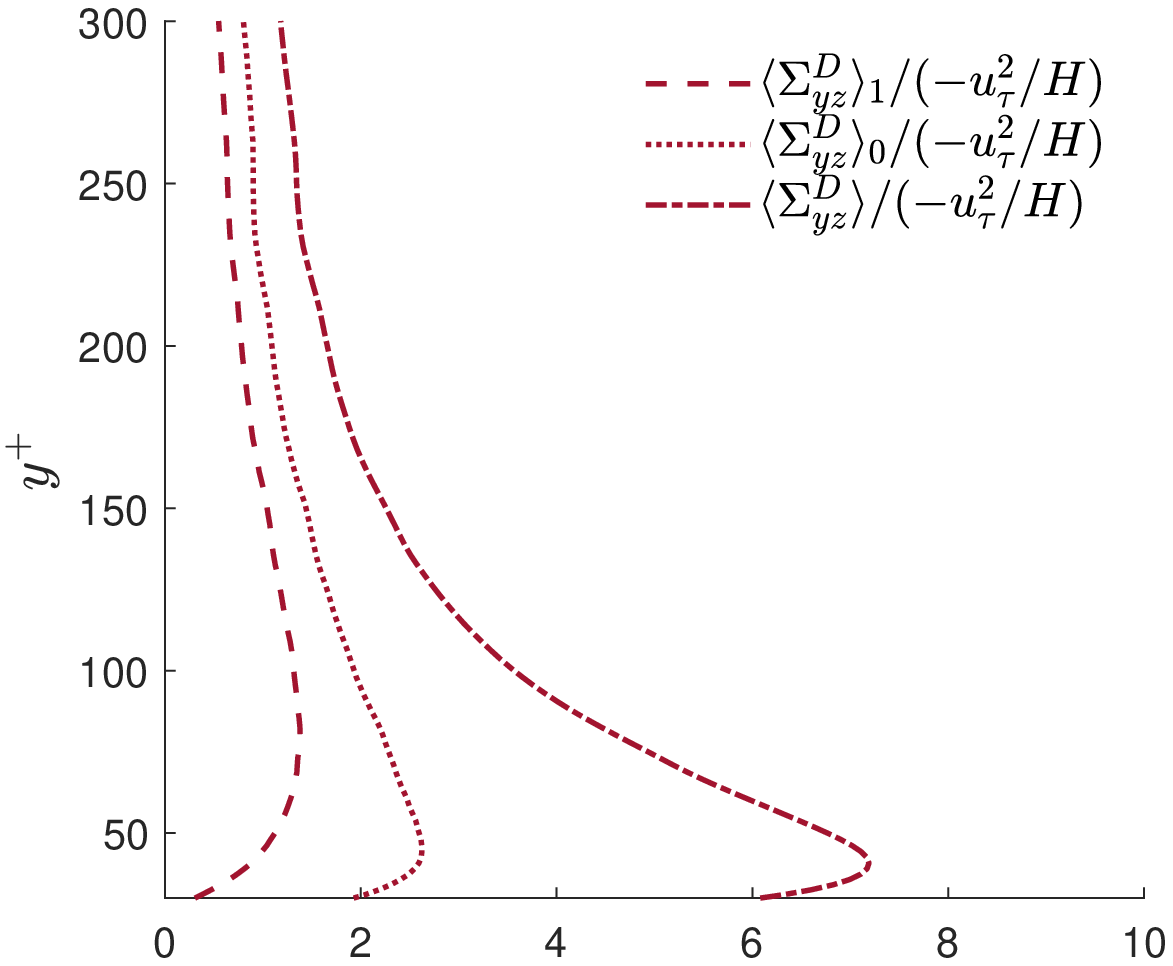}
         \caption{Vortical contributions to $\langle\Sigma_{yz}^D\rangle$. }
         \label{vhp}
     \end{subfigure}
     \hfill
     \begin{subfigure}[b]{0.329\textwidth}
         \centering
         \includegraphics[width=\textwidth]{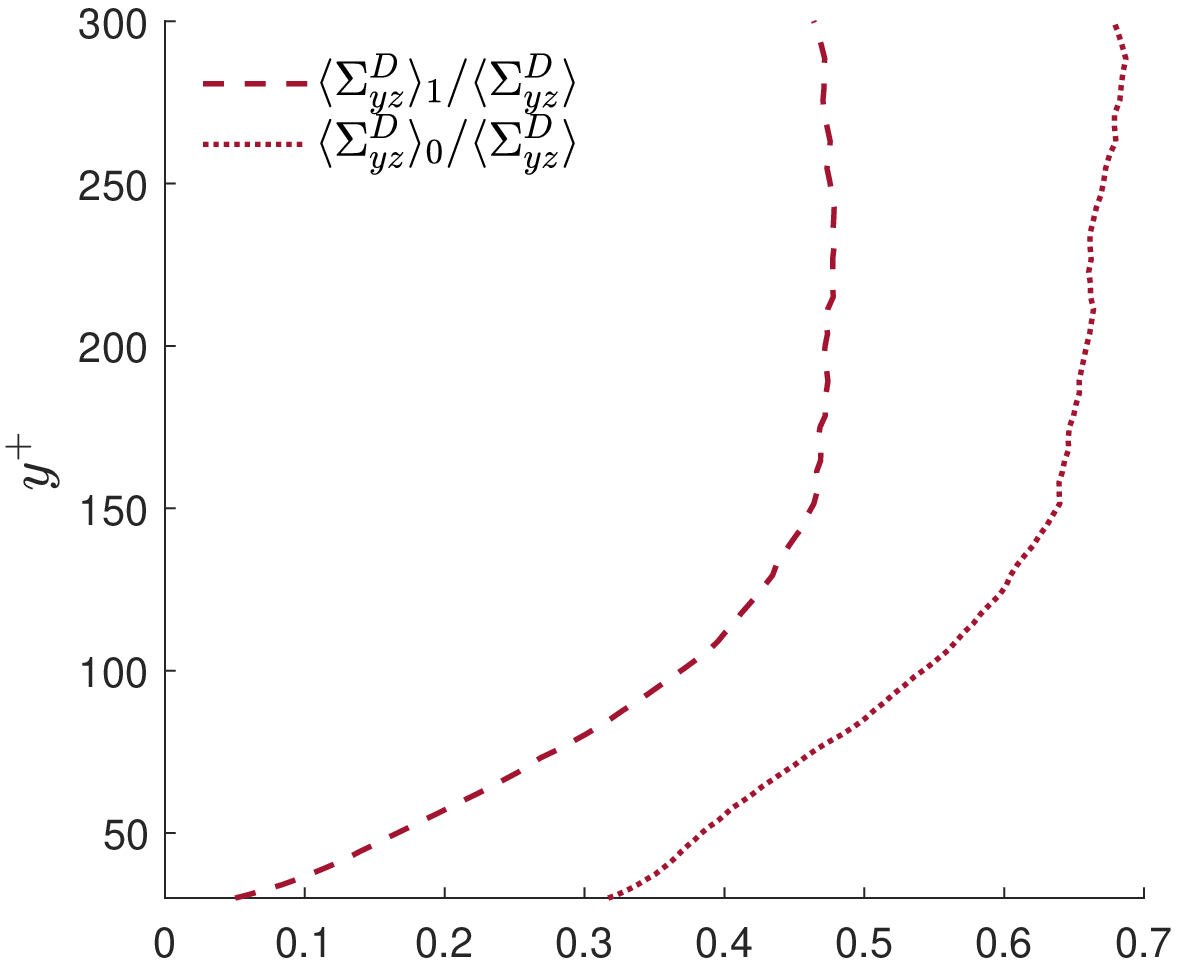}
         \caption{Fractional contributions to $\langle\Sigma_{yz}^D\rangle$. }
         \label{vhp_frac}
     \end{subfigure}
     \hfill
     \begin{subfigure}[b]{0.329\textwidth}
         \centering
         \includegraphics[width=\textwidth]{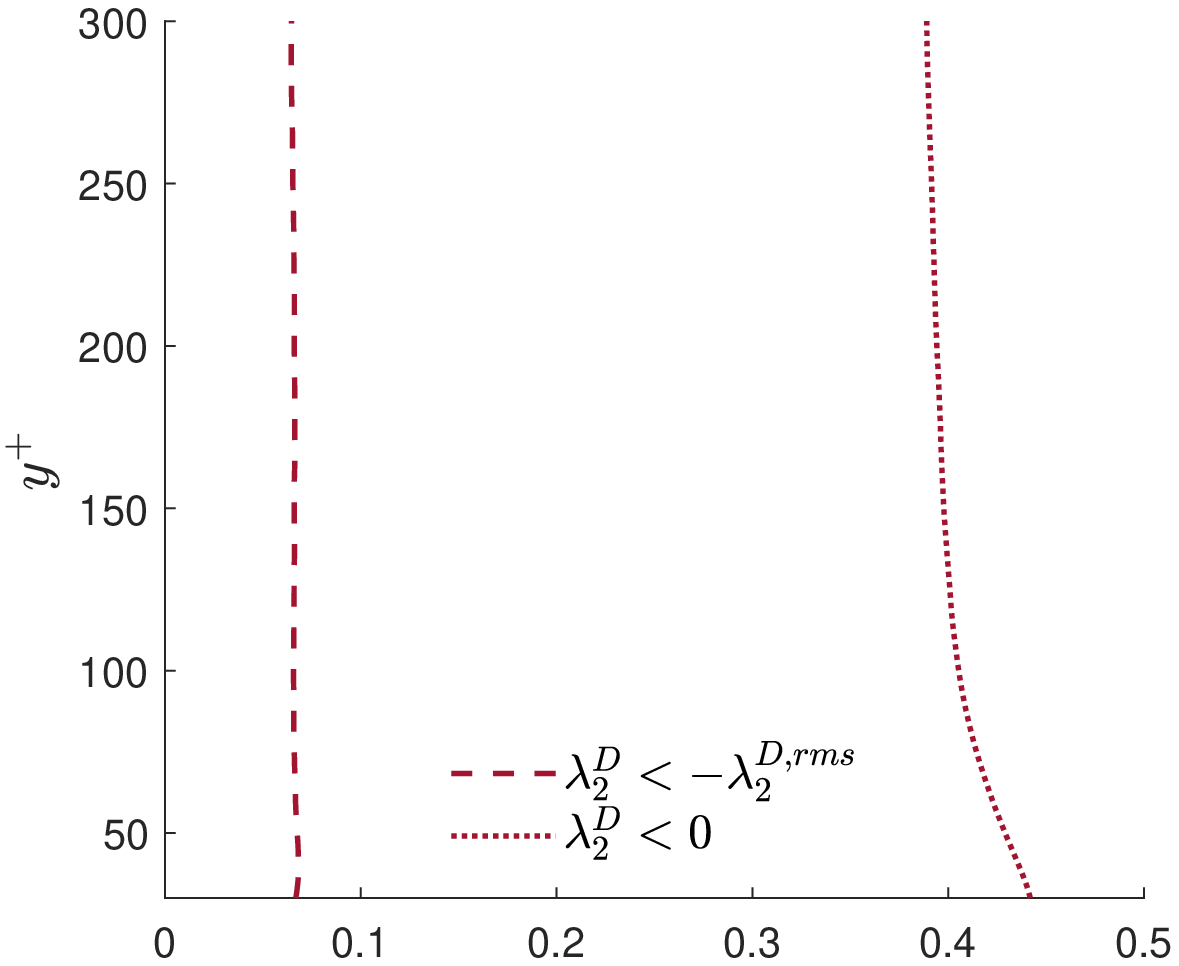}
         \caption{Area fraction occupied by $D$-type vortices. }
         \label{vhp_volfrac}
     \end{subfigure}
        \caption{(a) Mean down-gradient vorticity flux and its contributions from $D$-type coherent vortices.
        (b) Fractional contributions from $D$-type vortices. (c) Area fraction occupied by $D$-type vortices.}
        \label{vortcontriD} 
\end{figure}

The net vorticity flux of all D-type eddies is indeed down-gradient, however, as illustrated by Fig~\ref{vhp}
which plots $\langle\Sigma_{yz}^D\rangle$ normalized by $-u_\tau^2/H$ versus $y.$ Not only is the normalized 
flux positive but in fact $\langle\Sigma_{yz}^D\rangle<-u_\tau^2/H$ for all $y$ and it approaches almost $-7u_\tau^2/H$ 
in the buffer layer. These large values are possible because they (plus the viscous flux) are nearly cancelled 
by the opposing up-gradient flux supplied by the complementary field ${\bf u}^U.$ In addition to the
total $\langle\Sigma_{yz}^D\rangle,$ we can also calculate the partial average $\langle\Sigma_{yz}^D\rangle_\beta$
from the region of the vortex cores characterized by $\lambda_2^D<-\beta\lambda_2^{D,rms}.$ For $\beta=1$ this 
may be considered the direct contribution of the $D$-type vortices to the mean down-gradient flux. 
However, it is likely that this partial average on the cores underestimates the true contribution of the 
coherent vortices, which will also 
make an indirect contribution from a spatial neighborhood influenced by induced motions from Biot-Savart 
\citep{wu_christensen_2006,chen2018contributions}. As a crude estimate of this larger contribution from the region 
influenced by the coherent vortices,
we consider the partial average $\langle\Sigma_{yz}^D\rangle_0$ over the region with $\lambda_2^D<0.$ 
Both of these partial averages are plotted also versus $y$ in Fig~\ref{vhp}. The fractional 
contributions of the vortex cores and the vortex neighborhoods are furthermore plotted in Fig.~\ref{vhp_frac}. 
We can also calculate the area fractions of these two regions in the wall-parallel planes 
at each distance $y$ and these are plotted in Fig.~\ref{vhp_volfrac}.  
These plots show that the vortex cores contribute an increasing fraction of the down-gradient flux 
for increasing $y,$ starting from $\sim 5\%$ at $y^+=30,$  reaching $\sim 45\%$ at $y^+=150$, and 
remaining roughly constant then in the log layer, but vortex cores occupy only about  
$\sim 7\%$ of the area at every wall distance. The flux fraction from the vortex-dominated region 
behaves similarly but is even larger,  with $~\sim 30\% $ contribution at $y^+=30,$  $\sim 65\%$ at  $y^+=150$, 
and remaining roughly constant thereafter, while occupying only about 40\% of the area. 
Thus, by either of the measures, the $D$-type vortices make a contribution to the down-gradient flux 
out of proportion to their volume in the flow. 

Our results closely mirror previous works which have studied the effects of coherent vortices 
on the transport of spanwise vorticity. Following the pioneering work of \cite{wu_christensen_2006}
on spanwise vortex contributions to Reynolds stress, \cite{chen2014experimental} experimentally studied open 
channel flume flows at $Re_\tau=382\sim 740$ using particle image velocimetry (PIV) in 2D vertical planes, 
which gave access to the velocities $u(x,y)$ and $v(x,y).$ Identifying coherent vortices by the swirling 
strength with $\beta=1.5$, they could then analyze their contribution to the advective flux $v\omega_z.$ They found many such 
coherent vortices with almost 97\%  prograde at $y^+\simeq 50$ and this percentage declining with $y$ but 
still 65\% at $y^+=600,$ and it was conjectured that the prograde vortices were the heads of hairpin vortices.
\cite{chen2014experimental} observed as well a bipolar distribution of flux near the vortex cores, 
with $v'\omega_z'>0$ upstream and $v'\omega_z'<0$ downstream, which they explained also by local induced 
velocities. Finally, \cite{chen2014experimental} found for the region $100<y^+<0.9 Re_\tau$ that 
coherent vortex cores contribute about 45\% to $\langle v\omega_z\rangle$ while occupying only 9\% 
of the area. All of these results for the total velocity field in 2D planes are quite similar to ours for 
the high-pass filtered field component ${\bf u}^D$ in 3D.

The following paper of \cite{chen2018contributions} (see also \cite{chen2018comparison}) verified the results 
of \cite{chen2014experimental}, but in the channel-flow simulation of \cite{delalamo2004scaling} at 
$Re_\tau=934.$ Because of the availability of full 3D velocity fields, \cite{chen2018contributions} could 
apply the swirling strength criterion for various choices of $\beta$ in both 2D planes and in 3D and they 
found that comparable results were obtained for both, although the 3D criterion identified more coherent 
vortices than did the 2D criterion.
\cite{chen2018contributions} could also study the stretching flux $-w\omega_y$ and the full nonlinear flux
$v\omega_z-w\omega_y$ and then investigate the coherent vortex contribution to each of these. In fact, 
\cite{chen2018contributions} decomposed the entire space domain into four exclusive point sets consisting 
of ``filamentary vortices'' (FV), ``non-filamentary vortices'' (NFV), ``non-swirling vortex structures'' (NSVS),
and ``irrotational structures'' (IS), and all of these except the latter can contribute to vorticity flux. 
The FV structures correspond to roughly to our vortex cores, but identified by swirling strength
and a region growing algorithm, and the other structure types are precisely defined by \cite{chen2018comparison,chen2018contributions}. 
Thus, \cite{chen2018contributions} determined the fractional contributions of each of the first three
types of structures (FV, NFV, NSVS) to the three fluxes (advective, stretching, total nonlinear) and they 
emphasized the dominant role of the FV structures. For the region 
$y^+>300$ of their simulation where $\langle w\omega_y\rangle>0$ (down-gradient), FV structures 
contribute more than 80\% while occupying only 15\% of the volume. Likewise, for the region $y^+>50$
where $\langle v\omega_z-w\omega_y\rangle<0$ (down-gradient), FV structures contribute more than 60\%
while occupying area less than 15\%. They concluded that: ``Compared with the other three 
structures, FV play a very important role in velocity-vorticity correlations and the net force.''

From our perspective, however, a very important result of \cite{chen2018contributions} that was 
never explicitly mentioned by them is the fact that {\it the FV contributions to all three 
of the fluxes are negative (down-gradient) everywhere}, that is, $\langle v\omega_z\rangle_{FV}<0$, 
$-\langle w \omega_z\rangle_{FV} <0$ and $\langle v\omega_z- w \omega_z\rangle_{FV} <0$
for all $y$. These signs can be inferred from the data plotted in Fig.~2(b) and Fig.~9 of 
\cite{chen2018contributions}. Since however $-\langle w\omega_y\rangle>0$ for $y^+<300$  
and  $\langle v\omega_z-w\omega_y\rangle>0$ for $y^+<50$ in the simulation studied by 
\cite{chen2018contributions}, this means that the FV structures are not only not dominant 
in these regions but in fact give a contribution of the wrong sign! See Fig.~9(c)-(d) in
\cite{chen2018contributions}. Put another way, the FV structures do not account for the up-gradient 
vorticity transport in these regions. This is understandable if one can essentially identify the FV structures 
of \cite{chen2018contributions} with our $D$-type vortices, which contribute always a net down-gradient flux. 
This identification is plausible based on the evidence of visualizations, since the $D$-type vortices shown 
in our Fig.~\ref{vortices_hp} resemble quite closely the vortex structures detected by the 
swirling strength criterion in the full velocity field \citep{adrian2002observation,alfonsi2011hairpin}. 
In any case, the important result of \cite{chen2018contributions} not emphasized by them 
is that it is the NFV and NSVS structures which account for the up-gradient vorticity transport 
observed closer to the wall. Unfortunately, the works of \cite{chen2018comparison,chen2018contributions}
did not attempt to visualize the NFV and NSVS structures or to clarify their morphology and dynamics. 

\begin{figure}
         \centering
         \includegraphics[width=\textwidth]{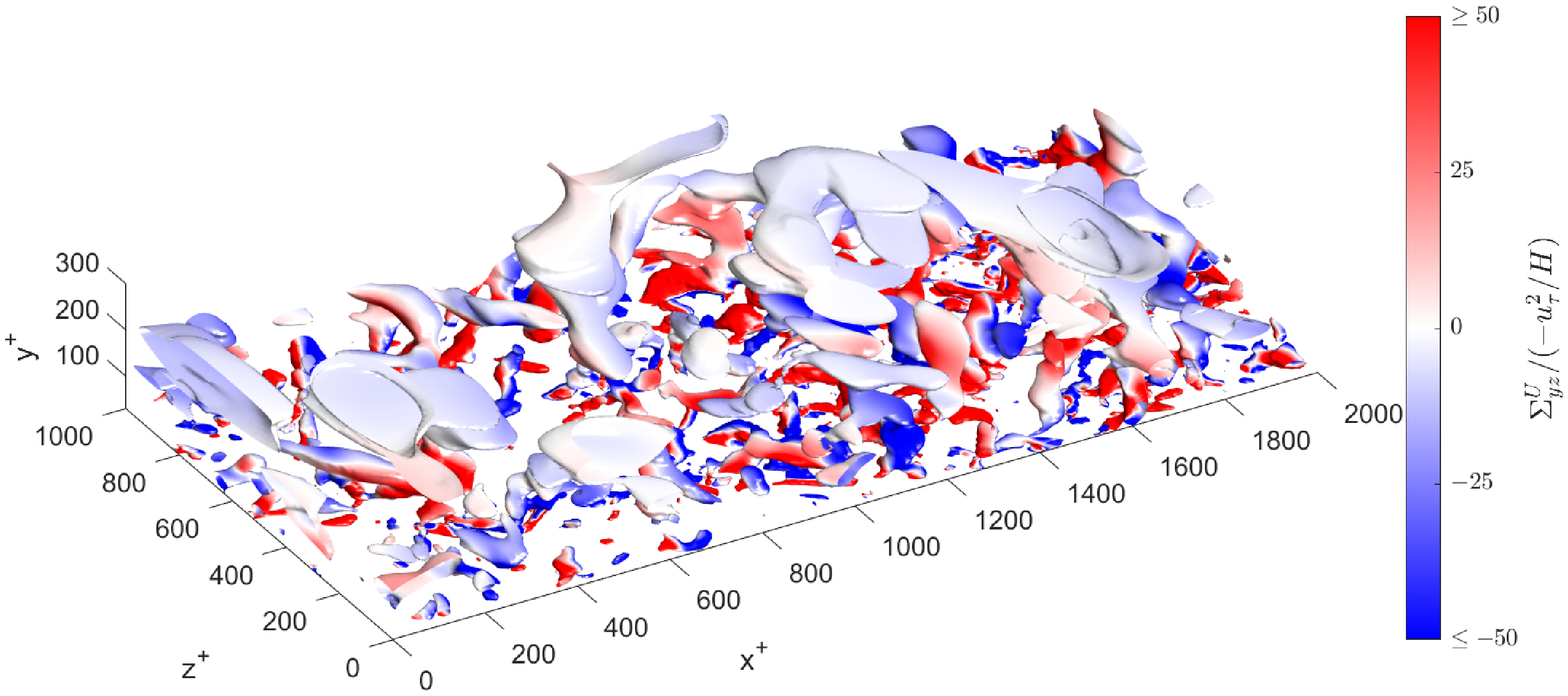}
        \caption{Vortices identified using the $\lambda_2$-criterion for the velocity field ${\bf u}^U$ filtered using $\mathcal{D}.$ Isosurfaces are plotted for $\lambda_2^U=-\lambda_2^{U,rms}$ and coloured by the nonlinear flux $\Sigma_{yz}^U$. The three dimensional figure is available at \href{https://protect-eu.mimecast.com/s/m66oCyrllIQ51r2fNX2V1?domain=cocalc.com}{http://cocalc.com/.../U-vortices.html}. The data and code to generate 3D vortices are available at \href{https://protect-eu.mimecast.com/s/M8UmCxvqqTgol1RSwC9Ij?domain=cocalc.com}{http://cocalc.com/.../3D-U-vortices/}. }
         \label{vortices_lp}
\end{figure}

We can now illuminate the nature of such structures by vortex visualizations for the low-pass field ${\bf u}^U$, 
using the discriminant function $\lambda_2^U$ and $\beta=1.$ These structures are plotted in Fig~\ref{vortices_lp} and shall 
be referred to here as ``$U$-type vortices'' since they arise from the field ${\bf u}^U$ which accounts for 
the up-gradient nonlinear vorticity flux toward the wall. As before, we visualize only the vortices in the log-layer 
of our simulation and the vortex surfaces are colored based on the values of the up-gradient flux $\Sigma_{yz}^U.$ 
The $U$-type vortices visualized in Fig~\ref{vortices_lp} have a pancake structure, vertically flattened 
and elongated along the streamwise direction but especially along the spanwise direction. 
Their characteristic shape becomes even clearer in the interactive 3D version of
Fig. \ref{vortices_lp} available at
\href{https://protect-eu.mimecast.com/s/m66oCyrllIQ51r2fNX2V1?domain=cocalc.com}{http://cocalc.com/../U-vortices.html}.
These are exactly 
the type of vortex structures one would imagine to arise from the \cite{lighthill1963} mechanism 
of correlated downflow and lateral stretching. In fact, the roughly twice longer extents spanwise 
than streamwise correspond well to Lighthill's remark that ``some longitudinal deformation is usually 
also present, which reduces the need for lateral deformation (perhaps, on average, by half).'' \black{Note, that these structures, ultimately based on the 2D co-spectra plotted in Fig.\ref{fig_2dcospec}, can be interpreted as evidence for a cascade process similar to that for momentum, since the vorticity transfer in space is observed to be carrried by a hierarchy of eddies whose size depends upon wall-distance.} 
However, there is no obvious self-similarity of these structures in scale and, according to the 
results presented in our Fig.~\ref{fig_1dcospec_z} and Table~\ref{tab:p}, one might expect the 
ratio $\lambda_x/\lambda_z$ to increase with scale. The $U$-type vortices show also no obvious attachment 
to the wall, being mainly horizontal to it. Indeed, an extension of the AEM by  
\cite{perry_marusic_1995,marusic_perry_1995} introduced in addition to the ``Type-A" attached 
eddies also ``Type-B" eddies to represent the wake flow in the outer layer and small-scale 
``Type C'' eddies to represent the Kolmogorov range. The Type-B eddies were viewed as detached 
vortex tubes which undulate in the spanwise direction, thus somewhat resembling our $U$-type eddies
(see Figure 3 of ~\cite{perry_marusic_1995}). More recently, \cite{hu_yang_zheng_2020} have attempted to decompose 
turbulent channel flow fields at $Re_\tau=5200$ into small-scale eddies, attached eddies, and detached eddies, 
and their $|u|$-isosurfaces for the detached eddies (see their Fig.~19(d)) have a similar pancake structure 
as our $U$-type vortices. We have also checked that the vorticity in the $U$-type vortices is predominantly 
spanwise and prograde  consistent with lateral stretching of pre-existing vorticity. This is demonstrated by Fig.~\ref{vortices_lp_omega} in the Supplementary Materials, 
where we colour these vortices by the cosine of the angle made by the vorticity vector $\boldsymbol{\omega}^U$ with the z-axis and find the prevalence of values close to -1, denoting prograde spanwise aligned vortices. Finally, we observe flux-bipolarity of the $U$-type vortices just as for the $D$-type, with red (down-gradient) generally on the upstream side of the vortex and blue (up-gradient) generally on the downstream side. This tendency is again more obvious in a 3D version of the figure available in the JFM Notebook.

\begin{figure}
    
      \begin{subfigure}[b]{0.329\textwidth}
         \centering
         \includegraphics[width=\textwidth]{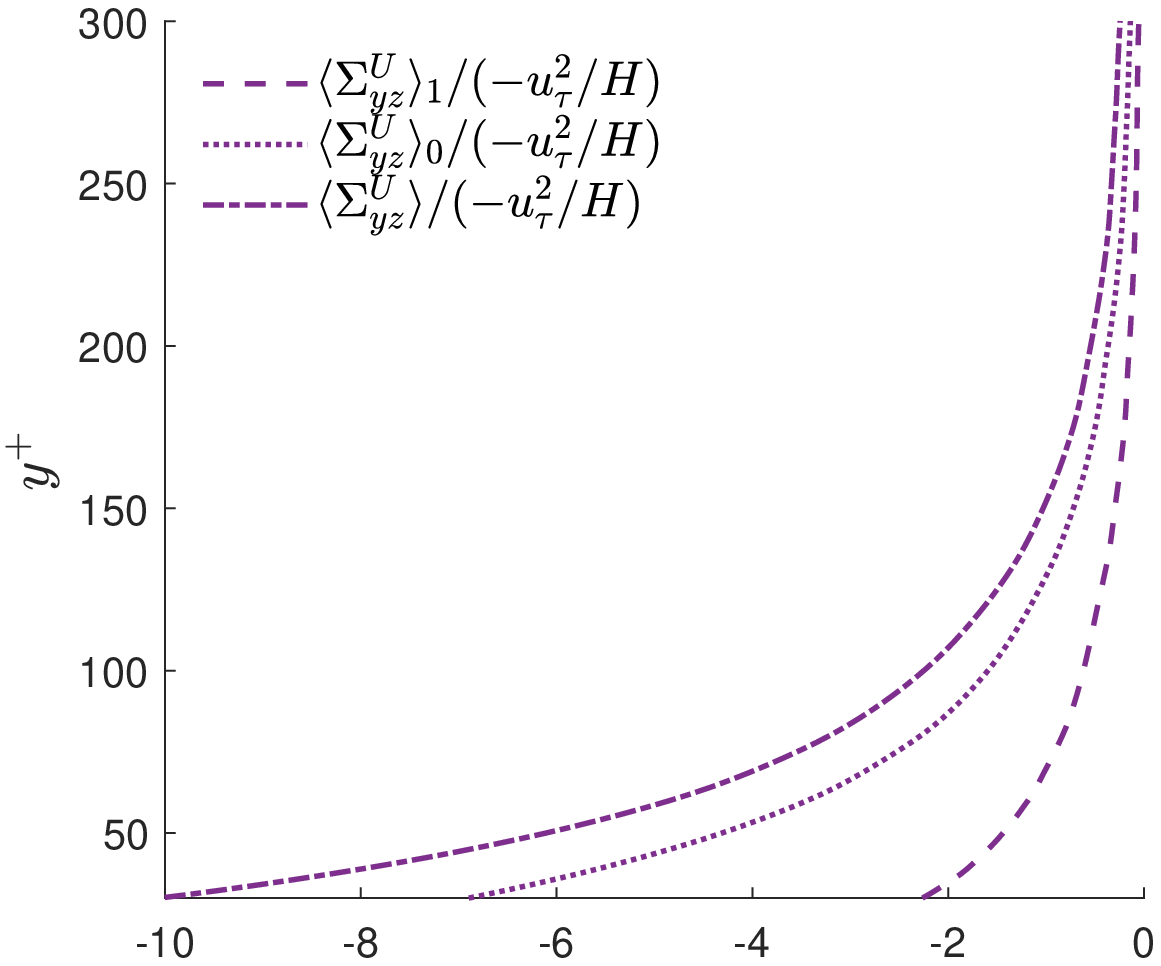}
        \caption{Vortical contributions to $\langle \Sigma_{yz}^U\rangle$}
         \label{vlp}
     \end{subfigure}
     \hfill
     \begin{subfigure}[b]{0.329\textwidth}
         \centering
         \includegraphics[width=\textwidth]{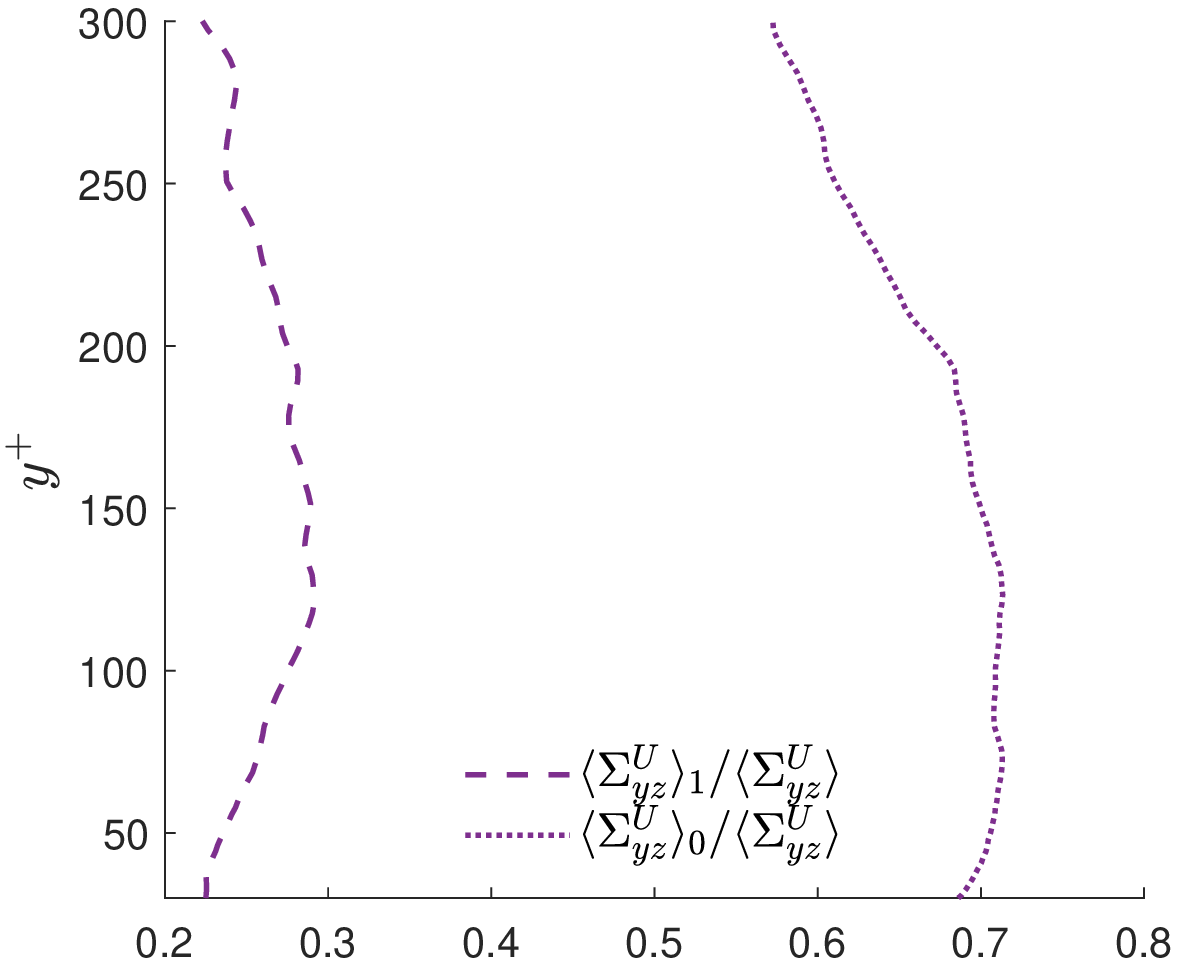}
        \caption{Fractional contribution to $\langle \Sigma_{yz}^U\rangle$}
         \label{vlp_frac}
          \end{subfigure}
    \hfill
     \begin{subfigure}[b]{0.329\textwidth}
         \centering
         \includegraphics[width=\textwidth]{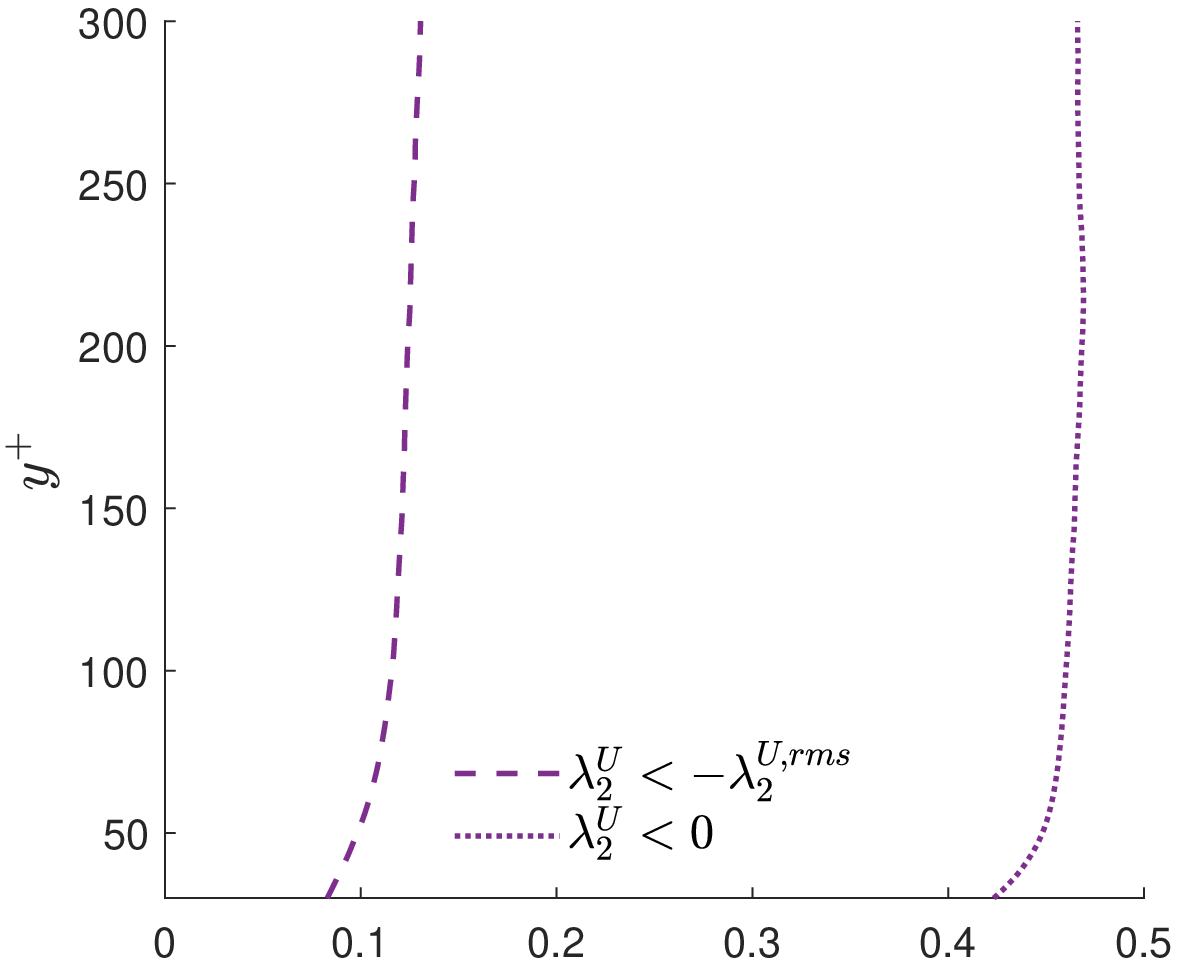}
        \caption{Area Fraction occupied by U-type vortices}
         \label{vlp_volfrac}
     \end{subfigure}
     
        \caption{(a) Mean down-gradient vorticity flux and its contributions from $U$-type coherent vortices.
        (b) Fractional contributions from $U$-type vortices. (c) Area fraction occupied by $U$-type vortices.}
        \label{vortcontriU}
        
\end{figure}

Despite large contributions to the vorticity flux of both signs, however, the net flux supplied by all 
$U$-type eddies is up-gradient. This is verified by the data in Fig~\ref{vlp} which plots $\langle\Sigma_{yz}^U\rangle$ 
normalized by $-u_\tau^2/H$ versus $y.$ Similar to the result for the $D$-type vortices, 
$\langle\Sigma_{yz}^U\rangle\gtrsim u_\tau^2/H$ across most of the log layer and approaches 
$10 u_\tau^2/H$ at the lower $y$ range where it must cancel most of the down-gradient flux from the 
$D$-type eddies and viscous diffusion. For the largest $y^+\sim 300$ in the log layer, 
$\langle\Sigma_{yz}^U\rangle\simeq 0.1 u_\tau^2/H$ and still cancels part of the 
contribution from the $D$-type eddies. In addition to the total $\langle\Sigma_{yz}^U\rangle,$ 
we can also calculate the partial averages $\langle\Sigma_{yz}^U\rangle_\beta$ for the 
condition $\lambda_2^U<-\beta \lambda_2^{U,rms}$ with $\beta=1$ (vortex cores) and 
$\beta=0$ (vortex neighborhoods). Both of these partial averages are plotted also versus $y$ 
in Fig~\ref{vlp}, the fractional contributions of the vortex cores and the vortex neighborhoods 
are plotted in Fig.~\ref{vlp_frac}, and the corresponding area fractions plotted in Fig.~\ref{vlp_volfrac}.   
We observe that the cores of the $U$-type vortices account for $\sim 20 \% -30\%$ of the up-gradient 
flux but occupy only about 10\% of the area,  while regions dominated by the $U$-type vortices provide 
$\sim 60 \% - 70\%$ of the flux but occupy only about 45\% of the area. The coherent $U$-type vortices 
thus contribute a percentage of the up-gradient flux roughly twice their area in the flow. 
This is not as outsize as the contribution of the $D$-type vortices to the down-gradient flux and 
the lower performance may be due to our rote application of the $\lambda_2$-criterion for vortex 
identification in ${\bf u}^U.$ This criterion was designed by \cite{jeong_hussain_1995} to detect 
rapidly swirling vortex tubes with low-pressure cores, whereas the $U$-type vortices clearly have 
a distinct structure. It is also possible that our filter function $\mathcal{D}$ does not have 
spectral support optimally chosen. More appropriate filter kernels and discriminant functions can 
probably be devised to characterize better the coherent $U$-type vortices that contribute most to 
the up-gradient vorticity transport.

\section{Discussion and Conclusions}\label{sec:conclusion} 

The main objective of this work has been to elucidate the dynamics involved in turbulent tranport of spanwise vorticity 
normal to a solid wall, motivated by the direct connection of this ``vorticity cascade'' to turbulent 
drag. We have carried out a numerical study for a canonical case of pressure-driven Poiseuille flow
in a channel with plane-parallel walls. We find that the mean vorticity transfer is the result of 
two intensely competing processes: an up-gradient transfer that concentrates spanwise vorticity
strongly near the wall and a slightly greater down-gradient transfer that disposes of the fresh spanwise vorticity 
generated at the wall by the mean pressure-gradient. This is exactly the picture suggested by 
\cite{lighthill1963}, who proposed also a concrete mechanism for up-gradient transport by 
inflow to the wall correlated with lateral stretching of vortex lines \black{and outflow correlated with lateral relaxation}. We have presented here detailed evidence 
for the validity of Lighthill's mechanism in the case of turbulent channel flow based upon:  
(i) correlations of wall-normal velocity with flux of spanwise vorticity, (ii) velocity-vorticity 
co-spectra that identify the eddies involved in nonlinear vorticity transport in the two directions, 
and (iii) visualization of the coherent vortex structures which contribute dominantly to the 
transport. All of the observations that we have accumulated are consistent with the proposed mechanism. \black{In addition, we have provided evidence to support the interpretation of this vorticity flux as a cascade similar to momentum cascade, in that the spatial transport of vorticity is associated also with a stepwise transfer in scale.}
This verification has consequences that extend far beyond channel flow, because \cite{lighthill1963}
suggested that the up-gradient transfer mechanism has very general validity for all turbulent flows  
interacting with solid walls, e.g. high-Reynolds flows around bluff bodies or airplane wings. In fact,
\cite{lighthill1963} proposed this mechanism to explain why turbulent boundary layers with concentrated vorticity 
form generally near solid walls. This is the necessary prelude to another fundamental vortex 
interaction with solid walls, the violent eruption of vorticity away from the wall 
\citep{doligalski1994vortex,smith1991dynamics,lenaers2012rare}, 
which apparently underlies phenomena such as boundary-layer separation behind bluff bodies and frequent ejections 
from the walls in turbulent flows over flat plates and in straight pipes.  

Our work exposes also a limitation of the ``attached eddy model'' of \cite{townsend1976structure} because 
we find considerable evidence that down-gradient vorticity flux out from solid walls is provided indeed 
by attached eddies but that the competing up-gradient flux into the walls is carried by detached eddies. 
Of course, many extensions of the AEM have been proposed previously. We have already mentioned the 
early work of \cite{perry_marusic_1995,marusic_perry_1995} and see also more recent papers of 
\cite{chandran2020spectral,hu_yang_zheng_2020}. The spectral aspects of the extension by 
\cite{perry_marusic_1995,marusic_perry_1995} focus on the streamwise spectrum $\phi_{uu}(k_x)$, 
as summarised in Figure 16 of~\cite{marusic_perry_1995}, reproduced in Figure 11 of~\cite{marusic2019attached}.
Our present cospectral results echo features of this model upon suitable interpretation.  
The model of \cite{perry_marusic_1995,marusic_perry_1995} has an overlap in the streamwise scales 
of Type-A and Type-B eddies, with the latter contribution dominating at larger streamwise scales. 
This type of behaviour is seen also in our 2D flux co-spectra in the log layer, plotted in 
Fig~\ref{fig_2dcospec} (d),(e),(f), where the $U$-type eddies dominate at large streamwise 
scales but overlap with the support of the $D$-type eddies at smaller streamwise scales. 
Note, however, that the effect of Type-B eddies in the AEM extension of Perry \& Marusic 
are supposed to diminish upon approaching the wall, whereas the vorticity transport effects 
of the $U$-type eddies in our work increase and dominate close to the wall. Unlike previous proposals 
to extend the AEM, our work has revealed that detached eddies play a fundamental dynamical 
role in the near-wall region, with direct importance to drag generation and reduction.  
The closest connection of our results are with those of \cite{chen2018contributions}, who 
found that ``non-filamentary vortices'' (NFV) and ``non-swirling vorticity structures'' (NSVS)
rather than ``filamentary vortices'' (FV) are responsible for the dominant up-gradient nonlinear vorticity transport near the wall.

In this paper we have exploited a database of a turbulent channel flow at $Re_\tau=1000$ and 
we have compared with related experiments and simulations at comparable Reynolds numbers. 
However, empirical data is available at a broader range of Reynolds numbers, both lower-Reynolds 
simulation data \citep{bernard1990turbulent,crawford1997reynolds} and field experiments at much 
higher Reynolds numbers \citep{priyadarshana2007statistical,morrill2013influences}. 
We do not expect the main conclusions of our work to be $Re$-dependent 
and we are aware of no data available at other Reynolds numbers which contradicts them. However,
some details of the story we have presented may change with $Re.$ For example, it is well-known 
that many predictions of the AEM are observable only for $Re\gg 1$ and, in particular, \cite{chandran2017two}
estimate that the similarity relation $\lambda_z\sim \lambda_x$ for attached eddies and clear 
$k^{-1}$ energy spectra for both $k_x$ and $k_z$ should be observable only for $Re_\tau>60,000.$
Thus, our conclusion in section \ref{sec:cospectra} that $D$-type and $U$-type eddies are 
non-self-similar might be Reynolds-number dependent, especially since self-similarity for 2D energy 
cospectra that is expected from attached eddies is not observed either at comparable $Re_\tau$
\citep{delalamo2004scaling,chandran2017two}. Thus, extending our analysis to higher Reynolds 
numbers is an important direction for future research. 


Another important direction is the investigation of turbulent vorticity dynamics by means of stochastic 
Lagrangian methods \citep{constantin2011stochastic,eyink2020stochastic}. The heuristic arguments of 
\cite{lighthill1963} are essentially Lagrangian and invoke the remarkable ``frozen-in'' properties enjoyed 
by vortex-lines in smooth ideal Euler flows. In a physical turbulent flow, however, these familiar 
properties of ideal vortex-lines suffer fundamental modifications by viscous diffusion, which can be 
exactly captured by the stochastic Lagrangian representation of vortex dynamics. These methods have already 
proved powerful to verify the validity of Lighthill's argument for origin of large magnitudes of wall-vorticity 
in a transitional zero pressure-gradient boundary layer \citep{wang2022origin}. The advantage of these 
methods compared with the Eulerian analysis in the present work is that they provide a complete and unambiguous 
account of the origin of the vorticity at any point in the flow, with precise and quantitative information 
about the physical mechanisms involved. Such stochastic Lagrangian methods have already been applied 
to ``ejections'' and ``sweeps'' in the buffer layer of the same $Re_\tau=1000$ turbulent channel flow 
studied in this paper, where it was demonstrated that the spanwise vorticity in those events is not assembled abruptly
from wall-vorticity but instead over many hundreds of viscous times \citep{eyink2020stochasticB}. It would be 
very illuminating to apply these methods in the log layer of the channel flow, reconstructing the spanwise vorticity 
under conditions of inflow and outflow and determining its origin unambiguously. The existing numerical schemes 
for the stochastic Lagrangian approach are quite inefficient in the log layer, however, because the Monte Carlo 
sampling errors grow exponentially in time. New algorithmic approaches are probably therefore required.  

Finally, the insights that we have obtained in this work about Eulerian vorticity dynamics described 
by the Huggins vorticity flux tensor \eqref{Sigma} can be exploited to understand drag generation 
and reduction via the detailed Josephson-Anderson (JA) relation \citep{Huggins1970a,huggins1994vortex,eyink2008,eyink2021}.
Such work is already in progress \citep{kumar2023turbulent}. We have thus intentionally 
omitted in the present paper any discussion of the work of \cite{yoon2016contribution} which directly relates
velocity-vorticity correlations to mean drag by a version of the so-called FIK identity \citep{fukagata2002contribution}.
This discussion requires a careful comparison with the JA-relation, which will be done by \cite{kumar2023turbulent}. 
The connections between these two approaches is indeed not straightforward, e.g. down-gradient nonlinear vorticity 
flux produces drag in the JA-relation but reduces drag in the identity of \cite{yoon2016contribution}! 
Here we just mention the principal difference that, whereas the identity of \cite{yoon2016contribution} represents 
the mean drag in a Reynolds averaging approach, the JA-relation connects the drag {\it instantaneously} in time to the vorticity 
flux throughout the flow volume. The shift away from ensemble flow statistics to recognize the dynamical heterogeneity and 
intermittency of drag has proved important, for example, in the problem of polymer drag reduction 
\citep{xi2019turbulent}. Our results here shed new light on the latter problem, because 
they imply that drag can be reduced instantaneously either by decreasing the down-gradient flux of spanwise 
vorticity or by increasing the up-gradient flux, or both. This will also be the subject of future work. 

\vskip 0.1in


\vskip 0.1in

\noindent{\bf Acknowledgements.} We thank Professors Nigel Goldenfeld, Bj\"orn Hof, Yves Pomeau and Tamer Zaki for fruitful discussions and the JHTDB team for their help and support of the public numerical turbulence laboratory.

\vskip 0.1in

\noindent{\bf Funding:}
We thank also the Simons Foundation for support of this work through the Targeted Grant No. MPS-663054, ``Revisiting the Turbulence Problem Using Statistical Mechanics'', and the National Science Foundation (Grant \# CSSI-210387) for support of JHTDB and of pursuing analysis of wall-bounded turbulence. 

\vskip 0.1in

\noindent{\bf Declaration of interests.}  The authors report no conflict of interest.

\vskip 0.1in

\noindent{\bf Author ORCIDs.} 

S. Kumar \href{https://orcid.org/0000-0002-6785-0072}{https://orcid.org/0000-0002-6785-0072}.

C. Meneveau \href{https://orcid.org/0000-0001-6947-3605}{https://orcid.org/0000-0001-6947-3605}.

G. Eyink \href{https://orcid.org/0000-0002-8656-7512}{https://orcid.org/0000-0002-8656-7512}.

\bibliographystyle{jfm}
\bibliography{jfm}
\newpage

\begin{center} 
{\large {\bf SUPPLEMENTARY MATERIALS}} 
\end{center}


\section{Quadrant contributions}\label{sec:quad} 
\begin{figure}
      \centering
     \begin{subfigure}[b]{0.32\textwidth}
         \centering
         \includegraphics[width=\textwidth]{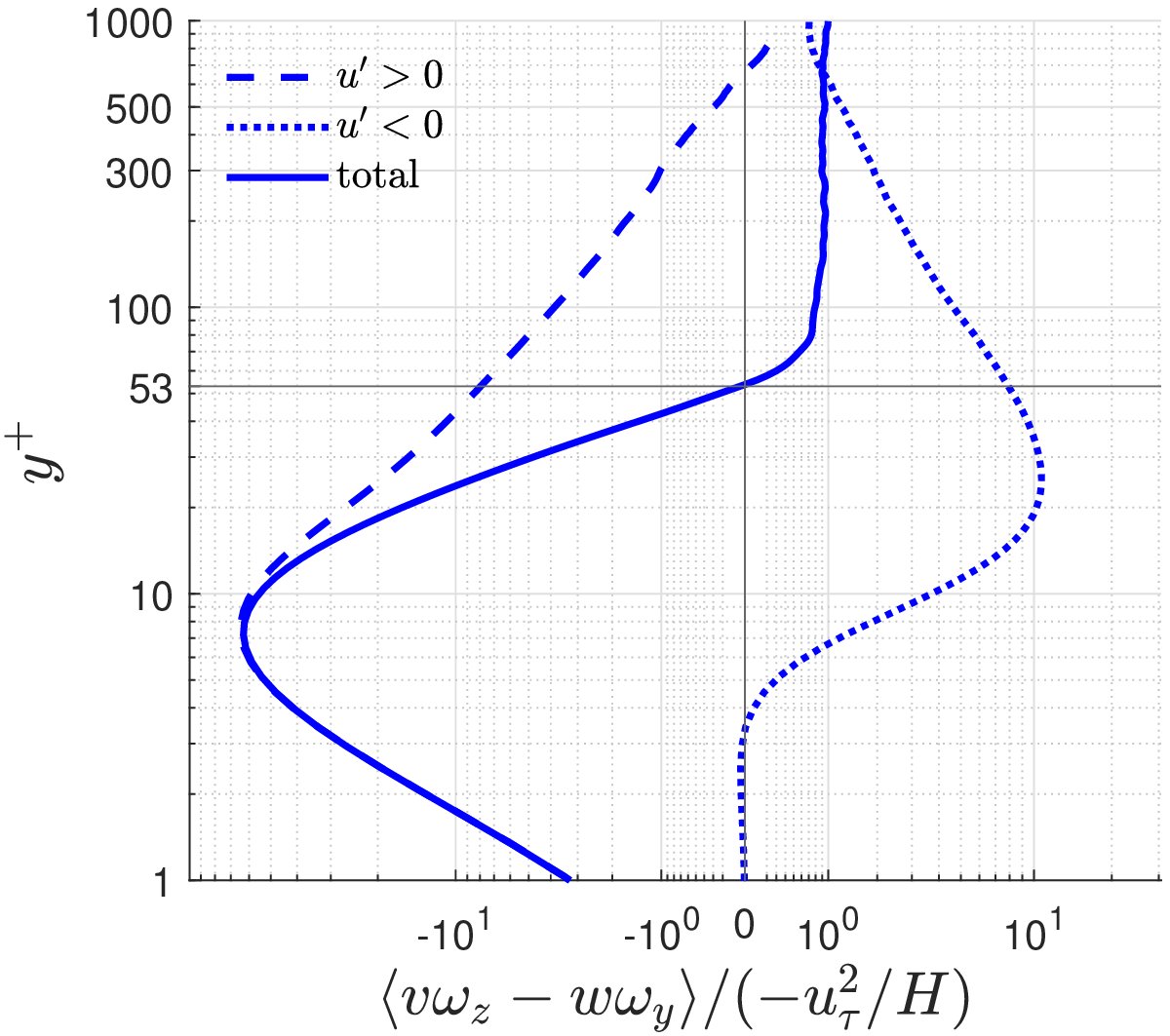}
        \caption{}
         \label{quad_nonlinear_u}
     \end{subfigure}
     \hfill
     \begin{subfigure}[b]{0.32\textwidth}
         \centering
         \includegraphics[width=\textwidth]{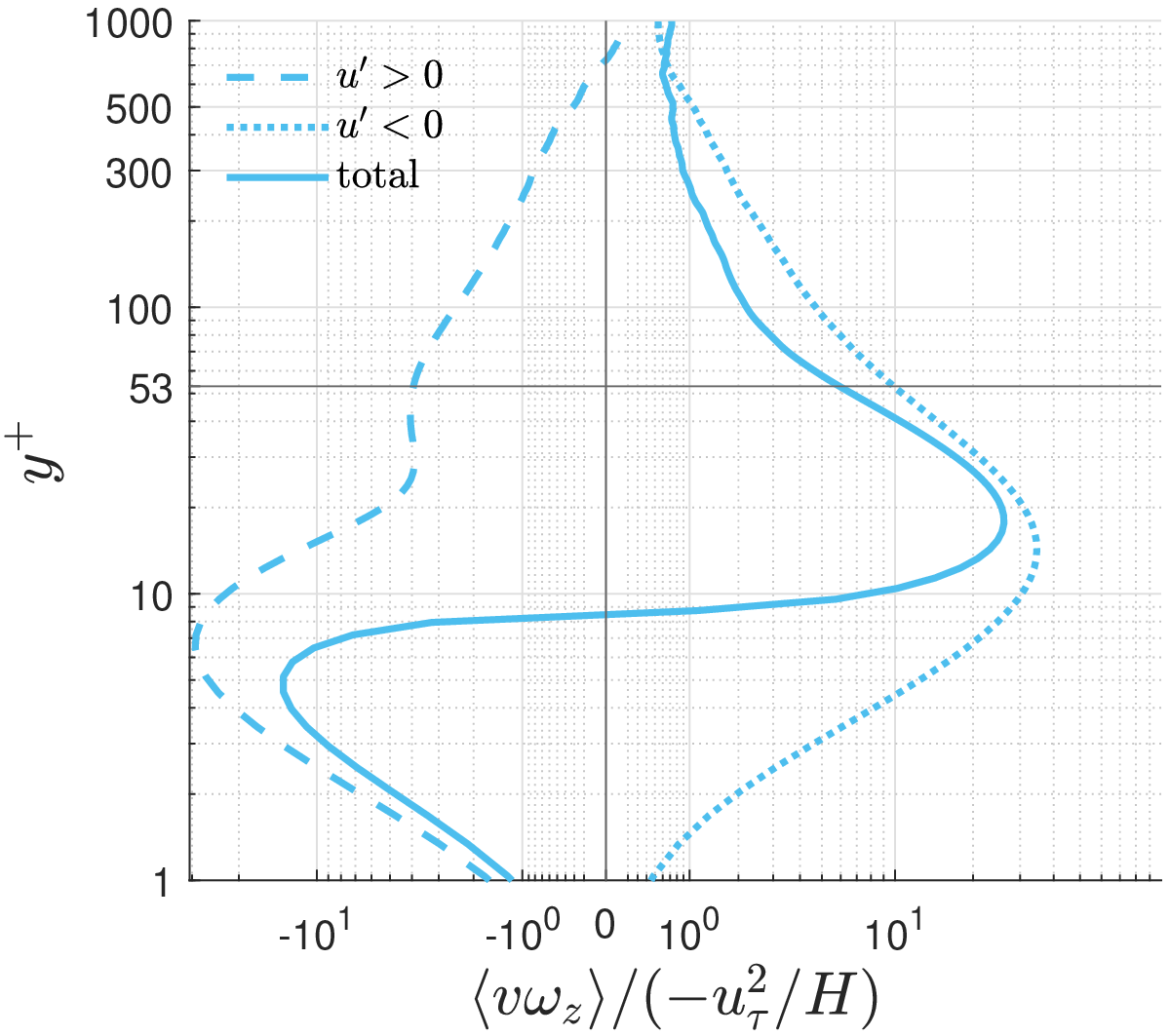}
         \caption{}
         \label{quad_voz_u}
     \end{subfigure}
     \hfill
     \begin{subfigure}[b]{0.32\textwidth}
         \centering
         \includegraphics[width=\textwidth]{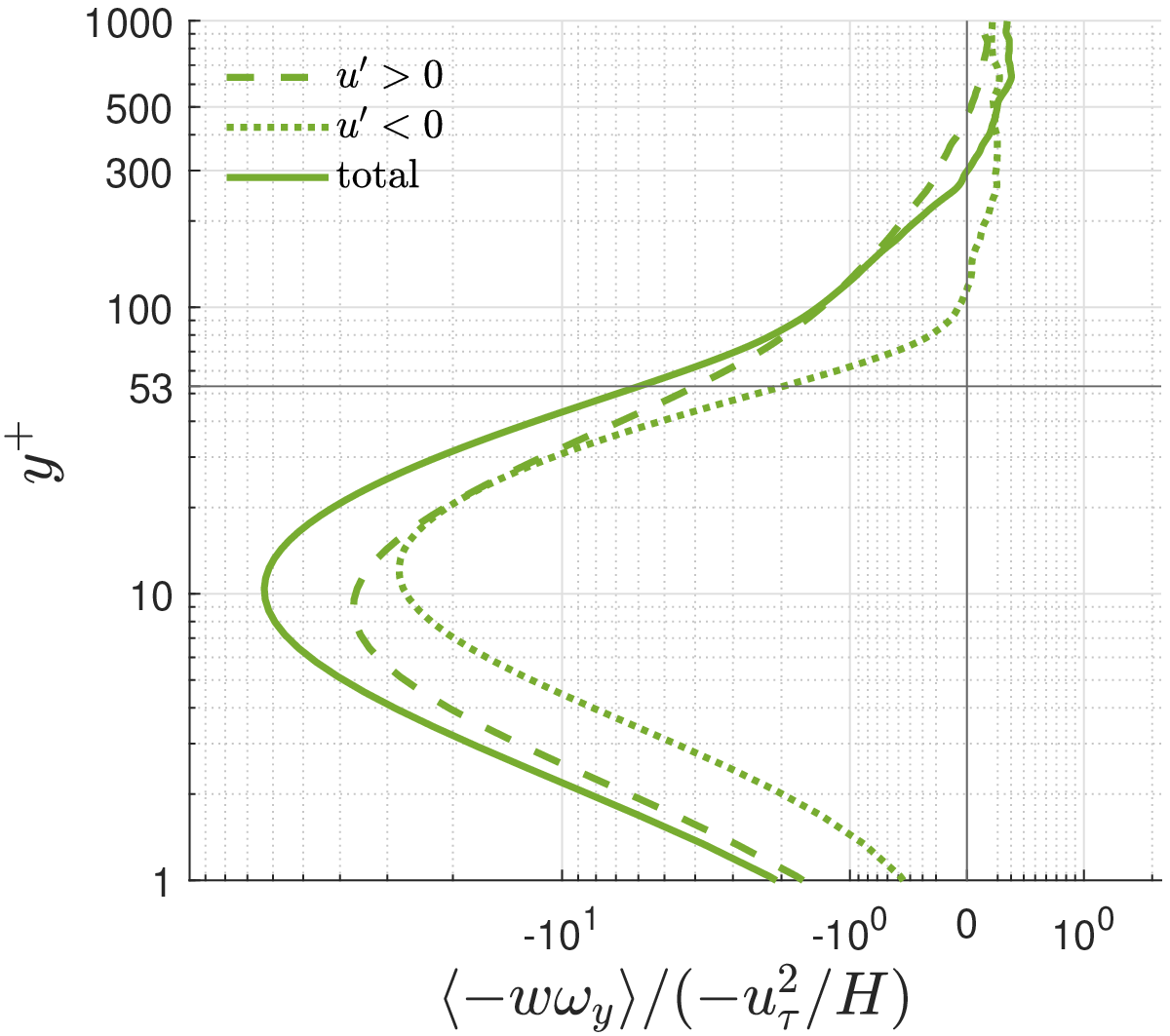}
         \caption{}
         \label{quad_woy_u}
     \end{subfigure}
        \caption{Contributions from high speed streaks $(u'>0)$ and low speed streaks $(u'<0)$, to the  (a) nonlinear flux, (b) convection/advection and (c) stretching/tilting, averaged over time and wall parallel planes, plotted as a function of wall distance.   }
        \label{streak_contri}
\end{figure}

Partial averages of the flux terms $v\omega_z -w\omega_y$, $v\omega_z$ and $-w\omega_y$ conditioned on ``low speed ($u'<0$)" and ``high speed ($u'>0$)" events are shown in Fig~\ref{quad_nonlinear_u}, ~\ref{quad_voz_u} and \ref{quad_woy_u} respectively. The plot in Fig \ref{quad_woy_u} shows that the stretching/tilting term ($-w\omega_y$) is agnostic to the sign of $u'$ for $y^+\lesssim100$, where both low speed and high speed streaks produce up-gradient contributions. For $100\lesssim y^+\lesssim500$ low speed streaks make down-gradient contributions while high speed streaks make up-gradient contributions to this stretching term. Close to the centerline ($y^+\gtrsim500$), both contributions are down-gradient.
The convective term (shown in Fig ~\ref{quad_voz_u}), on the other hand, shows strongly opposing behaviours for low speed and high speed streaks across nearly the entire channel (for $y^+ \lesssim 700$), with low speed streaks making  down-gradient contributions but high speed streaks up-gradient contributions. 
By contrast, both contributions to the convective flux are down-gradient close to the centerline ($y^+\gtrsim 700$).
The total nonlinear flux (shown in Fig ~\ref{quad_nonlinear_u}), is dominated by the convective term and behaves similarly across most of the channel ($5 \lesssim y^+\lesssim700$), with low-speed streaks being down-gradient and high-speed streaks being up-gradient. Within the viscous sublayer($y^+\lesssim5$), low speed streaks make no contributions to the flux and the entire flux is due to high speed streaks. Close to the centerline ($y^+\gtrsim 700$) both contributions are down-gradient. The observed correlations of the separate 
flux terms with $u'$ are plausibly explained as a consequence of the primary correlation with $v'$ due to Lighthill's mechanism and the secondary correlation of $v'$ with $u'.$ 

\begin{figure}
      \centering
     \begin{subfigure}[b]{0.32\textwidth}
         \centering
         \includegraphics[width=\textwidth]{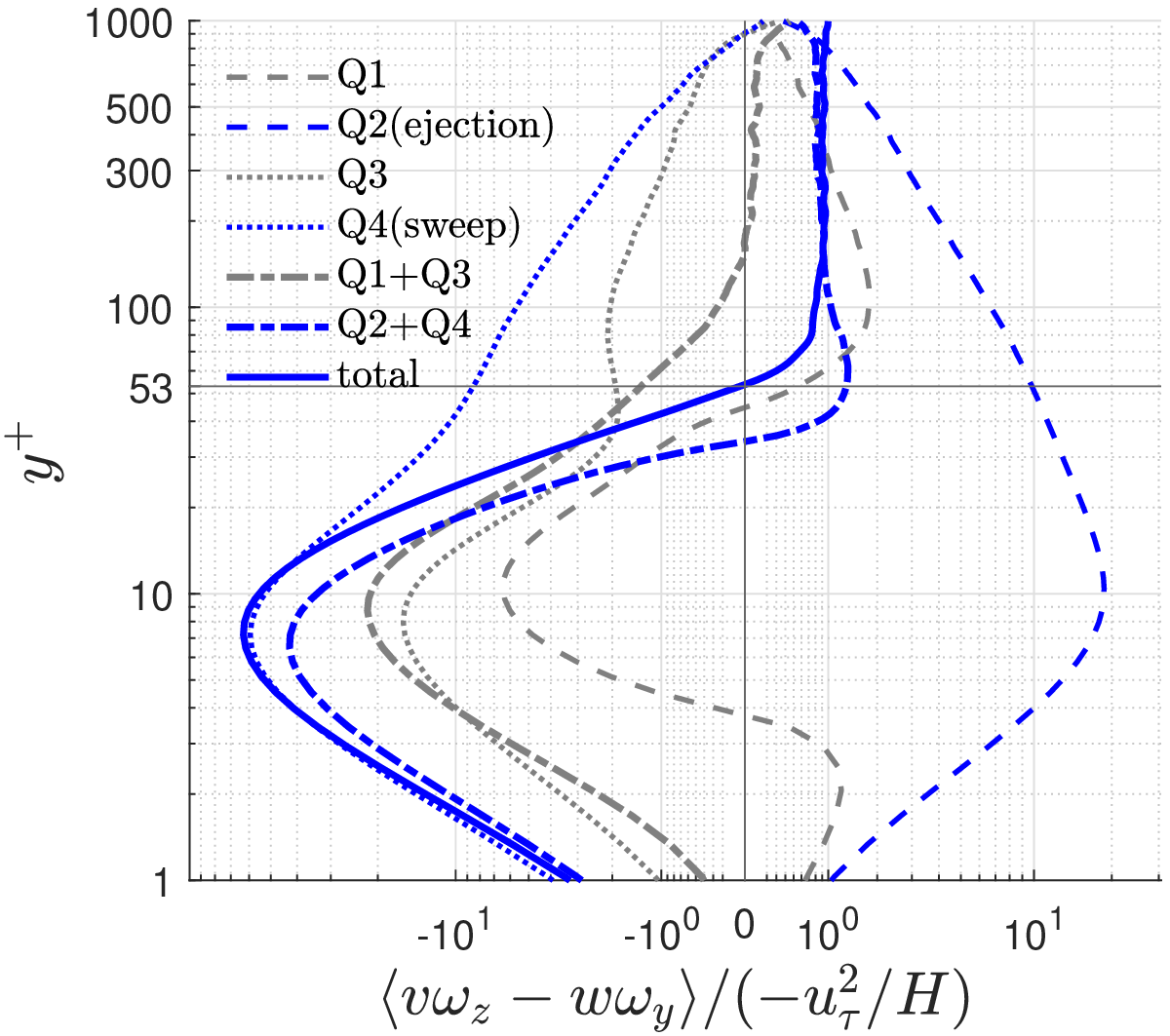}
        \caption{}
         \label{quad_nonlinear}
     \end{subfigure}
     \hfill
     \begin{subfigure}[b]{0.32\textwidth}
         \centering
         \includegraphics[width=\textwidth]{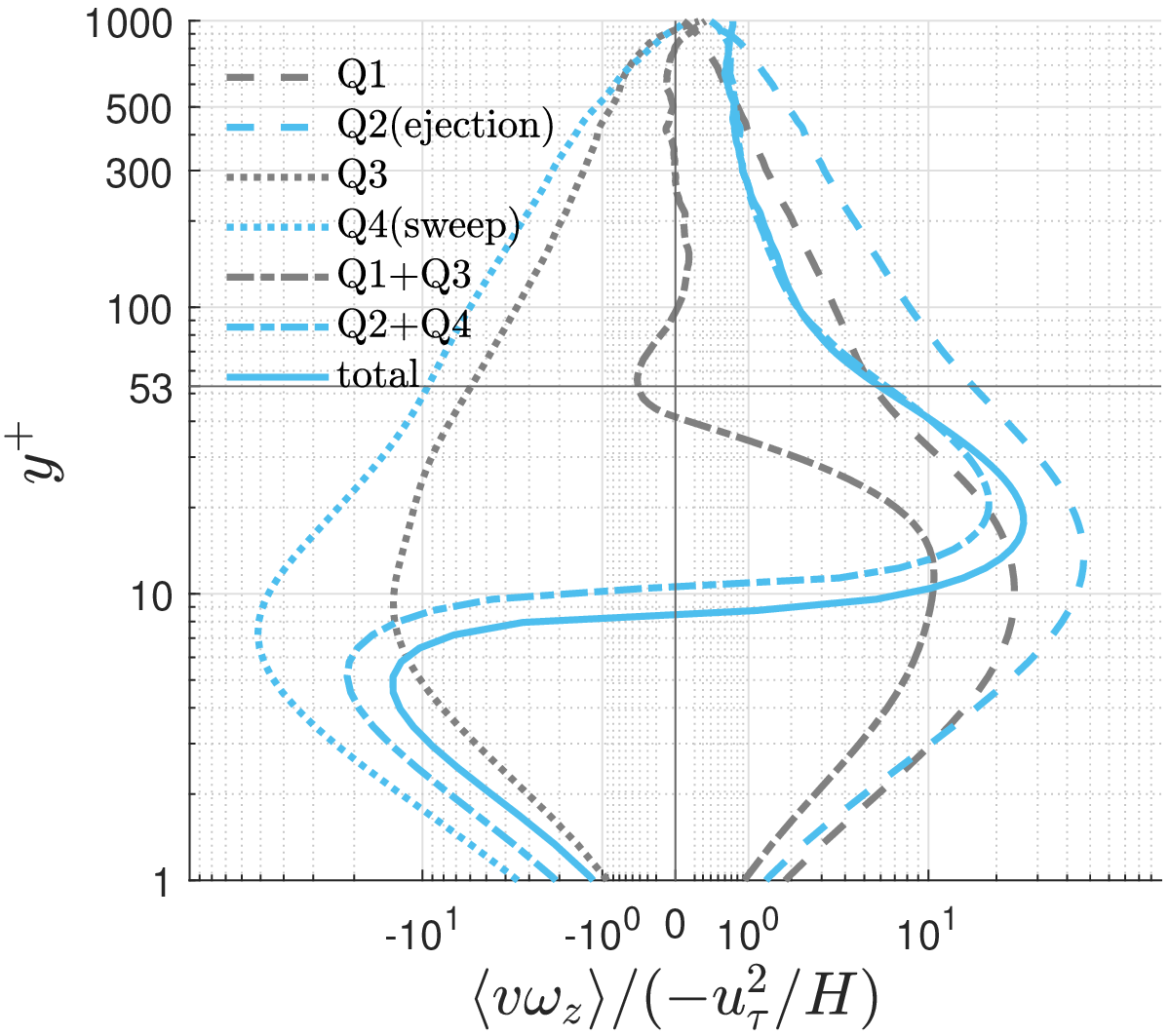}
         \caption{}
         \label{quad_voz}
     \end{subfigure}
     \hfill
     \begin{subfigure}[b]{0.32\textwidth}
         \centering
         \includegraphics[width=\textwidth]{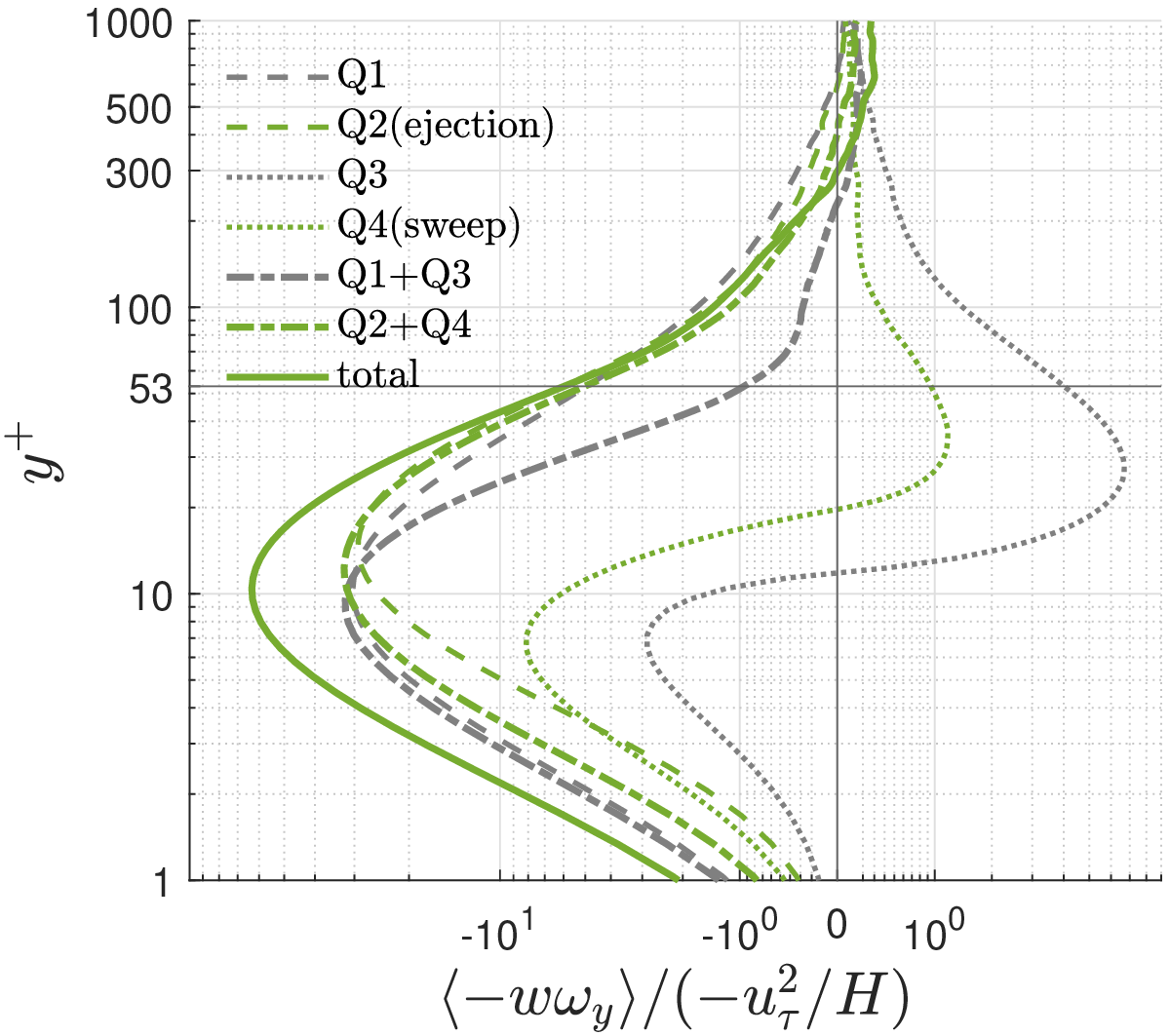}
         \caption{}
         \label{quad_woy}
     \end{subfigure}
        \caption{Contributions from quadrants to the nonlinear flux (a), convection/advection (b) and stretching/tilting (c), averaged over time and wall parallel planes, plotted as a function of wall distance.   }
        \label{quad_contri}
\end{figure}

This idea is illuminated by the quadrant correlations, discussed next.  
The contributions from the four individual quadrants of the $u'-v'$ plane (see ~\cite{pope2000turbulent}) are shown for  the total nonlinear flux (Fig ~\ref{quad_nonlinear}), the convection/advection term (Fig ~\ref{quad_voz}) and the stretching/tilting term (Fig ~\ref{quad_woy}). Contributions from ``active (Q2+Q4)" and ``inactive (Q1 +Q3)" motions are plotted as well. The latter show that active motions contribute nearly the entire flux for the convective term, while inactive motions make a much a smaller contribution (Fig ~\ref{quad_voz}). The stretching/tilting term is nearly agnostic to active/inactive motions for $y^+\lesssim 30$ but also dominated by active motions for $y^+\gtrsim 30$
(Fig ~\ref{quad_woy}). On the whole, the net nonlinear flux (Fig ~\ref{quad_nonlinear}) is dominated by contributions from active motions, with inactive motions making a decidedly smaller contribution, and this effect is mainly through the convection term. These observations are consistent with our explanation above that the observed correlations 
of the flux contributions are due to the primary correlation with $v'$ and the strong anti-correlation between 
$u'$ and $v'$ in $Q2+Q4$

Further evidence for this picture is provided by the separate quadrant contributions. From 
Fig ~\ref{quad_voz} for the convective term it may be seen 
that Q1 and Q2 where $v'>0$ both make down-gradient contributions,
while Q3 and Q4 where $v'<0$ both make up-gradient contributions across the entire channel. 
On the other hand, the stretching/tilting term in Fig ~\ref{quad_woy} exhibits opposite flux directions across most of the channel, with Q1 and Q2 up-gradient and Q3 and Q4 down-gradient. Furthermore, for both convection and stretching terms, the $Q1$ correlations while similar to the $Q2$ correlations are smaller in magnitude,and likewise the $Q3$ correlations while similar to the $Q4$ correlations are smaller. This suggests again that the primary correlation is with $v',$ but that the dominant contribution 
arises from the ``active'' quadrants
$Q2+Q4$ where $u'$ and $v'$ are anti-correlated.  

Altogether, these results support our claim that the correlation most relevant to the physics is that between the flux and regions of outflow ($v'>0$) and inflow ($v'<0$), as shown in Fig ~\ref{quad_contri_v} in the main text. 
The dominance of the ``active'' regions produces a secondary correlation of vorticity flux with $u'.$

We note that contributions to vorticity flux from the four quadrants $Q1-Q4$ were calculated previously by ~\cite{vidal_etal_2018}, but 
for duct flow with sidewalls (both straight and curved) at two constant $z$ planes. We cannot compare our results 
with theirs, not only because of the differences in the simulated flows but also because they considered 
products of fluctuating terms $v'\omega_z'$ and $w'\omega_y'.$ Since $w\omega_y=w'\omega_y',$ our results 
for this term agree well with theirs for $z$ away from sidewalls, but our results for $v\omega_z$ differ
considerably from theirs for $v'\omega_z'.$

\newpage 


\clearpage 

\section{Comparison with data from ~\cite{delalamo2004scaling}}
\begin{figure}
\centering
     \begin{subfigure}[b]{0.75\textwidth}
         \centering
         \includegraphics[width=\textwidth]{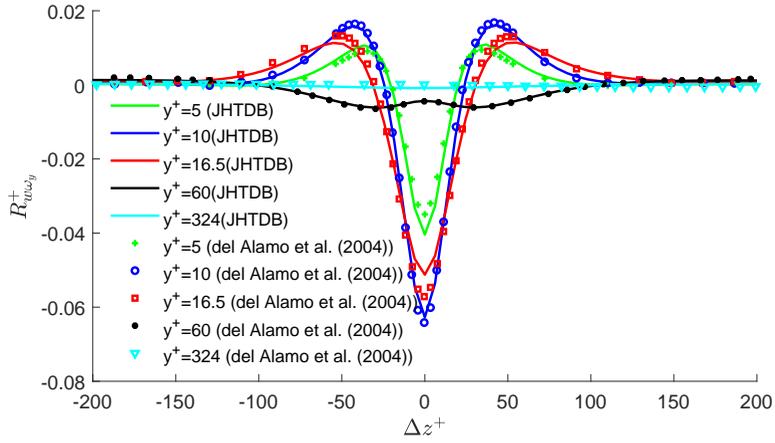}
        \caption{}
         \label{corwoy}
     \end{subfigure}
     \vfill
     \begin{subfigure}[b]{0.75\textwidth}
         \centering
         \includegraphics[width=\textwidth]{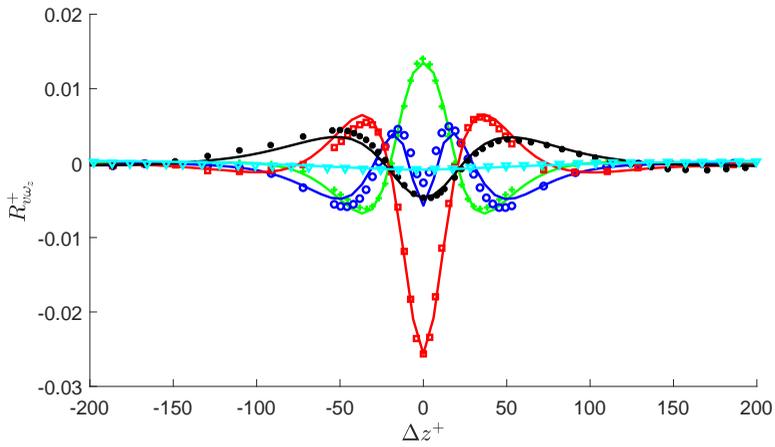}
         \caption{}
         \label{corvoz}
     \end{subfigure}
        \caption{Spanwise two-point correlation of (a) spanwise velocity and wall normal vorticity $(R_{w\omega_y}^+)$, (b) spanwise vorticity and wall normal velocity $(R_{v\omega_z}^+)$, computed from channel flow data at $Re_{\tau}=1000$ from JHTDB(~\cite{jhtdb_channel}) and from earlier simulation data of channel flow at $Re_{\tau}=934$ by ~\cite{delalamo2004scaling} reported in ~\cite{monty2011characteristics}.}
        \label{cor_comp}
\end{figure}

We compare the spanwise two point velocity-vorticity correlations computed from channel flow data at $Re_{\tau}=1000$ from the Johns Hopkins Turbulence Database ~\cite{jhtdb1,jhtdb_channel} and at $Re_{\tau}=934 $ from ~\cite{delalamo2004scaling} reported in ~\cite{monty2011characteristics} in Fig~
\ref{cor_comp}. The correlations are related to the respective spanwise co-spectra as follows:
\begin{align}
R_{w\omega_y}^+(\Delta z)&=\frac{R_{w\omega_y}(\Delta z)}{u_{\tau}^2/\delta_{\nu}}=\frac{1}{u_{\tau}^2/\delta_{\nu}}\int_0^{\infty} \phi_{w\omega_y}(k_z)e^{ik_z \Delta z}dk_z\\
R_{v\omega_z}^+(\Delta z)&=\frac{R_{v\omega_z}(\Delta z)}{u_{\tau}^2/\delta_{\nu}}=\frac{1}{u_{\tau}^2/\delta_{\nu}}\int_0^{\infty} \phi_{v\omega_z}(k_z)e^{ik_z \Delta z}dk_z
\end{align}
We observe good agreement between correlations from both datasets.

\newpage 

\clearpage

\section{Velocity-vorticity co-spectra }\label{spec2d}
The co-spectrum of nonlinear flux is given by $\phi_{v\omega_z}-\phi_{w\omega_y}$ with the spanwise co-spectra shown in Fig~\ref{fig_1dcospec_z} and the streamwise cospectrum in Fig.~\ref{fig_1dcospec} of the main text. 
{The latter streamwise ``net force spectra'' have been the subject of detailed study in prior works of ~\cite{guala_hommema_adrian_2006,balakumar_adrian_2007,wu_baltzer_adrian_2012}. The wall-normal derivative of the Reynolds shear stress is characterized in these works as producing retardation of the mean flow above $y_p$ and acceleration below, associated with a negative and a positive sign respectively. This retarding force is produced by a down-gradient flux of spanwise vorticity while an accelerating force results from an up-gradient flux, as discussed in Section ~\ref{intro}. 
The detailed study by ~\cite{wu_baltzer_adrian_2012} found large positive (accelerating) values for the streamwise net force spectrum concentrated below $y^+=20$, and observed that below the top of the buffer layer (at $y^+=30$), all scales except the very smallest ($\lambda_x<0.15R$, $R^+=685$) accelerate the mean flow (or contribute an up-gradient flux). Conversely, for $y>0.2R$, they found negative (decelerating or contributing a down-gradient flux) values for all scales. In the wall-normal region where $y^+>20$ and $y<0.2R$, they found a complicated $y$ variation of the spectra with negative (decelerating or with a down-gradient flux) values sandwiched between positive (decelerating or with an up-gradient flux) values, each occupying a varying range of scales. These observations mirror our own, as illustrated particularly by our Fig~\ref{fig_1dcospec} }. 

In this section, we look at the constituent co-spectra, i.e., $\phi_{v\omega_z}$ and $-\phi_{w\omega_y},$
both spanwise and streamwise. All of the mean features of these 1D spectra can be inferred from the corresponding 
2D cospectra plotted in Section \ref{sec:separate2D}. However, we present the 1D cospectra here for completeness.  


\newpage 

\begin{figure}
     \centering
     \begin{subfigure}[b]{0.32\textwidth}
         \centering
         \includegraphics[width=\textwidth]{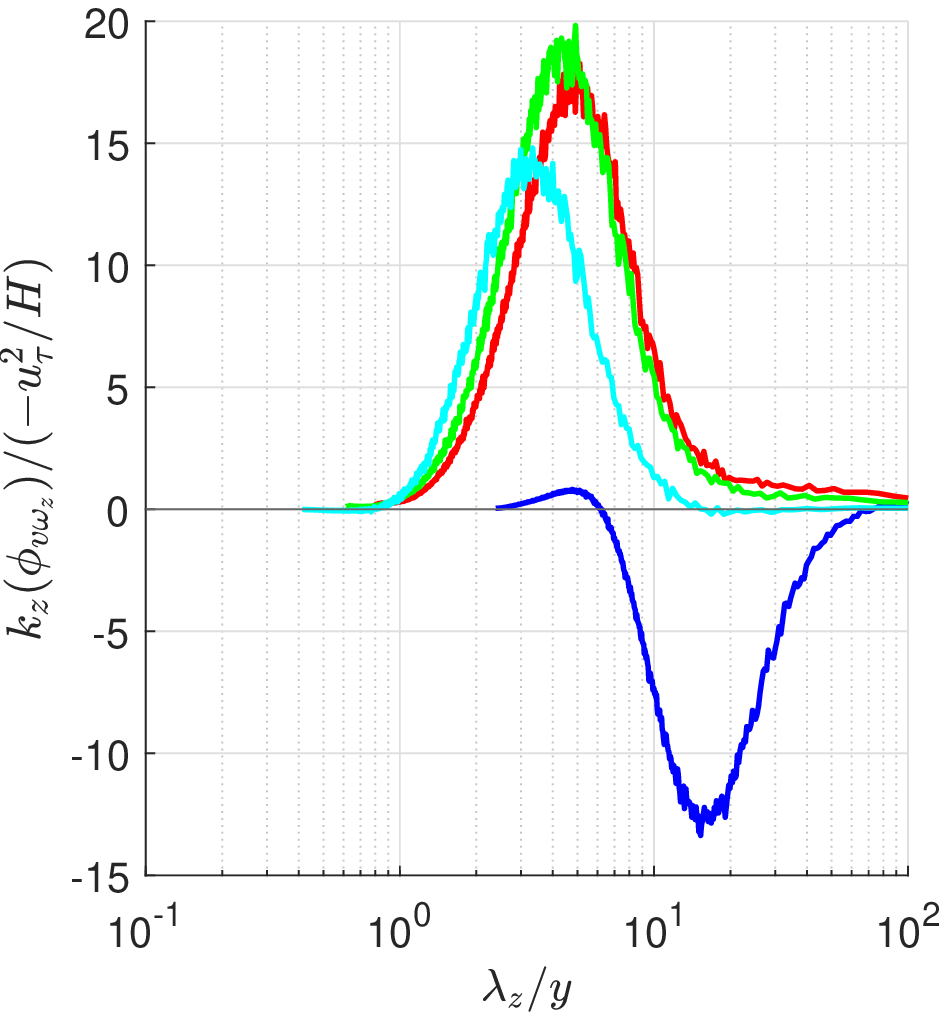}
        \caption{Spanwise co-spectra: $\mbox{Viscous \& Buffer Layers}$}
         \label{fig_bufferzvoz}
     \end{subfigure}
     \hfill
     \begin{subfigure}[b]{0.32\textwidth}
         \centering
         \includegraphics[width=\textwidth]{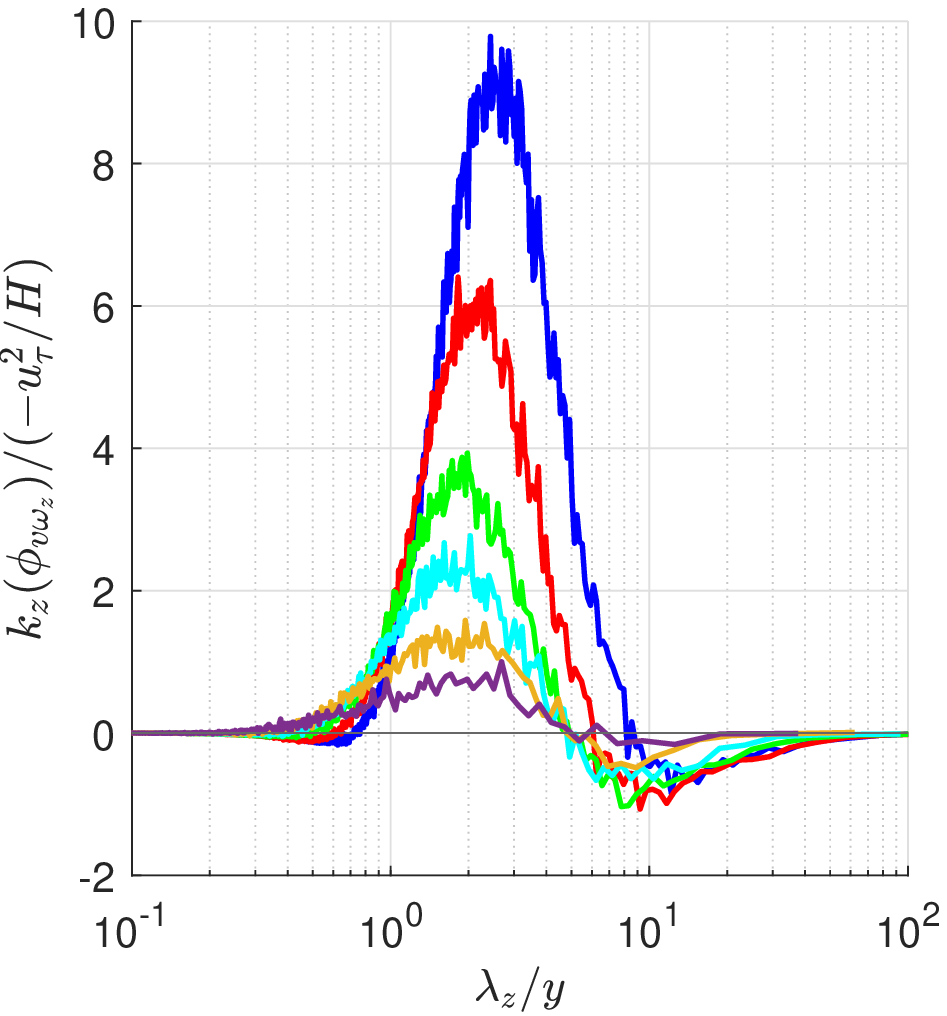}
         \caption{Spanwise co-spectra: Log Layer}
         \label{fig_inertialzvoz}
     \end{subfigure}
     \hfill
     \begin{subfigure}[b]{0.32\textwidth}
         \centering
         \includegraphics[width=\textwidth]{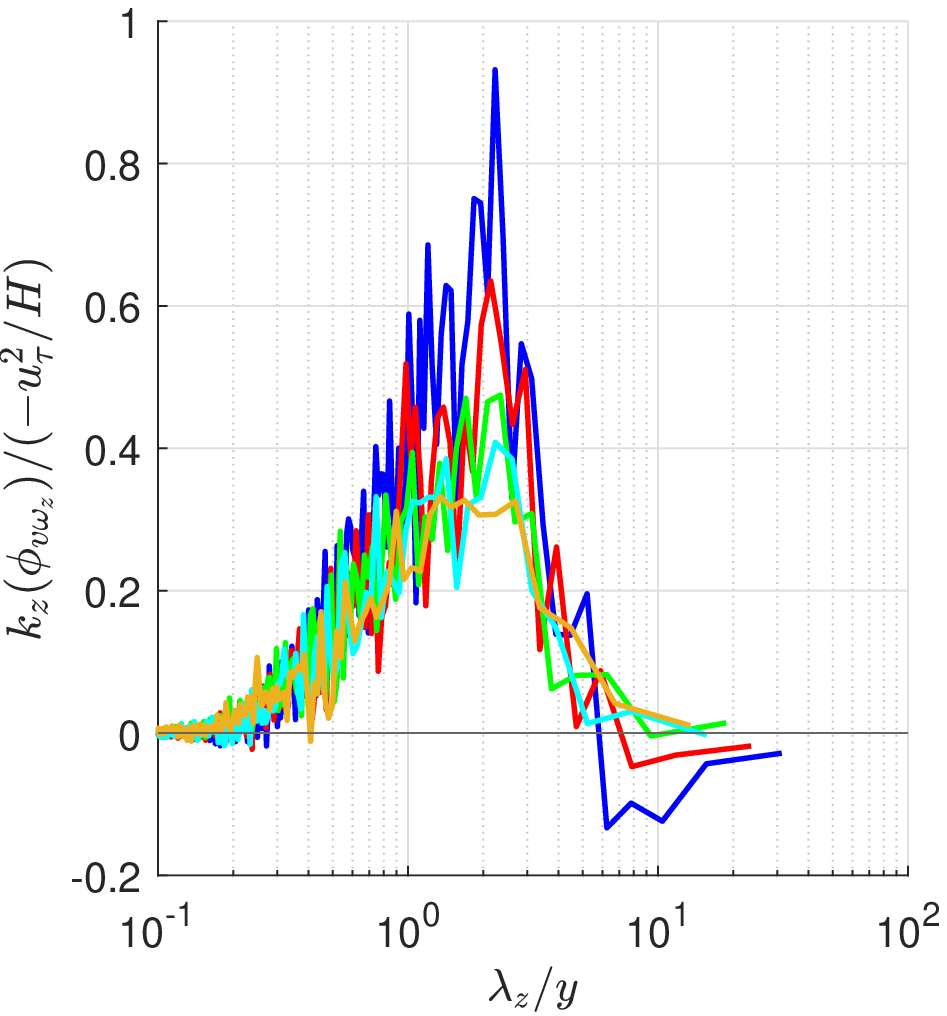}
         \caption{Spanwise co-spectra: Outer Layer}
         \label{fig_outerzvoz}
     \end{subfigure}
             \caption{Normalized spanwise cospectra of wall normal velocity- spanwise vorticity ($\phi_{v\omega_z} $), in the (a) viscous \& buffer layers, (b) log layer and (c) outer layer. Curves have the same meaning as in corresponding plots in Fig~\ref{fig_1dcospec_z}.}
        \label{fig_1dcospec_voz}
\end{figure}
\begin{figure}
     \centering
     \begin{subfigure}[b]{0.32\textwidth}
         \centering
         \includegraphics[width=\textwidth]{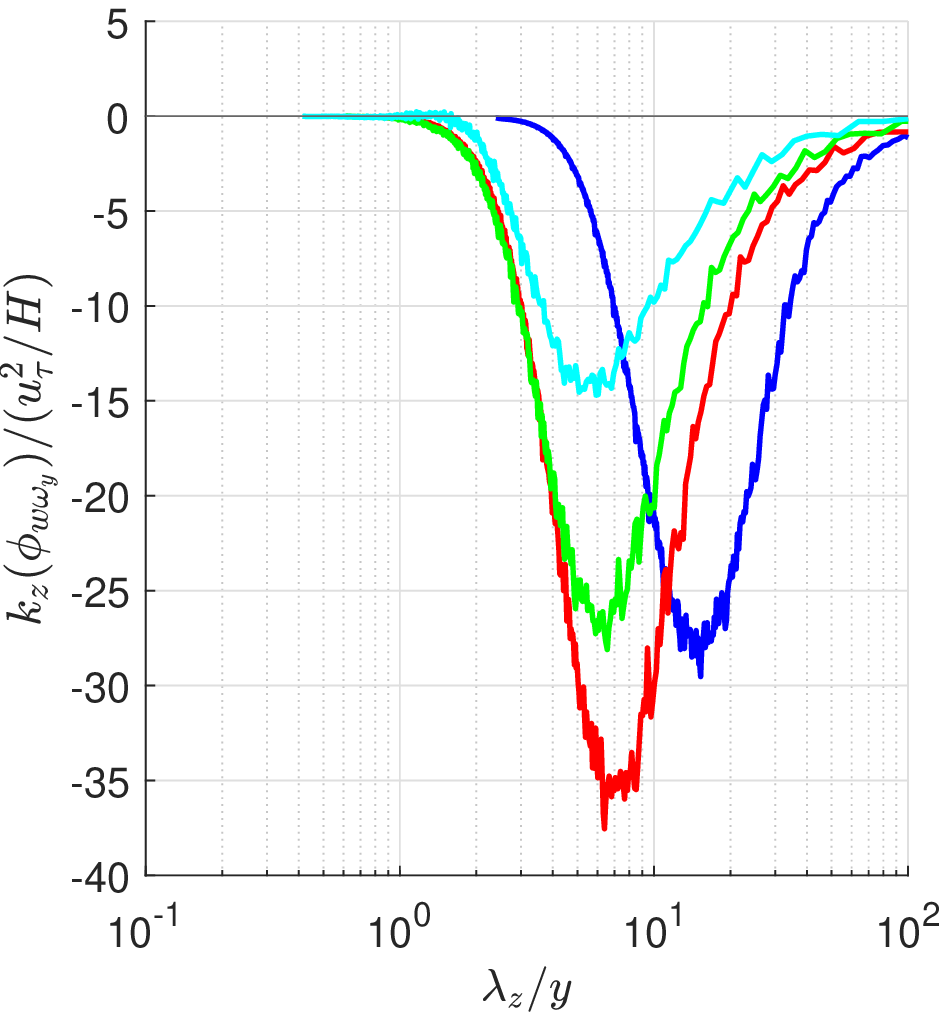}
        \caption{Spanwise co-spectra: $\mbox{Viscous \& Buffer Layer}$}
         \label{fig_bufferzwoy}
     \end{subfigure}
     \hfill
     \begin{subfigure}[b]{0.32\textwidth}
         \centering
         \includegraphics[width=\textwidth]{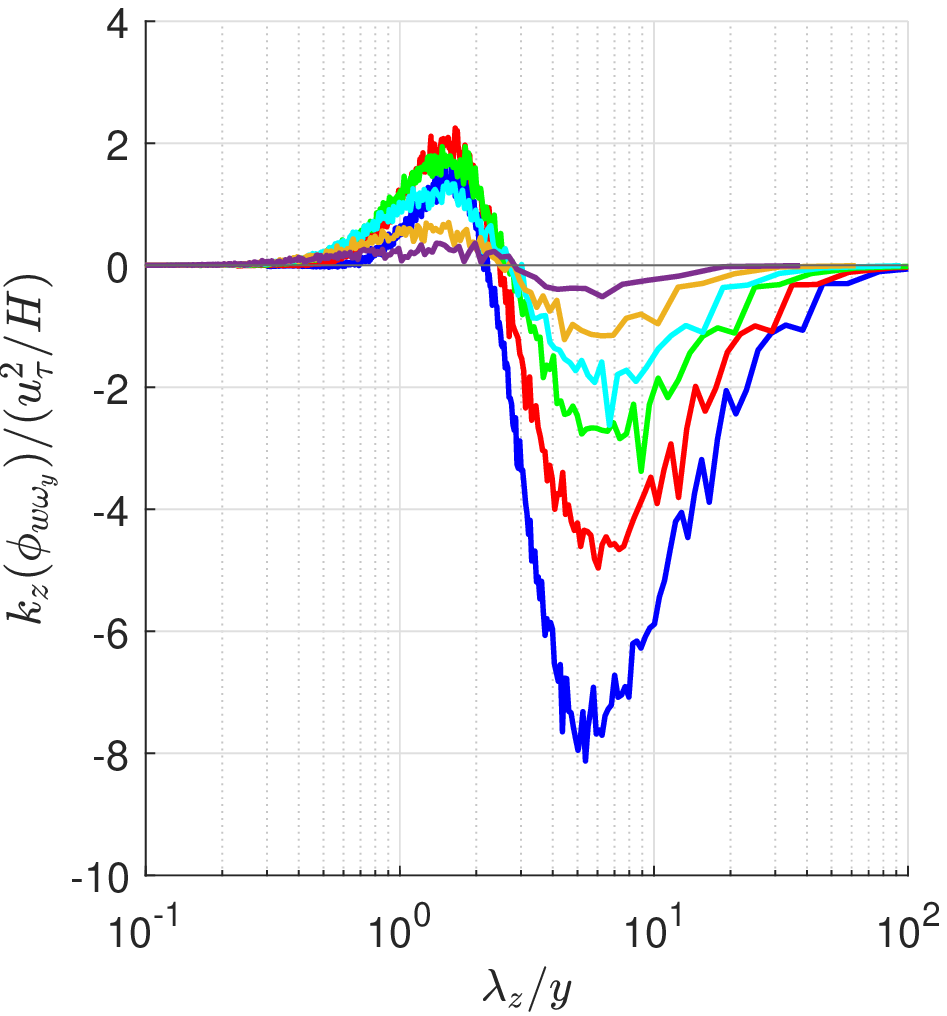}
         \caption{Spanwise co-spectra: Log Layer}
         \label{fig_inertialzwoy}
     \end{subfigure}
     \hfill
     \begin{subfigure}[b]{0.32\textwidth}
         \centering
         \includegraphics[width=\textwidth]{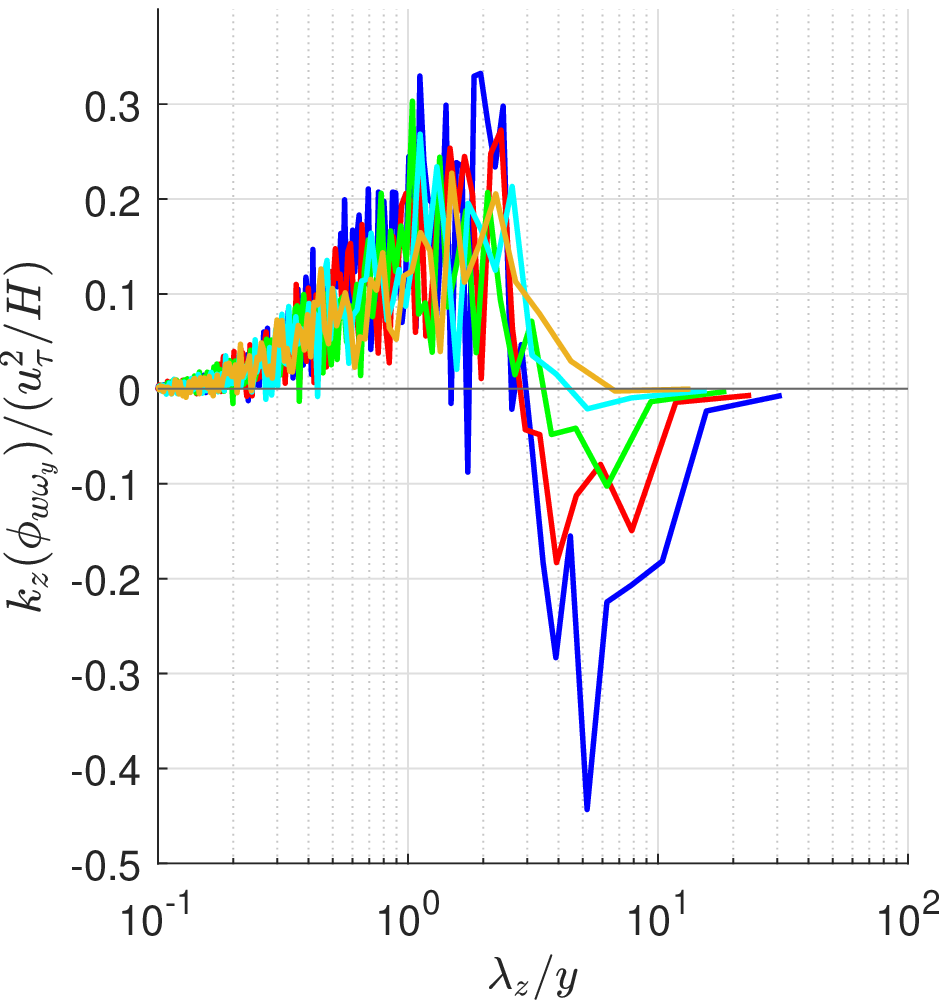}
         \caption{Spanwise co-spectra: Outer Layer}
         \label{fig_outerzwoy}
     \end{subfigure}
             \caption{Normalized spanwise cospectra of (negative of) the spanwise velocity - wall normal vorticity ($-\phi_{w\omega_y}$), in the (a) viscous \& buffer layers, (b) log layer and (c) outer layer. Curves have the same meaning as in corresponding plots in Fig~\ref{fig_1dcospec_z}.}
        \label{fig_1dcospec_woy}
\end{figure}
\begin{figure}
     \centering
     \begin{subfigure}[b]{0.32\textwidth}
         \centering
         \includegraphics[width=\textwidth]{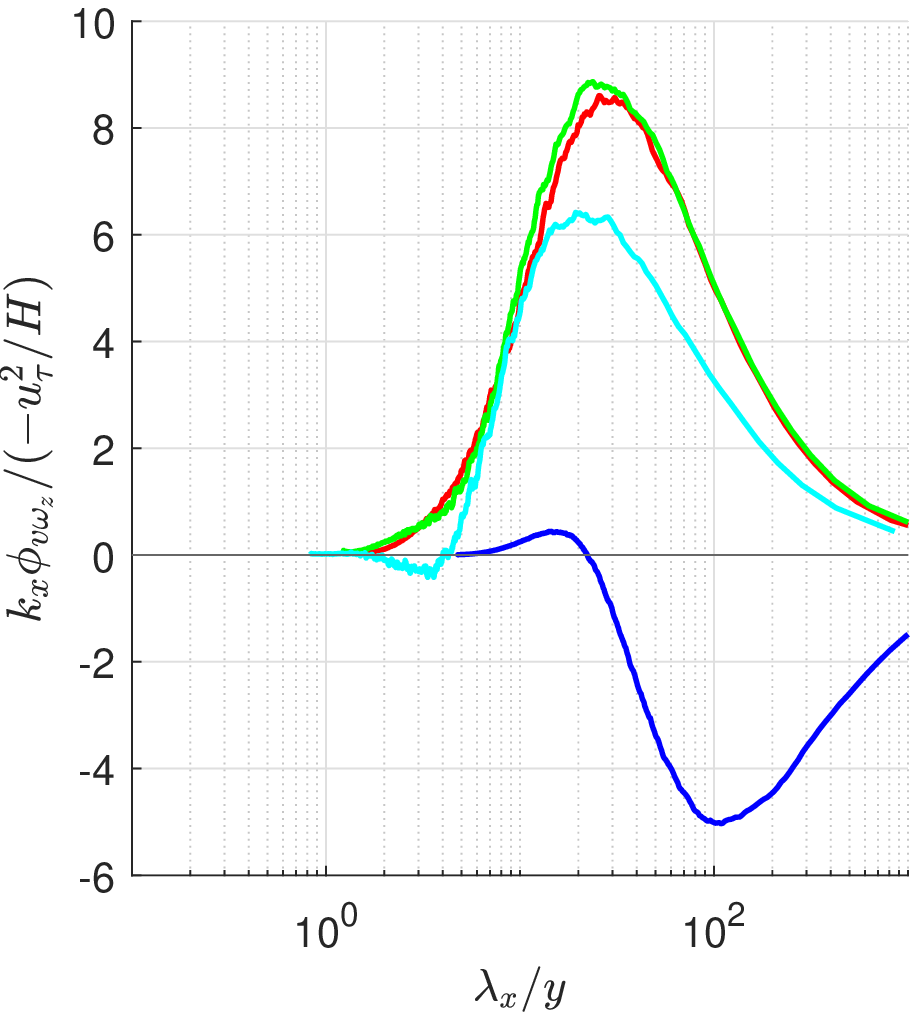}
        \caption{Streamwise co-spectra: $\mbox{Viscous \& Buffer Layers}$}
         \label{fig_bufferxvoz}
     \end{subfigure}
     \hfill
     \begin{subfigure}[b]{0.32\textwidth}
         \centering
         \includegraphics[width=\textwidth]{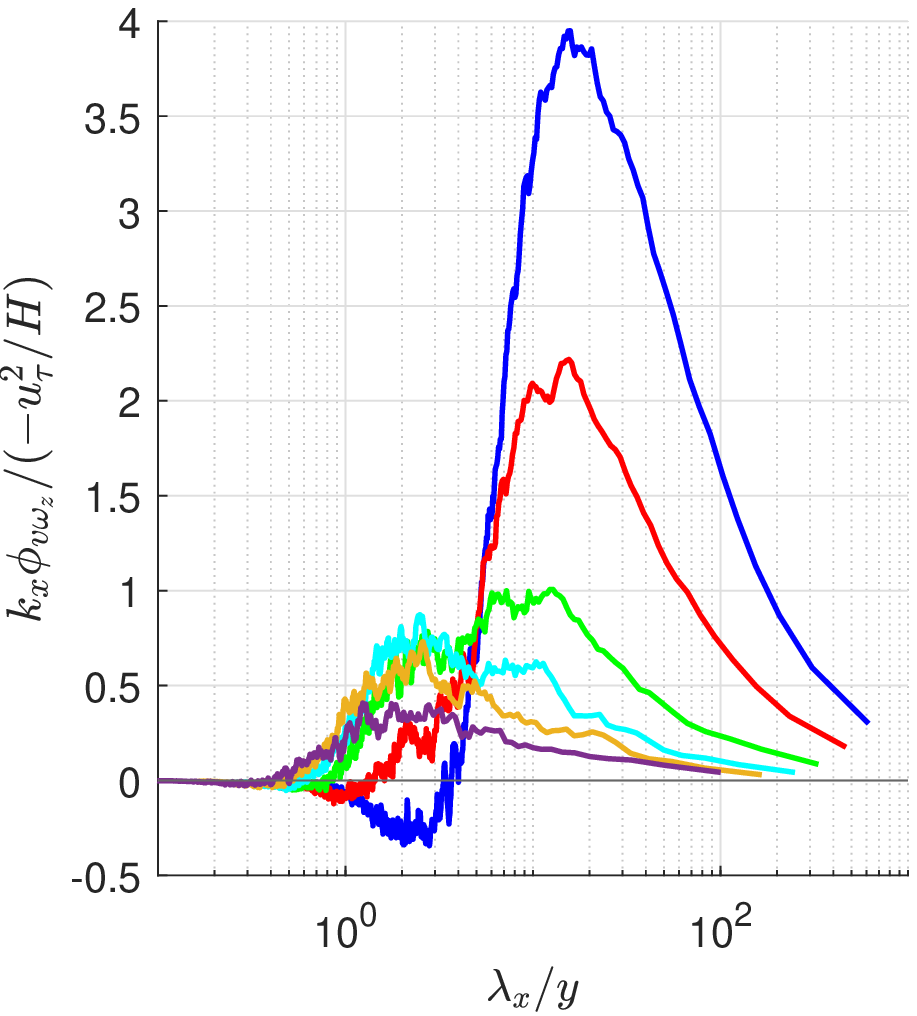}
         \caption{Streamwise co-spectra: Log Layer}
         \label{fig_inertialxvoz}
     \end{subfigure}
     \hfill
     \begin{subfigure}[b]{0.32\textwidth}
         \centering
         \includegraphics[width=\textwidth]{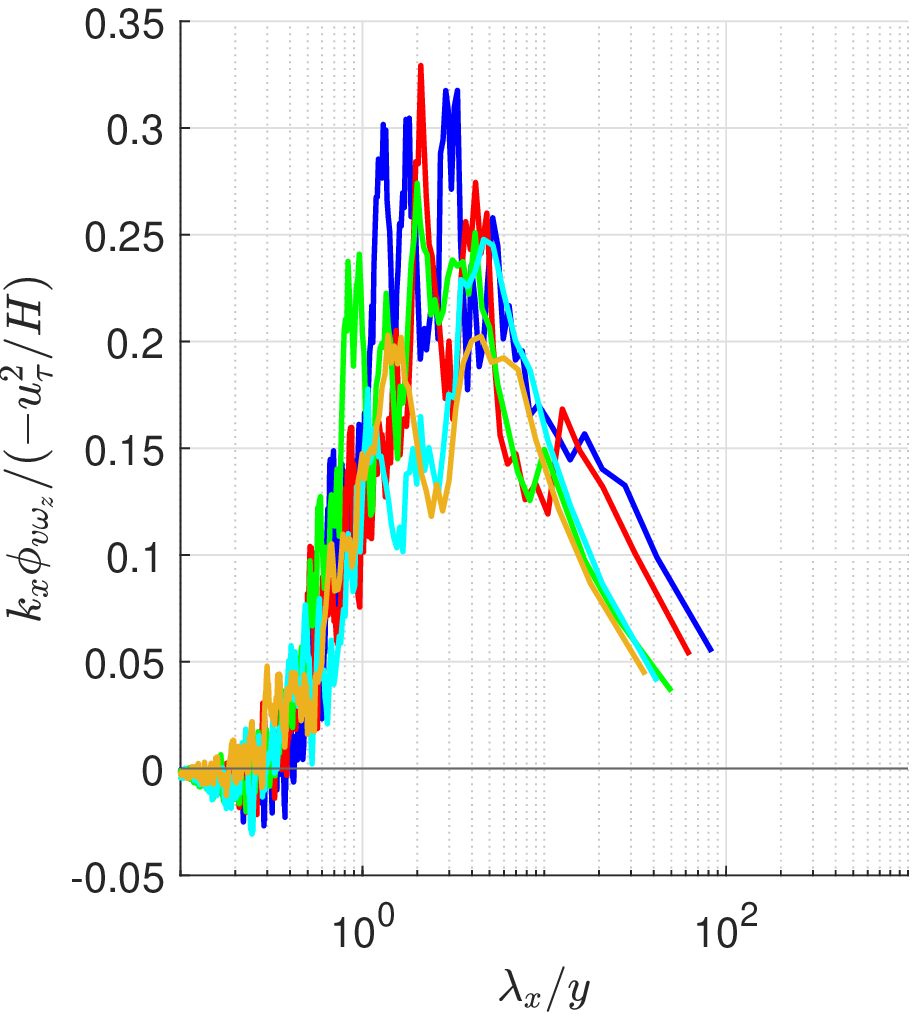}
         \caption{Streamwise co-spectra: Outer Layer}
         \label{fig_outerxvoz}
     \end{subfigure}
             \caption{Normalized streamwise cospectra of wall normal velocity- spanwise vorticity ($\phi_{v\omega_z} $), in the (a) viscous \& buffer layers, (b) log layer and (c) outer layer. Curves have the same meaning as in corresponding plots in Fig~\ref{fig_1dcospec_z}.}
        \label{fig_1dcospecx_voz}
\end{figure}
\begin{figure}
     \centering
     \begin{subfigure}[b]{0.32\textwidth}
         \centering
         \includegraphics[width=\textwidth]{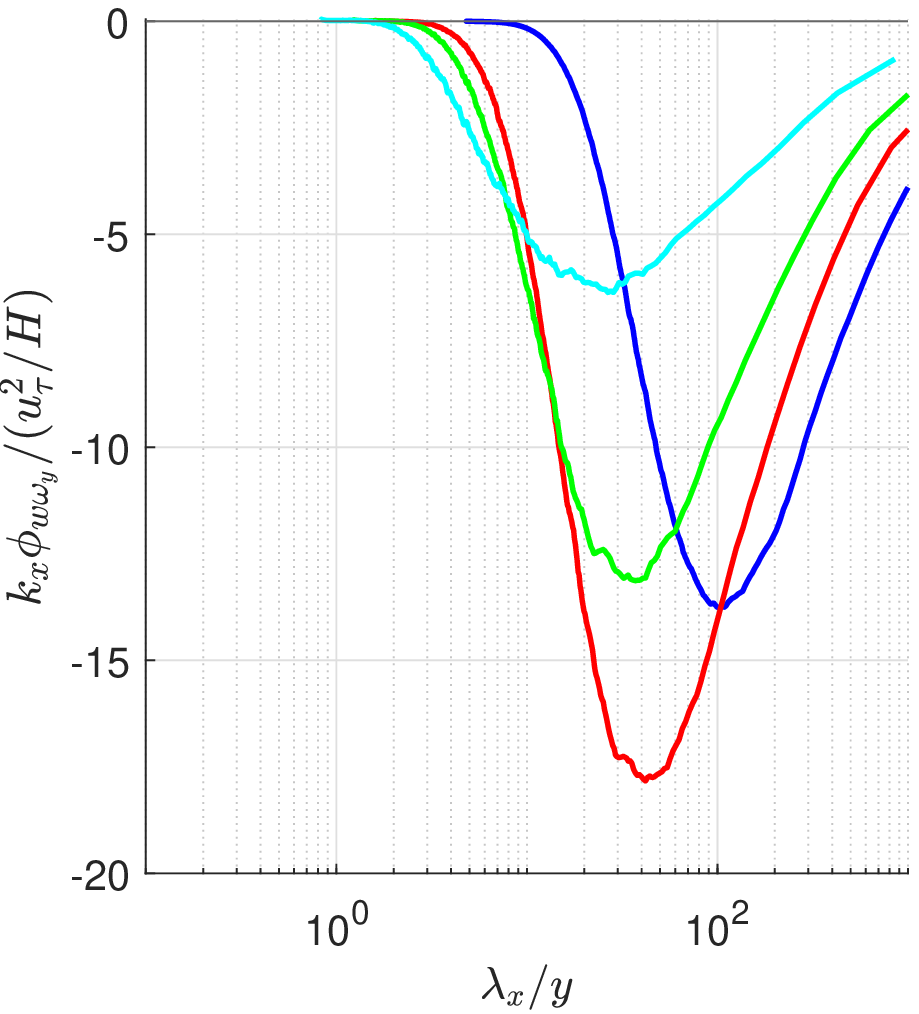}
        \caption{Streamwise co-spectra: $\mbox{Viscous \& Buffer Layers}$}
         \label{fig_bufferxwoy}
     \end{subfigure}
     \hfill
     \begin{subfigure}[b]{0.32\textwidth}
         \centering
         \includegraphics[width=\textwidth]{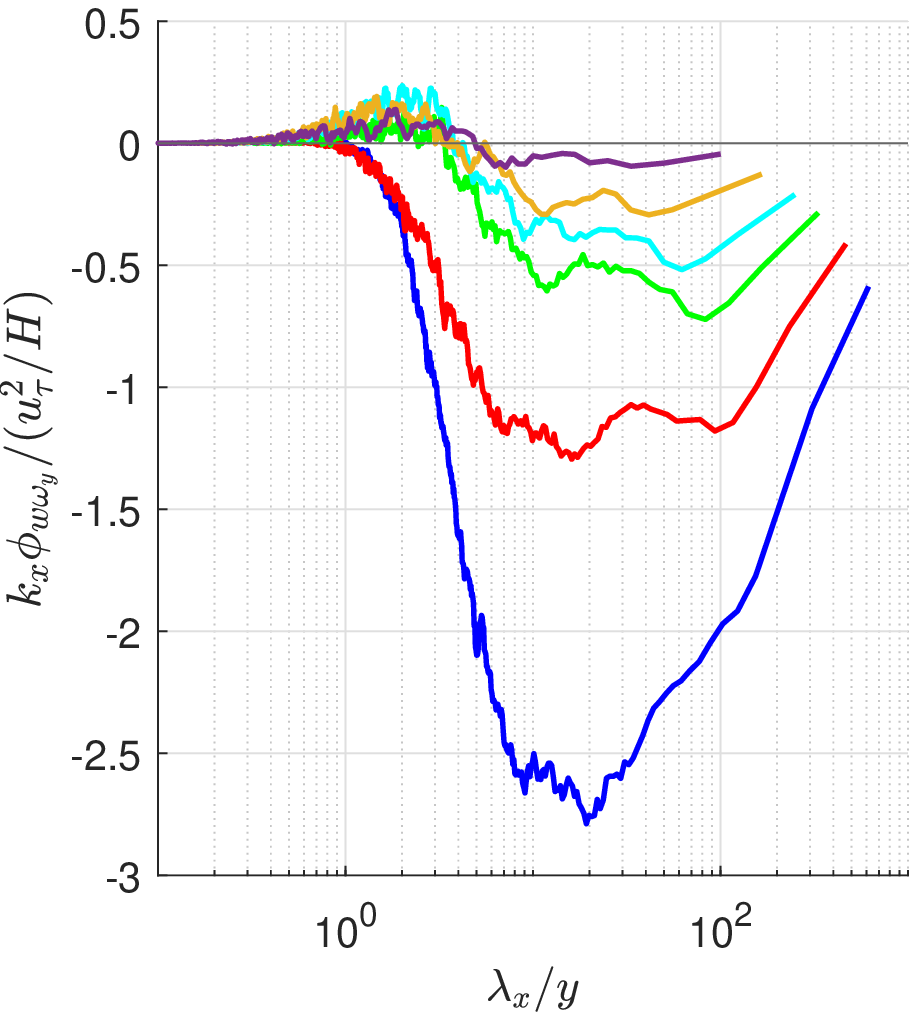}
         \caption{Streamwise co-spectra: Log Layer}
         \label{fig_inertialxwoy}
     \end{subfigure}
     \hfill
     \begin{subfigure}[b]{0.32\textwidth}
         \centering
         \includegraphics[width=\textwidth]{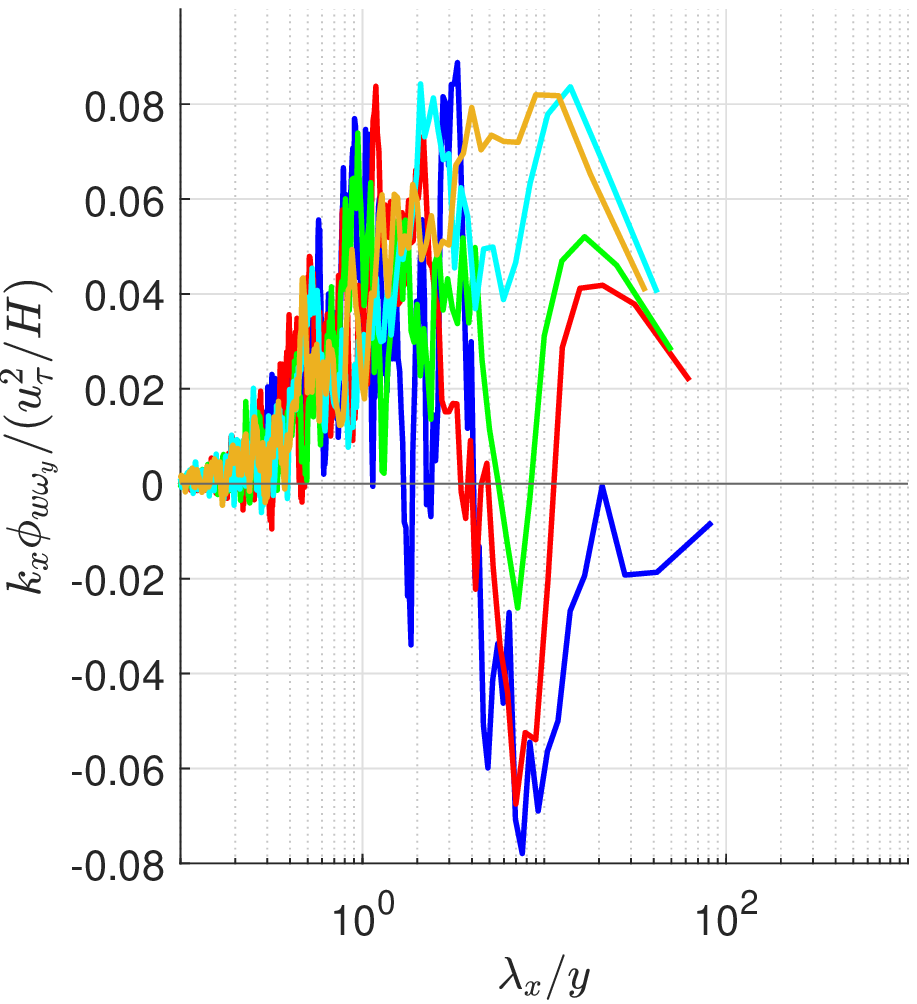}
         \caption{Streamwise co-spectra: Outer Layer}
         \label{fig_outerxwoy}
     \end{subfigure}
             \caption{Normalized streamwise cospectra of (negative of) the spanwise velocity - wall normal vorticity ($-\phi_{w\omega_y}$), in the (a) viscous \& buffer layers, (b) log layer and (c) outer layer. Curves have the same meaning as in corresponding plots in Fig~\ref{fig_1dcospec_z}.}
        \label{fig_1dcospecx_woy}
\end{figure}

\clearpage 

\section{Smoothing of 2D Spectra}\label{smoothing}
Since the 2D cospectra in this study were obtained by averaging over only 38 snapshots, we smooth the 2D co-spectra by a simple running average in Fourier space. Given that the streamwise and spanwise domain size is $L_x$ and $L_z$, and the number of corresponding grid points are $N_x$ and $N_z$ (assuming both are even), the streamwise and spanwise wavenumbers are given by $k_i=2\pi i/L_x$ and $k_j=2\pi j/L_z$ where $i,j\in \mathbb{Z}$. We demonstrate the smoothing procedure by showing its application to obtain $\varphi_{v\omega_z} (k_i,k_j)$, where $i,j\in\{0,1,2,... \}$ (shown in Fig~\ref{fig_2dvoz}). We start by defining the relevant 2D Fourier transforms and the cospectrum as, 
\begin{align}
&\hat{v}(k_i,k_j,y)=FFT_{2D}[v(x,y,z)],  \ \hat{\omega}_z(k_i,k_j,y)=FFT_{2D}[\omega_z(x,y,z)], \ \text{and} \nonumber \\
&\Phi_{v\omega_z}(k_i,k_j,y):=\langle\hat{v}\hat{\omega}_z^*\rangle, \ \text{where},\ i=\left \{ -{N_x}/{2}+1, -{N_x}/{2}+2,...-1,0,1,..{N_x}/{2}-1\right \}, \nonumber \\
 &j=\left \{ -{N_z}/{2}+1, -{N_z}/{2}+2,...-1,0,1,...{N_z}/{2}-1\right \}. 
 \end{align}
 We extend the co-spectrum to the full wavenumber space by defining,
 \begin{align}
\Phi_{v\omega_z}(k_i,k_j,y)&:=0, \ \forall \  |i| \geq \frac{N_x}{2}, |j| \geq \frac{N_z}{2}.  
\end{align}
This spectrum satisfies the property, 
\begin{align}
\sum_{i=-\infty}^{\infty} \sum_{j=-\infty}^{\infty} \Phi_{v\omega_z}(k_i,k_j,y) \Delta k_x \Delta k_z &= \langle v\omega_z\rangle(y) , \ where, \ \Delta k_x=\frac{2\pi}{L_x}, \Delta k_z=\frac{2\pi}{L_z}.
\end{align}
We now introduce the smoothed co-spectrum, with streamwise window size $\delta k_x=2b_x\Delta k_x$ and spanwise window size $\delta k_z= 2 b_z \Delta k_z$ as, 
\begin{align}
\Phi_{v\omega_z}^{b_x,b_y}(k_i,k_j,y):= \frac{1}{(2b_x+1)(2b_z+1)}\sum_{m=-b_x}^{b_x} \sum_{n=-b_z}^{b_z} \Phi_{v\omega_z}(k_{i+m},k_{j+n},y).
\end{align}
This smoothing maintains the value of the integral over the full wavenumber space. 
We then add contributions reflected in the $x$- and $z$-axes so that the spectra depend only on wavenumber magnitudes $k_x\geq 0,$ $k_z\geq 0$, yielding,
\begin{align}
&\varphi_{v \omega_z}(k_i,k_j,y):=  \Phi_{v\omega_z}^{b_x,b_z}(k_i,k_j,y)+ \Phi_{v\omega_z}^{b_x,b_z}(-k_i,k_j,y) 
+ \Phi_{v\omega_z}^{b_x,b_z}(-k_i,-k_j,y)+ \nonumber \\
& \Phi_{v\omega_z}^{b_x,b_z}(k_i,-k_j,y), \ 
 i=\left \{0,1,2,...,\frac{N_x}{2}+b_x-1  \right \}, \ j=\left \{0,1,2,...,\frac{N_z}{2}+b_z-1  \right \}.
 \end{align}
This single quadrant co-spectrum satisfies the relation,
 \begin{align}
 \sum_{i=0}^{\infty} \sum_{j=0}^{\infty} \varphi_{v\omega_z}(k_i,k_j,y) \Delta k_x \Delta k_z = \langle v\omega_z\rangle(y).
\end{align}

To choose the appropriate window size $b_x=b_z=b$ we use the Principle of Minimal Sensitivity \citep{Stevenson1981}. 
For this purpose, we calculate the $L^2$ distances between cospectra filtered with consecutive window sizes $(||\Phi^{b+1,b+1}_{v\omega_z-w\omega_y}-\Phi^{b,b}_{v\omega_z-w\omega_y}||_2, b=0,1,2,...)$ and plot these 
versus $b$ in Fig~\ref{specl2}. We find that the distance is least sensitive to window size for $2 \geq b \geq 4$, 
so that we keep the window size at $b=3$ for all 2D cospectra plotted in the main text. 
Raw co-spectra, as well as those smoothed with two window sizes, $b=3$ and $b=6$, are plotted in Fig ~\ref{fig_2dw}. We observe that smoothing the co-spectra removes some of the high wavenumber noise present in the un-smoothed spectrum ( Fig~\ref{unfiltered}). Increasing the window size beyond $b=3$ (Fig~\ref{w3}) does not lead to any appreciable 
noise reduction but begins to smear out larger scale features (Fig~\ref{w6})

\newpage

{ \begin{figure}
 \centering
\includegraphics[width=0.4\textwidth]{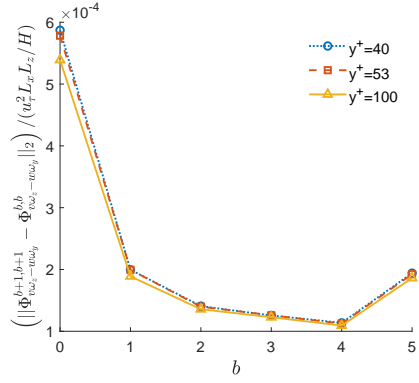}
\caption{The $L^2$ distance between co-spectra filtered with consecutive window sizes. We select $w_x=w_z=3$ for all 2D co-spectra, based on the Principle of minimal sensitivity (see ~\cite{Stevenson1981}). }
\label{specl2}
\end{figure}}

\begin{figure}
     \centering
      \begin{subfigure}[b]{0.329\textwidth}
         \centering
         \includegraphics[width=\textwidth]{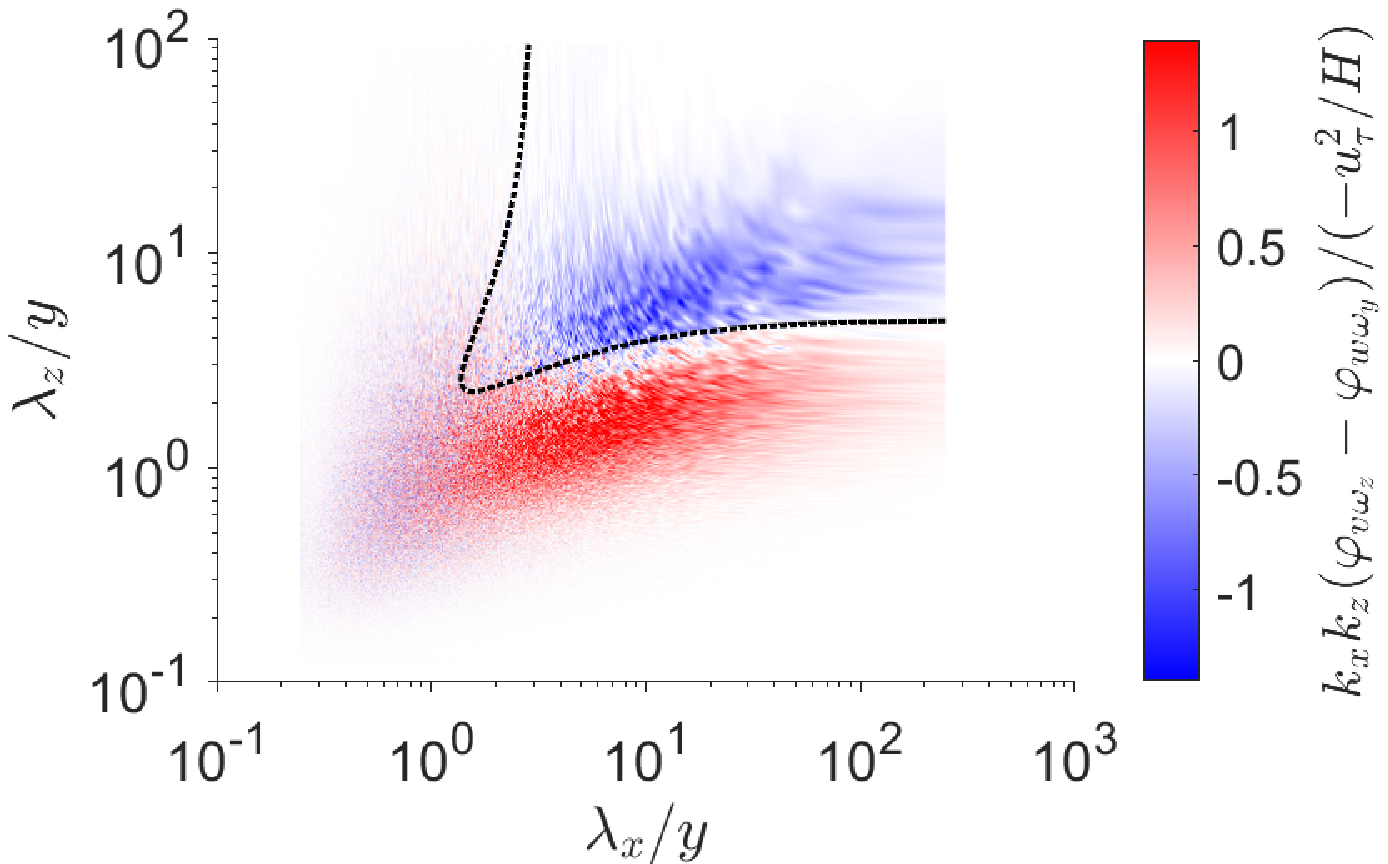}
        \caption{Without smoothing}
         \label{unfiltered}
     \end{subfigure}
     \hfill
     \begin{subfigure}[b]{0.329\textwidth}
         \centering
         \includegraphics[width=\textwidth]{fig11f.eps}
         \caption{$b_x=b_z=3$}
         \label{w3}
     \end{subfigure}
     \hfill	
     \begin{subfigure}[b]{0.329\textwidth}
         \centering
         \includegraphics[width=\textwidth]{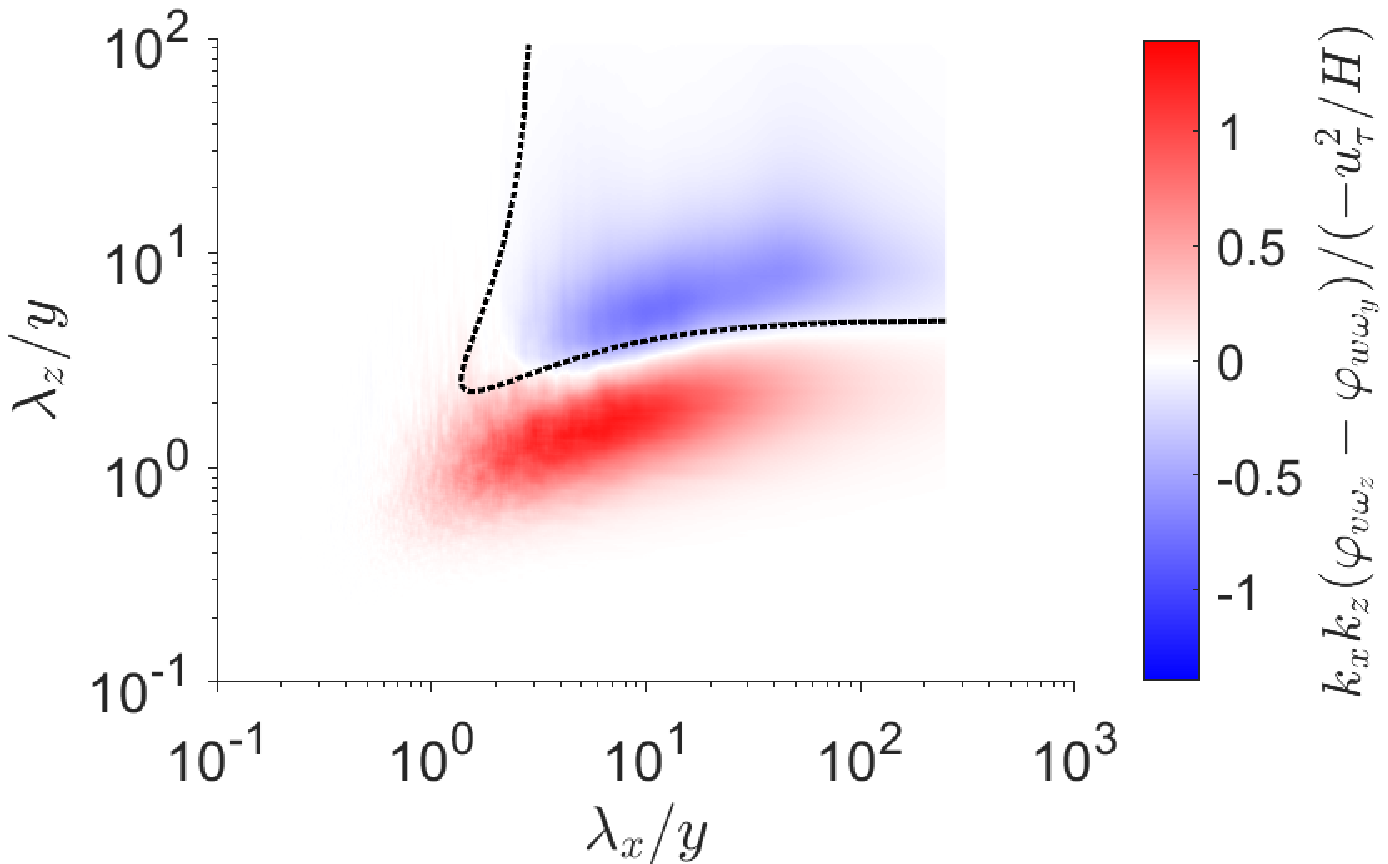}
        \caption{$w_x=w_z=6$}
         \label{w6}
     \end{subfigure}
\caption{Normalized 2D co-spectra of the nonlinear flux  ($\varphi_{v\omega_z}-\varphi_{w\omega_y}$) at $y^+=100$. The co-spectra ahown are (a) unsmoothed , smoothed with window size (b) $b_x=b_z=3$, and (c) $b_x=b_z=6$. The black dashed curves mark iso-contour of the filter $\mathcal{D}(k_x,k_z,y)=0.5$, described in Appendix~\ref{dragonfly}. Smoothing removes some of the high wavenumber noise seen in (a).  }

\end{figure}

\clearpage

\section{Individual Velocity-Vorticity 2D Cospectra}\label{sec:separate2D}  

\begin{figure}
     \centering
      \begin{subfigure}[b]{0.329\textwidth}
         \centering
         \includegraphics[width=\textwidth]{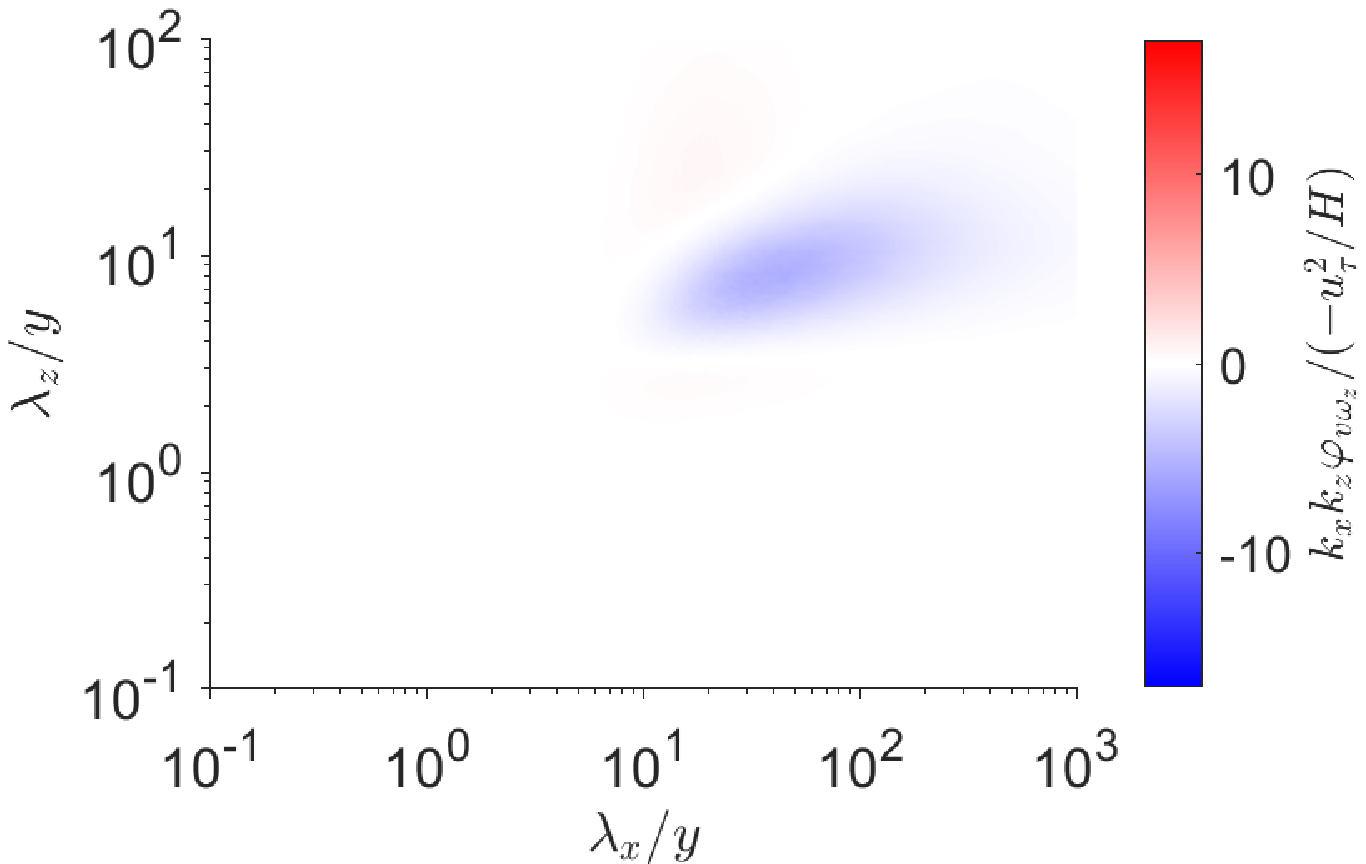}
        \caption{$y^+=5$}
         \label{y5voz}
     \end{subfigure}
     \hfill
     \begin{subfigure}[b]{0.329\textwidth}
         \centering
         \includegraphics[width=\textwidth]{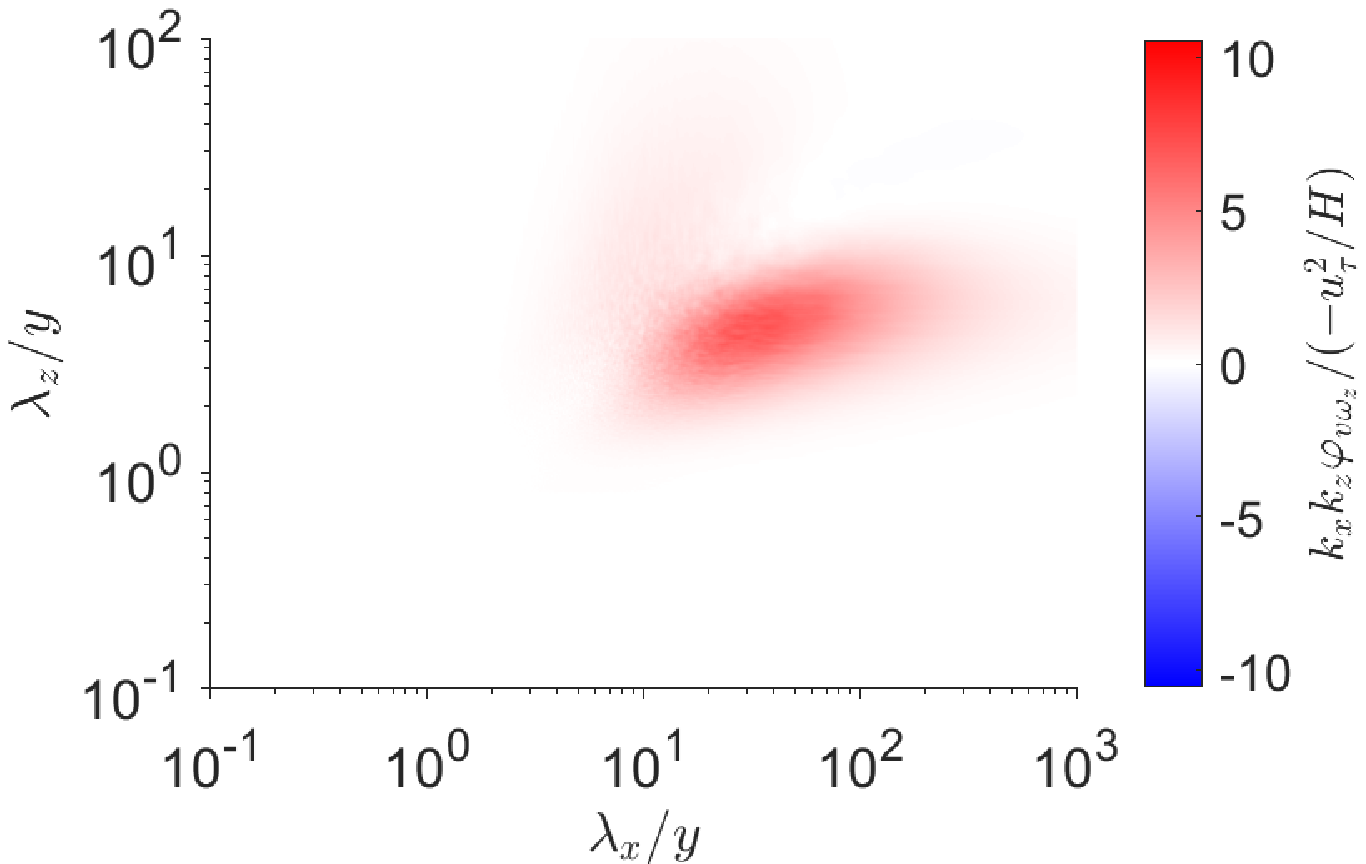}
         \caption{$y^+=15$}
         \label{y15voz}
     \end{subfigure}
     \hfill	
     \begin{subfigure}[b]{0.329\textwidth}
         \centering
         \includegraphics[width=\textwidth]{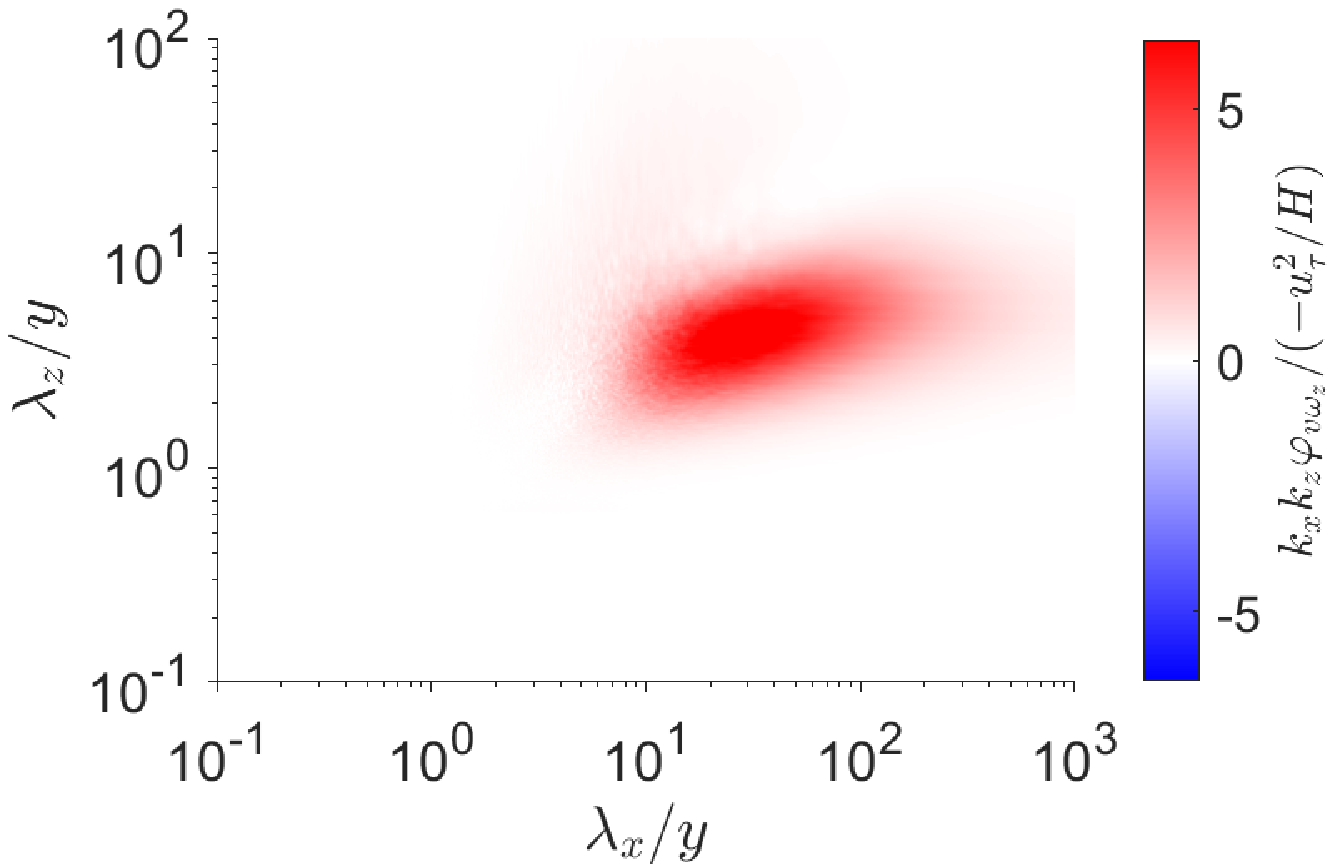}
        \caption{$y^+=20$}
         \label{y20voz}
     \end{subfigure}
     \hfill
          \begin{subfigure}[b]{0.329\textwidth}
         \centering
         \includegraphics[width=\textwidth]{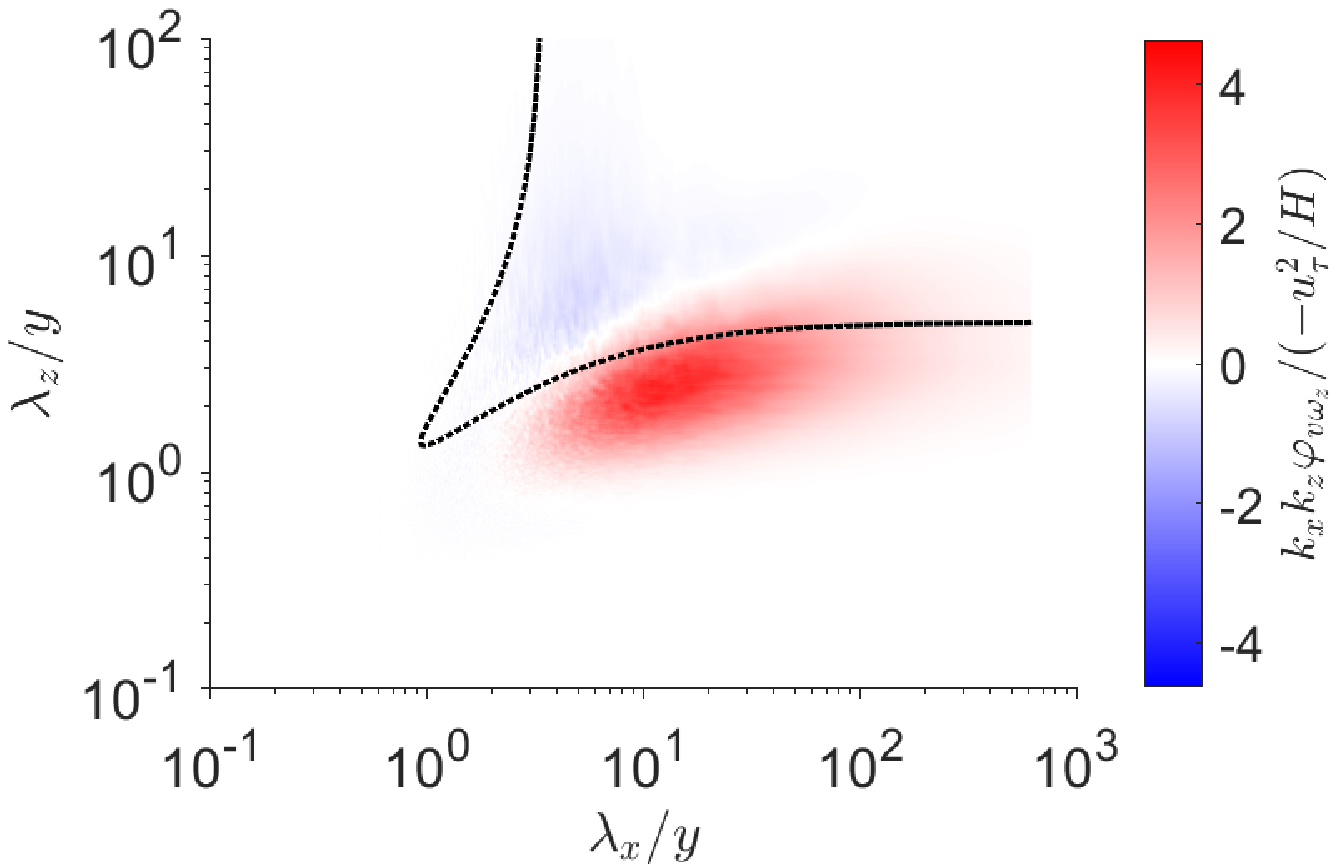}
        \caption{$y^+=40$}
         \label{y40voz}
     \end{subfigure}
     \hfill
    \begin{subfigure}[b]{0.329\textwidth}
         \centering
         \includegraphics[width=\textwidth]{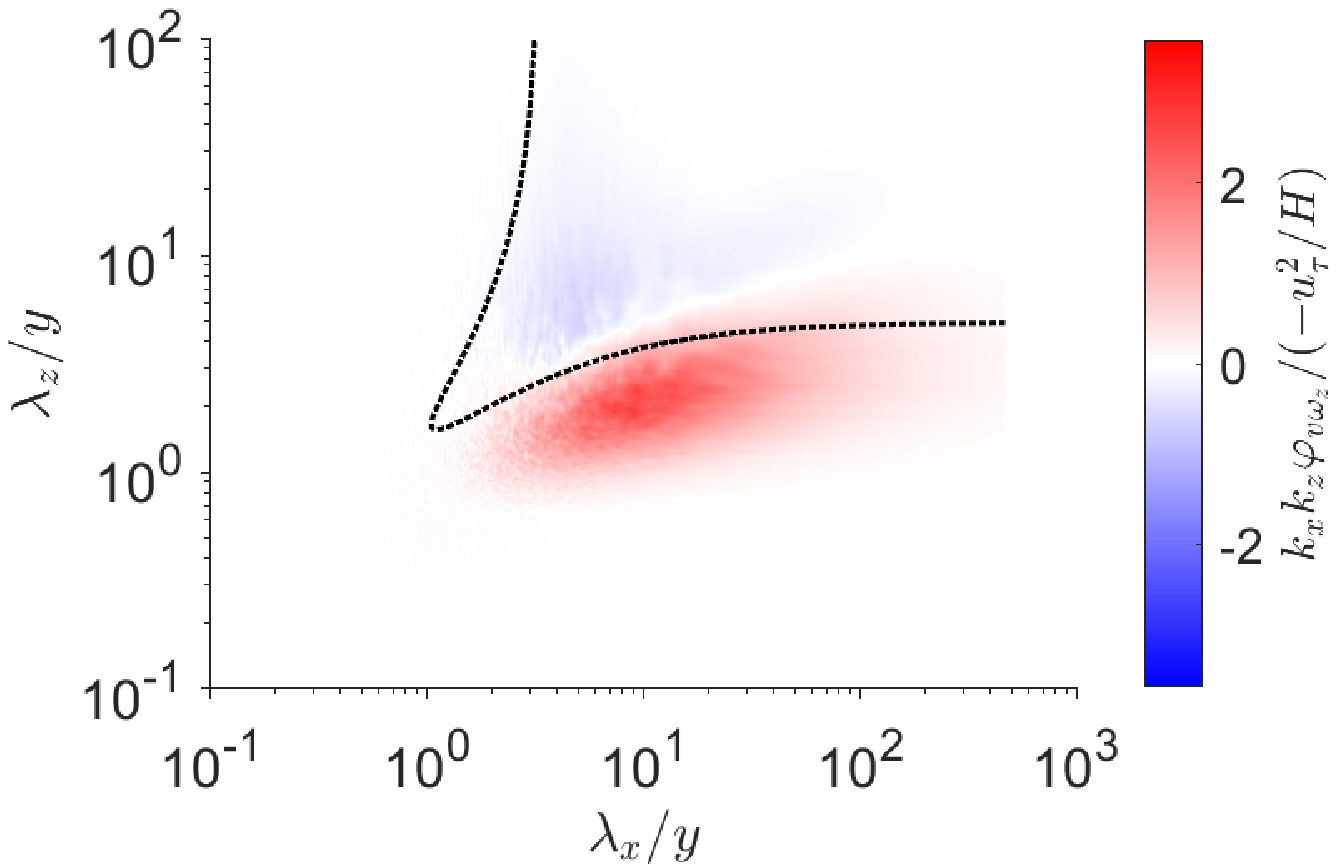}
         \caption{$y^+=53$}
         \label{y53}
     \end{subfigure}
     \hfill
     \begin{subfigure}[b]{0.329\textwidth}
         \centering
         \includegraphics[width=\textwidth]{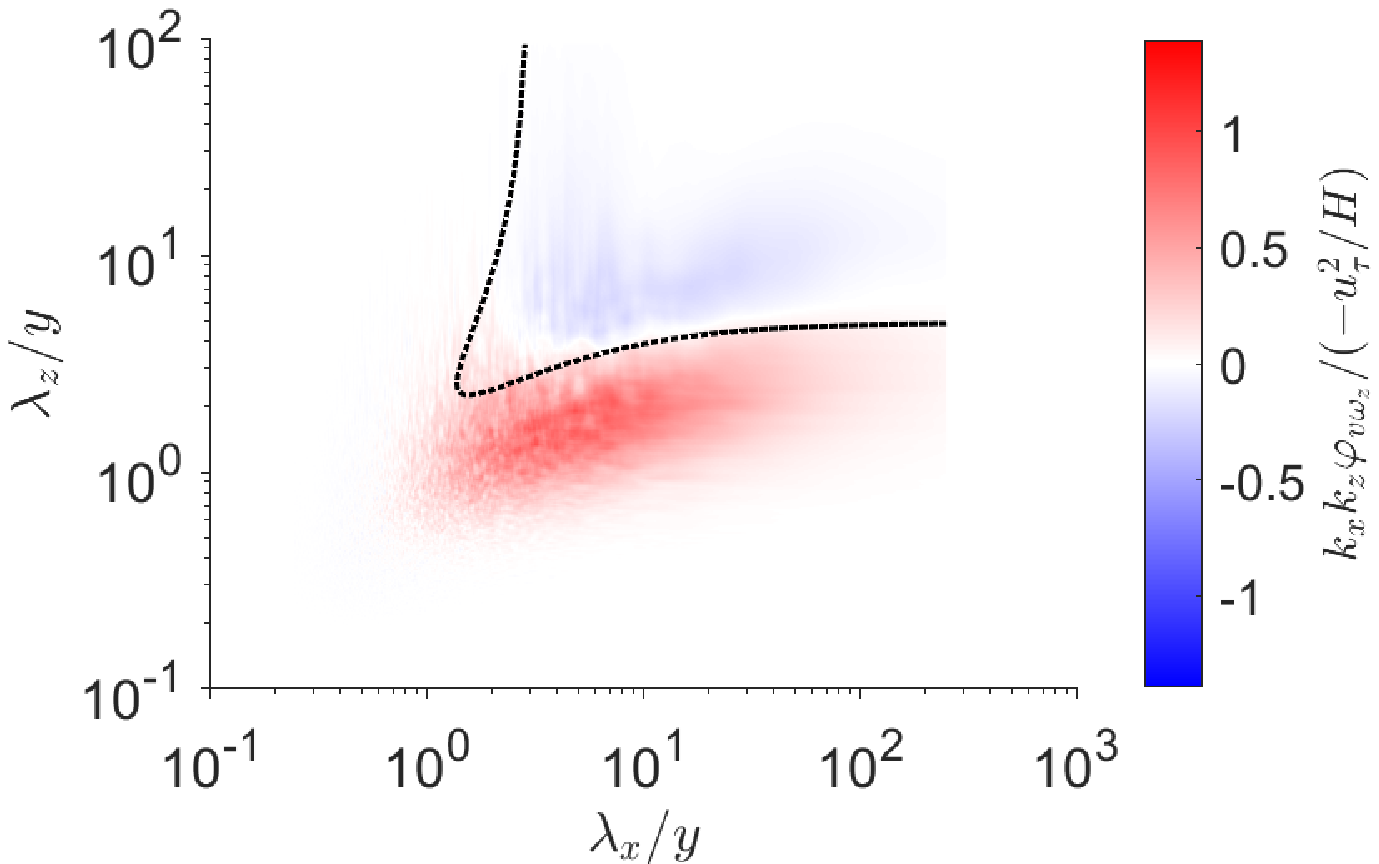}
         \caption{$y^+=100$}
         \label{y100voz}
     \end{subfigure}
     \hfill
      \begin{subfigure}[b]{0.329\textwidth}
         \centering
         \includegraphics[width=\textwidth]{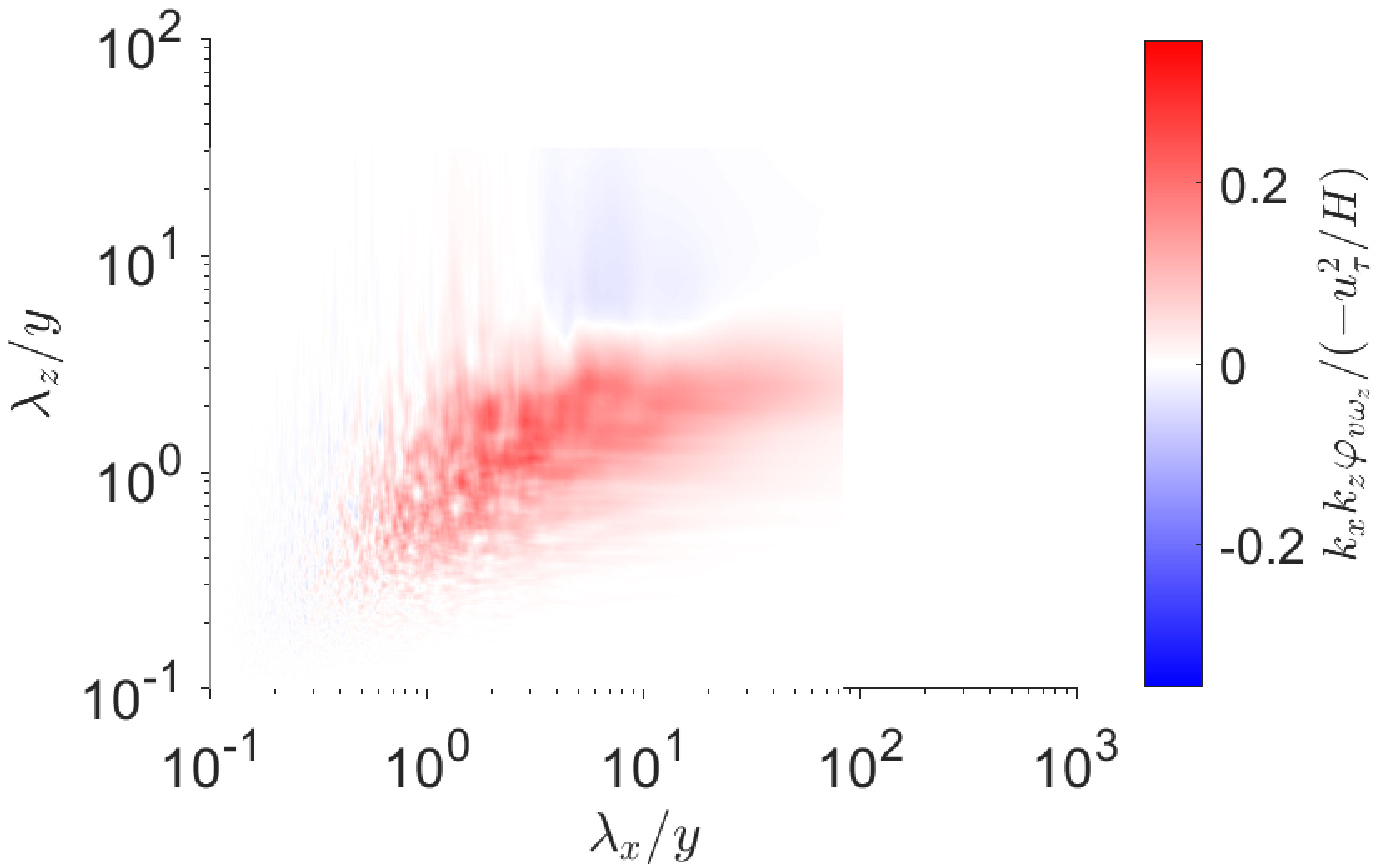}
        \caption{$y^+=300$}
         \label{y300voz}
     \end{subfigure}
      \hfill
     \begin{subfigure}[b]{0.329\textwidth}
         \centering
         \includegraphics[width=\textwidth]{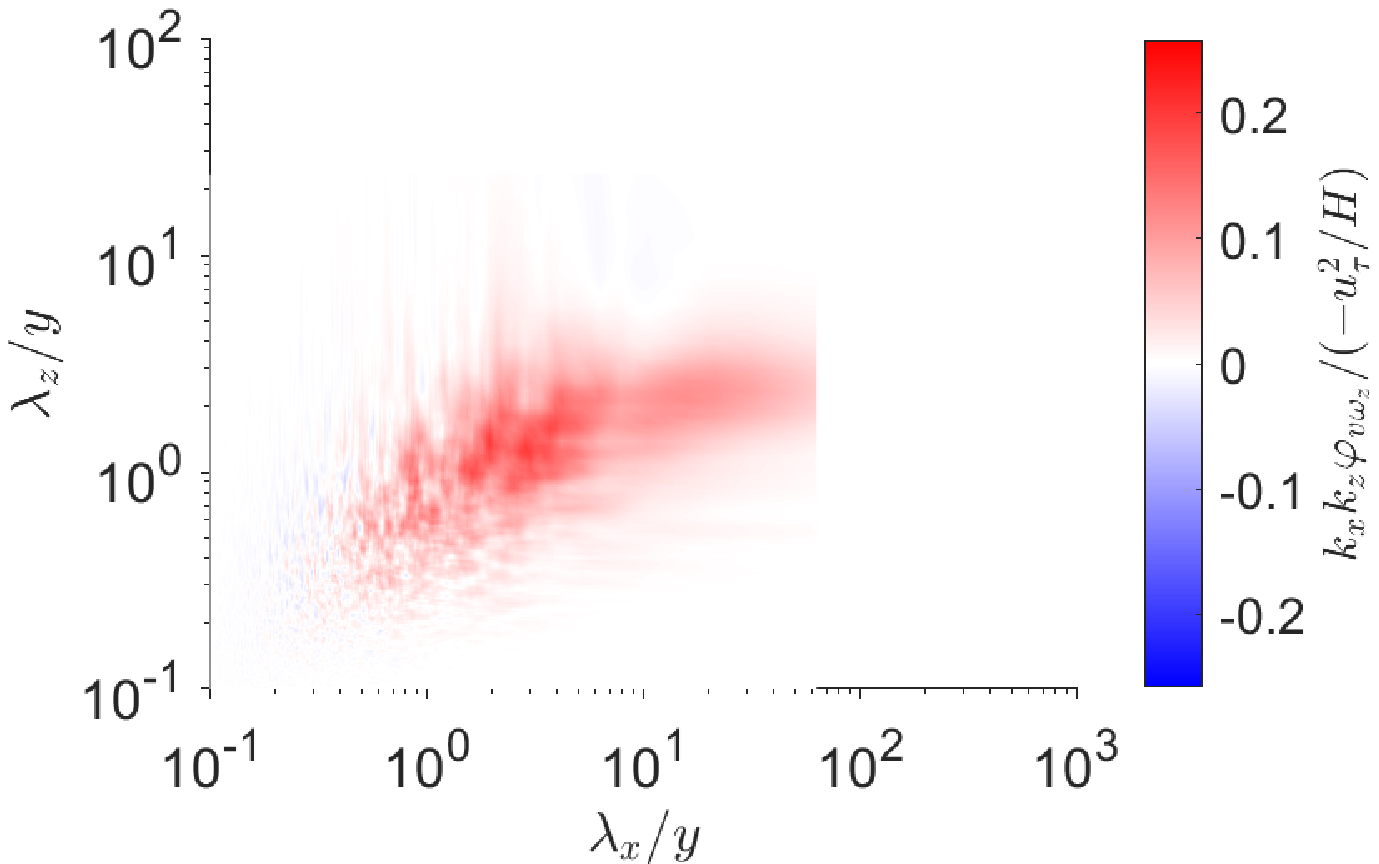}
         \caption{$y^+=400$}
         \label{y400voz}
     \end{subfigure}
     \hfill
     \begin{subfigure}[b]{0.329\textwidth}
         \centering
         \includegraphics[width=\textwidth]{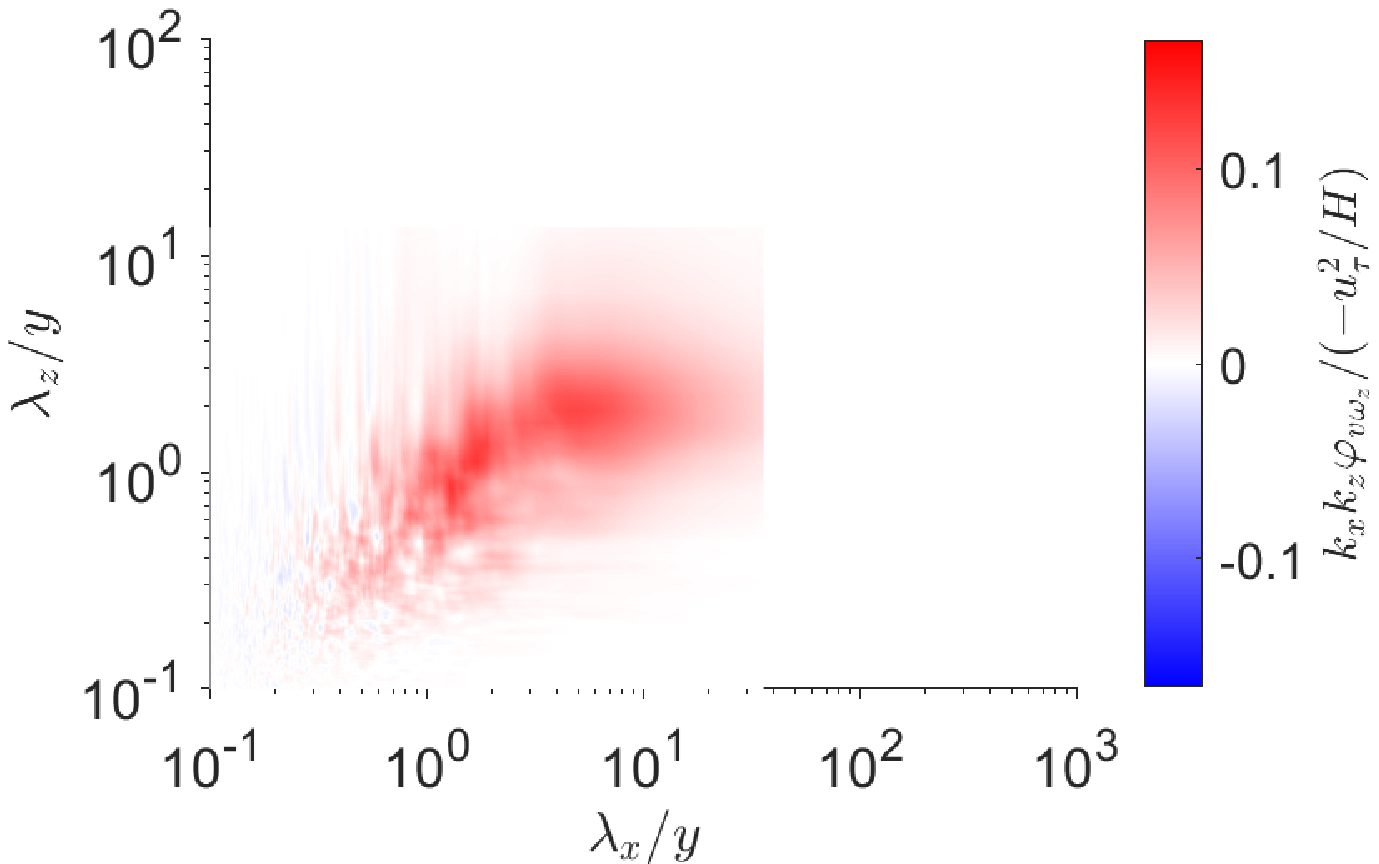}
        \caption{$y^+=700$}
         \label{y700voz}
     \end{subfigure}
     
        \caption{Normalized 2D co-spectra of the convective term ($\varphi_{v\omega_z}$), in the viscous \& buffer layers (a,b,c), log layer (d,e,f) and outer layer (g,h,i).The black dashed curves mark iso-contour of the filter $\mathcal{D}(k_x,k_z,y)=0.5$, described in Appendix~\ref{dragonfly}.}
        \label{fig_2dvoz}
        
\end{figure}

\begin{figure}
     \centering
      \begin{subfigure}[b]{0.329\textwidth}
         \centering
         \includegraphics[width=\textwidth]{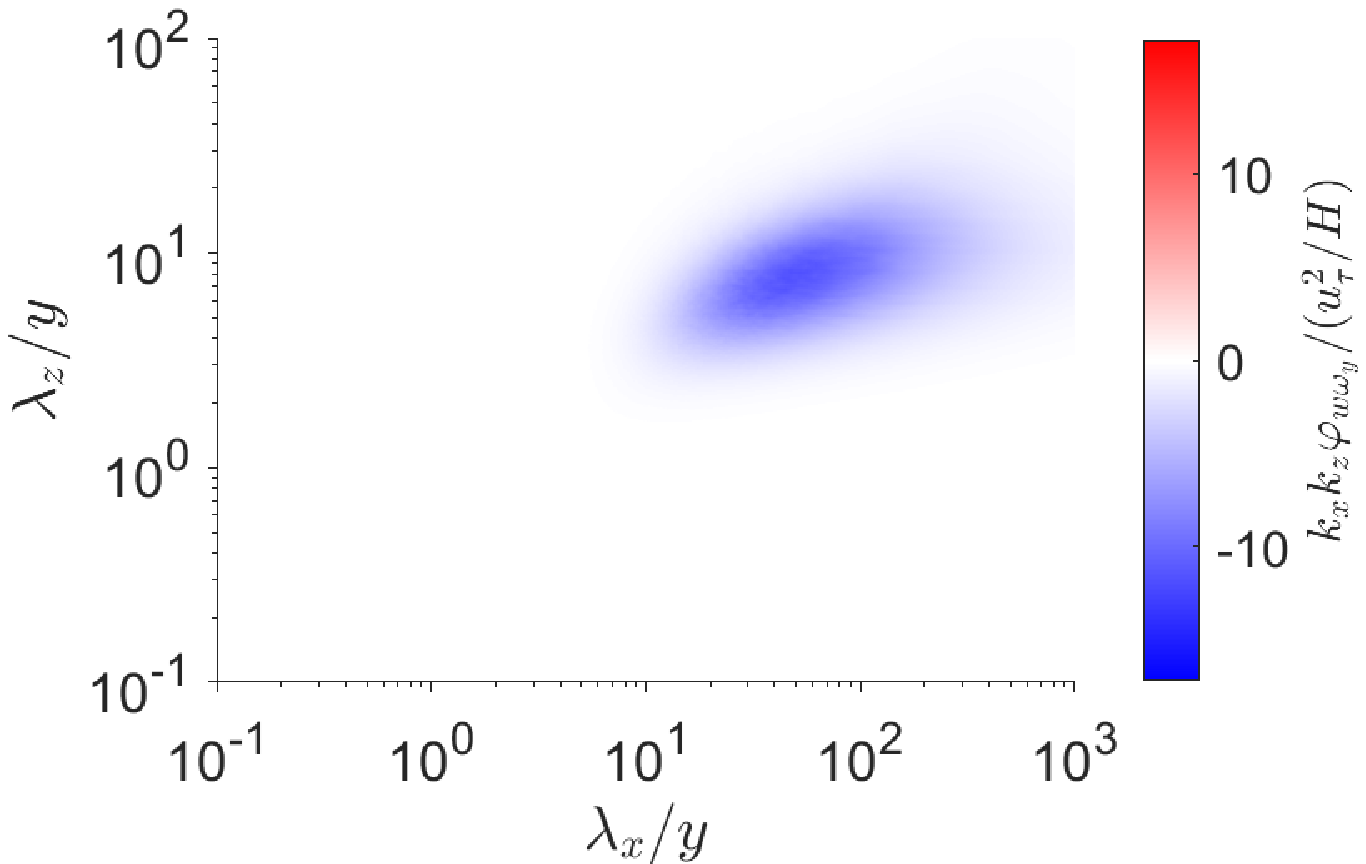}
        \caption{$y^+=5$}
         \label{y5woy}
     \end{subfigure}
     \hfill
     \begin{subfigure}[b]{0.329\textwidth}
         \centering
         \includegraphics[width=\textwidth]{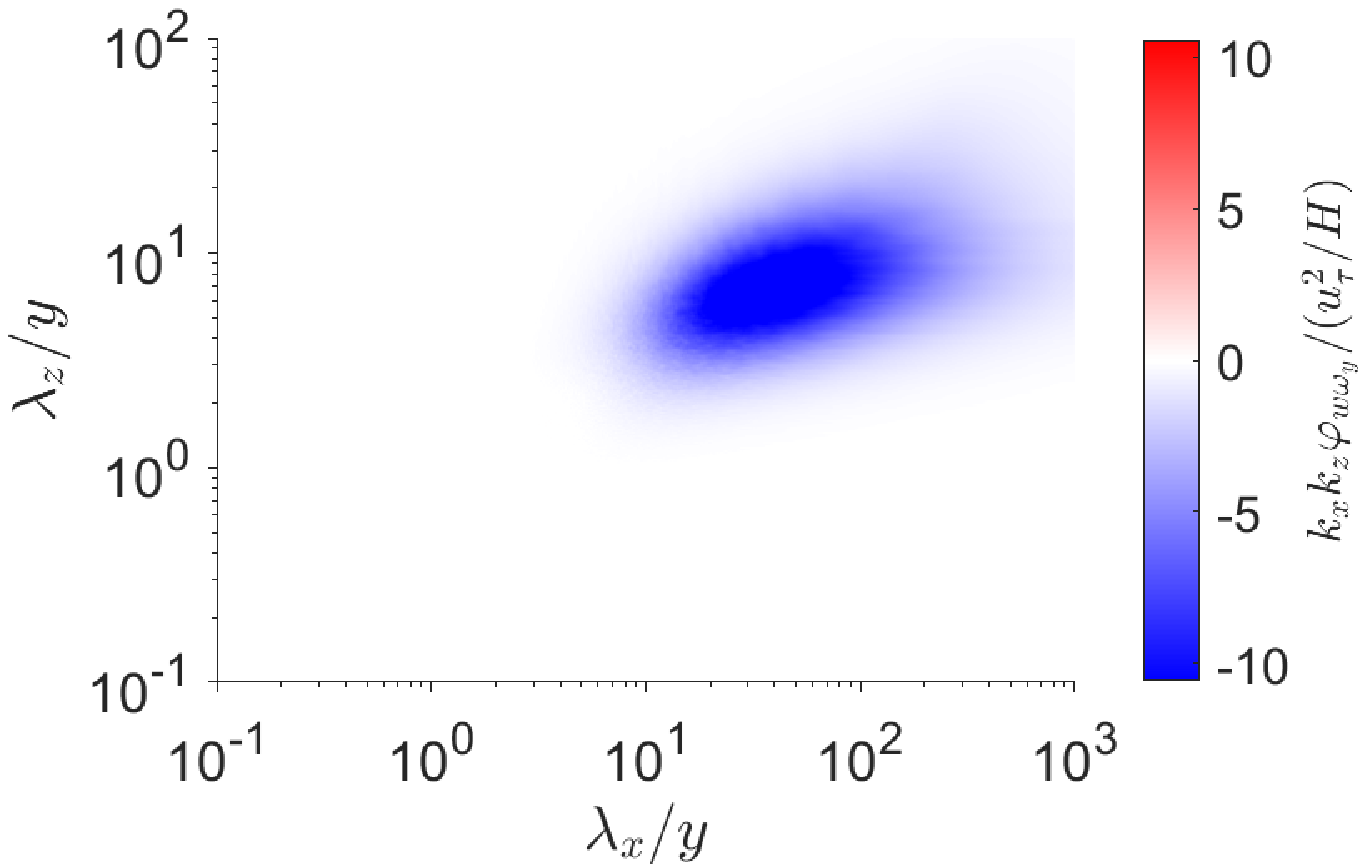}
         \caption{$y^+=15$}
         \label{y15woy}
     \end{subfigure}
     \hfill	
     \begin{subfigure}[b]{0.329\textwidth}
         \centering
         \includegraphics[width=\textwidth]{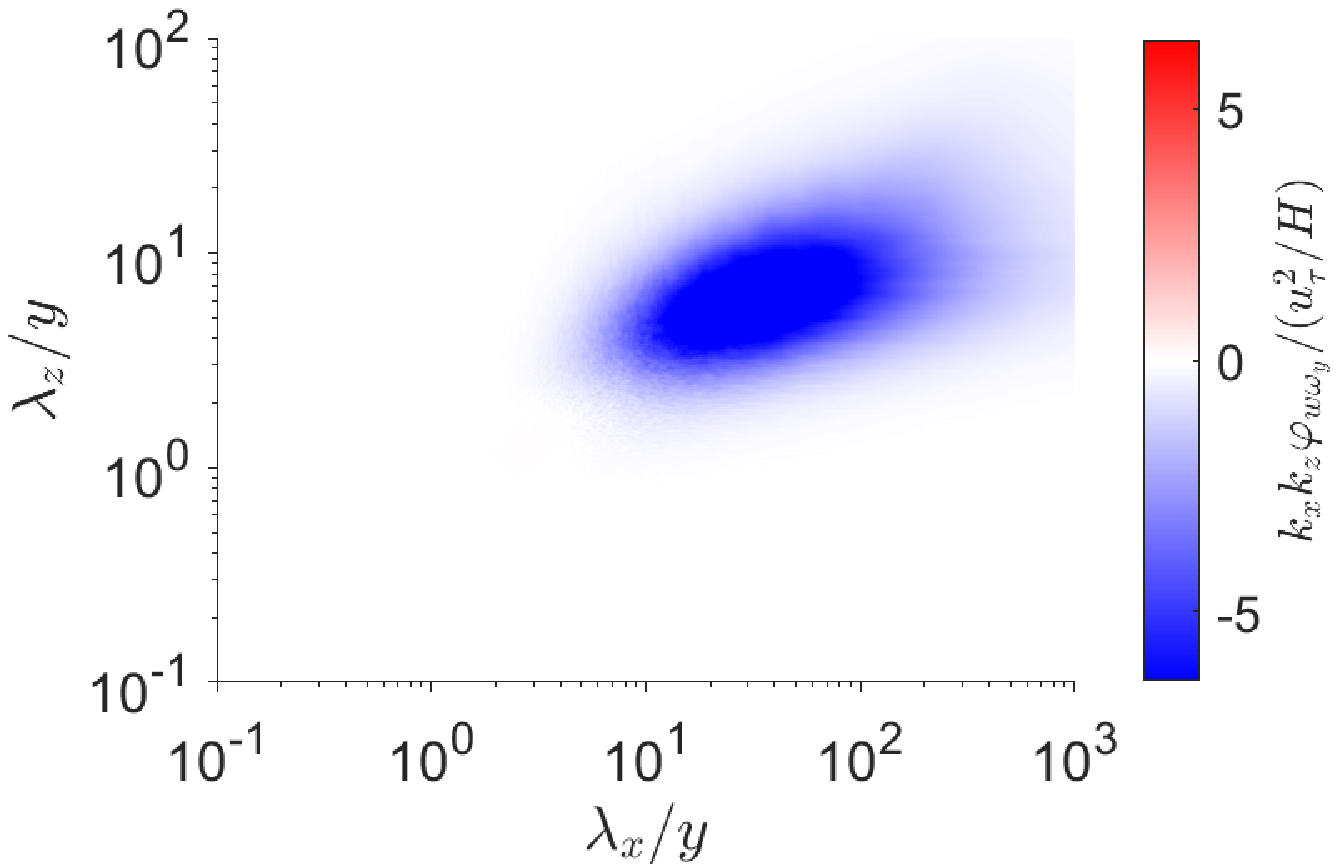}
        \caption{$y^+=20$}
         \label{y20woy}
     \end{subfigure}
    \vfill
    \begin{subfigure}[b]{0.329\textwidth}
         \centering
         \includegraphics[width=\textwidth]{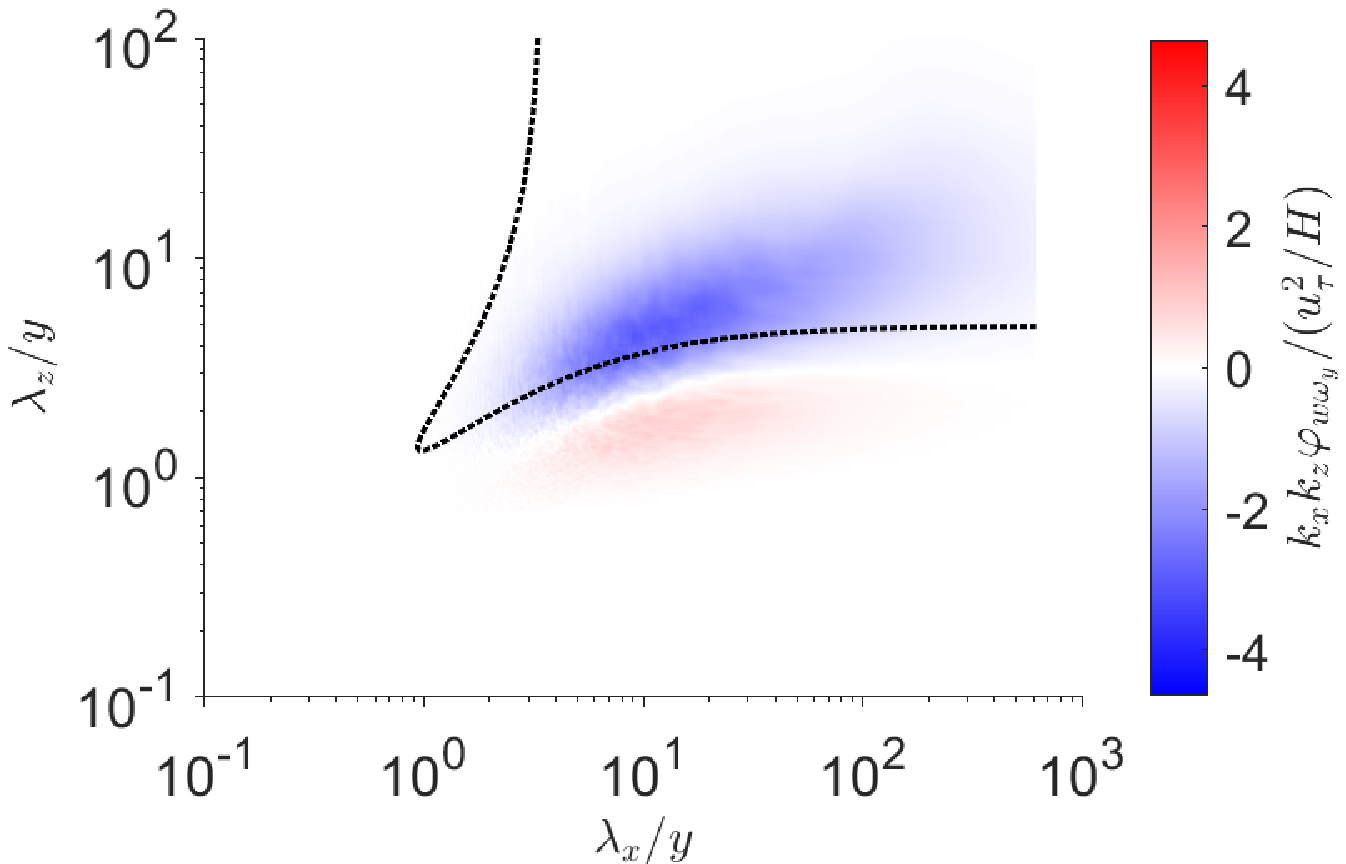}
        \caption{$y^+=40$}
        \label{y40woy}
    \end{subfigure}
    \hfill
     \begin{subfigure}[b]{0.329\textwidth}
         \centering
         \includegraphics[width=\textwidth]{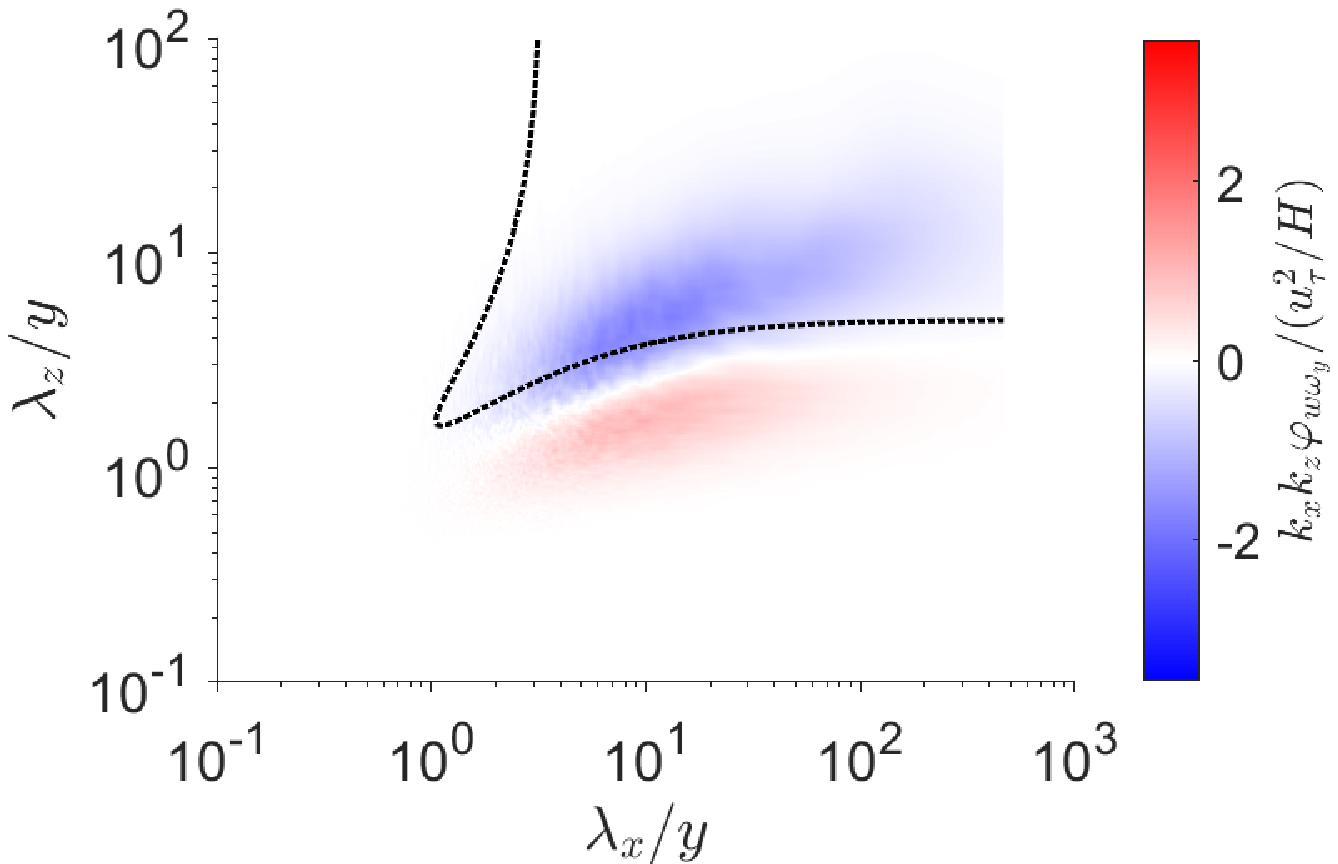}
        \caption{$y^+=53$}
        \label{y53woy}
    \end{subfigure}
    \hfill
    \begin{subfigure}[b]{0.329\textwidth}
        \centering
        \includegraphics[width=\textwidth]{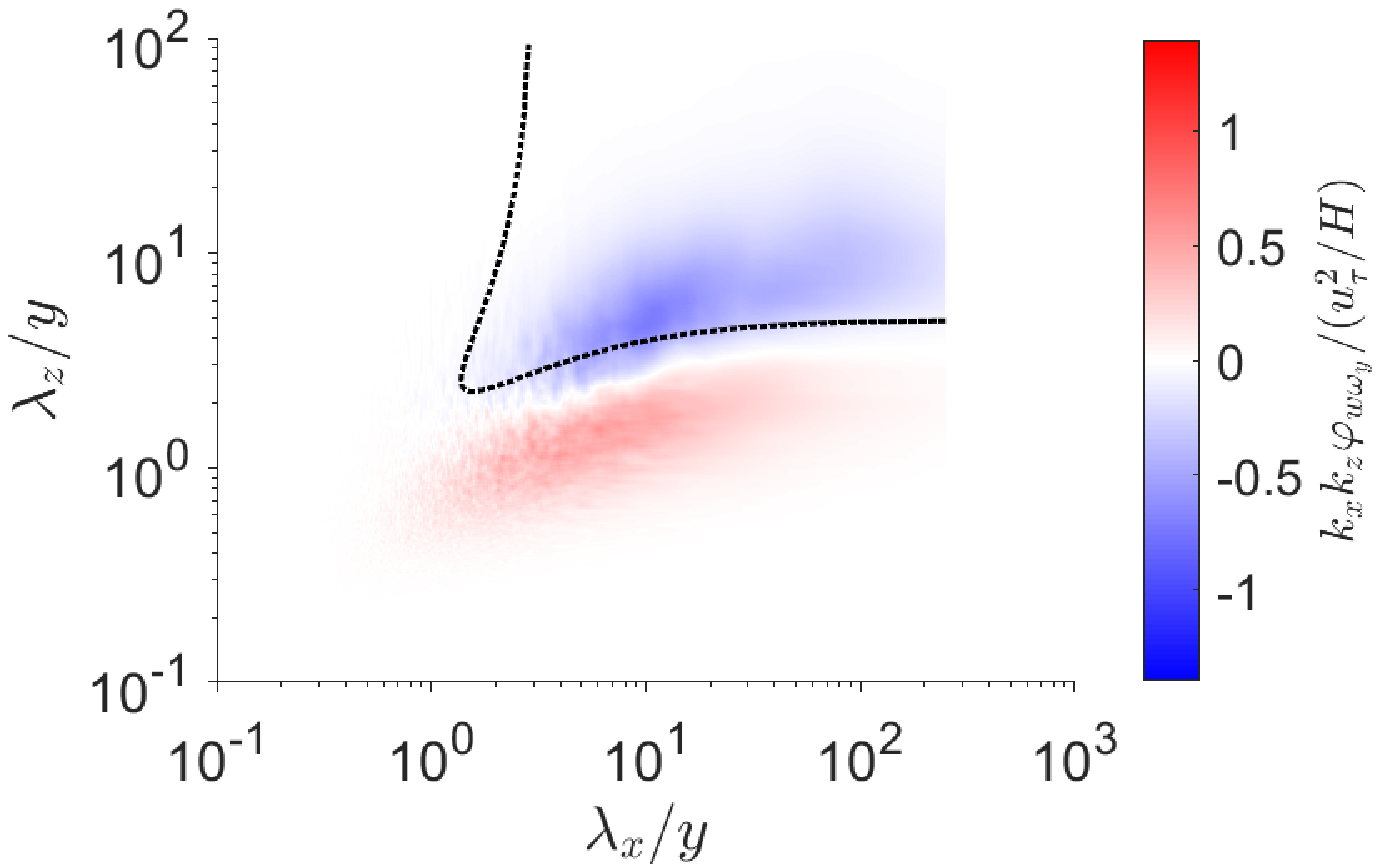}
         \caption{$y^+=100$}
         \label{y100woy}
     \end{subfigure}
     \vfill 
      \begin{subfigure}[b]{0.329\textwidth}
         \centering
         \includegraphics[width=\textwidth]{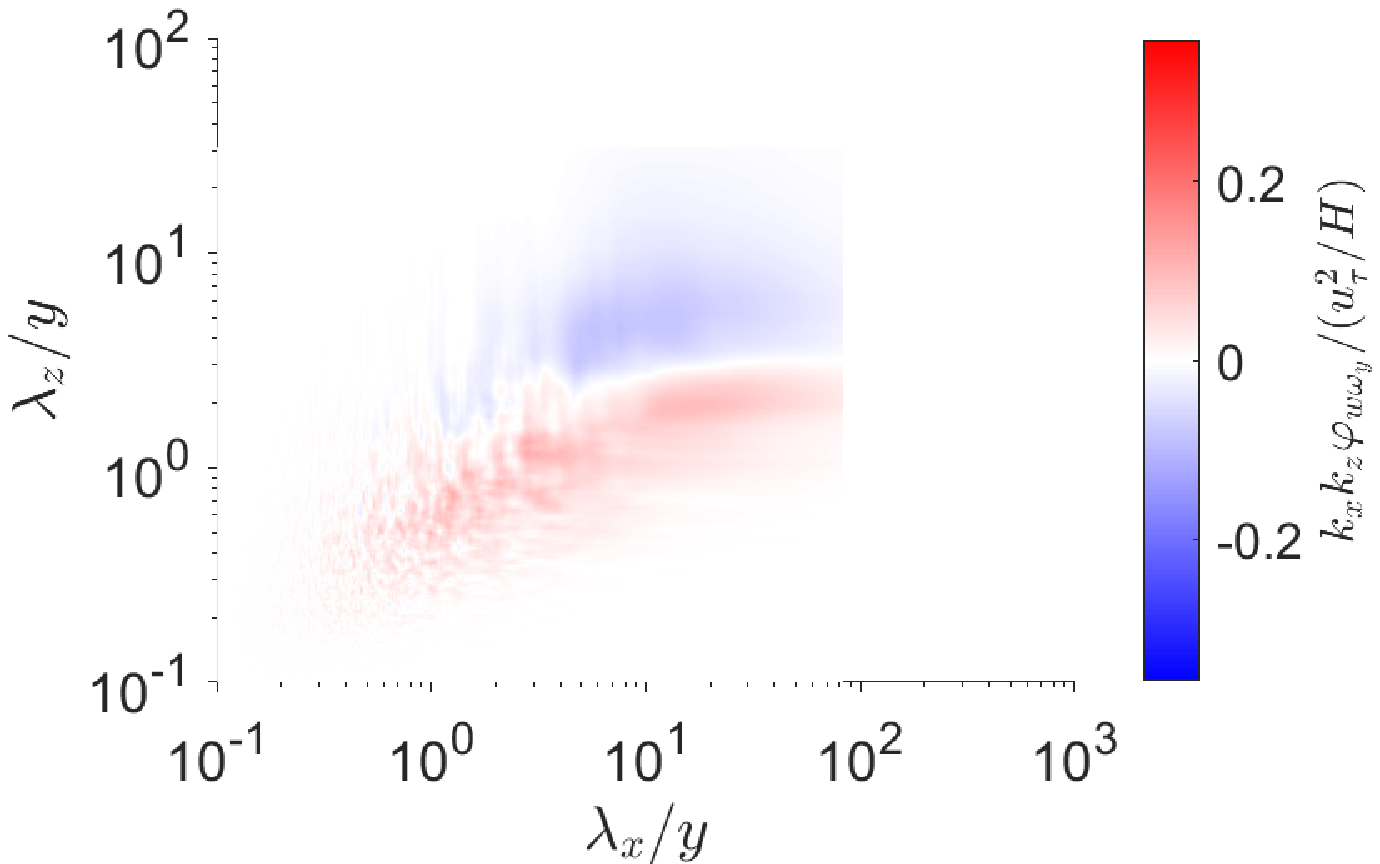}
        \caption{$y^+=300$}
         \label{y300woy}
     \end{subfigure}
      \hfill
     \begin{subfigure}[b]{0.329\textwidth}
         \centering
         \includegraphics[width=\textwidth]{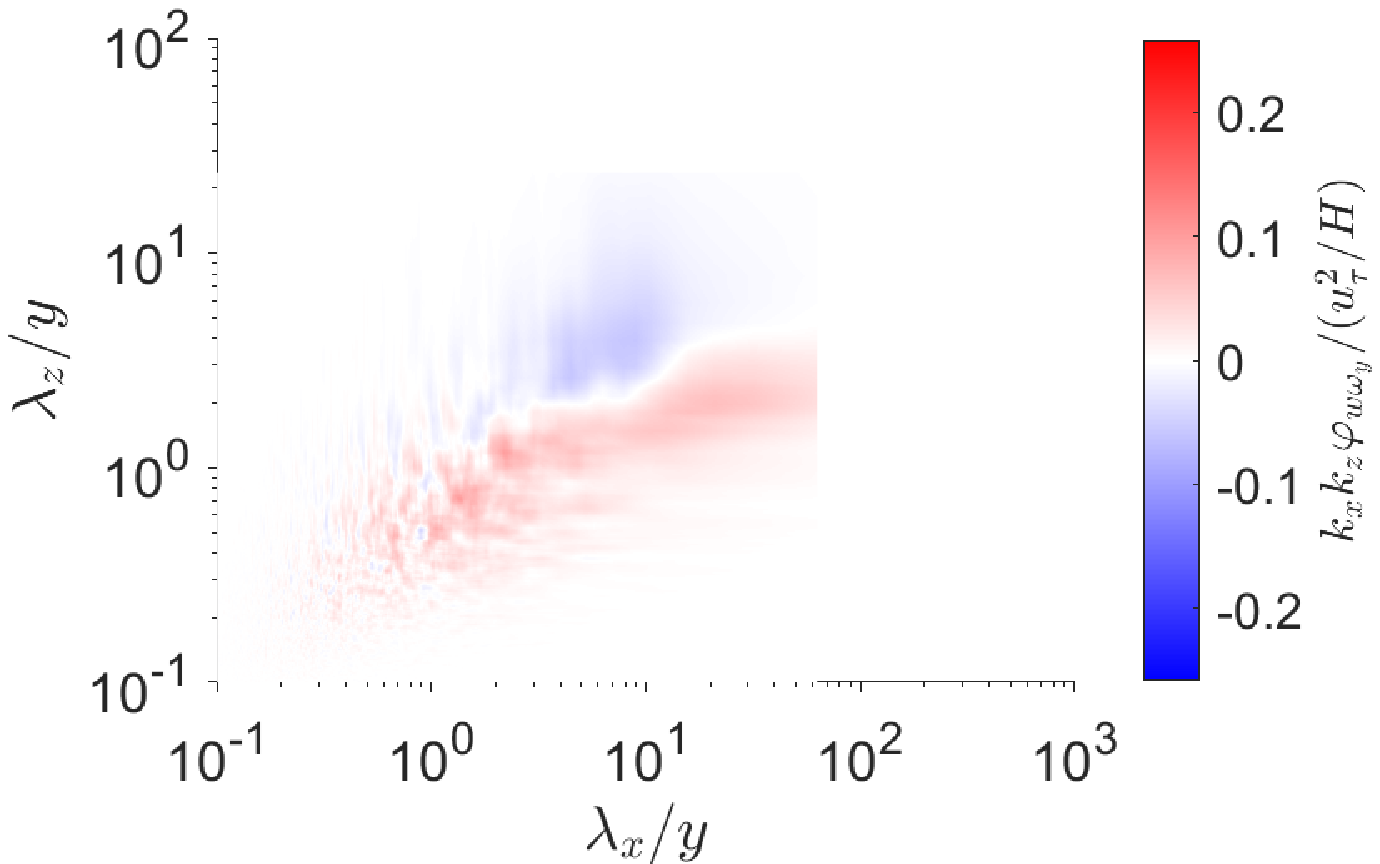}
         \caption{$y^+=400$}
       \label{y400woy}
    \end{subfigure}
    \hfill
    \begin{subfigure}[b]{0.329\textwidth}
        \centering
         \includegraphics[width=\textwidth]{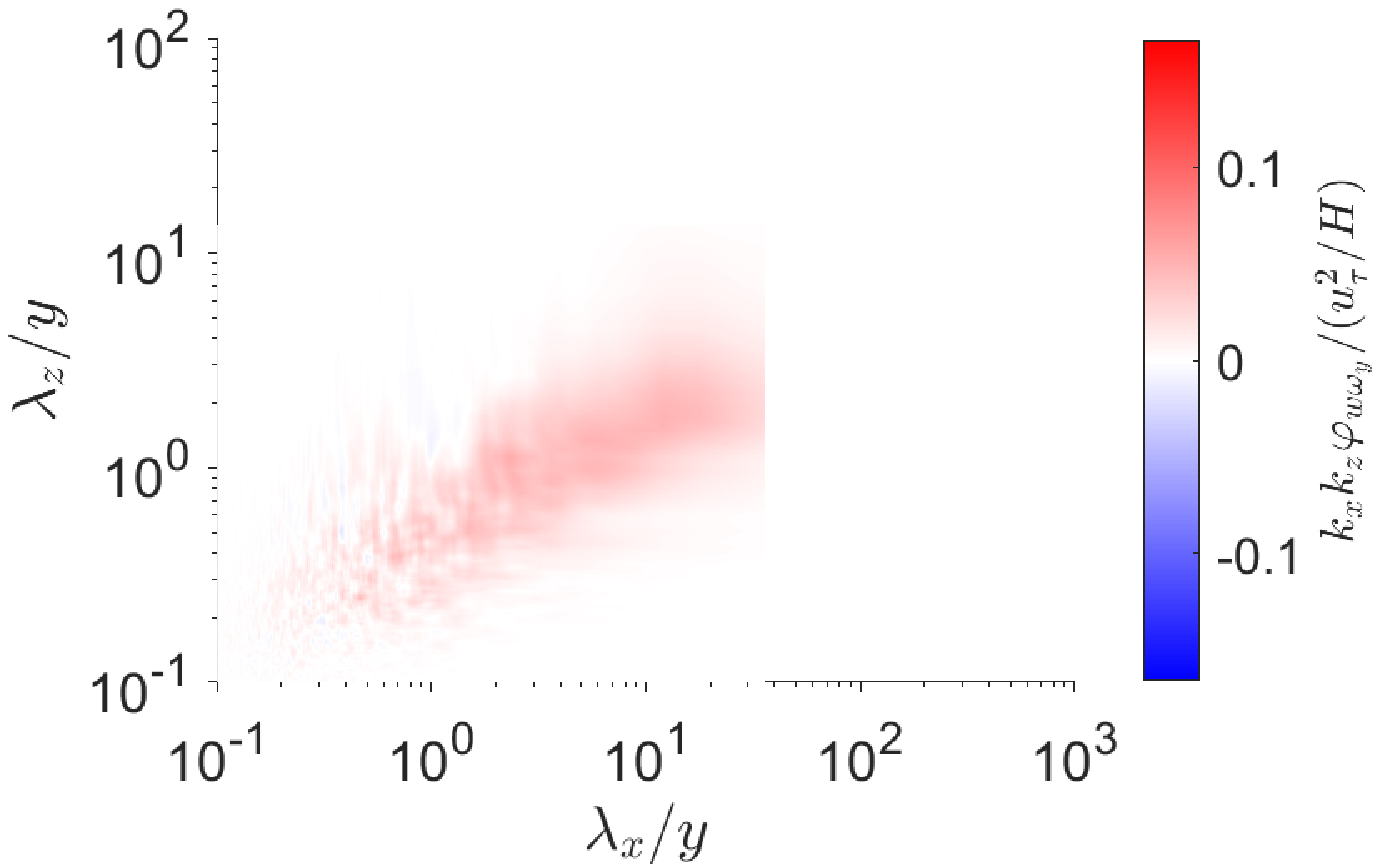}
         \caption{$y^+=700$}
         \label{y700woy}
     \end{subfigure}
     
        \caption{Normalized 2D co-spectra of the stretching/tilting term ($-\varphi_{w\omega_y}$) , in the viscous \& buffer layers (a,b,c), log layer (d,e,f) and outer layer (g,h,i).The black dashed curves mark iso-contour of the filter $\mathcal{D}(k_x,k_z,y)=0.5$, described in Appendix~\ref{dragonfly}.}
        \label{fig_2dwoy}
        
\end{figure}

Corresponding to the 2D flux cospectra shown in Fig ~\ref{fig_2dcospec} of the main text, we plot 
the separate 2D cospectra for the advective flux ($\varphi_{v\omega_z}(k_x,k_z,y)$) and stretching flux ($-\varphi_{w\omega_y}(k_x,k_z,y)$), in Fig~\ref{fig_2dvoz} and Fig~\ref{fig_2dwoy}, respectively. These cospectra have a bipartite structure similar to the total nonlinear flux cospectra plotted in Fig~\ref{fig_2dcospec}. However, the advective flux cospectra (Fig~\ref{fig_2dvoz}) make a largely down-gradient contribution, while the stretching flux cospectra (Fig~\ref{fig_2dwoy}) make a largely up-gradient contribution to the total nonlinear flux. An exception to this trend is marked by the cospectra at $y^+\lesssim 10$, where both contributions are up-gradient and $y^+\gtrsim500$ where both are down-gradient. Therefore, we can say that, by and large, the advective flux makes a down-gradient contribution while the stretching flux makes an up-gradient contribution to the nonlinear flux co-spectra, for $10\lesssim y^+ \lesssim 500$.   

\newpage

\clearpage 

\section{Dragonfly Filter}\label{dragonfly}
\begin{figure}
     \centering
      \begin{subfigure}[b]{0.329\textwidth}
         \centering
         \includegraphics[width=\textwidth]{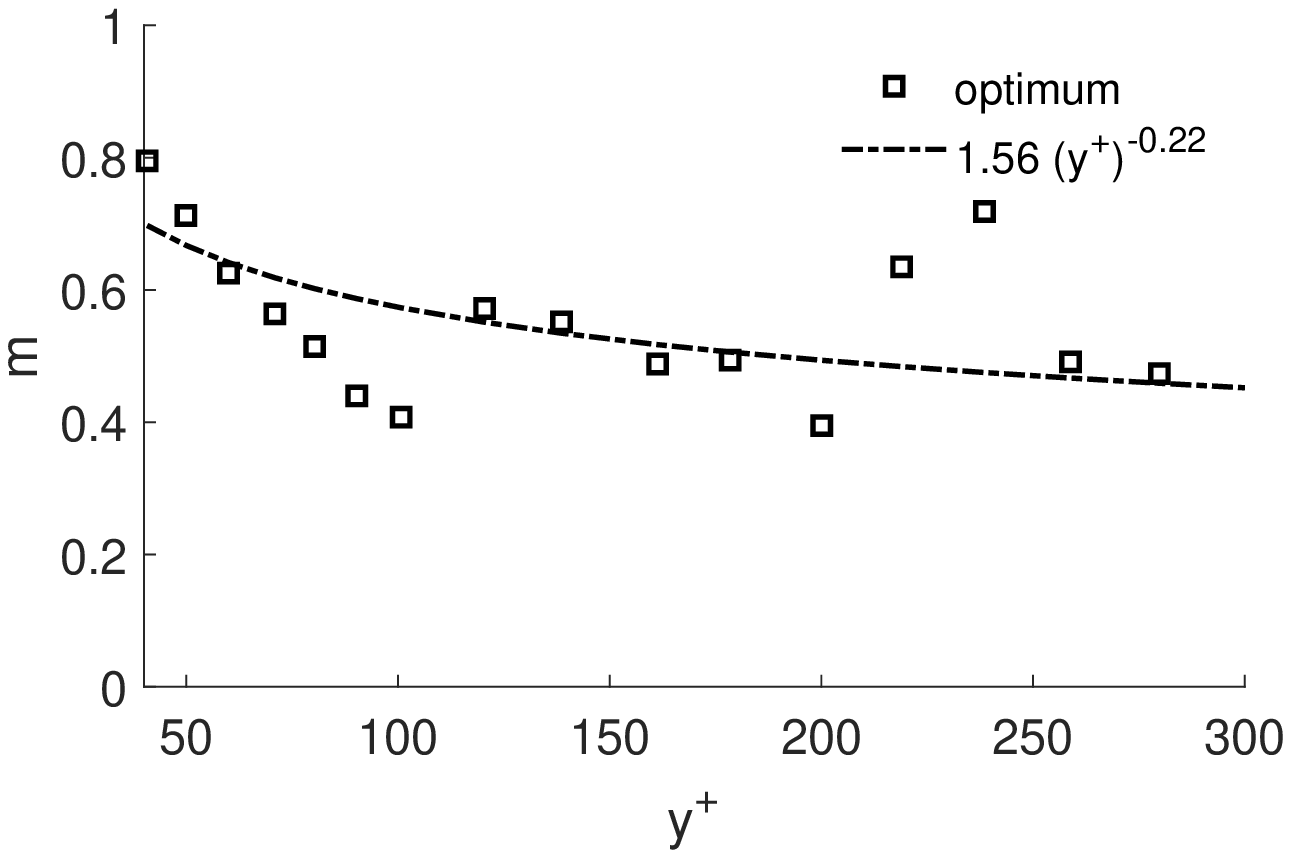}
        \caption{}
         \label{m}
     \end{subfigure}
     \hfill
     \begin{subfigure}[b]{0.329\textwidth}
         \centering
         \includegraphics[width=\textwidth]{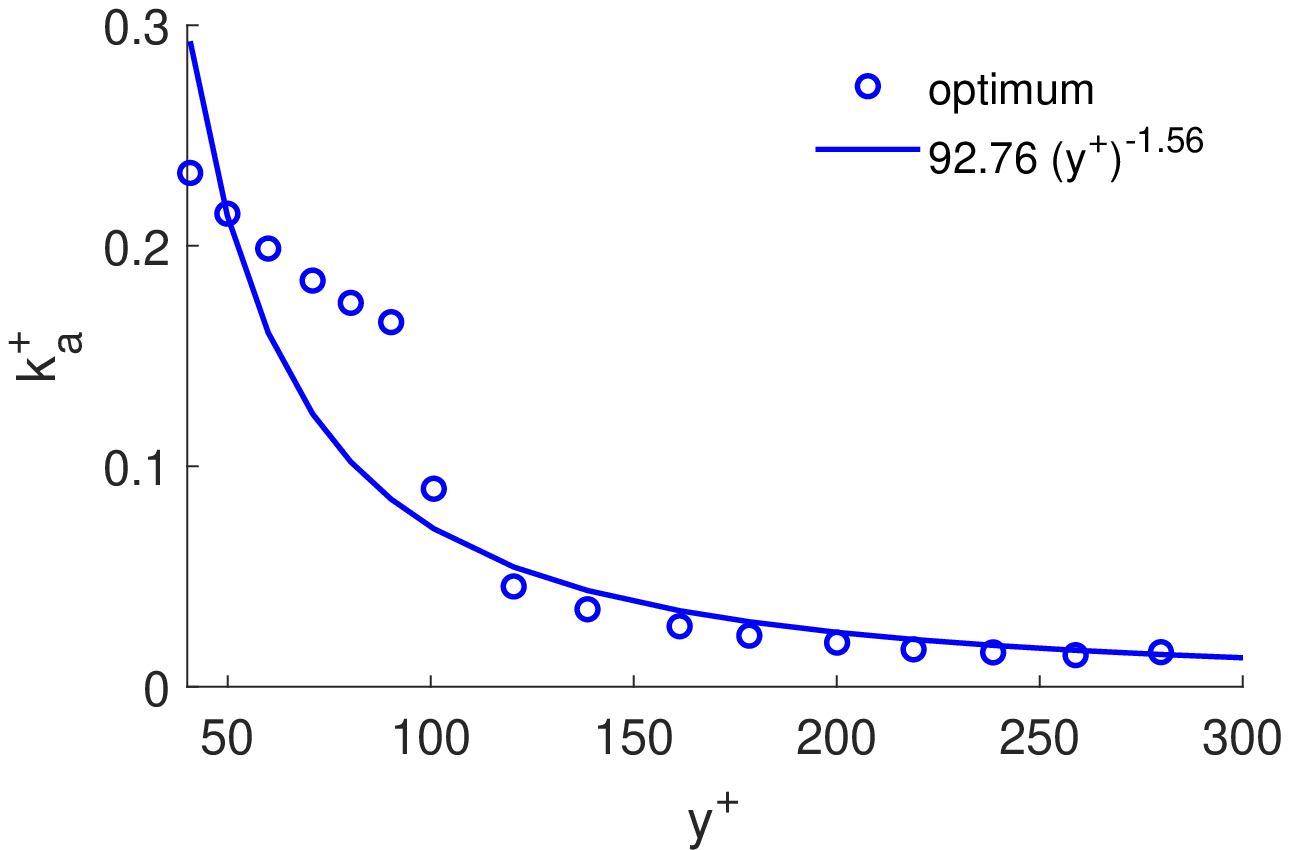}
         \caption{}
         \label{ka}
     \end{subfigure}
     \hfill	
     \begin{subfigure}[b]{0.329\textwidth}
         \centering
         \includegraphics[width=\textwidth]{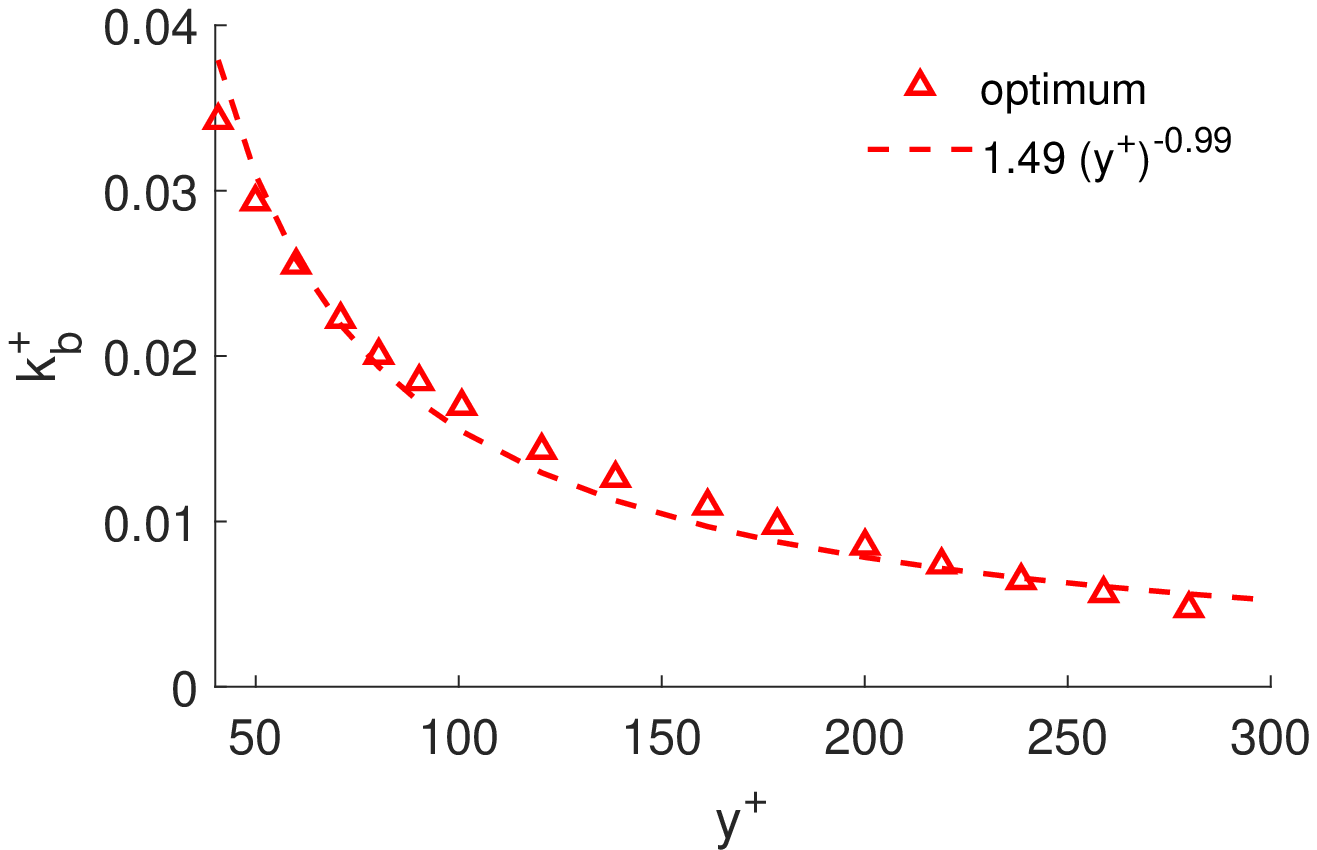}
        \caption{}
         \label{kb}
     \end{subfigure}
\caption{ Parameters of the Dragonfly filter $\mathcal{D}$ in the log layer. Points mark optimum values calculated by minimizing $\Sigma_{yz}^U$. Curves mark power law fits, which have been subsequently used to calculate the filter.}
        \label{filterfit}
        
\end{figure}

The 2D nonlinear flux cospectra shown in Fig~\ref{fig_2dcospec}, particularly in the log layer, possess a natural ``boundary" in wave number space separating regions of down-gradient and up-gradient transport . In this section, we propose a simple filter that allows us to distinguish the two competing scales in the log layer. We chose the filter kernel 
to be graded in order to reduce Gibbs-type oscillations in the spatial filtered fields. A simple choice which we dub the 
``dragonfly" filter ($\mathcal{D}$) is a product of two Gaussian filters. To capture the spectral region 
of interest, the two Gaussians are chosen to have elliptical level curves centered at the origin with principal 
axes of respective slopes $\pm m:$
\begin{equation} 
\mathcal{D}(k_x,k_z,y):=\exp\left(-\left[ \frac{(|k_x|+m|k_z|)^2}{k_a^2} +\frac{(|k_z|-m|k_x|)^2}{k_b^2} \right] \right)
\end{equation} 
We then define also a complement filter ($\mathcal{D}^c:=1-\mathcal{D}$). To choose the optimum 
parameters $m,k_a^+$ and $k_b^+,$ we minimize the flux value $\Sigma_{yz}^U$ (largest negative value)  
separately for each $y^+=40,60,80,...300.$ However, for computational convenience, it is useful to 
have an explicit representation of these optimum parameters as functions of $y.$ The optimum values are 
shown in Fig~\ref{filterfit} and may be reasonably described by power laws. In fact,  the optimum $k_b$ fits a $y^{-1}$ power law (shown in Fig~\ref{kb}) very well. Parameters $m$ and $k_a$ are not represented as well by power laws and 
show a ``kink'' around $y^+=100$, which can be a subject of further investigation. The best fits by power laws yield 
\begin{align} 
m&=1.56(y^+)^{-0.22},\\
k_a^+&=92.76(y^+)^{-1.56},\\
k_b^+&=1.49(y^+)^{-0.99}.
\end{align}
which are plotted also in  Fig~\ref{filterfit}. These power-law relations were deemed adequate and have 
been used for the results presented in the paper. 

\begin{figure}
     \begin{subfigure}[b]{0.49\textwidth}
         \centering
         \includegraphics[width=\textwidth]{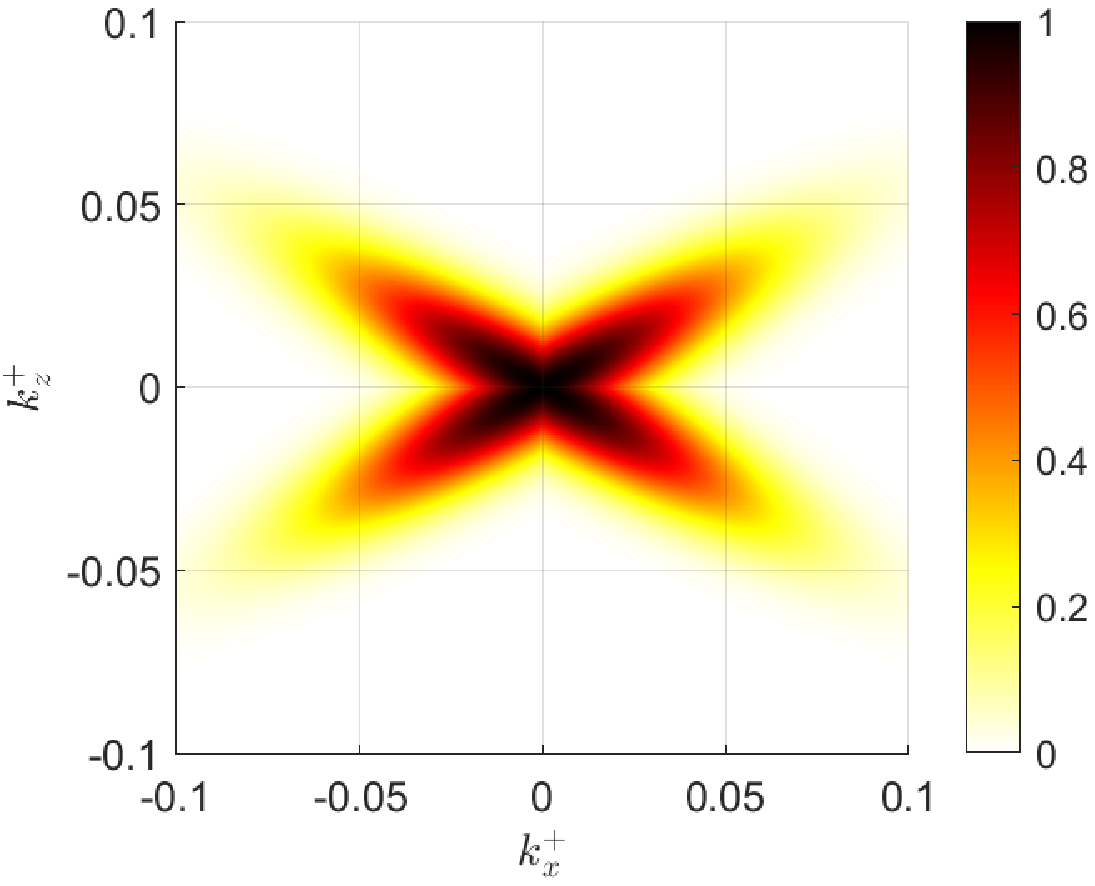}
        \caption{Dragonfly filter $\mathcal{D}$ at $y^+=100$ }
         \label{flower}
     \end{subfigure}
     \hfill
     \begin{subfigure}[b]{0.49\textwidth}
         \centering
         \includegraphics[width=\textwidth]{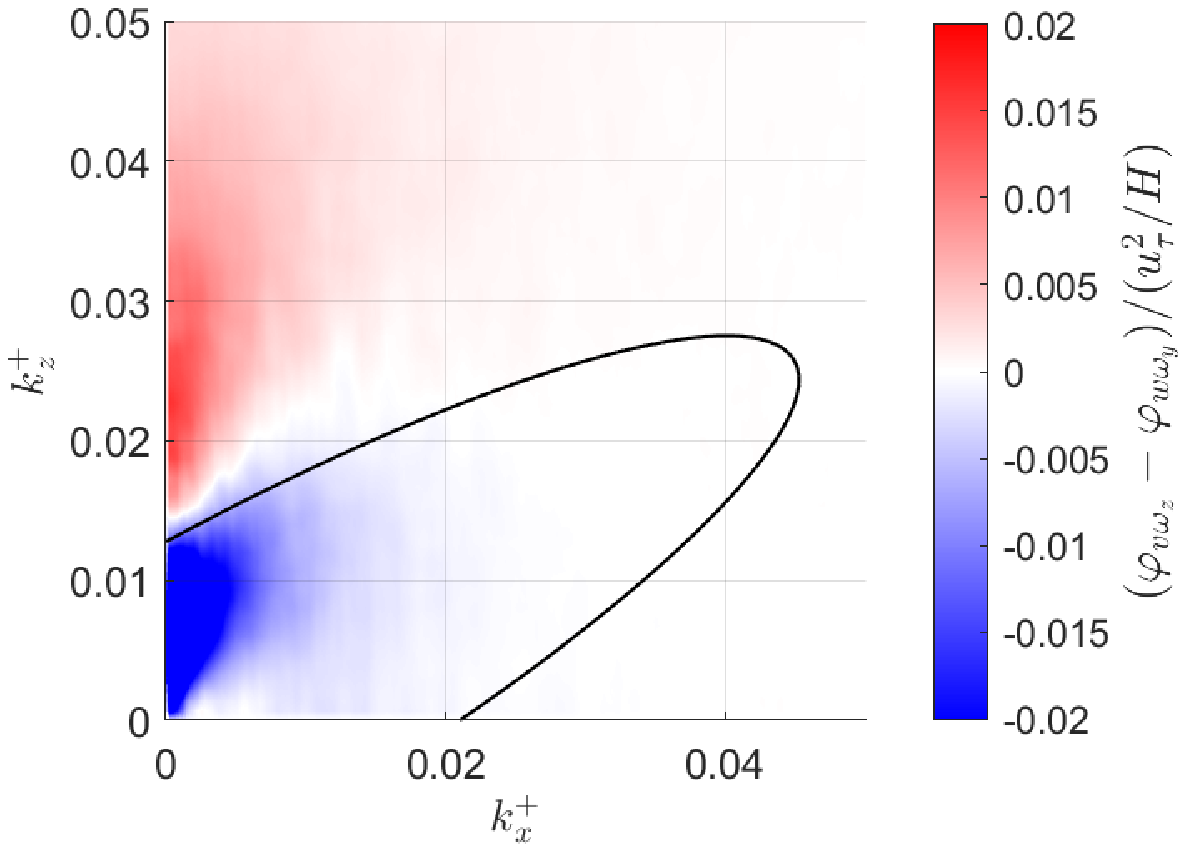}
        \caption{Co-spectra at $y^+=100$, the contour marks $\mathcal{D}=0.5$}
         \label{fig:y equals x}
     \end{subfigure}
     \hfill
     \begin{subfigure}[b]{0.49\textwidth}
         \centering
         \includegraphics[width=\textwidth]{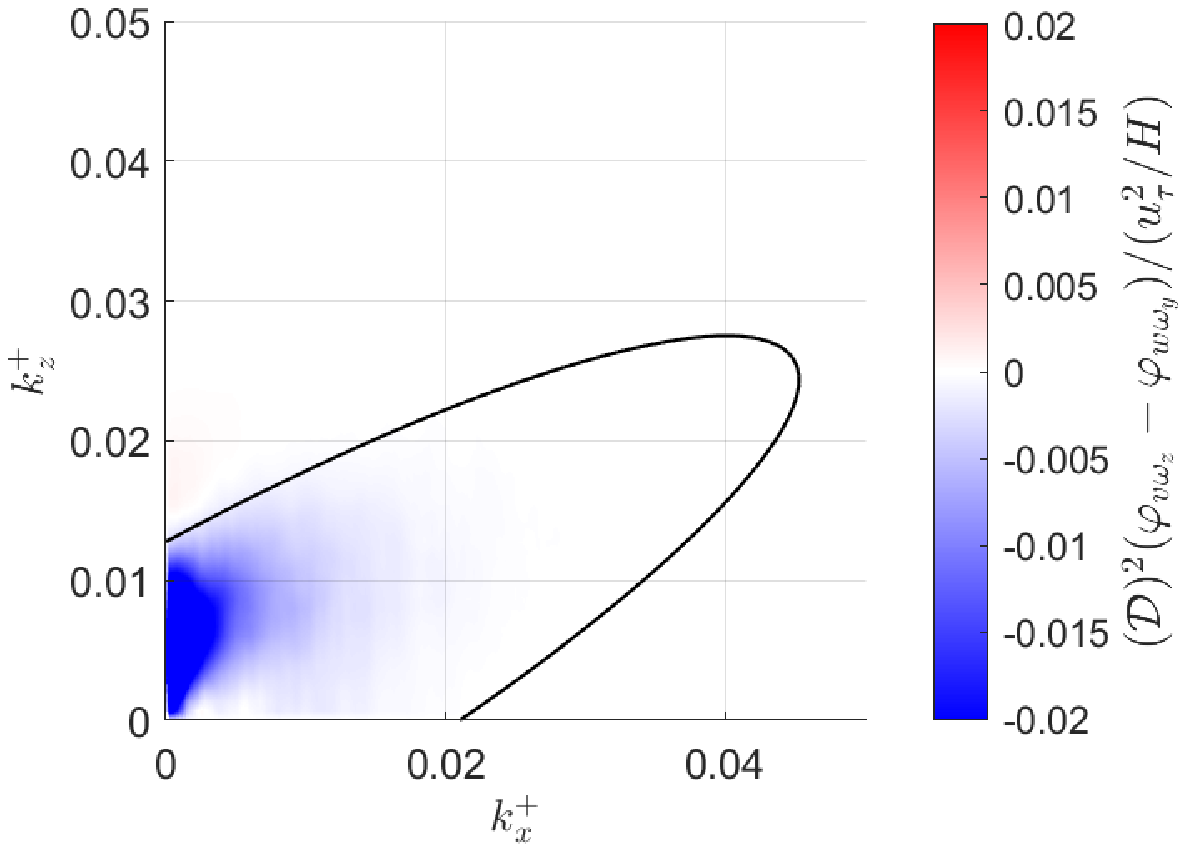}
        \caption{Filtered co-spectra at $y^+=100$, the contour marks $\mathcal{D}=0.5$}
         \label{flowerD}
     \end{subfigure}
     \hfill
     \begin{subfigure}[b]{0.49\textwidth}
         \centering
         \includegraphics[width=\textwidth]{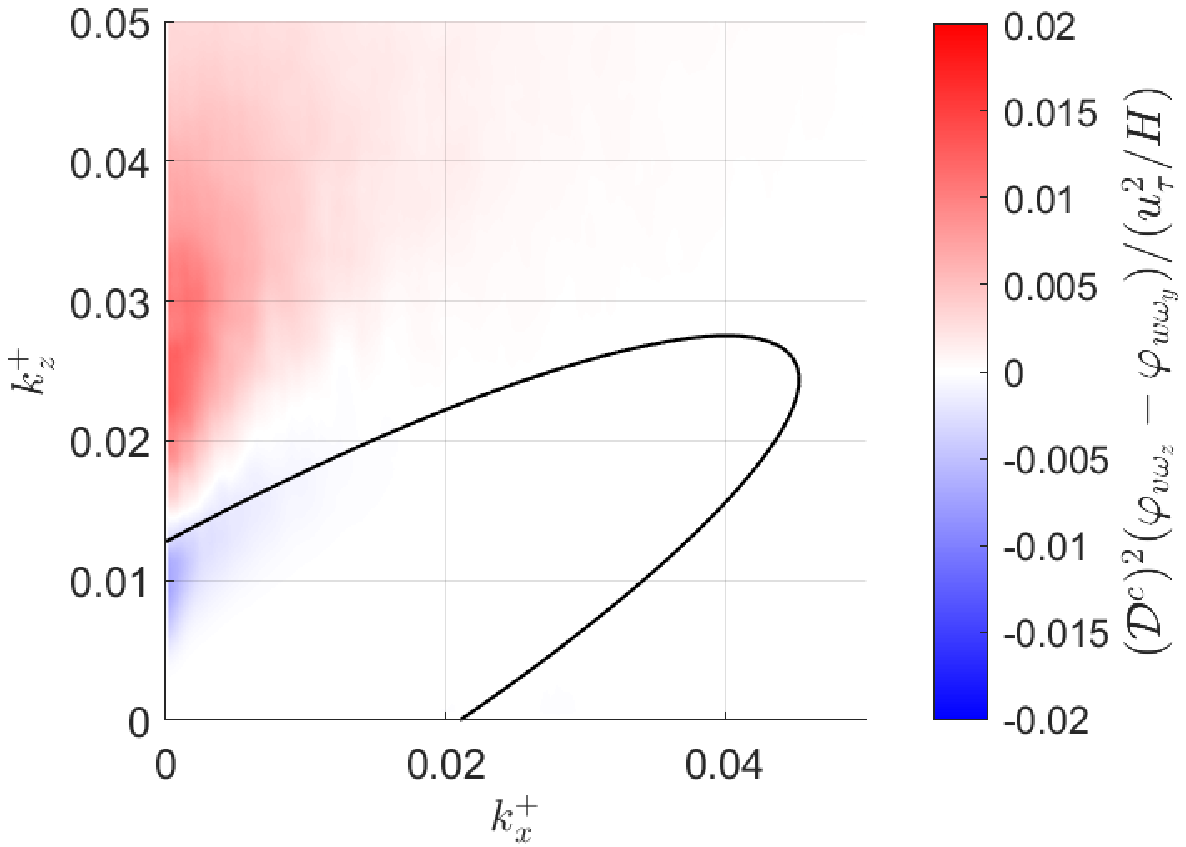}
         \caption{Filtered co-spectra at $y^+=100$, the contour marks $\mathcal{D}^c=0.5$}
         \label{flowerDc}
     \end{subfigure}
        \caption{Dragonfly filter and its application to the velocity-vorticity co-spectrum at $y^+=100$}
        \label{fig:three graphs}
\end{figure}

The velocity and vorticity fields are filtered using $\mathcal{D}(k_x,k_z,y)$ and $\mathcal{D}^c(k_x,k_z,y)$ yielding the up-gradient and down-gradient parts of the fields respectively. The procedure to obtain the filtered fields ($q^U$ and $q^D$) from an unfiltered field $q$, at a given wall distance $y$, is as follows:
 \begin{align}
\hat{q}(k_x,k_z,y)&=FFT_{2D}[q(x,y,z)] \\
q^U(x,y,z)&=iFFT_{2D}[\mathcal{D}(k_x,k_z,y)\hat{q}(k_x,k_z,y)],\label{eqlp}\\
q^D(x,y,z)&=iFFT_{2D}[\mathcal{D}^c(k_x,k_z,y)\hat{q}(k_x,k_z,y)]\label{eqhp}.
 \end{align}
 Filtering with $\mathcal{D}$ selects low-wavenumber (large lengthscale) up-gradient scales and results in the nonlinear flux plotted in Fig~\ref{vlp}. The complement $\mathcal{D}^c$ selects high wavenumber (small lengthscale) down-gradient scales that result in the nonlinear flux plotted in Fig~\ref{vhp}. 
We plot $\mathcal{D}$ for $y^+=100$ in Fig~\ref{flower}. Co-spectra resulting from filtering with $\mathcal{D}$ and with $\mathcal{D}^c$ are shown in Fig~\ref{flowerD} and Fig~\ref{flowerDc}, respectively. These plots illustrate that the 
constructed filters separate the cospectrum into mainly down-gradient and up-gradient parts. The separation is 
not perfect, because of the graded nature of the filter kernel, but it was deemed sufficient for our analysis. 

\newpage

\clearpage 

\section{Orientation of $U$-type vortices}\label{orientation}
\begin{figure}
         \centering
         \includegraphics[width=\textwidth]{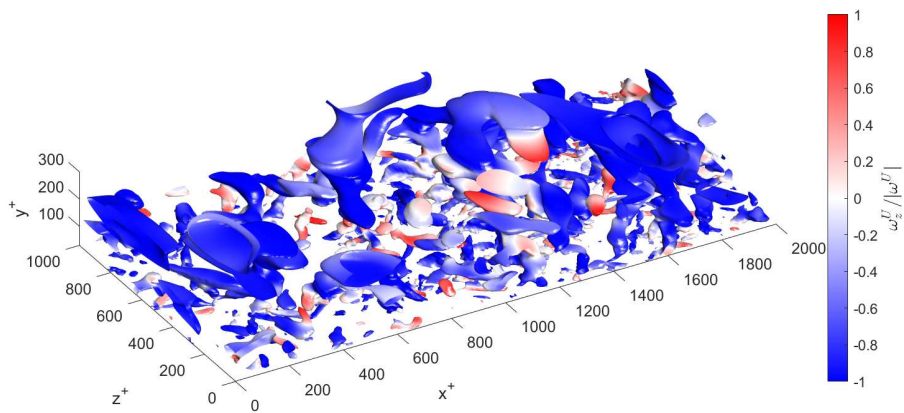}
        \caption{Vortices identified using the $\lambda_2$-criterion for the velocity field ${\bf u}^U$ filtered using $\mathcal{D}.$ Isosurfaces are plotted for $\lambda_2^U=-\lambda_2^{U,rms}$ and colored by cosine of the angle made by the vorticity vector $\boldsymbol{\omega}^U$ with the z-axis, given by $\omega_z^U/|\boldsymbol{\omega}^U|$. }
         \label{vortices_lp_omega}
\end{figure}

We here present evidence that vorticity vector orientation within $U$-type vortices is predominantly 
spanwise and prograde, consistent with lateral stretching of pre-existing vorticity. This is demonstrated by Fig.~\ref{vortices_lp_omega}, which plots the same vortices visualized by the $\lambda_2$-criterion 
in Fig.~\ref{vortices_lp} in the main text but coloured now by the cosine of the angle between vorticity vector $\boldsymbol{\omega}^U$ and the z-axis. We observe a prevalence of values smaller than -0.7, denoting prograde vortices forming angles smaller than $\pi/4$ with the z-axis. We note also the presence of a few retrograde vortices (shown in red) and a few which are not spanwise aligned (white). 

\newpage 

\end{document}